\documentclass[preprint,number,sort&compress]{elsarticle}
\usepackage[utf8]{inputenc}
\usepackage{color}
\usepackage[
scale=0.7, marginratio={1:1, 2:3}, ignoreall,
]{geometry}
\usepackage{amsmath}
\usepackage{amssymb}
\usepackage[linktocpage=true,colorlinks=true,linkcolor=blue,anchorcolor=black,citecolor=magenta,filecolor=black,menucolor=black,runcolor=black,urlcolor=blue]{hyperref}
\usepackage{tabularx}
\usepackage{booktabs}
\usepackage{units}
\usepackage{url}
\usepackage{aasmacros}
\usepackage{chronosys}
\usepackage{comment}

\newcommand{\mrm}[1]{_{\rm #1}}
\renewcommand{\d}{{\rm d}}
\newcommand{\etal}{\textit{et al.}}
\newcommand{\BBbar}{\epsilon_{{\rm B}\overline{\rm B}}}
\newcommand{\rate}[1]{$R < #1\,$pc$^{-3}\cdot$yr$^{-1}$}
\begin{document}
	
	\begin{center}
		{\Large
		Primordial black hole constraints with Hawking radiation - a review}
	\end{center}

	\begin{center}
		J\'er\'emy Auffinger$^{a,}$\footnote{\url{j.auffinger@ipnl.in2p3.fr}}
	\end{center}

	\begin{center}
		\textit{$^a$Univ. Lyon, Univ. Claude Bernard Lyon 1, CNRS/IN2P3, IP2I Lyon, UMR 5822, F-69622, Villeurbanne, France}
	\end{center}
	
	\hrule
	\vspace{0.3cm}
	
		Primordial black holes are under intense scrutiny since the detection of gravitational waves from mergers of solar-mass black holes in 2015. More recently, the development of numerical tools and the precision observational data have rekindled the effort to constrain the black hole abundance in the lower mass range, that is $M < 10^{23}$g. In particular, primordial black holes of asteroid mass $M \sim 10^{17}-10^{23}\,$g may represent 100\% of dark matter.
		While the microlensing and stellar disruption constraints on their abundance have been relieved, Hawking radiation of these black holes seems to be the only detection (and constraining) mean.
		Hawking radiation constraints on primordial black holes date back to the first papers by Hawking. Black holes evaporating in the early universe may have generated the baryon asymmetry, modified big bang nucleosynthesis, distorted the cosmic microwave background, or produced cosmological backgrounds of stable particles such as photons and neutrinos. At the end of their lifetime, exploding primordial black holes would produce high energy cosmic rays that would provide invaluable access to the physics at energies up to the Planck scale.
		In this review, we describe the main principles of Hawking radiation, which lie at the border of general relativity, quantum mechanics and statistical physics. We then present an up-to-date status of the different constraints on primordial black holes that rely on the evaporation phenomenon, and give, where relevant, prospects for future work. In particular, non-standard black holes and emission of beyond the Standard Model degrees of freedom is currently a hot subject.
	
	\vspace{0.3cm}
	\hrule
	
	\tableofcontents
	
	\section{Introduction}
	
	Dark matter (DM) is currently one of the deepest problems of the cosmological standard model. Indeed, whereas the precision era of cosmology has probed the bases of this model, many unanswered questions remain. Even if the concept of DM---an invisible matter which nevertheless has visible gravitational effects---is well established, there are numerous solutions to explain its nature (see \textit{e.g.}~\cite{2021PrPNP.11903865A} for a recent review and~\cite{2018RvMP...90d5002B} for an historical account). Primordial black holes (PBHs) were one of the candidates proposed to explain DM, in the early 1970's. PBHs are black holes (BHs) that formed in the early universe from the collapse of overdensities resulting from quantum fluctuations, and are predicted also by other mechanisms such as phase transitions.
	
	PBHs are not the result of stellar collapse, thus their mass can span a very wide range of masses, in fact from the Planck mass to the mass of the universe enclosed by the Hubble horizon today. On the other hand, their abundance is tightly constrained by the numerous effects they have on the cosmological history. Heavy PBHs, with mass $M > 10^{23}\,$g, may act as gravitational lenses, disrupt binaries, accrete matter or coalesce. This gives corresponding microlensing, stellar surveys, X-ray observations and gravitational wave (GW) constraints. Overall, PBHs with $M > 10^{23}\,$g cannot represent 100\% of DM and the constraints will certainly tighten with the upcoming GW instruments (LISA, Einstein telescope) or microlensing surveys (Roman, Rubin). For a recent review of the PBH constraints as a DM candidate, see~\cite{2020ARNPS..70..355C,2021JPhG...48d3001G}, for a review of the general prospects on PBH DM, we refer the reader to the Snowmass white paper~\cite{2022arXiv220308967B}, and for a dedicated review of GW perspectives see~\cite{2018CQGra..35f3001S}.
	
	The lightest PBHs are more interesting as their abundance is yet unconstrained in the window $10^{17}-10^{23}\,$g, so that they could constitute 100\% of DM. These PBHs are light enough so that the phenomenon of Hawking radiation (HR) becomes important. HR is a semi-classical process predicted by Hawking, that makes BHs lose their mass by emitting all the quantum fields with a quasi-thermal spectrum. Theorized in the 1970's, this effect remains unobserved, mainly because it is so faint for heavy BHs: the temperature of the Hawking spectrum is inversely proportional to the BH mass. PBHs in the ``HR window'' $10^{17}-10^{23}\,$g however would radiate huge amounts of particles of all types, which could be the only way of assessing their abundance. The ``HR window'' coincides with MeV astronomy, a quite unexplored domain that has become the stake of many instrument proposals (AMEGO, ASTROGRAM)~\cite{2022arXiv220307360E}. Prediction of the PBH signals in this mass window would then allow to close the last possibility for PBH DM~\cite{2022arXiv220308967B}.
	
	Finally, PBHs with mass $M < 5\times10^{15}\,$g would have already evaporated by now, hence they would not participate in the DM density. While their final explosion is searched for in $\gamma$-ray and even neutrino experiments~\cite{2021JCAP...12..051C}, more promising prospects come from the signature they may have left in the past cosmological eras. The smallest PBHs, with mass $M < 10^{9}\,$g, would have evaporated before big bang nucleosynthesis (BBN). They may have produced the baryon asymmetry or generated the particle DM abundance: BHs evaporate in all (beyond) Standard Model ((B)SM) particles. Slightly heavier PBHs with mass $M \sim 10^9-10^{12}\,$g evaporate during BBN, and their energetic evaporation products enter the chain of nuclear reactions that created the light elements; their abundance is then constrained by the successful predictions of standard BBN. PBHs with $M \sim 10^{12}-10^{17}\,$g modify the recombination history of the universe. As the cosmic microwave background (CMB) is one of the better known cosmological probe, with precision data from the Planck satellite and prospective ones (CMB-S4, PIXIE), any exotic energy injection is limited by observation. In particular, the EDGES collaboration has cleared the way for surveys of the Dark Ages after recombination, observing the $21\,$cm sky for the first time. This may result in the strongest constraints on PBHs up to $M \sim 10^{18}\,$g, with future instruments such as SKA. PBHs emit particles during their whole lifetime, and the spectrum of those backgrounds today encodes some of the thermal history of the universe, on top of the genuine characteristics of HR. The photon background, the easiest to observe, goes back to the CMB ($t\sim 380\,000\,$yr); the neutrino background would have been emitted since neutrino decoupling ($t\sim 1\,$s) and would then constitute a complementary probe of the early universe with BBN; the putative graviton spectrum would have propagated unperturbed since inflation, which would represent the ultimate probe of the first instants of the universe. Most of these topics are reviewed in~\cite{2021RPPh...84k6902C}, but the constraints evolve constantly and no review focusing on HR limits has yet been published.
	
	With the development of precision observations, it further becomes more and more important to predict the precise shape of the PBH signal, which is the aim of several recent numerical tools such as \texttt{BlackHawk}~\cite{2019EPJC...79..693A,2021EPJC...81..910A}, \texttt{CosmoLED}~\cite{2022arXiv220111761F}, \texttt{nuHawkHunter}~\cite{2022arXiv220314979B} and \texttt{ULYSSES}~\cite{2022PhRvD.105a5022C,2022PhRvD.105a5023C}. These tools rely on particle physics codes that compute the interactions of the Hawking radiated particles from the MeV to the Planck scale. Moreover, intense work is performed to obtain the signals (and constraints) on non-standard (P)BHs, such as regular solutions to the Einstein equations, or BSM particles, such as interacting DM. Realistic PBH formation mechanisms are now taken into account when constraining their abundance, with \textit{e.g.}~extended mass distributions or spinning BHs.
	
	In this review, we present first the historical development of the PBH (as DM) (Section~\ref{sec:PBHs}) and HR (Section~\ref{sec:HR}) ideas, with a recall of the fundamental equations required to compute the HR spectra. We briefly review the constraints on (heavy) PBHs not linked to evaporation (Section~\ref{sec:non-evap}). Then, and this is the main focus of the present paper, we give a complete account of the different constraints, in cosmological chronological order: baryogenesis (Section~\ref{sec:baryogenesis}), BBN (Section~\ref{sec:BBN}), CMB (Section~\ref{sec:CMB}), diffuse backgrounds of photons, neutrinos and gravitons (Section~\ref{sec:backgrounds}), cosmic rays such as electrons/positrons and antiprotons (CRs, Section~\ref{sec:CRs}), and the final burst of PBHs (Section~\ref{sec:final_burst}). In the last part (Section~\ref{sec:prospects}), we review the recent developments regarding extended mass functions, spinning PBHs and non-standard BHs/BSM degrees of freedom. We finally conclude with general prospects (Section~\ref{sec:conclusion}). A master figure summarizing all the constraints is given in Fig.~\ref{fig:master}. Throughout the review, we use the natural system of units defined as $G = \hbar = k\mrm{B} = c = 4\pi\epsilon_0 = 1$.

	\section{Primordial black holes}
	\label{sec:PBHs}
	
	PBHs are BHs formed in the early universe. In this Section, we first go through a brief historical overview of the PBH ideas, and then sketch the main formation mechanisms and mass ranges expected for PBHs, with particular focus on formation during a dust phase, and extended mass and angular momentum distributions. 
	
	\subsection{Historical overview}
	
\begin{figure}[!t]
		\centering
		\includegraphics{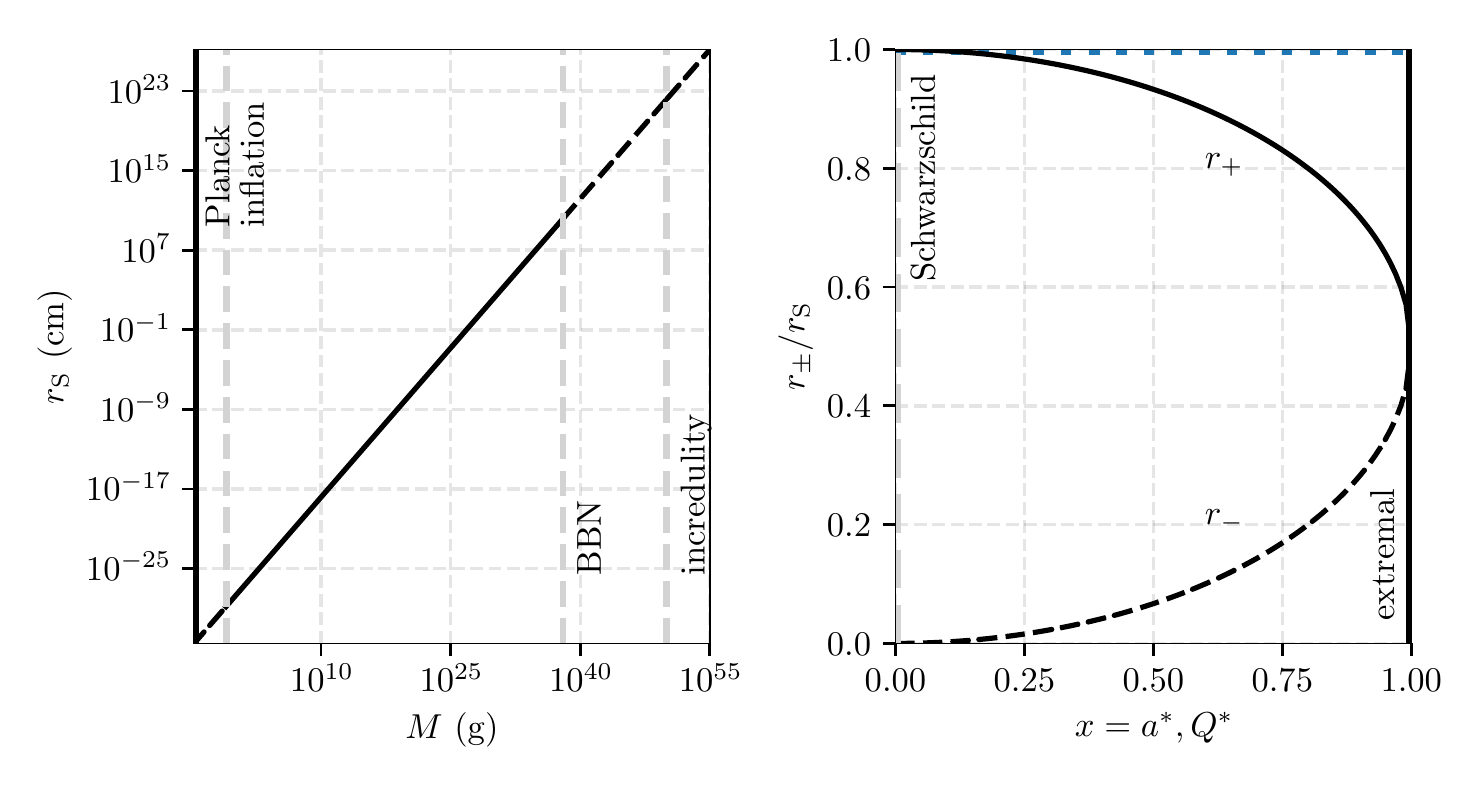}
		\caption{\textbf{Left:} Radius of a SBH as a function of its mass. The vertical solid line at $M = M\mrm{Pl}$ represents the classical limit, the vertical dashed line at $M \sim 0.1\,$g represents the inflation limit, the vertical dashed line at $M \sim 10^{38}\,$g represents the BBN limit and the vertical dotted line at $M \sim 10^{50}\,$g represents the ``incredulity limit'' (see text). \textbf{Right:} Event (solid) and Cauchy (dashed) horizons radii $r_\pm$ normalized to the Schwarzschild radius (dotted) $r\mrm{S}$ for a KBH ($x = a^*$) and RNBH ($x = Q^*$). The vertical line at $x = 0$ represents the Schwarzschild limit and the vertical solid line at $x = 1$ represents the extremal limit (see text).}
		\label{fig:radii}
	\end{figure}

	Using only Newtonian mechanics, Michell and Laplace independently predicted at the end of the XVII$^{\rm th}$ century that if there were an object sufficiently dense that the escape velocity at its surface was faster than the speed of light, it would be totally black and absorb anything that would come gravitationally bound to it. That is precisely what a ``black hole'' would do. Schwarzschild was the first one to derive the solution of the spherically symmetric and static Einstein equations around a pointlike mass $M$ in 1916~\cite{1916SPAW.......189S} (hereafter SBH, for Schwarzschild BH), right after Einstein published the general relativity framework
	\begin{equation}
		\d s^2 = -G(r)\d t^2 + \frac{1}{F(r)}\d r^2 + H(r)\d\Omega\,,
	\end{equation}
	with
	\begin{equation}
		\left\{\begin{array}{ll}
			F(r) = G(r) = 1 - \dfrac{r\mrm{S}}{r}\,,  \\
			H(r) = r^2\,,
		\end{array}\right.
	\end{equation}
	and $r\mrm{S} \equiv 2M$ is the Schwarzschild radius, illustrated in Fig.~\ref{fig:radii} (left panel). At space infinity, that is $r\gg r\mrm{S}$, the metric becomes asymptotically flat, and reduces to the Minkowsky metric
	\begin{equation}
		\d s^2 = -\d t^2 + \d r^2 + r^2 \d\Omega^2\,,
	\end{equation}
	which is the empty and flat solution to the Einstein equations. There is a singularity of the metric at $r = 0$, corresponding to a real curvature singularity as the Ricci scalar $R$ goes to infinity, and a fake singularity ($R$ is finite) at $r = r\mrm{S}$, which can be resolved by a change of coordinates. The singularity is an intricate problem, but the general relativity calculations show that the worldlines of free-falling objects toward the singularity exchange time and space characteristics precisely at $r = r\mrm{S}$, denoted as the BH horizon of events. Hence, any signal emitted by a probe that would have crossed the horizon would never reach an observer at infinity but remain enclosed in the BH. The interior of the BH horizon is causally separated from the exterior by the event horizon, which effectively hides away the metric singularity and satisfies the ``cosmic censorship conjecture''~\cite{2021FoPh...51...42L} (CCC). Interestingly, ``invisible collapsed objects'', (\textit{i.e.}~BHs), were proposed very early as a potential candidate for DM~\cite{1969Natur.224..891V,1970ApJ...161..419W}.
	
	The Maxwell equations were included in that framework to obtain the metric around a pointlike mass $M$ with electric charge $Q$ by Reissner \& Nordstr\"om~\cite{1916AnP...355..106R,1918KNAB...20.1238N} (hereafter RNBH, for Reissner-Nordstr\"om BH) with the metric coefficients
	\begin{equation}
		\left\{\begin{array}{ll}
			F(r) = G(r) =  1 - \dfrac{r\mrm{S}}{r} + \dfrac{r\mrm{Q}^2}{r^2}\,, \\
			H(r) = r^2\,,
		\end{array}\right.
	\end{equation}
	where $r_Q \equiv Q^2$. This metric exhibits one real singularity at $r = 0$ and two fake coordinate singularities at
	\begin{equation}
		r_\pm \equiv r\mrm{S}\dfrac{1 \pm \sqrt{1-Q^{*2}}}{2}\,,
	\end{equation}
	where $r_-$ is a Cauchy horizon and $r_+$ is the BH event horizon. $Q^* \equiv Q/M$ is the reduced (dimensionless) BH charge.
	
	The case of a rotating BH with angular momentum $a \equiv J/M$, which is only axisymmetric, is more mathematically involved and was solved by Kerr in 1963~\cite{1963PhRvL..11..237K} (hereafter KBH, for Kerr BH). The metric is here
	\begin{equation}
		\d s^2 = \big(\d t - a \sin^2\theta \d\phi\big)^2\, \frac{\Delta}{\Sigma} - \left(\frac{\d r^2}{\Delta} + \d\theta^2\right) \Sigma - \big((r^2+a^2)\d\phi-a \d t\big)^2 \, \frac{\sin^2\theta}{\Sigma}\,,
	\end{equation}
	where $\Sigma(r) \equiv r^2 + a^2 \cos^2\theta$ and $\Delta(r) \equiv r^2 - 2Mr + a^2$. The Cauchy and event horizons are located at
	\begin{equation}
		r_\pm \equiv r\mrm{S}\dfrac{1 \pm \sqrt{1 - a^{*2}}}{2}\,,
	\end{equation}
	where $a^* \equiv a/M$ is the reduced (dimensionless) angular momentum or ``BH spin''.
	
	The combination of these two solutions gives the Kerr--Newman metric around a charged, rotating mass~\cite{1965JMP.....6..915N,1965JMP.....6..918N}, with horizons located at $r_\pm = r\mrm{S}(1 \pm \sqrt{1 - Q^{*2} - a^{*2}})/2$. A comparison of the radii of a SBH with those of a KBH or RNBH for different values of $a^*$ and $Q^*$ are given in Fig.~\ref{fig:radii} (right panel).
	
	One immediately remarks that when $a^*,Q^* \rightarrow 0$, the SBH is recovered, while in the opposite limit $a^*,Q^* \rightarrow 1$, the Cauchy and event horizons radii collapse to the value $r\mrm{S}/2$. In this case, denoted as ``extremal'', the BH is ill-defined as the coordinate singularity at the BH center $r = 0$ is not hidden behind a horizon---the CCC is violated; this limit is shown as a black vertical line on the right panel of Fig.~\ref{fig:radii}.
	
	BHs result in general from the collapse of some energy density inside its own Schwarzschild radius. At the end of its evolution, a star shrinks because most hydrogen has burned into helium and the gravitational forces win against the failing nuclear reactions radiation pressure. For a certain parameter space of stellar conditions, the star can collapse into a ``stellar'' BH (sometimes denoted as ``astrophysical'' BH). Tolman, Oppenheimer \& Volkoff~\cite{1939PhRv...55..364T,1939PhRv...55..374O} (TOV) identified an inferior mass for these BHs, with modern studies giving $M\mrm{TOV} \sim 2\,M_\odot$~\cite{1996A&A...305..871B}. Upper limits on their mass are more dubious but it is generally not expected that genuine stellar mass BHs of more than $\sim 100\,M_\odot$ can form.
	
	Nowhere else in the present universe is there matter denser than the interior of stars, where BHs could form. However, the universe was order of magnitudes denser in the past. Then, it is not unimaginable that in the early universe, patches of the universe which were overdense relatively to the average density due to some statistical fluctuation collapsed into early non-stellar BHs. These are generally denoted as ``primordial black holes''. The very first mention of PBHs dates back to 1966: Zel'dovich \& Novikov, in a paper entitled ``The hypothesis of cores retarded during expansion and the hot cosmological model''~\cite{1967SvA....10..602Z}, speak about ``retarded cores''~\cite{1965SvA.....8..857N} whose evolution, because of gravitational collapse, would escape the overall expansion of the universe, and describe these cores to have radius $R < 2M$ which is precisely the radius of a SBH.
	
	This original idea seems to have gone quite unnoticed until Hawking published the paper ``Gravitationally collapsed objects of very low mass''~\cite{1971MNRAS.152...75H}. The idea is fundamentally the same as Zel'dovich \& Novikov, \textit{i.e.}~large perturbations collapsing after inflation inside their Schwarzschild radius to give BHs (explicitly named here). The collapse is described as a classical process, hence a lower limit is imposed on the mass of the BH formed $M > M\mrm{Pl} \sim 2 \times 10^{-5}\,$g for general relativity to be reliable. Hawking concludes that paper by stating that:
	\begin{quote}
		\textit{[\dots] it is tempting to suppose that the major part of the mass of the Universe is in the form of collapsed objects. This extra density could stabilize clusters of galaxies which, otherwise, appear mostly not to be gravitationally bound.} \cite{1971MNRAS.152...75H}
	\end{quote}
	
	Some years later, Carr \& Hawking~\cite{1974MNRAS.168..399C} show that the concerns raised by Zel'dovich \& Novikov~\cite{1967SvA....10..602Z} about catastrophic early PBH accretion are not established, however:
	\begin{quote}
		\textit{The obvious place to look for such giant black holes} [that would have accreted rapidly] \textit{would be in clusters of galaxies where they might provide the missing mass necessary to bind clusters gravitationally.} \cite{1974MNRAS.168..399C}
	\end{quote}
	Meszaros~\cite{1974A&A....37..225M,1975A&A....38....5M} finally argues that galaxy formation could benefit from the existence of PBHs, the latter seeding early structures and giving the ``hidden mass in clusters required by virial theorem arguments.''
	These claims, together with that of~\cite{1969Natur.224..891V,1970ApJ...161..419W} concerning astrophysical BHs, constitute the prelude of the ``DM = PBHs'' scenario.
	
	\subsection{Formation mechanisms}
	\label{sec:PBH_formation}
	
	Carr \& Hawking~\cite{1974MNRAS.168..399C} first quantitatively described the process of PBH formation, which was numerically computed in~\cite{1978SvA....22..129N} (for a recent review, see~\cite{2022Univ....8...66E}). A region of locally enhanced density collapses if its radius is smaller than the associated Jeans length. The collapse must supplant the pressure forces (which depend on the equation of state of the cosmological fluid) and the universe expansion. BH formation then depends on the amplitude of the perturbation, which must be superior than some threshold $\Delta\rho / \rho > \delta$.
	
	At time $t$, the collapsing region must be of order of the particle horizon size $M\mrm{hor} \sim H(t)^{-1}$ for causal reasons. This is the \textit{standard scenario}, which relies on the assumption that there was a spectrum of primordial inhomogeneities. Modern calculations show that a PBH formed during radiation domination (RD) has an initial mass $M\mrm{i}$ linked to its formation time $t\mrm{f}$ by~\cite{2021RPPh...84k6902C}
	\begin{equation}
		M\mrm{i}(t\mrm{f}) = \dfrac{\gamma}{2H(t\mrm{f})} \sim 10^{15}\,\text{g} \left( \dfrac{t\mrm{f}}{10^{-23}\,{\rm s}}\right),\label{eq:formation_time}
	\end{equation}
	where $\gamma$ is a parameter linked to the collapse mechanism. For spherical collapse during RD, $\gamma \sim 0.2$~\cite{2021RPPh...84k6902C}. In the usual model of PBH formation, PBHs form after the inflation period, hence there is an inferior limit on the size of the particle horizon derived from CMB observations $H(t\mrm{f})^{-1}$~\cite{2020A&A...641A..10P} translating into an inferior limit on the mass of PBHs
	\begin{equation}
		M\mrm{min} \sim 0.1\,\text{g}\,.
	\end{equation}
	However, this constraint applies only on conventional inflationary scenarios, \emph{i.e.}~the standard slow-roll models of inflation with Einstein gravity (see \emph{e.g.}~the review~\cite{2011pls..conf..523B} and references therein). In more sophisticated scenarios,
	the scale of inflation can not be determined by CMB observations.
	An absolute lower limit on the PBH mass is the Planck mass $M\mrm{Pl}$ at which quantum gravity effects should be sizeable and our understanding of general relativity breaks down. PBHs with size smaller than the Planck length $\ell\mrm{Pl} \sim 10^{-35}\,$m would have formed at times before the Planck time $t\mrm{Pl}\sim 10^{-43}\,$s which is forbidden by the trans-Planckian conjecture~\cite{2020JHEP...09..123B}.
	
	An upper limit on the mass of PBHs at formation is derived from the fact that they should form before the onset of BBN not to spoil the baryon-to-photon ratio
	\begin{equation}
		M\mrm{BBN} \sim M(t\mrm{f} \approx 1\,\text{s}) \sim 10^5\,M_\odot\,.
	\end{equation}
	There are also trivial ``incredulity'' reasons to believe that there is no giant PBH of $M\gtrsim M\mrm{incr} \sim 10^{50}\,$g occupying most of the Hubble sphere today, first because there are other complex structures in the sky, second because if there were a giant PBH, we would be falling onto it due to its gravitational attraction and would see a strong Doppler dipole feature in the CMB~\cite{2018MNRAS.478.3756C,2021MNRAS.501.2029C}.
	The limits $M\mrm{Pl}$, $M\mrm{min}$, $M\mrm{BBN}$ and $M\mrm{incr}$ are represented by vertical lines on the left panel of Fig.~\ref{fig:radii}.
	
	The list of PBH formation scenarios is extensive, and we refer the reader to the recent reviews~\cite{2021RPPh...84k6902C,2021JPhG...48d3001G} for a complete description. They include:
	\begin{itemize}
		\item collapse of primordial overdensities (the \textit{standard scenario});\footnote{This scenario seems deprecated due to the tremendous increase of the fluctuation spectrum needed at small scales compared to the CMB, see the very explicit Fig.~1 of~\cite{2021JPhG...48d3001G}. This has motivated more exotic formation scenarios, and refutes the usual claim that PBH DM is compelling because it does not require new physics.}
		\item collapse of inflaton fields;
		\item collapse of topological defects;
		\item collapse from bubble collision due to an early first order phase transition.
	\end{itemize}
	In particular, some formation mechanisms (\textit{e.g.}~grand unified theory---GUT---phase transition) explain both PBH formation and later cosmological events (\textit{e.g.}~PBH baryogenesis mediated by GUT bosons~\cite{1982PhRvD..26.2681H}). In general, an alternative cosmological model results in very different PBH constraints~\cite{1998PhR...307..125L}. Let us detail some formation mechanisms related to EMDE, or which result in extended PBH mass or spin distributions.
	
	For example, formation of PBHs during an early matter-dominated era (EMDE) was reviewed in detail in~\cite{2010RAA....10..495K}. It is shown that PBH formation is easier in this case due to reduced pressure forces. The relationship between a PBH mass and its formation time, and thus the PBH constraints then depend on the EMDE parameters and are in general weaker than the corresponding RD constraints~\cite{1997PhRvD..56.7559G}.
	Modern studies of PBH formation during an EMDE have been performed by~\cite{2015IJMPD..2430022K,2016ApJ...833...61H,2016PhRvD..93l3523G,2017JHEP...09..138G,2018PhRvD..98l3024K,2019JCAP...11..014G,2019PhRvD.100l3544M}.
	
	\subsubsection{Extended mass distribution}
	
	One prominent feature of the first review of PBH constraints by Carr~\cite{1975ApJ...201....1C} is that they use the Press-Schechter formalism designed for scale-invariant perturbations in the early universe and the formation of structures. Carr predicts that the mass spectrum of PBHs is a power-law with exponent $\d n \propto M^{-f(w)} \d M$ where $f(w)$ is some function of the equation of state parameter $P = w\rho$ and where the collapse into a PBH happens once the density perturbation exceeds some threshold $\Delta\rho/\rho > \delta$. In the RD era, one obtains $f(w) = 5/2$. This exponent should be modified for PBH formation during an EMDE.
	
	Hence, the mass spectrum of PBHs was originally believed to be \textit{extended} and to span a rather large range of masses (from $M\mrm{Pl}$ upward). This link between primordial scale-invariant fluctuations of density and the PBH abundance shows that:
	\begin{quote}
		\textit{PBHs are unique since they alone could be expected to survive the dissipative effects which erase all other imprints of conditions in the first second of the universe.} \cite{1975ApJ...201....1C}
	\end{quote}
	Hence, PBH constraints are sometimes converted into primordial inhomogeneities (see \textit{e.g.}~Fig.~19 of~\cite{2021RPPh...84k6902C}). The monochromatic distribution of PBHs often used to derive abundance constraints is then only a convenient approximation to a more realistic extended mass distribution. This poses the \textit{mathematical} question of how to convert the constraints obtained for monochromatic mass distributions to extended ones?
	
	Yokoyama~\cite{1998PhRvD..58j7502Y} is the first, to our knowledge, to try and use an analytical procedure of conversion. This follows the discovery of the critical behaviour of the PBH collapse process by Niemeyer \& Jedamzik~\cite{1998PhRvL..80.5481N,1999PhRvD..59l4013N}. The critical collapse mechanism leads to the formation of PBHs with mass $M < M\mrm{hor}$; hence at each time, a distribution of PBHs masses would arise. The total distribution would then be the sum of these instantaneous formations. The rough computation by Yokoyama shows that using an extended mass function may \textit{broaden} the excluded PBH parameter space.
	
	At about the same time, calculations showed that a monochromatic peak in the primordial fluctuation power spectrum results in fact in a log-normal PBH mass distribution, centered around the characteristic horizon scale at formation~\cite{1993PhRvD..47.4244D}. This distribution was further refined using peak theory~\cite{2008PhRvD..78b3004T,2019PhRvL.122n1302G}. In fact, a Gaussian log-normal function can mimic any peak in the PBH distribution resulting from a particular mechanism of formation~\cite{2016PhRvD..94f3530G}. The most common PBH mass distributions are described \textit{e.g.}~in~\cite{2017PhRvD..96b3514C}.
	
	Concerning conversion methods, pioneering analytical and numerical work has been done in \cite{2017PhRvD..96d3504I,2017PhRvD..95h3508K,2017PhRvD..96b3514C,2018JCAP...01..004B}. Refs.~\cite{2017PhRvD..96d3504I,2017PhRvD..95h3508K} propose to use a sufficiently fine binning procedure to separate the extended distribution into independent Dirac functions, and Ref.~\cite{2018JCAP...01..004B} implicitly computes the equivalent monochromatic amplitude resulting in the same constraint as the extended distribution at some mass $M$. The most commonly used method is that of Carr \etal~\cite{2017PhRvD..96b3514C} (hereafter the ``Carr method'') which decomposes any measure as an infinite sum over \textit{correlated} PBH contributions from different masses; the difficulty is to find the $n$-th order correlation functions. These works confirm that the constraints on an extended distribution should be more stringent than the expected constraint resulting from the addition of monochromatic ones. The precision of the conversion has been examined in some papers, showing mitigated results~\cite{2019arXiv190706485P}. For HR constraints, we advocate the use of a numerical procedure such as \texttt{BlackHawk} which can directly compute the full HR spectrum for any PBH distribution \textit{before} applying the constraints.
	
	In the following, PBH mass distributions at formation are denoted as $\d n/\d M$ with the total number of PBHs $N\mrm{PBH}$ and their density $\rho\mrm{PBH}$ given by
	\begin{equation}
		N\mrm{PBH} = \int_{M\mrm{min}}^{M\mrm{max}} \dfrac{\d n}{\d M}\,\d M\,, \quad \text{and} \quad \rho\mrm{PBH} = \int_{M\mrm{min}}^{M\mrm{max}} M\,\dfrac{\d n}{\d M}\,\d M\,.
	\end{equation}
	
	\subsubsection{Spin distribution}
	
	PBHs formed during RD are believed to have negligible spin~\cite{2019JCAP...05..018D}, which has also been proven for the critical collapse mechanism~\cite{2017PTEP.2017h3E01C}. On the other hand, PBHs formed during an EMDE could have sizeable to near-extremal spin~\cite{2016ApJ...833...61H,2017PhRvD..96h3517H,2017PhRvD..96j3002C,2021OJAp....4E...1A}. In any case, \textit{some} of the PBHs should form with a significant spin~\cite{2021PhRvD.104h3018C}, and PBHs can also be spun up either through early accretion processes~\cite{2020JCAP...04..052D} or through hierarchical mergers~\cite{2017ApJ...840L..24F}.
	
	The problem is then the same as for extended mass distributions. The constraints on SBHs could be converted to KBHs using some analytical procedure; or realistic extended spin distributions could be reduced to their peak/average value (\textit{e.g.}~\cite{2020arXiv200400618H}). We instead advocate for the use of the \texttt{BlackHawk} capacity to simulate extended distributions of \textit{both} mass and spin to compute accurate constraints, as was done for the first time in~\cite{2021PhRvD.103l3549A}.
	
	In the following, the distribution of PBH secondary parameter $x$ ($x$ can be the PBH spin $a^*$, electric charge $Q^*$, \textit{etc.}) is denoted as $\d\tilde{n}/\d x$ and is normalized as
	\begin{equation}
		\int_{x\mrm{min}}^{x\mrm{max}} \dfrac{\d \tilde{n}}{\d x}\,\d x = 1\,.
	\end{equation}
	A joint distribution may be denoted as $\d^2 n/\d M\d x$.
	
	\section{Hawking radiation}
	\label{sec:HR}
	
	In this Section, we review the physics of Hawking radiation, with historical ideas and papers. We describe the thermodynamics and quantum mechanics bases, then detail the HR calculation and finally give the basic formulas, with analytical and numerical results.
	
	\subsection{Historical overview}
	
	\subsubsection{Thermodynamics aspects}
	
	In the 1970's, some work was done in order to conciliate BH mechanics with known theories of thermodynamics. Indeed, BHs seemed to violate the 2$^{\rm nd}$ law of thermodynamics as they can swallow a great amount of information with only modification of mass $M$, charge $Q$ and angular momentum $J$. As BHs are described only by these three quantities, as stated by the ``no-hair'' theorem set out by Israel~\cite{1967PhRv..164.1776I,1968CMaPh...8..245I,1971PhRvL..26..331C}, there is a loss of entropy known as the ``information paradox''~\cite{1993PhRvL..71.3743P}. Thus, Bekenstein~\cite{1973PhRvD...7.2333B} searched for a generalized second law for BHs, and associated their area to an effective entropy that should not decrease during BH evolution. This claim was further supported by an analog version of the first law of thermodynamics
	\begin{equation}
		\d M = \Theta \d A + \vec{\Omega}\cdot\vec{\d L} + \Phi \d Q\,,
	\end{equation}
	where $\vec{\d J}$ (resp.~$\d Q$) is the change in angular momentum (resp.~charge) of the BH while $\vec{\Omega}$ (resp.~$\Phi$) plays a role analog to angular frequency (resp.~electric potential). Thus, the term $\Theta$ in front of the entropy change would be identified as an effective temperature whose expression depends on the surface gravity $\kappa$ of the BH.
	
	Bardeen, Carter \& Hawking~\cite{1973CMaPh..31..161B} go one step further by giving explicitly the 4 laws of BH mechanics:
	\begin{itemize}
		\item[] \textbf{0$^{\rm th}$ law:} the temperature of a BH $T = \kappa/2\pi$ is constant over its event horizon;
		\item[] \textbf{1$^{\rm st}$ law:} the infinitesimal evolution of the BH mass is given by $\d M = T \d A/4 + \vec{\Omega}\cdot\vec{\d J} + \Phi \d Q$;
		\item[] \textbf{2$^{\rm nd}$ law:} the total entropy composed of that of a black hole $S = A/4$ plus that of the rest of the universe can only increase;
		\item[] \textbf{3$^{\rm rd}$ law:} the temperature of a BH cannot be decreased to absolute 0 by a finite number of operations.
	\end{itemize}
	In this paper, the entropy is found to be $S = A/4$ in the first law, so that the effective temperature is identified as
	\begin{equation}
		T = \dfrac{\kappa}{2\pi}\,. \label{eq:temp_historical}
	\end{equation}
	The third law flows from the CCC. Then follows some noticeable statement in adequacy with the views at that time:
	\begin{quote}
		\textit{In fact the temperature of a black hole is absolute zero} [because it] \textit{cannot be in equilibrium with blackbody radiation at any non-zero temperature.} \cite{1973CMaPh..31..161B}
	\end{quote}
	precisely because a BH can only \textit{accrete} matter and radiation. For a complete modern review of the thermodynamics of BHs, we refer the reader to~\cite{2018SHPMP..64...52W,2017arXiv171002725W}.
	
	\subsubsection{Quantum mechanics aspects}
	
	At about the same epoch, theoretical work was pursued in order to check whether BHs were stable objects regarding quantum mechanics, that is, if BHs could develop diverging perturbations that would challenge their survival. This was particularly interesting for KBHs as these can experience stimulated emission of bosonic fields through ``superradiance'' effects~\cite{1972PhRvL..28..994M}, which would efficiently extract mass and angular momentum from the black hole and could lead to ``black hole bombs''~\cite{1972Natur.238..211P}.\footnote{For a complete review of the superradiance effect we refer the interested reader to~\cite{2015LNP...906.....B}.}
	Teukolsky \& Press published a series of 3 papers~\cite{1973ApJ...185..635T,1973ApJ...185..649P,1974ApJ...193..443T} in which they study the stability of the Kerr metric against a spin $s$ equal to 1, \nicefrac{1}{2} and 2 wave scattering. The separated equations for the scalar case were derived earlier by Carter~\cite{1968CMaPh..10..280C}.\footnote{This problem was also explored by Starobinskii \& Churilov for bosonic fields in~\cite{1973JETP...37...28S,1974JETP...38....1S} and Unruh for fermionic field in~\cite{1973PhRvL..31.1265U}.} They obtained the fundamental separated radial and angular equations for the propagation of a wave of spin $s$ and energy $E$ outside a BH of mass $M$, in semi-classical general relativity (\textit{i.e.}~classical metric and quantum fields). The radial equation is therefore named the ``Teukolsky equation'' and the angular part is solved by spin-weighted spheroidal harmonics~\cite{1977JMP....18.1849F}. Resolution of the Teukolsky equation with the correct boundary conditions yields the transmission (absorption) and reflection coefficients of a wave over a BH horizon and thus the superradiant amplification of incident bosonic waves, while fermions are not superradiant.
	
	\subsubsection{Hawking papers}
	
	Confronted with the possible existence of PBHs with size and mass way below the classical stellar processes---in fact in the quantum mechanics regime---Hawking wondered how quantum mechanics effects would come into play. This led to the famous discovery that BHs are not \textit{black} but radiate a steady flux of particles like thermal bodies. This was first proposed in a paper published in Nature in 1974 entitled ``Black hole explosions?''~\cite{1974Natur.248...30H}. In this paper, Hawking claims that:
	\begin{quote}
		\textit{It seems that any black hole will create and emit particles such as neutrinos or photons at just the rate as one would expect if the black hole was a body with a temperature of} [$T = \kappa/2\pi$]. \cite{1974Natur.248...30H}
	\end{quote}
	which is precisely the temperature \eqref{eq:temp_historical} encountered when reconciling BH mechanics with thermodynamics. Hawking immediately suspected a deep fundamental link between this temperature (blackbody emission) and the effective temperature in BH thermodynamics (entropy). The link was identified in the context of the microcanonical ensemble in~\cite{1975PhRvD..12.3077B,1976PhRvD..13..191H}. Follows:
	\begin{quote}
		\textit{As a black hole emits this thermal radiation one would expect it to lose mass. This in turn would increase the surface gravity} [$\kappa \propto 1/M$] \textit{and so increase the rate of emission.} \cite{1974Natur.248...30H}
	\end{quote}
	Hence, any BH would have a finite lifetime that can be estimated if its luminosity follows the blackbody law $L \sim A T^4$. Straightforward integration shows that a PBH with the critical mass $M_*\sim 10^{15}\,$g would have evaporated by today if formed just after inflation.\footnote{Given the roughness of this estimation and the uncertainty concerning the age of the universe at that time, it is remarkable that the estimation by Hawking falls within 50\% of the modern value $M_* = 5\times 10^{14}\,$g.} Due to the inverse power-law dependency of the emissivity with the mass, a very powerful ``explosion'' is to be expected at the end of the BH evolution.
	
	The details of HR are given in a subsequent longer paper from 1975 entitled ``Particle creation by black holes''~\cite{1975CMaPh..43..199H}. The calculations are based on quantum mechanics in curved spacetime. We don't give them here but refer the reader to~\cite{2000mmp..conf..180T} which follows step by step the original Hawking derivation in a comprehensive way. Basically, the flux of particle originates in the fact that an observer close to the BH is freely falling with an acceleration given by the surface gravity $\kappa$, while an observer far away from the BH horizon is at rest in the BH frame. Thus, they define different local bases and vacuum quantum states and the conversion between these bases results in a net thermal flux of particles at infinity with rate
	\begin{equation}
		\dfrac{\d^2 N}{\d t \d E} = \dfrac{1}{2\pi} \dfrac{\Gamma_{Eslm}}{e^{E/T} - (-1)^{2s}}\,, \label{eq:HR_historical}
	\end{equation}
	where $\Gamma_{Eslm}$ is precisely the absorption coefficient encountered by Teukolsky \& Press, here interpreted as a spontaneous emission coefficient. It is often denoted as ``greybody'' (or ``graybody'') factor (GF) as it encodes the departure of the BH from a pure blackbody: not all radiation is absorbed but some part is reflected (or, equivalently, spontaneously emitted).
	
	The rate of emission is given for massless scalar fields from a SBH but extension to higher spin $s$ fields for a Kerr--Newmann BH is straightforwardly obtained by adding the corresponding angular velocity and electric potential into the Boltzmann factor
	\begin{equation}
		\dfrac{\d^2 N}{\d t \d E} = \dfrac{1}{2\pi}\dfrac{\Gamma_{Eslm}(\Omega,\Phi)}{e^{(E - m\Omega - q\Phi)/T(M,a^*,Q)} - (-1)^{2s}}\,,\label{eq:HR_KN}
	\end{equation}
	where $\Omega = 4\pi a/A$ and $\Phi = 4\pi Q r_+ /A$~\cite{1976PhRvD..13..198P}.
	The tendency of BHs to emit aligned momentum and same-charge particles is apparent from the form of the effective chemical potential in Eq.~\eqref{eq:HR_KN}.
	
	For the emission of a particle of non-zero rest mass $\mu$ to be kinematically allowed, the energy of emission must be $E > \mu$. However, the rate of emission is exponentially suppressed at high energies, hence:
	\begin{quote}
		\textit{As the temperature rose, it would exceed the rest mass of particles such as the electron and the muon and the black hole would begin to emit them also.} \cite{1975CMaPh..43..199H}
	\end{quote}
	Therefore, if the number of kinematically allowed degrees of freedom (dofs) of emission increases monotonically with energy, such as in the Hagedorn model, then the latest emission would indeed be explosive.
	
	The evolution of BHs by HR is considered as quasi-static, so that backreaction can be ignored, and the successive particles are emitted independently. Furthermore, to avoid problems with Planck scale naked singularities, Hawking assumed that when reaching this scale the system shall ``disappear altogether'' in a final flash of energy.
	Many interpretations were subsequently given as for this spontaneous emission. Hawking proposed that it could be grasped considering the spontaneous creation of pairs of particles and antiparticles just outside the horizon with a Boltzmann distribution given by the temperature of the BH, with one half escaping to infinity with positive energy and the second half falling inside the BH with negative energy, therefore decreasing its mass. Another analogy proposed is the spontaneous decay of the BH because of particles tunneling out of the horizon~\cite{1976PhRvD..13.2188H} (see in particular the discussion by Parikh \& Wilczek~\cite{2000PhRvL..85.5042P}). For other HR derivation and interpretations, we refer the interested reader to~\cite{2018SHPMP..64...52W}.
	
	\subsection{Theoretical aspects}
	\label{sec:theory}
	
	The Hawking process was thus firmly established on statistical mechanics, quantum mechanics and general relativity grounds. That new process and the corollary finite lifetime of BHs has immediate consequences on PBH constraints. PBHs with mass $M < M_*$ cannot participate in the DM density today, and their evaporation may leave observational signatures, as discussed in the first ever PBH review by Carr~\cite{1975ApJ...201....1C}.\footnote{A review was also published on the Soviet side~\cite{1976SvPhU..19..244F} but it went almost unnoticed.} In fact, the evaporation of small PBHs may be the only reasonable way to detect them.
	
	Very early arguments by Carter and Gibbons~\cite{1974PhRvL..33..558C,1975CMaPh..44..245G} showed that even if formed with non-zero charge and angular momentum, a BH would lose them more rapidly than its mass and end up as a simple SBH for most of its lifetime $\tau$. For rotating PBHs, even if the temperature $T\rightarrow 0$ as the angular momentum approaches the extremal limit $a^*\rightarrow 1$, enhanced emission rates (wrongly associated to superradiance) should compensate so that the lifetime $\tau(a^* \lesssim 1) \sim \tau(a^* = 0)$. Hence, if formed initially rotating, a PBH should keep a sizeable angular momentum for most of its lifetime; this in fact contradicts the above statement. The situation is very different for charged BHs since a rapid discharge process called the Schwinger effect~\cite{1951PhRv...82..664S} neutralizes the BH independently of its HR rate in a timescale much shorter than the mass loss, due to spontaneous pair-production of charged particles (see also~\cite{1974Natur.247..530Z}). In fact, it is estimated that only supermassive BHs formed with appreciable charge would still be charged today---they are protected from the Schwinger effect by the non-zero mass of the electron. Hence, for quite a long time, HR by PBHs was considered only for SBHs. This was further supported by the simple formation mechanism proposed for PBHs; they originate from the collapse of neutral radiation dominated material.
	
	The GFs represented the complicated part of the Hawking process analysis. Apart from some high- and low-energy limits, the calculation of GFs requires numerical tools. Page was the first to compute them numerically in 1976-1977 \cite{1976PhRvD..13..198P,1976PhRvD..14.3260P,1977PhRvD..16.2402P}, using the programs of Teukolsky \& Press \cite{1973ApJ...185..635T,1973ApJ...185..649P,1974ApJ...193..443T}. The first paper deals with SBHs and defines the ``Page coefficients'' as the fundamental functions for computing the BH evolution, integrating over the mass (and spin) loss through HR. The second paper deals with KBHs and Page remarks that depending on the set of accessible dofs, the spin evolution may be drastically different. A coupling between the fundamental dofs spin and the BH angular momentum is identified, which causes enhanced emission of higher spin particles for rapidly spinning BHs. The third paper deals with charged BHs and massive particles, with the conclusion that a non-zero rest mass $\mu$ acts more or less as a kinematic cut-off on HR at $E > \mu$.
	
	Once the ``primary'' emission rates of fundamental particles are known, remains the difficulty of their subsequent interactions and decays. This is deeply related to the model of particle physics considered, and has dramatic consequences on the PBH constraints, as noted in the second review by Carr~\cite{1976ApJ...206....8C}. The theory of HR for coloured particles and gauge bosons~\cite{1977PhLB...71..234P} was rapidly implemented. However, the relative independence of emitted particles remained a subject of debate until MacGibbon \& Webber \cite{1990PhRvD..41.3052M,1991PhRvD..44..376M} (hereafter MG\&W) settled the fundamental theory of BH ``secondary'' HR, with the hypothesis that BHs radiate mostly independent particles that subsequently behave like the products of e$^+$-e$^-$ collisions in accelerators.\footnote{This work is based on earlier discussion by Carter~\etal~\cite{1976A&A....52..427C} and in particular by Oliensis \& Hill~\cite{1984PhLB..143...92O}.} The first paper deals with the instantaneous emission of particles and the hadronization of the coloured ones; use is made of the particle physics code \texttt{HERWIG}~\cite{1988NuPhB.310..461M} to obtain the secondary spectra. The second paper deals with the total emission integrated over the BH lifetime, with the conclusion that most of the energy radiated by BHs is in the form of QCD jets originating in the hadronization and decay of quarks and gluons. The composition of the primary and secondary PBH HR products gives therefore privileged access to particle physics at \textit{any} energy up to the Planck limit $E \sim 10^{19}\,$GeV.
	
	The MG\&W model was challenged by others, mainly in the 1990's, with very different PBH constraint predictions:
	\begin{itemize}
		\item Cline \& Hong~\cite{1992ApJ...401L..57C} argue that the correct model of HR must lie between the two extreme Hagedorn and Gell-Mann models for hadrons (or equivalently between the ``elementary'' and ``composite'' particle models);
		\item Heckler~\cite{1997PhRvD..55..480H,1997PhRvL..78.3430H} claims that taking into account number non-conserving interactions of the $2 \rightarrow 3$ kind, the density of an expanding plasma around an exploding PBH can grow very fast (see also the computations by Daghigh \& Kapusta~\cite{2003PhRvD..67d4006D});
		\item Belyanin \etal~\cite{1996MNRAS.283..626B} predict that a full magneto-hydrodynamic plasma should develop around evaporating PBHs, based or earlier ideas from Rees \& Blandford~\cite{1977Natur.266..333R,1977MNRAS.181..489B};
		\item Nagatani~\cite{1999PhRvD..59d1301N} argues that symmetry restoration is possible in a sphere around an evaporating BH when its temperature overcomes the associated threshold.\footnote{Hawking had suspected just the contrary in~\cite{1981CMaPh..80..421H}.}
	\end{itemize}
	All of these competing models rely on the formation of some kind of ``photosphere'' or ``fireball'' around PBHs, due to non-negligible interactions between emitted particles~\cite{1999PhRvD..59f3009C}. They predict very different HR signatures in the astronomical observations, in particular concerning the PBH final bursts. The party has been definitely settled by MacGibbon \etal~\cite{2008PhRvD..78f4043M} in 2008 in a detailed study of all the relevant interactions in the SM framework. The MG\&W model is now commonly used by all HR papers, and we stick to it hereafter, with comments on other models at relevant places.
	
	\subsection{Basic formulas}
	\label{sec:basic_formulas}
	
\begin{figure}[!t]
		\centering
		\includegraphics{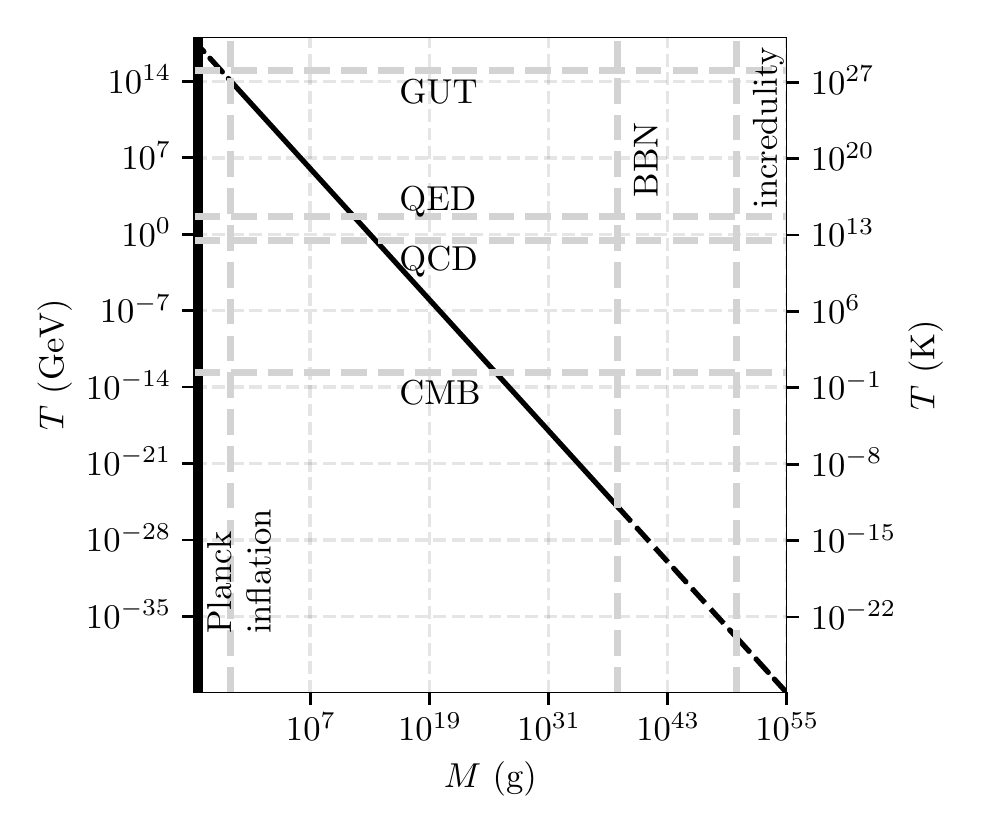}
		\caption{Temperature of a SBH, compared to the energy scales of the CMB ($T\sim 2.7\,$K), the QCD ($\Lambda\mrm{QCD}\sim 150\,$MeV) and QED ($\Lambda\mrm{QED}\sim 100\,$GeV) phase transitions and the putative GUT scale ($E\mrm{GUT} \sim 10^{14}-10^{16}\,$GeV). The mass scales are the same as in Fig.~\ref{fig:radii}.}
		\label{fig:temperature}
	\end{figure}

	In this Section we present the basic formulas used to compute the HR rates in the MG\&W model. Recall that the fundamental equation for the emission rate of a single spin $s$ dof with angular momentum numbers $(l,m)$, rest mass $\mu$, electric charge $q$ and energy $E$ and for a BH metric described by the mass $M$ and the set of secondary parameters $\{x_j\}$ ($\{x_j\}$ can contain the BH spin, its electric charge, \textit{etc.}) is
	\begin{equation}
		\dfrac{\d^2 N}{\d t\d E} = \dfrac{1}{2\pi}\dfrac{\Gamma_{Esqlm}(M,\{x_j\})}{e^{E^\prime/T(\{x_j\})} - (-1)^{2s}}\Theta(E - \mu)\,. \label{eq:HR_fundamental}
	\end{equation}
	The GFs $\Gamma$ and the BH temperature $T$ depend in general of the BH metric. In the high-energy limit (``geometrical optics'' limit or GO), they are the same for all particle spins and in the Schwarzschild case one obtains
	\begin{equation}
		\Gamma\mrm{GO} (E) \underset{\lim E\rightarrow+\infty}{=} \sum_{l,m} \Gamma_{Eslm} = 27 E^2 M^2\,. \label{eq:GO}
	\end{equation}
	The GO limit for general spherically symmetric BHs is given by~\cite{2021PhRvD.104h4016A}. The temperature is obtained in general from $T = \kappa/2\pi$, which becomes in the KBH case
	\begin{equation}
		T\mrm{K} = \dfrac{1}{2\pi}\left( \dfrac{r_+ - M}{r_+^2 + a^2} \right) \quad \underset{a^* \rightarrow 0}{\longrightarrow} \quad T\mrm{S} = \dfrac{1}{8\pi M}\,. \label{eq:temperature}
	\end{equation}
	The temperature of a SBH, compared to some typical energy scales, is given in Fig.~\ref{fig:temperature}, while the temperature of a general spherically symmetric BH is obtained in~\cite{2021PhRvD.103j4010A}.
	
	The energy of the particle may be corrected by effective chemical potentials due to \textit{e.g.}~spin or charge coupling: $E^\prime = E - m\Omega - q\Phi$. The rate of emission of a primary particle $i$ with internal multiplicity $g_i$ (color, helicity) is
	\begin{equation}
		\dfrac{\d^2 N_i}{\d t\d E} = g_i \sum_{l,m} \dfrac{\d^2 N}{\d t\d E}\,. \label{eq:HR_primary}
	\end{equation}
	The emission of a secondary particle $j$ must be computed by convolving the primary spectra with particle physics branching ratios $\text{Br}_{i\rightarrow j}(E,E^\prime)$
	\begin{equation}
		\dfrac{\d^2 N_j}{\d t\d E} = \int_0^{+\infty} \sum_i \text{Br}_{i\rightarrow j}(E,E^\prime)\,\dfrac{\d^2 N_i}{\d t\d E^\prime}\,\d E^\prime\,. \label{eq:HR_secondary}
	\end{equation}
	Hence, a distribution of BHs produces the following instantaneous flux of secondary particles
	\begin{equation}
		\dfrac{\d^2 n_j}{\d t\d E} = \int \d \{x_j\} \int_{M\mrm{min}}^{M\mrm{max}} \d M\,\dfrac{\d^2 n}{\d M \d \{x_j\}}\,\dfrac{\d^2 N_j}{\d t\d E}\,. \label{eq:HR_instantaneous}
	\end{equation}
	This flux may be integrated over some period of cosmological time $(t_1,t_2)$, taking the universe redshift dilution $\mathfrak{r} \equiv a(t_2)/a(t)$ into account, through
	\begin{equation}
		\dfrac{\d n_j}{\d E}(t_2) = \int_{t_1}^{t_2} \mathfrak{r}(t) \left.\dfrac{\d^2 n_j}{\d t\d E^\prime}\right|_{E^\prime = E/\mathfrak{r}(t)}\,\d t\,, \label{eq:HR_integrated}
	\end{equation}
	where the time evolution of the PBH parameters should also be taken into account.
	
	The evolution of KBHs is encoded in the ``Page coefficients''
	\begin{equation}
		\begin{pmatrix} -f(M,a^*)/M^2 \\ -g(M,a^*)a^*/M \end{pmatrix} \equiv \dfrac{\d}{\d t} \begin{pmatrix} M \\ J \end{pmatrix} =  \int_0^{+\infty} \sum_i \dfrac{\d^2 N_i}{\d t\d E} \begin{pmatrix} E \\ m \end{pmatrix}\d E\,,\label{eq:Page_coefs}
	\end{equation}
	where one should remember that the emission rates depend non-trivially on the particle angular momentum (an analog coefficient could be defined for \textit{e.g.}~the BH electric charge). The following differential equations are obtained straightforwardly
	\begin{equation}
		\left\{\begin{array}{l}
			\dfrac{\d M}{\d t} = -\dfrac{f(M,a^*)}{M^2}\,, \\
			\dfrac{\d a^*}{\d t} = \dfrac{a^*(2f(M,a^*) - g(M,a^*))}{M^3}.
		\end{array}\right.
		\label{eq:evolution_equations}
	\end{equation}
	and the lifetime of a BH is computed as
	\begin{equation}
		\tau(M,a^*) = \int_0^{M} \frac{M^2}{f(M,a^*)}\d M\,.\label{eq:lifetime_general}
	\end{equation}
	Assuming constant $f(M,a^*)$ raises $\tau \sim M^3/(3f(M,a^*))$. With the SM emission only, one obtains $\tau(M_*) = t_0$ the age of the universe for the critical mass $M_* \sim 5\times 10^{14}\,$g.
	
	\subsection{Numerical recipes}
	\label{sec:numerical_recipes}
	
	There are several methods to compute the quantities given by Eqs.~\eqref{eq:HR_fundamental}--\eqref{eq:HR_integrated}. This can be done analytically under some assumptions, \textit{e.g.}~in the geometrical optics (GO) high-energy limit for HR rates~\cite{2021EPJP..136..261A}. One can also compute the numerical rates and fit them with heuristic analytical formulas, see \textit{e.g.}~\cite{2022PhRvD.105a5022C}. As the GFs are not analytically calculable for general BH metrics at all energies, numerical resolution is required. In general, the Teukolsky equation is complicated to solve numerically as such. It can be reduced to a Sch\"odinger-like wave equation with short-ranged potentials, as shown in the general spherically symmetric case by~\cite{2021PhRvD.103j4010A} or in the Kerr case by~\cite{1975RSPSA.345..145C,1976RSPSA.349..217D,1976RSPSA.350..165C,1977RSPSA.352..325C}. This equation may be solved numerically in the WKB approximation~\cite{2000PhRvL..85.5042P}, through a path-integral formalism~\cite{2015arXiv151205018G} or by brute force computing.
	
	Several numerical public codes are available to compute the HR rates. Historically, the first programs were released in the context of HR by small higher-dimensional BHs evaporating in the LHC detectors, and took the form of event generators. These include \texttt{Charybdis}~\cite{2003JHEP...08..033H,2009JHEP...10..014F}, \texttt{Catfish}~\cite{2007CoPhC.177..506C}, \texttt{BlackMax}~\cite{2008PhRvD..77g6007D,2009arXiv0902.3577D,2019CoPhC.236..285D} and \texttt{QBH}~\cite{2010CoPhC.181.1917G}. The present author (together with A.~Arbey) released the \texttt{C} code \texttt{BlackHawk}~\cite{2019EPJC...79..693A,2021EPJC...81..910A} which is more dedicated to PBH studies with the possibility to simulate general mass and spin distributions and to study the cosmological evolution of the HR emission. More recently, the PBH evaporation in the context of DM production by PBH evaporation and higher-dimensional PBHs were implemented in respectively the \texttt{Python} code \texttt{ULYSSES}~\cite{2022PhRvD.105a5022C,2022PhRvD.105a5023C} and \texttt{CosmoLED}~\cite{2022arXiv220111761F}. Comparison of the PBH codes, when possible, shows that they are in reasonable quantitative agreement with each other.
	
\begin{figure}[!t]
		\centering
		\includegraphics{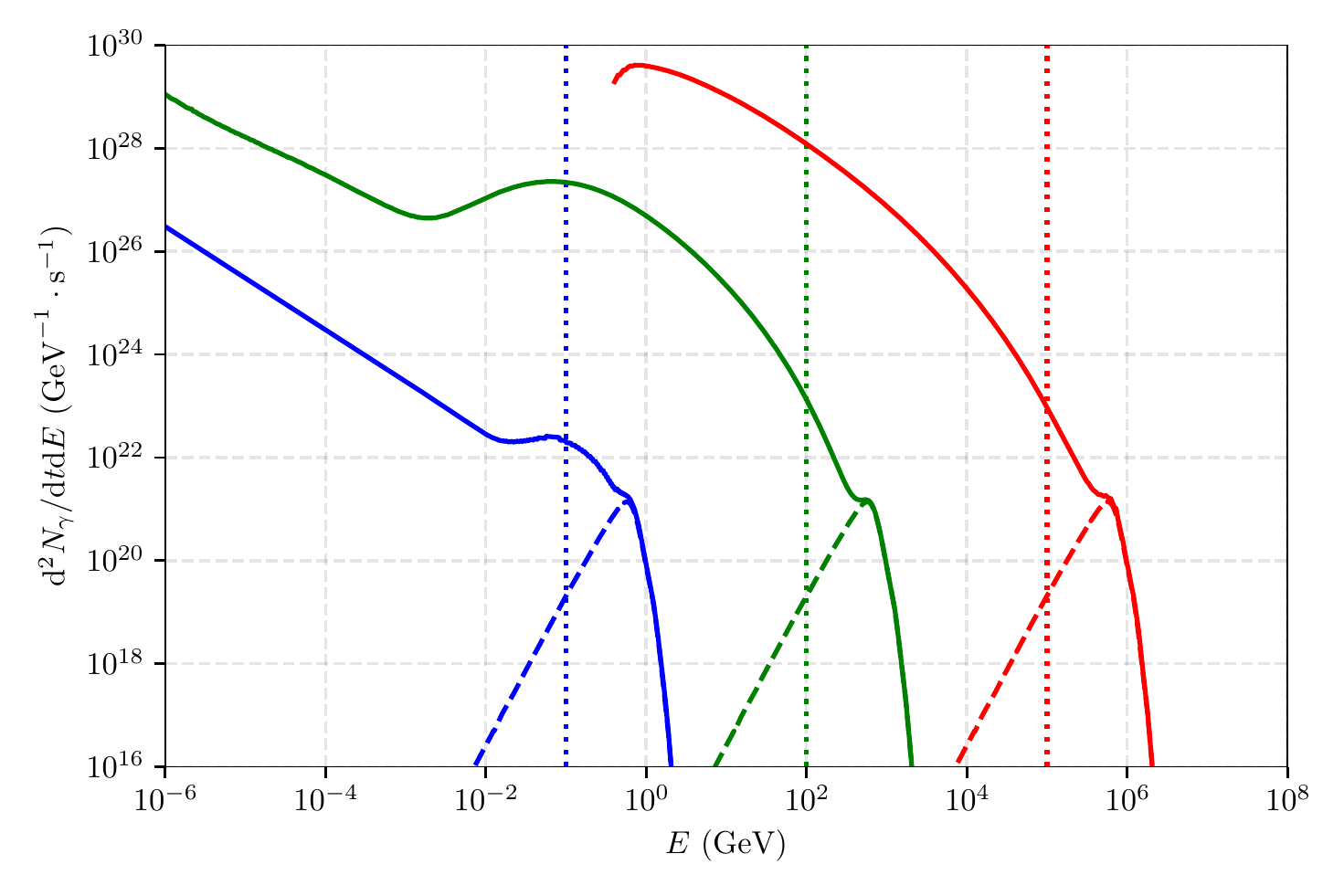}
		\caption{Photon primary (dashed) and secondary (solid) spectra from a SBH with temperature $T = 100\,$MeV (left, \texttt{Hazma}), $100\,$GeV (center, \texttt{PYTHIA}) and $100\,$TeV (right, \texttt{HDMSpectra}) as computed with \texttt{BlackHawk}. The vertical lines represent the BH temperature.}
		\label{fig:spectra}
	\end{figure}

	The secondary spectra are computed by particle physics codes, which differ following the energy scale considered. For example, the public code \texttt{BlackHawk}, available at \url{https://blackhawk.hepforge.org/}, relies on the \texttt{Python} package \texttt{Hazma}~\cite{2020JCAP...01..056C} at low energies ($E \lesssim$ GeV), \texttt{PYTHIA}~\cite{2022arXiv220311601B} or \texttt{HERWIG}~\cite{2020EPJC...80..452B} at LHC energies ($E \sim $ GeV$-$TeV) and on \texttt{HDMSpectra}~\cite{2021JHEP...06..121B} at high energies (TeV upward). An example of photon spectra is given in Fig.~\ref{fig:spectra} for each of these cases. On that figure, the bumpy feature in the secondary spectra at $\sim 100\,$MeV is due to the primary pion decays. The codes are not reliable for $E\mrm{sec}/E\mrm{prim} \ll 10^{-6}$, hence the \texttt{HDMSpectra} line stops at $\sim 1\,$GeV. As predicted by MG\&W, most of the emission is realized at energies $E \ll T$ due to the jet hadronization and decay. In the context of DM studies, Cirelli \& collaborators constructed tabulated branching ratios for DM decay that can be directly transported into PBH secondary spectrum calculations~\cite{2011JCAP...03..051C,2011JCAP...03..019C}. The material is available at \url{http://www.marcocirelli.net/PPPC4DMID.html} and is generally denoted as PPPC4DMID.
	
	\section{Non-evaporated primordial black holes}
	\label{sec:non-evap}
	
	In this Section, we provide a very brief account of the constraints on the PBH density which are not based on HR. This is only a brief summary as complete recent accounts can be found \textit{e.g.}~in~\cite{2020ARNPS..70..355C,2021JPhG...48d3001G}. Theses reviews focus on PBHs as a DM candidate. Indeed, PBHs are non-baryonic and non-relativistic, while not being a new object of particle physics like WIMPs or axions. They appear therefore as a convenient DM candidate and the constraints on their abundance today $\Omega\mrm{PBH} \equiv \rho\mrm{PBH} / \rho\mrm{c}$ (where $\rho\mrm{c}$ is the critical density) are often set as the fraction of DM $f\mrm{PBH} \equiv \Omega\mrm{PBH}/\Omega\mrm{DM}$ they can represent, with the upper limit $f\mrm{PBH} < 1$ not to overclose the universe. A more universal parameter is the fraction of the universe density into PBHs at the time of their formation $\beta \equiv \Omega\mrm{PBH}(t\mrm{f})/\Omega\mrm{tot}(t\mrm{f})$, with the particular combination $\beta^\prime$ that could encode non-standard cosmologies~\cite{2021RPPh...84k6902C}; we stick to $\beta^\prime$ hereafter to express the PBH constraints. PBHs with mass over the evaporation limit $M > M_*$ are constrained mostly by their gravitational effects, such as gravitational lensing, dynamics in structures, accretion and gravitational waves. The constraints that rely on PBH evaporation are presented in detail in the next Sections.
	
	\subsection{Lensing constraints}
	
	PBHs of any mass behave as gravitational lenses that deviate light rays due to their strong gravitational field (for a review of this effect and its implications in cosmology, see~\cite{2010CQGra..27w3001B}). PBHs should have a sensible proper motion with respect to the sources they lense the light of. Their radius is so small that they only behave as very transient lenses, that is called millilensing, microlensing, femtolensing or picolensing depending on the duration of the effect. 
	
	The visible effect on Earth is that the light from remote sources is suddenly brightened by the passing of a PBH due to the focusing effect~\cite{2010CQGra..27w3001B}. This applies to the light of quasars, supernovae, stars, and $\gamma$-ray bursts (GRBs). The most famous studies are that of MACHO~\cite{2000ApJ...542..281A}, EROS~\cite{2007A&A...469..387T},\footnote{The names of these two collaborations are obviously irrespective.} OGLE~\cite{2019PhRvD..99h3503N}\footnote{OGLE might have observed microlensing events compatible with planetary mass PBH origin~\cite{2017Natur.548..183M}.} and Subaru-HSC~\cite{2020PhRvD.101f3005S}, which exclude $f\mrm{PBH} = 1$ in the range $\sim 10^{23}-10^{35}\,$g. The original Subaru-HSC constraint was initially much more stringent and excluded PBHs as DM down to $\sim 10^{21}\,$g~\cite{2019NatAs...3..524N} but it was shown recently that this was due to a simplistic treatment of the source spatial extension~\cite{2020PhRvD.101f3005S}.
	
	There is a disputed constraint on GRB femtolensing~\cite{2012PhRvD..86d3001B,2018JCAP...12..005K} in the $10^{17}-10^{19}\,$g mass range which would close a good deal of the HR window, and an ongoing search for fast radio burst lensing~\cite{2020MNRAS.496..564K}. Microlensing constraints are the ones with the most expected improvement in the Snowmass PBH white paper~\cite{2022arXiv220308967B}.
	
	\subsection{Dynamical constraints}
	
	PBHs are massive objects, and thus they exert gravitational attraction on the nearby material. This results: $i)$ in gravitational friction that make them settle down in compact objects that they can subsequently absorb or destruct, $ii)$ in gravitational disturbance of bound systems like stellar binaries, and $iii)$ in gravitational heating of the interstellar medium (ISM) gas. The most severe constraints in the $10^{36}-10^{47}\,$g come from gas heating and galaxies, globular clusters and stellar binaries stability~\cite{1999ApJ...516..195C,2014ApJ...790..159M} (see~\cite{2018MNRAS.478.3756C} for a review). ``Stupendously'' large PBHs would also accrete surrounding material with too much efficiency if formed in the early universe; depleting structure formation~\cite{2018MNRAS.478.3756C}. There could be constraints on the disruption of neutron stars and white dwarfs (the most compact non-BH objects) in the range $10^{18}-10^{23}\,$g~\cite{2013PhRvD..87b3507C,2013PhRvD..87l3524C} (effectively closing all the HR window), but these were recently debunked~\cite{2014arXiv1402.4671C,2014PhRvD..90j3522D,2019JCAP...08..031M} and await more rigorous theoretical treatment.
	
	\subsection{Accretion constraints}
	
	PBHs may accrete gas and dust material into disks like any other celestial object. These disks would radiate in the X-ray band and heat the surrounding ISM or the CMB~\cite{2017JCAP...10..034I,2020PhRvR...2b3204S}, or the PBHs could be so massive that their gravitational potential distorts the CMB~\cite{2018PhRvD..97d3525N} ($\mu$-distortions). These constraints exclude all DM into PBHs in the super-Solar mass range $M \gtrsim M_\odot$. Accretion indirectly takes a part in all other constraints as it modifies the PBH mass spectrum.
	
	\subsection{Gravitational wave constraints}
	
	As GWs propagate in space with no measurable alteration, they are the most precise probes of gravitational events back to the early universe. The stochastic GW background encompasses the GWs generated by PBH formation, by the statistical distribution of PBHs (scalar-induced GWs), by their sudden evaporation and modification of the equation of state, and by their continuous mergers since the early era. This may be the set of constraints totaling the greatest number of publications since 2015 and the first observation of the merger of BHs by the LIGO instrument~\cite{2016PhRvL.116f1102A,2016ApJ...833L...1A}. It was immediately suggested that the BH components of the binary merger were PBHs~\cite{2016PhRvL.116t1301B,2016PhRvL.117f1101S}, and the background of GWs from PBH mergers was predicted in~\cite{2017JCAP...09..037R}. The associated constraints are model-dependent but exclude PBHs in the $10^{30}-10^{34}\,$g mass range \cite{2019PhRvD.100b4017A,2020PhRvL.124y1101C} (observations by the LIGO/VIRGO instruments and the NANOGrav experiment). For a review about the GW perspectives of PBH studies, please refer to~\cite{2018CQGra..35f3001S}.\footnote{For a more general review about GWs and the early universe, see~\cite{2022arXiv220307972C}.} GWs from mergers are also one of the only positive evidence for PBHs, as even if they represent just a percent of the DM density, their mergers are still in the range of detectability at LIGO/VIRGO and future detectors~\cite{2021arXiv210503349F}.
	
	Peculiar GW events such as very heavy $M\gtrsim 100\,M_\odot$ or light $M\lesssim M\mrm{TOV}$ BH binary mergers (\textit{e.g.}~\cite{2020JCAP...08..039C,2022PhRvD.105f4063C}) would be smoking-guns towards PBH origin (see however~\cite{2021PhRvL.126n1105D}). A stochastic GW background should also be generated by PBH formation~\cite{2020PhRvD.101l3533I,2021PhLB..82336722D,2021PhRvL.126e1303V}, and most interestingly it depends on the spin of the PBHs, giving access to a double-check of the PBH origin of BH mergers~\cite{2020EPJC...80..243K}. It could be the only way of constraining the PBH abundance in the remote $M \lesssim 10^9\,$g mass range for which PBHs evaporate before BBN. The forthcoming LISA instrument is expected to shed some light on these yet unmeasured features~\cite{2022arXiv220405434A}.
	
	\section{Baryogenesis}
	\label{sec:baryogenesis}
	
	The universe is not symmetric in (anti)baryon content. The baryon-to-photon ratio $\eta \sim 6\times 10^{-10}$~\cite{2020JCAP...03..010F}, as measured at BBN or CMB epochs shows that if initially produced in the same quantities, most of the baryons and antibaryons should have annihilated into photons. However, a slight overabundance of baryons of the order of $\eta$ must have been produced, so that after annihilation some baryon content was still present. This baryon-antibaryon asymmetry resulted in baryogenesis, the genesis of the baryon content of the universe. For baryogenesis to take place, the three conditions of Sakharov~\cite{1967JETPL...5...24S} must be fulfilled:
	\begin{enumerate}
		\item[] \textbf{Cond.~1:} C and CP symmetry violation;
		\item[] \textbf{Cond.~2:} baryonic number violation;
		\item[] \textbf{Cond.~3:} thermodynamic equilibrium violation.
	\end{enumerate}
	In the standard cosmological model with only SM fields, these conditions are not met with sufficient amplitude to explain baryogenesis. Thus, extensions to the standard model must assumed. One possibility resides in the capability of BHs to fulfill conditions 1 and 3 immediately. Condition 1 is satisfied because as a PBH explosion manifestly violates the T symmetry, it should violate C/CP as well in order to conserve the CPT symmetry resulting from the global Lorentz invariance~\cite{1976ApJ...206....8C}. CP symmetry violation is also present in the SM as seen \textit{e.g.}~in neutral kaon decay~\cite{2018PhRvD..98c0001T}. Furthermore, when BHs evaporate, the temperature of the quasi-thermal flux they produce is uncorrelated to the background radiation temperature. Therefore, they are sources of out-of-equilibrium processes, which fulfills condition 3.
	
	Condition 2 is trickier: baryonic number violation is present in the SM but not with a magnitude sufficient to explain the value of $\eta$. Furthermore, any asymmetry generated in the early universe must survive until BBN and CMB epochs, thus it must not be diluted away by other mechanisms, such as wash-out or entropy injection~\cite{1976JETPL..24..571Z,2021JETP..133..552C}. In general, it is assumed that baryogenesis should occur after the background temperature has cooled to $T\mrm{EW} \sim 100\,$GeV, for the electroweak (EW) interactions to freeze out. In the standard RD model, this translates to a time $t\mrm{EW} \sim 10^{-12}\,$s which is the lifetime of a $M \sim 10^{5}\,$g PBH.	PBH participation in baryogenesis may proceed by five main scenarios:
	\begin{itemize}
		\item the ``Hawking model'' where baryogenesis is the direct result of an asymmetry in the GFs (\textit{e.g.}~an effective chemical potential);
		\item the ``Weinberg model'' where heavy (GUT) bosons emitted by HR of PBHs decay asymmetrically into (anti)baryons;
		\item the ``Dolgov model'' where a mass asymmetry between (anti)baryons cause asymmetric re-absorption probability by PBHs;
		\item the ``Nagatani model'' where an EW symmetry-restored region exists around high-temperature PBHs and baryon asymmetry develops by sphaleron processes;
		\item the ``Fujita model'' where the heavy decaying particles are right-handed neutrinos (RHNs) instead of GUT bosons.
	\end{itemize}
	These five models fulfill condition 2, as showed below. We have listed them in chronological order of appearance.
	
	The natural mass range for baryogenesis is $M \ll 10^9\,$g so that PBHs evaporate before BBN and their abundance is not constrained. Indeed, the initial abundance of PBHs would have to be very large in order to produce the correct asymmetry~\cite{1979PhRvD..19.1036T}, PBHs may even have dominated the energy density of the universe before their evanescence~\cite{1980MNRAS.192..427B}, an era denoted as BHD (for black hole domination). However, a very extended PBH distribution like the original scale-invariant one poses immediate difficulty as it would conflict with the BBN constraints at $M > 10^9\,$g~\cite{1980PhLB...94..364G}.
	
	In general, the asymmetry proceeds from a baryon non-conservation process either in the PBH evaporation directly or in the PBH evaporation product decay. Thus, the generated asymmetry is related to the PBH abundance by a proportionality factor $\BBbar$, which depends on the PBH mass and on the asymmetry mechanism
	\begin{equation}
		\eta \propto \left\{\begin{array}{ll}
			\BBbar(M)\, n\mrm{PBH}\,, & \text{RD} \\
			\BBbar(M)\,, & \text{BHD}.
		\end{array}\right.
	\end{equation}
	Reviews about PBH baryogenesis can be found \textit{e.g.}~in~\cite{1980SHEP....1..183B,1981NuPhB.181..461B,1992PhR...222..309D}.
	
	\subsection{Hawking model}
	
	In the seminal paper~\cite{1975CMaPh..43..199H}, Hawking expressed the idea that PBHs may be formed from symmetric (anti)baryonic content, and evaporate preferentially into baryons rather that antibaryons, due to \textit{e.g.}~a chemical potential in the GFs leading to asymmetric branching ratios. On the other hand, PBHs could be formed by collapsing baryonic material, but as they have no memory of their progenitor composition due to the no-hair theorem, they would carry no baryonic charge. Hence, PBHs would radiate (anti)baryons in equal numbers and contribute only in the radiation density, making the baryon-to-photon ratio effectively diminish. These two mechanisms fulfill condition 2.
	
	Barrow~\cite{1980MNRAS.192..427B} shows first that the correct PBH baryogenesis model depends on the capability of PBHs to dominate the energy density of the universe. If they do, then baryogenesis can result from the ``Hawking model''. In this case, the asymmetry does not depend on the initial PBH abundance because they are at the origin of a second reheating (the first reheating is the decay of the putative inflaton field) and of all the radiation content of the universe. If they do not, then it should proceed through \textit{e.g.}~the ``Weinberg model''. There is a region of overlap that depends greatly on the assumptions on $\BBbar$.
	
	Hook~\cite{2014PhRvD..90h3535H} proposes a realization of the ``Hawking model'' of baryogenesis via a chemical potential in the HR rate that is not the same for (anti)particles. This chemical potential is assumed to be present everywhere in the universe in~\cite{2014PhRvD..90h3535H}, but it may also be located in the BH vicinity due to modified EW interactions in curved spacetime~\cite{2015PhRvD..92d6008F,2015arXiv150500472B} and it could depend on the BH mass~\cite{2017PTEP.2017c3B02H}.
	
	The chemical potential could also be generated by baryon/lepton number interactions with the space-time curvature~\cite{2021PhRvD.103h3504B,2022JCAP...03..013S}. Ref.~\cite{2021PhRvD.103h3504B} is the first to consider baryogenesis with an extended mass function originating in the critical collapse mechanism.
	
	\subsection{Weinberg model}
	
	The first complete scenario of baryogenesis based on PBH evaporation in a GUT may also be the first appearance of the idea that PBHs can emit particle dofs that have not yet been discovered in astronomical and laboratory experiments. It was presented by Turner \& Schramm in~\cite{1979Natur.279..303T} and assumes that PBHs can evaporate into a GUT-scale ($\sim 10^{16}\,$GeV) boson $X$ (that could be the Higgs boson), whose decay would strongly violate CP. The subsequent asymmetric X/$\overline{\text{X}}$ decays into B/$\overline{\text{B}}$ would generate a net baryon density $n\mrm{B}$ (fulfilling condition 2) which would then be diluted by the emission of the rest of the PBH mass under the form of radiation (entropy) down to the observed value of $\eta$. This is based on the idea by Weinberg~\cite{1979PhRvL..42..850W,1979PhRvD..20.2484N} but provides a natural origin for the GUT bosons in the evaporation of PBHs. The first numerical calculation of the relevant parameter space in the PBH mass/GUT boson characteristics was done in~\cite{1979PhLB...89..155T}.
	
	The GUT ``Weinberg model'' was greatly improved by Barrow \& Ross~\cite{1980MNRAS.192..427B,1980MNRAS.192P..19B,1981NuPhB.181..461B}. They remark that if the PBH spatial distribution is not homogeneous, then inhomogeneous baryon-to-photon ratio can develop. This could be at the origin of structure formation. They also discuss the monopole problem: PBHs could produce dramatic magnetic monopole abundance by HR, which is strictly constrained. This would result in a strong reduction of the available parameter space for PBH baryogenesis.
	
	10 years later, Barrow \etal~\cite{1991PhRvD..43..977B,1991PhRvD..43..984B} review the PBH GUT baryogenesis mechanism. They present a version of it where the heavy GUT boson is the Higgs boson. Baumann \etal~\cite{2007hep.th....3250B} spotted a typo in the numerical calculations of Barrow \etal~and shrink accordingly the available parameter space in \cite{2007hep.th....3250B}. The inclusion of this scenario in the full MSSM was done in~\cite{2021PhRvD.103d3504H}.
	
	\subsection{Dolgov model}
	
	This model is the reverse of the ``Hawking model'' and it does not really rely on HR mechanics, even if absorption and emission coefficients are not independent; it is given here for completeness. Toussaint~\etal~\cite{1979PhRvD..19.1036T} proposed that the \textit{absorption} of surrounding particles by PBHs could be asymmetric: if B/$\overline{\text{B}}$ have different interactions with the gravitational field (\textit{e.g.}~different rest masses), then their absorption coefficients are different and an initially symmetric medium would become asymmetric because of PBH absorption.
	
	Dolgov's model~\cite{1981PhRvD..24.1042D} is based on the ``Weinberg model''. Suppose that the PBH radiated heavy bosons decay into mass-asymmetric mesons. These are subsequently recaptured by PBHs with a different rate due to different coupling with the PBH metric. The baryonic asymmetry is computed analytically thanks to a resolution of the Teukolsky equation with an approximate potential for the massive GUT fields in~\cite{1980JETP...52..169D}.
	
	Ambrosone \etal~\cite{2022PhRvD.105d5001A} note that the ``Dolgov model'' can in fact be complementary to the other baryogenesis models because the very high PBH fraction necessitated by those may result in effective re-absorption of the asymmetry produced. If this re-absorption is asymmetric, the baryon asymmetry is preserved.
	
	\subsection{Nagatani model}
	
	A new model of baryogenesis by PBHs is introduced by Nagatani in 1999~\cite{1999PhRvD..59d1301N}. They argue that the if PBHs dominated the early universe, their energetic radiation might heat up the PBH environment so much that the EW (Higgs) symmetry might be restored in a spherical region around PBHs. Then, a domain wall should form between the restored symmetry patch and the outer broken symmetry universe. Baryogenesis would occur inside that domain wall (or in the restored region) via sphaleron processes~\cite{1985PhLB..155...36K} which fulfill condition 2. With very fine tuning, this model could also explain reheating and DM~\cite{2007hep.th....3070A}. The ``Nagatani model'' seems deprecated since PBHs may not form symmetry-restoring fireballs~\cite{1981CMaPh..80..421H,2008PhRvD..78f4043M}.
	
	\subsection{Fujita model}
	
	Baryogenesis through PBH evaporation into asymmetric RHNs is considered by Fujita \etal~\cite{2014PhRvD..89j3501F} as a mean to explain the coincidence between the (dark) matter cosmological densities. This baryogenesis model is based on an old tentative to avoid recourse to \textit{ad hoc} GUT bosons~\cite{1986PhLB..174...45F}. The parameter space for baryogenesis depends on the ``light'' ($T>\mu\mrm{\nu}$) or ``heavy'' ($T<\mu\mrm{\nu}$) mass of RHNs and is incompatible with DM generation except if entropy non-conservation is allowed (even so, very fine-tuning is needed~\cite{2020JCAP...08..045B}).
	
	Recent numerical calculations done with the public code \texttt{ULYSSES}~\cite{2022PhRvD.105a5022C,2022PhRvD.105a5023C} have shown that interplay between PBH generated RHNs and thermal ones from Higgs decay complicate the scenario~\cite{2021PhRvD.104j3021P}. The parameter space depends further on the neutrino model~\cite{2021JCAP...11..019D,2021JCAP...08..021D}. Dilution of the asymmetry by entropy injection may be avoided with very light PBHs $M \gtrsim 10\,$g~\cite{2022arXiv220308823B,2022JCAP...03..031B} but in this case the wash-out problem is still present.
	
	\subsection{Other ideas and prospects}
	
	A small amount of asymmetry could also originate in the randomness of the HR process, with a fluctuation of the baryon number of the emitted flux of the same form as that of the emitted charge~\cite{1976ApJ...206....8C}.
	
	Lindley~\cite{1981MNRAS.196..317L,1982MNRAS.199..775L} argued that PBH baryogenesis models are un-natural since they would require ``entirely unjustified'' fine-tuning in the PBH abundance and mass in order to obtain the correct asymmetry. They argue rather that the PBHs could be used as a second-hand source of asymmetry, while the first-hand source should remain \textit{e.g.}~the thermal GUT mechanism.
	
	Majumdar~\etal~\cite{1995IJMPD...4..517M,1999PhRvD..60f3513U} were the first to study the impact of PBH early accretion on the baryogenesis mechanisms. The overall conclusion is that the initial increase in the PBH mass delays their evanescence, which would help them decay \textit{after} the EW freeze-out and thus avoid the wash-out problem. However, this effect can be sizeable only for $M > 10^3\,$g PBHs~\cite{2003PAN....66..476B}.
	
	If PBHs generate the baryon asymmetry of the universe, this constitutes a major positive evidence for their existence. The distinctive signature of such models is that they predict a stochastic GW background of multiple origins: PBH evaporation into gravitons, or PBH formation/Poisson distribution (scalar-induced GWs)~\cite{2020PhRvD.101l3533I}.
	
	Overall, the PBH baryogenesis scenario is still perfectly viable, with the downside that fine-tuning is required in every model. As all these processes rely on the existence of BSM physics or non-standard BH behavior, they provide no direct constraint on the PBH fraction $\beta^\prime$ in the early universe. Constraints could come from the GW background (see Section~\ref{sec:bckg_graviton}), from the density of PBH remnants (see Section~\ref{sec:PMRs}) or from the concomitant production of DM (see Section~\ref{sec:DM_production}); these could allow observational discrimination of the different models. Note that the full GFs for HR of heavy particles (Weinberg or Fujita models) are never used, with the remarkable exception of Ref.~\cite{2021PhRvD.104j3021P}. Other PBH baryogenesis scenarios would benefit from precision calculations of HR.
	
	\section{Big bang nucleosynthesis}
	\label{sec:BBN}
	
	BBN corresponds to the epoch when the light elements where built up by fusion and fission of nucleons into nuclei (by order of importance H, $^3$He, $^4$He, $^7$Li and $^6$Li). It began when the universe was $t \sim 2\,$min old, and lasted effectively until it was $t \sim 20\,$min old. The complete BBN mechanisms extend out of this time window with the freezing of the neutron-to-proton ratio at $t \sim 1\,$s when p$\,\longleftrightarrow\,$n inter-conversion by neutrino capture became ineffective and the dissociation or decay of some light elements (\textit{e.g.}~$^3$H, $^7$Be) into stable elements at $t \gg 20\,$min. These light elements enter in the composition of the first generation of stars where they are further processed into heavier elements up to Fe.
	
	The standard BBN scenario works remarkably well, and is the more robust probe of the very early universe so far. The light element abundances were observed in distant (old) gas clouds, and correspond perfectly to the prediction of numerical codes such as \textit{e.g.}~the public one~\texttt{AlterBBN}~\cite{2012CoPhC.183.1822A,2020CoPhC.24806982A} that compute the nuclear reaction chains taking the universe expansion into account. The only free parameter of the standard scenario is the baryon-to-photon ratio, and if set to the CMB value of $\eta \sim 6\times 10^{-10}$, all the relevant abundances are obtained. This success remains however mitigated by the $^7$Li problem, namely the discrepancy of a factor of a few between the predicted $^7$Li abundance and its value as deduced from the Spite plateau in metal-poor stellar abundances. Recent element abundances and discussions on BBN can be found in the review~\cite{2020JCAP...03..010F}. PBHs modify the BBN in several ways, described in the following:
	\begin{itemize}
		\item modification of the neutron-to-proton ratio prior to BBN;
		\item dissociation of elements by injection of hadrons;
		\item dissociation of elements by injection of photons;
		\item modification of the expansion history by injection of entropy.
	\end{itemize}
	As the BBN yields are finely tuned by the different reactions at stake, they tolerate only minimal exotic interference. This results in very stringent limits on PBHs with mass $M\sim 10^9-10^{13}\,$g.
	
\begin{figure}[!t]
		\centering
		\includegraphics{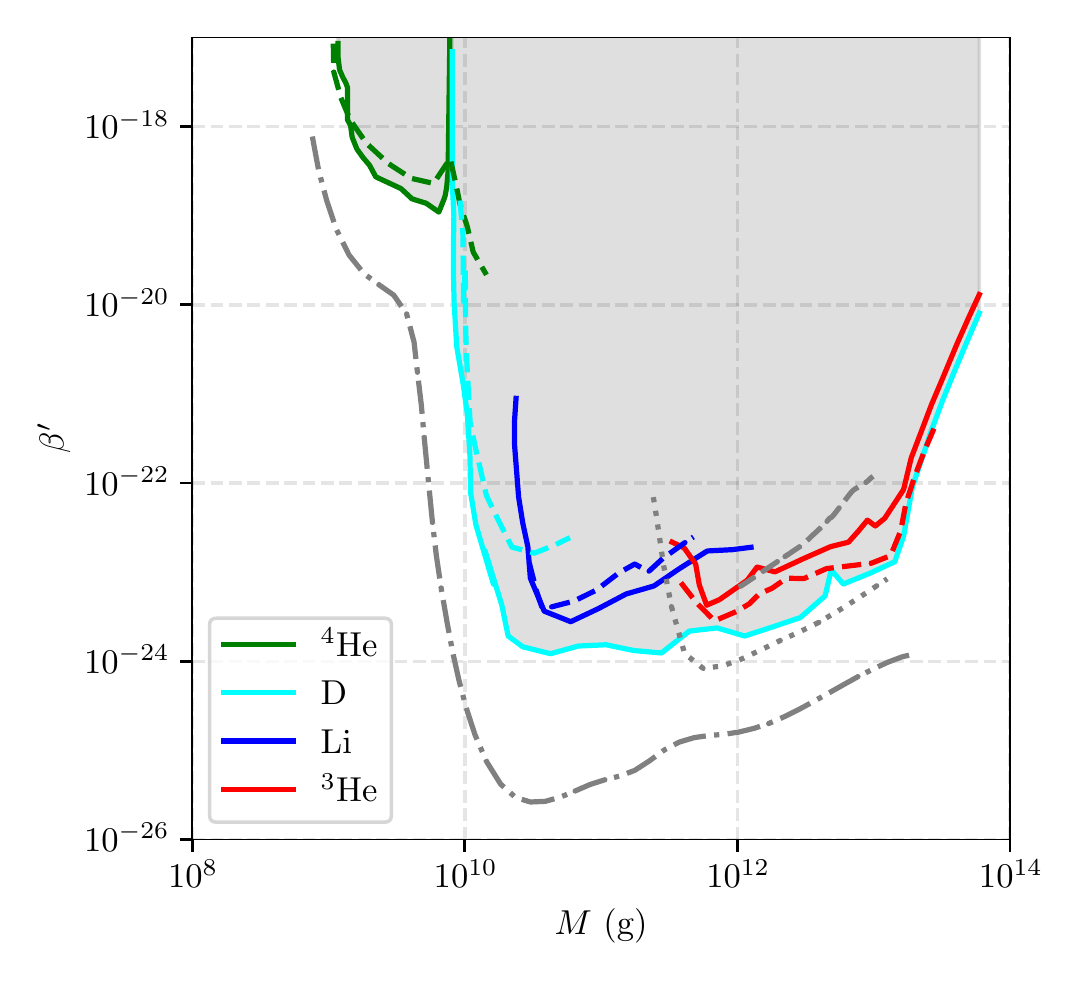}
		\caption{BBN constraints on PBHs. The coloured solid lines are that of Ref.~\cite{2021RPPh...84k6902C}, compared to the older limits of Ref.~\cite{2010PhRvD..81j4019C} (coloured dashed lines). For completeness, we also plot in grey the photodissociation limits obtained by Refs.~\cite{2020JCAP...06..018A} (dotted) and \cite{2021JCAP...05..042L} (dashed), and the limit of \cite{2020PhRvD.102j3512K} (dot-dashed) obtained by conversion of DM constraints. The shaded area is robustly excluded.}
		\label{fig:BBN}
	\end{figure}

	The BBN constraints from PBHs long remained subject to large uncertainties because of the precision of the nuclear reaction rates, the need for full numerical Boltzmann equation resolution as the PBH products are non-thermal, and the uncertainties of the high-energy PBH spectrum. It is only with the MG\&W model in the 1990's that studies have correctly taken into account the hadronic injection in the form of QCD jets. A summary of the modern constraints is provided in Fig.~\ref{fig:BBN}.
	
	\subsection{Nucleon inter-conversion}
	
	PBHs of $M \gtrsim 10^9\,$g can produce energetic neutrinos and hadrons \textit{after} the neutron-to-proton ratio has frozen out. Hence, they modify the initial conditions of BBN. This has a sizeable effect on the final yields as all the supplementary neutrons that are produced will ultimately end up in helium, with a corollary reduction of the deuterium abundance.
	
	Vainer \& Nasel'skii were the first to examine the effect of neutrino injection in 1977~\cite{1977SvAL....3...76V,1978SvA....22..138V} (followed by Zel'dovich \etal~\cite{1977SvAL....3..110Z} for the nucleon injection). They rely on the Wagoner code~\cite{1967ApJ...148....3W,1969ApJS...18..247W,1973ApJ...179..343W} to compute the light element abundances. All these effects modify the abundance of D at the end of BBN, which is tightly constrained by observations.
	
	Kohri \& Yokoyama recomputed the nucleon interconversion constraints in 1999~\cite{1999PhRvD..61b3501K}. Their paper is inspired by the work performed for particle DM decay during BBN by Dimopoulos and collaborators~\cite{1988PhRvL..60....7D,1988ApJ...330..545D,1989NuPhB.311..699D}. They used new observational data on the D, $^4$He and $^7$Li abundances, and on the neutron lifetime $\tau\mrm{n}$. They studied the alteration of (early) BBN by hadron injection by $10^8-3\times10^{10}\,$g PBHs at $t\sim 10^{-3}-10^{4}\,$s. As the timescale of particle interaction in the BBN plasma is $\mathcal{O}(10^{-8})\,$s, they considered all hadrons with that lifetime in their cascade reactions: charged pions $\pi^{\pm}$, charged kaons K$^\pm$, the long neutral kaon K$_{\rm L}$ and the (anti)nucleons. They estimated the rate of emission of these particles from jets in the MG\&W model, for the first time with unstable particles. The other electromagnetic (EM) particles are assumed to decay or thermalize with the plasma. They showed in particular that the constraints are highly dependent of the baryon-to-photon ratio $\eta$, which is left free in their study (as well as $M$ and $n\mrm{PBH}$), and on the choice of the D data, which suffers from high uncertainties.
	
	\subsection{Hadro-dissociation}
	
	Hadrodissociation means dissociation of light elements by hadronic injection, namely light mesons (pions, kaons) and nucleons (protons, neutrons) radiated by PBHs.
	
	Zel'dovich \etal~\cite{1977SvAL....3..110Z} analytically studied the effect of PBHs on the abundance of D. They find that energetic hadronic injection can destroy helium and overproduce deuterium. Numerical calculations are performed with the Wagoner code by Miyama \& Sato~\cite{1978PThPh..59.1012M}. Vainer \etal~\cite{1978SvAL....4..185V} further discussed hadrodissociation of light nuclei by energetic neutrons from PBHs and conclude that the effect is globally an increase in the production of D and a decrease of $^4$He. The energetic protons, due to their non-zero electric charge, would thermalize in the vicinity of the PBHs.
	
	Sedel'nikov returned to the problem of standard BBN alterations by evaporating PBHs into energetic hadrons in 1996~\cite{1996AstL...22..797S}. They took into account the fact that the PBH p/$\overline{\rm p}$ yield comes uniquely from jet decays, following the MG\&W model. They obtained constraints on the abundance of PBHs slightly more stringent that what was obtained before in the $10^9-10^{13}\,$g mass range. They also discussed the effect of adding BSM dofs on the evaporation of the PBHs: it mainly reduces their lifetime and their branching ratio into p/$\overline{\rm p}$.
	
	\subsection{Photo-dissociation}
	
	Vainer \& Nasel'skii~\cite{1977SvAL....3...76V} argued that the very high energy $\gamma$-rays emitted at the end of PBH evaporation could break $^4$He into D by photo-dissociation. Lindley~\cite{1980MNRAS.193..593L} further showed that high energy photons can trigger EM cascades and then produce more high energy photons, leading to the late photodissociation of D after the BBN yields have frozen out.
	
	The photo-dissociation constraints have been revisited by Acharya \& Khatri~\cite{2020JCAP...06..018A}. They considered only EM cascades from $10^{11}-10^{16}\,$g PBHs, and followed the procedure of~\cite{2019JCAP...12..046A} for DM decay. Full GFs are taken into account, secondary particles are computed with \texttt{PYTHIA} and results are presented for the first time for a log-normal mass distribution of PBHs.
	
	The photo-dissociation constraints were studied by Luo \etal~\cite{2021JCAP...05..042L} with a full spectrum Boltzmann treatment (based on the DM study~\cite{2015PhRvD..91j3007P}). They computed the constraints for critical collapse mass function, for the first time. \texttt{BlackHawk} is used to compute the precise spectrum of secondary particles and the evolution of PBHs. The Kawano code is used with the REACLIB database for the BBN computations. Only a very narrow mass range around $10^{12}-10^{13}\,$g is constrained.
	
	\subsection{Lithium problem}
	
	Jedamzik~\cite{2004PhRvD..70f3524J} studied the general impact of ``something'' decaying during BBN (DM, PBHs), with a modified version of the more Kawano code. Under some circumstances, namely fine-tuned neutron injection \textit{during} BBN, the $^7$Li problem is resolved. As can be expected, this is more complicated with full Boltzmann injection of un-thermalized particles generated by PBH evaporation. In fact, it has been proven that a monochromatic distribution of Schwarzschild PBHs cannot solve the lithium problem because their temperature is unequivocally related to their lifetime, forbidding fine-tuning~\cite{2021RPPh...84k6902C}.\footnote{Unpublished work performed by the author with J.~Pradler reached the same conclusions.} An extended mass distribution may alleviate that conclusion by destroying $^7$Li without overproducing D.
	
	\subsection{Modern studies}
	
	Carr \etal~\cite{2010PhRvD..81j4019C} were the first to compute the BBN constraints from PBHs in detail in the MG\&W model, taking into account all above phenomena. This work is a direct continuation of Ref.~\cite{1999PhRvD..61b3501K}. The new constraints rely on a modified version of the Kawano code \texttt{v4.1} to include PBH injection and the relevant interactions are borrowed from DM studies (\textit{e.g.}~\cite{2005PhRvD..71h3502K} and references therein). The full energy spectrum of the PBH products are taken into account and convoluted with the nuclear chains with an iterative method.
	
	The BBN constraints are revisited by Keith \etal~\cite{2020PhRvD.102j3512K}, based on the latest Planck measurements~\cite{2020JCAP...03..010F} and of direct conversion of constraints from the more recent particle decay theory~\cite{2018PhRvD..97b3502K,2020JCAP...12..048K}; however only pure blackbody HR is used. They discuss the impact of a dark sector: the evaporation rate is faster, and the branching ratio into SM particles is weaker. The expansion history is modified by PBHs and their (dark) decay products, which is taken into account using \texttt{AlterBBN}.
	
	In their most recent review, Carr \etal~\cite{2021RPPh...84k6902C} update their BBN constraints from Ref.~\cite{2010PhRvD..81j4019C} and discuss the results of~\cite{2020JCAP...06..018A,2020PhRvD.102j3512K} which obtain stronger limits. They argue that the hadron and photon effects must be taken into account together as they may cancel each other, as explained \textit{e.g.}~in \cite{2018PhRvD..97b3502K,2020JCAP...12..048K} on which they base their calculation. Their results is thus slightly less stringent while they use the same recent data from Planck. Their constraints are given as solid colored lines in Fig.~\ref{fig:BBN}, as we consider they are the most robust to date. Older constraints from~\cite{2010PhRvD..81j4019C} are shown for comparison, together with that of Refs.~\cite{2020JCAP...06..018A,2021JCAP...05..042L,2020PhRvD.102j3512K}.
	
	Overall, BBN is a very well understood process that strongly constrains the light PBH abundance. It does not seem that PBHs can have any positive impact on that cosmological era, thus we expect that constraints will become more and more stringent with additional light element data, in particular deuterium observations.
	
	\section{Cosmic microwave background}
	\label{sec:CMB}
	
	The measure of the CMB by Penzias \& Wilson in 1965~\cite{1965ApJ...142..419P} was one of the chief evidence in favor of the hot Big Bang model, together with the BBN yields. The CMB has been observed by the satellites COBE~\cite{1990ApJ...354L..37M}, WMAP~\cite{2003ApJS..148....1B,2003ApJS..148..135H} and more recently Planck~\cite{2016A&A...594A..11P,2020A&A...641A...5P}, to reach an extraordinary sensitivity. Future measures with the CMB-S4 experiments should increase further the understanding on the CMB epoch~\cite{2016arXiv161002743A,2018JCAP...08..029B,2019arXiv190210541H}.
	
	The isotropic CMB light corresponds to the first light emitted by atoms when nuclei and electrons combined and photons decoupled form the plasma. The CMB had a pure blackbody distribution at $E = 13.6\,$eV originally, redshifted today to a $T \sim 2.7\,$K background. This light was emitted at $t\sim 380\,000\,$yr. The decoupling process encompasses all the $\Lambda$CDM physical ingredients: the spectrum of the CMB is extremely sensitive to the composition of the universe, the distribution of the energy density between (dark) matter and radiation. The spatial fluctuations of the CMB measured by Planck give stringent constraints on the amount of energy injection, as well as the departure from pure Planckian spectrum measured by COBE/FIRAS. The CMB epoch is followed by the Dark Ages, namely the period before the burning of the first stars and galaxies. During this period, some reionization process should have occurred, with disputed origin. Another signature is the $21\,$cm background which contains information about an hyperfine transition of the hydrogen atom. The reionization fraction of the universe and the $21\,$cm ``temperature'' $T_{21}$ give access to the Dark Ages period.	CMB physics has benefited from numerous public computational tools, such as \texttt{CLASS}~\cite{2011arXiv1104.2932L,2011JCAP...07..034B,2011arXiv1104.2934L,2011JCAP...09..032L}, and many modules adapted for non-standard cosmology, including PBHs (\texttt{HyRec}~\cite{2011PhRvD..83d3513A,2020PhRvD.102h3517L}, \texttt{DarkHistory}~\cite{2020PhRvD.101b3530L}).
	
	The simplest idea linking PBHs and the CMB would be that PBHs of temperature $T \sim 2.7\,$K distributed isotropically generate the observed background with steady state radiation. Nartikar \& Rana~\cite{1979PhLA...72...75N} showed that this is unrealistic. However, evaporating PBHs modify the genuine CMB in several ways, described chronologically in the following (see~\cite{2020JCAP...02..026L} for a review):
	\begin{itemize}
		\item energy injection may distort the CMB spectrum and make it depart from pure blackbody;
		\item energetic EM particles may reionize the universe, leading to:
		\item damping of the CMB anisotropies;
		\item heating of the $T_{21}$ temperature (modified optical depth of Lyman-$\alpha$ signal).
	\end{itemize}
	All these processes rely on energy injection in the form of EM particles, that is photons, charged leptons and charged mesons (see the seminal work by Nasel'skii~\cite{1978Ap.....14...82N,1978SvAL....4..209N}). The other particles are either unstable on the CMB timescales or not produced by PBHs that survive until then. Hence, PBHs that can modify the CMB must have initial mass $M \gtrsim 10^{11}\,$g. The fact that CMB disturbance relies most on the electron injection makes these constraints competitive only for $M < 10^{18}\,$g.
	
	\subsection{CMB distortions}
	
	The CMB is characterized by a power spectrum very close to a pure blackbody of initial temperature $13.6\,$eV corresponding to the ionization energy of the hydrogen atom. After the redshift due to the expansion of the universe from $t\sim 380\,000\,$yr to $t_0 \sim 13.9\,$Gyr, the CMB is measured today as a microwave background of temperature $T\mrm{CMB}\sim 2.7\,$K~\cite{2020A&A...641A...6P}. This light is due to the recombination of electrons onto protons to form neutral hydrogen atoms. Exotic EM injection, \textit{e.g.}~by evaporating PBHs, contributes to or scatters on the CMB photons. Hence, the pure blackbody can be \textit{distorted} in several ways~\cite{2020JCAP...02..026L}:
	\begin{itemize}
		\item $g$ distortions: a shift of the blackbody temperature away from $13.6\,$eV;
		\item $\mu$ distortions: an effective chemical potential in the CMB spectrum;
		\item $y$ distortions: a departure from equilibrium of the Compton scattering process.
	\end{itemize}
	These are \textit{thermal} distortions, but the CMB can be distorted by \textit{non-thermal} mechanisms, such as PBH injection of energy. In this case a dedicated treatment is required. Ref.~\cite{2020JCAP...02..010A} hence distinguishes ``\textit{yim}'' distortions from ``\textit{ntr}'' ones. All of these are severely constrained by CMB observations, the most precise to date being COBE/FIRAS. As the CMB distortions would be caused \textit{at the time of} recombination, it concerns PBHs of mass $M\sim 10^{11}-10^{13}\,$g.
	
\begin{figure}[!t]
		\centering
		\includegraphics{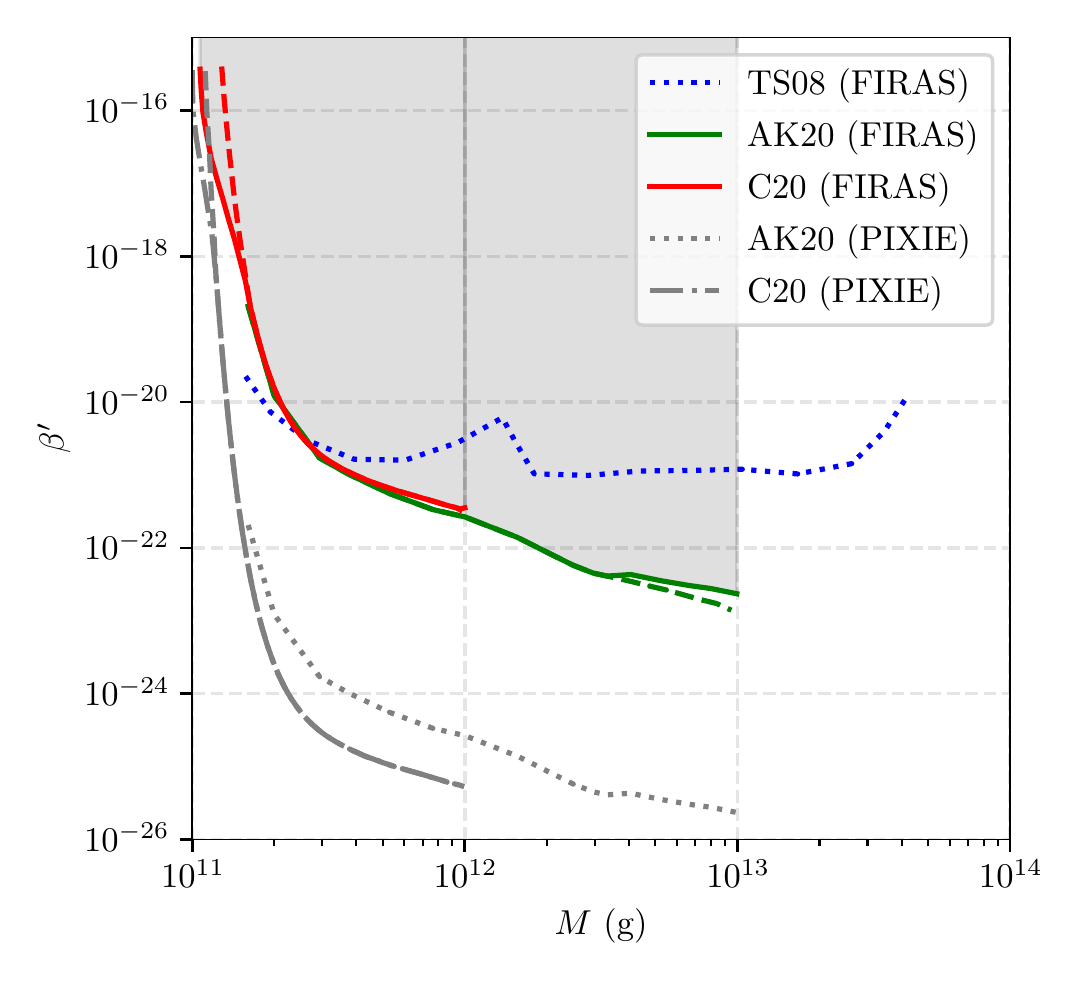}
		\caption{CMB distortion constraints on PBHs. The solid lines represent the more robust constraints by Acharya \& Khatri~\cite{2020JCAP...02..010A} (AK20, solid green is the \textit{ntr} limit, while dashed green is the \textit{yim} limit) and Chluba \etal~\cite{2020MNRAS.498..959C} (C20, solid red is the high-injection corrected version of dashed red). The older constraints from Tashiro \& Sugiyama~\cite{2008PhRvD..78b3004T} (TS08) is given in dashed blue for comparison, while the prospective limits from PIXIE are shown as discontinuous grey lines (the PIXIE design is not definitive, which explains the differences between Refs.~\cite{2020JCAP...02..010A,2020MNRAS.498..959C}, resp.~AK20 and C20).}
		\label{fig:CMB_distortions}
	\end{figure}

	Zel'dovich \etal~\cite{1976JETPL..24..571Z} were the first to estimate PBH constraints from CMB distortions. PBHs evaporating after the temperature of the universe cools down to $T\sim m\mrm{e}$, emit energetic radiation that cannot reach thermal equilibrium with the plasma, resulting in a distortion of the CMB. Nasel'skii \& Shevelev~\cite{1978Ap.....14..386N} then considered the Boltzmann equations describing the inverse Compton scattering of e$^\pm$ on the CMB and the induced distortion, but they were limited by numerical capabilities at that time.
	Tashiro \& Sugiyama~\cite{2008PhRvD..78b3004T} studied the $\mu$ and $y$ distortions of the CMB by PBHs in the $10^{11}-10^{13}\,$g range, using the WMAP data, considering only the primary photon spectrum.
	The constraints were revisited by Acharya \& Khatri~\cite{2020JCAP...02..010A}, based on the DM study~\cite{2019PhRvD..99l3510A} and the theory on EM cascades in~\cite{2019PhRvD..99d3520A} (see also~\cite{2021arXiv211206699A}). They used COBE/FIRAS data that result in constraints less stringent than the BBN ones but predict major improvement with future PIXIE measures. The PBH mass evolution during evaporation is taken into account and secondary particles are also considered thanks to the PPPC4DMID package.
	The CMB distortions have been corrected by Chluba \etal~\cite{2020MNRAS.498..959C} in the case of very large energy injection that depart from simple linear analysis. This results in mild modifications of the PBH constraints from~\cite{2020JCAP...02..010A}. The constraints from Refs.~\cite{2020JCAP...02..026L,2020JCAP...02..010A} and the corrections from Ref.~\cite{2020MNRAS.498..959C} are shown in Fig.~\ref{fig:CMB_distortions}.
	
	\subsection{Reionization}
	
	In a series of 3 papers, Gibilisco~\cite{1996IJMPA..11.5541G,1997IJMPA..12.2855G,1997IJMPA..12.4167G} computed the reionization of the universe due to EM energy injection by PBHs. The FIRAS/COBE data impose strong constraints on the fraction of the universe reionized after the CMB. In the first paper, they used the old Page spectra for the injected energy and obtained very strong limitations $\Omega\mrm{PBH} \lesssim 10^{-8}-10^{-12}$ in the $M\sim 10^{14}\,$g range. They considered PBH remnants in the second paper. In the third one, they finally used the MG\&W model to compute the total amount of energy injected into the CMB: both direct EM and QCD decays are taken into account. Interestingly, the constraints are presented with a dependency on the Hubble parameter. Indeed, the mass of PBHs evaporating at the reionization epoch depends on the age of the universe.
	Reionization by PBHs was reviewed by Belotsky \etal~\cite{2015JCAP...01..041B} (see also~\cite{2014MPLA...2940005B,2015IJMPD..2445005B}). They computed the reionization due to $10^{16}-10^{17}\,$g PBHs, saturating the PBH bounds from $\gamma$-rays. The distinctive signature would be that such a reionization should be homogeneous, whereas that from the first stars would be inhomogeneous. They proposed a way of alleviating the constraints on PBHs by using their idea of PBH clusters~\cite{2015PAN....78..387B}. Clustering would allow to ``disregard'' the Galactic $\gamma$-ray limit. The monochromatic and power-law mass functions considered result in a negligible reionization, while the two-peaked function of~\cite{2017IJMPD..2650102B} provides enhanced effect.
	Reionization by axions produced by PBHs is studied by Schiavone \etal~\cite{2021JCAP...08..063S}. They obtain the first constraints on axions by PBHs with the \texttt{RecFast} code by adapting the study of~\cite{2016JCAP...05..006E}.
	
	Reionization has two major observational consequences. The first one is to dampen the CMB anisotropies, the second one is to modify the Lyman-$\alpha$ optical depth, and thus the $21\,$cm signal. Both are used below to obtain constraints on PBHs.
	
	\subsection{CMB anisotropies}
	
	CMB anisotropy damping proceeds from EM energy injection at the \textit{late} CMB epoch, it then concerns PBHs with $M > 10^{13}\,$g. In fact, it is a direct corollary of the reionization mechanism as the free electrons interact with the CMB photons and dampen their power spectrum~\cite{2017JCAP...03..043P}. Any EM energy injection that alters the ionization fraction sufficiently late hence modifies the CMB anisotropy spectrum.
	
	Dorosheva \& Nasel'skii~\cite{1986Ap.....24..321D,1987SvA....31....1D}, and Nasel'skii \& Polnar\"ev~\cite{1987SvAL...13...67N} assessed how the CMB anisotropies could help constrain the DM candidates. They presented a complete computation of the $\Delta(T)/T$ anisotropy damping of the CMB, related to non-standard reionization by either decaying massive particles or PBHs, both envisaged as decaying ``hidden mass'' candidates. Carr \etal~\cite{2010PhRvD..81j4019C} were the first to apply this constraint to PBHs, for which they obtained a limit in the $M \sim 10^{13}-10^{14}\,$g mass range adapted from that of decaying particles~\cite{2007PhRvD..76f1301Z}, based on WMAP data.
	
\begin{figure}[!t]
		\centering
		\includegraphics{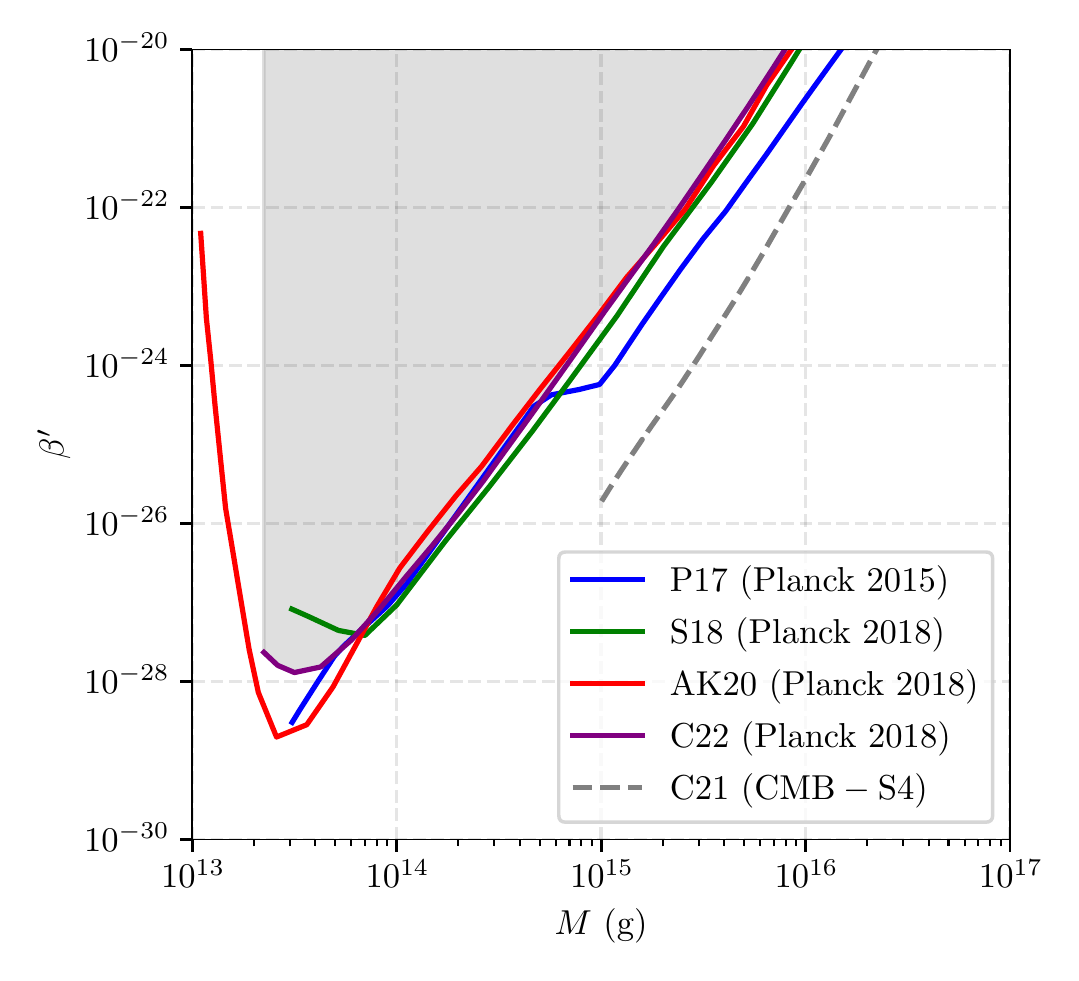}
		\caption{CMB anisotropy constraints on PBHs. The solid lines represent the results of Poulin \etal~\cite{2017JCAP...03..043P} (P17, blue), St\"ocker \etal~\cite{2018JCAP...03..018S} (S18, green), Acharya \& Khatri~\cite{2020JCAP...06..018A} (AK20, red) and Cang \etal~\cite{2022JCAP...03..012C} (C22, purple). The prospective CMB-S4 limit is from Ref.~\cite{2021JCAP...05..051C} (C21, grey dashed).}
		\label{fig:CMB_anisotropies}
	\end{figure}

	This first detailed numerical study was performed by Poulin \etal~\cite{2017JCAP...03..043P}. They used the time-dependent evaporation products of PBHs and the code \texttt{CLASS} to study the effect of exotic energy injection. The bounds are competitive with the EGXB ones in the $10^{14}-10^{17}\,$g mass range with the Planck 2015 data (see also~\cite{2019MNRAS.486.4569Y}). However, the HR spectrum was simplified to the extreme by considering the GO limit at $E>3T$ and no emission below. The constraints are alleviated if the $\Lambda$CDM parameters are left free during the constraining procedure~\cite{2017PhRvD..95h3006C,2019arXiv190706485P}. The CMB constraint in the $10^{15}-10^{17}\,$g mass range was recomputed by Poulter \etal~\cite{2019arXiv190706485P} using Planck 2018 data.\footnote{See also~\cite{2020JCAP...02..026L} which relied on the ``on-the-spot'' approximation, meaning immediate thermalization and resulting in stronger constraints; but used complete secondary estimation with \texttt{PYTHIA}.} Explicit extended mass functions (uniform and log-normal) are used for the first time but only primary particles were considered.\footnote{Interestingly, they compared the results obtained with these genuine mass functions to the conversion method of Ref.~\cite{2017PhRvD..96b3514C} and a good agreement for log-normal mass function is found, whereas the uniform mass function shows $\mathcal{O}(1)$ discrepancy.} St\"ocker \etal~\cite{2018JCAP...03..018S} used the same simple primary HR parametrization as~\cite{2017JCAP...03..043P}, however, they computed the secondary particles with \texttt{PYTHIA}, with particular attention given to the QCD phase transition at $E \sim \Lambda\mrm{QCD}$. The CMB constraints were revisited by Acharya \& Khatri~\cite{2020JCAP...06..018A}, based on the particle DM study~\cite{2019JCAP...12..046A}. The constraints of Refs.~\cite{2017JCAP...03..043P,2018JCAP...03..018S,2020JCAP...06..018A} are shown on Fig.~\ref{fig:CMB_anisotropies}.
	
	Prospective CMB anisotropy constraints on PBHs were computed by Cang \etal~\cite{2021JCAP...05..051C} using \textit{e.g.}~the CMB-S4 designs (see Fig.~\ref{fig:CMB_anisotropies}). Several kinds of extended mass functions were considered and full HR rates were extracted from literature. \texttt{HyRec} was used to produce mock data for these instruments and extended limits are obtained from the Carr method. Finally, the case of spinning PBHs was treated in the subsequent paper~\cite{2022JCAP...03..012C} thanks to \texttt{BlackHawk}, with extended mass functions as well (Carr method). They are the first to use numerical GFs, thus we have chosen to represent their constraint in Fig.~\ref{fig:CMB_anisotropies} as the robust limit. CMB-S4 is predicted to improve the constraints a little in the higher mass range.
	
	\subsection{21 cm signal}
	\label{sec:21cm}
	
	PBH late EM energy injection provokes heating of the intergalactic medium (IGM) during the Dark Ages. This in turn affects the optical depth of hydrogen as the equilibrium of the hyperfine states HI and HII depends on the IGM temperature. Hence, the redshifted $21\,$cm data contain the whole thermal history of the hydrogen gas after recombination, during the Dark Ages. This is a unique probe of an otherwise dark period. Other processes impact the $21\,$cm temperature, such as particle DM~\cite{2021MNRAS.508.3446H}, baryon cooling~\cite{2021PhRvD.103f3044H} and X-ray heating by stars~\cite{2022JCAP...03..030M}.
	
	At the time of Gibilisco's papers~\cite{1996IJMPA..11.5541G,1997IJMPA..12.2855G,1997IJMPA..12.4167G}, no Lyman-$\alpha$ data were available, only upper limits. The $21\,$cm signal as a constraint on PBHs was evaluated by Mack \& Wesley~\cite{2008arXiv0805.1531M} with first comparison to prospective instrument sensitivity. They noted that:
	\begin{quote}
		\textit{[\dots] future 21$\,$cm observations can provide better constraints on PBHs than are currently available} [due] \textit{to a coincidence between the Hawking temperature of PBHs that evaporate during the Dark Ages and a window of low optical thickness of the IGM to photon absorption.} \cite{2008arXiv0805.1531M}
	\end{quote}
	This explains why the $21\,$cm signal has received so much attention lately. Ref.~\cite{2008arXiv0805.1531M} modified \texttt{RecFast} to include the full PBH spectrum interaction with the IGM temperature and the reionization process, and obtained constraints on PBHs from prospective $21\,$cm surveys: LOFAR, SKA, EDGES, 21CMA, MWA.
	
	The first EDGES data were finally released in 2018~\cite{2018Natur.555...67B}, with a trough in the $21\,$cm temperature deeper than expected from standard reionization history. Clark \etal~\cite{2018PhRvD..98d3006C} were the first to derive the PBH constraints in the $10^{15}-10^{17}\,$g mass range. They took only primary photons and electrons into account and discussed prospective data from PRIZM, HERA, LEDA and SKA.
	Yang~\cite{2020PhRvD.102h3538Y} recomputed the $21\,$cm constraints on PBHs with the code \texttt{CAMB}, based on work done for particle DM decay~\cite{2021JCAP...10..033H}. The primary and secondary EM particles were taken into account and the lower mass range $10^{13}-10^{15}\,$g was explored.
	
	Spinning PBH constraints have been obtained by~\cite{2022JCAP...03..012C,2022MNRAS.510.4236N} using \texttt{BlackHawk}, with typically more stringent limits due to the enhanced HR rates.
	
	Saha \& Laha \cite{2021arXiv211210794S} noted that the SARAS 3 experiment seems to mitigate the EDGES data. Therefore, they only considered putative $21\,$cm measurements and conservatively produced prospective constraints, taking the PBH spin into account and using \texttt{BlackHawk} to compute the EM energy injection and \texttt{DarkHistory} to compute the $21\,$cm signal for the first time. Full EM cascades in the IGM plasma and PBH evolution were finally considered by Cai \etal~\cite{2022JCAP...03..012C}. Indeed, using the full redshift dependency of the $21\,$cm trough rather than just the $21\,$cm temperature allows to place more stringent constraints~\cite{2022JCAP...03..030M}.
	
\begin{figure}[!t]
		\centering
		\includegraphics{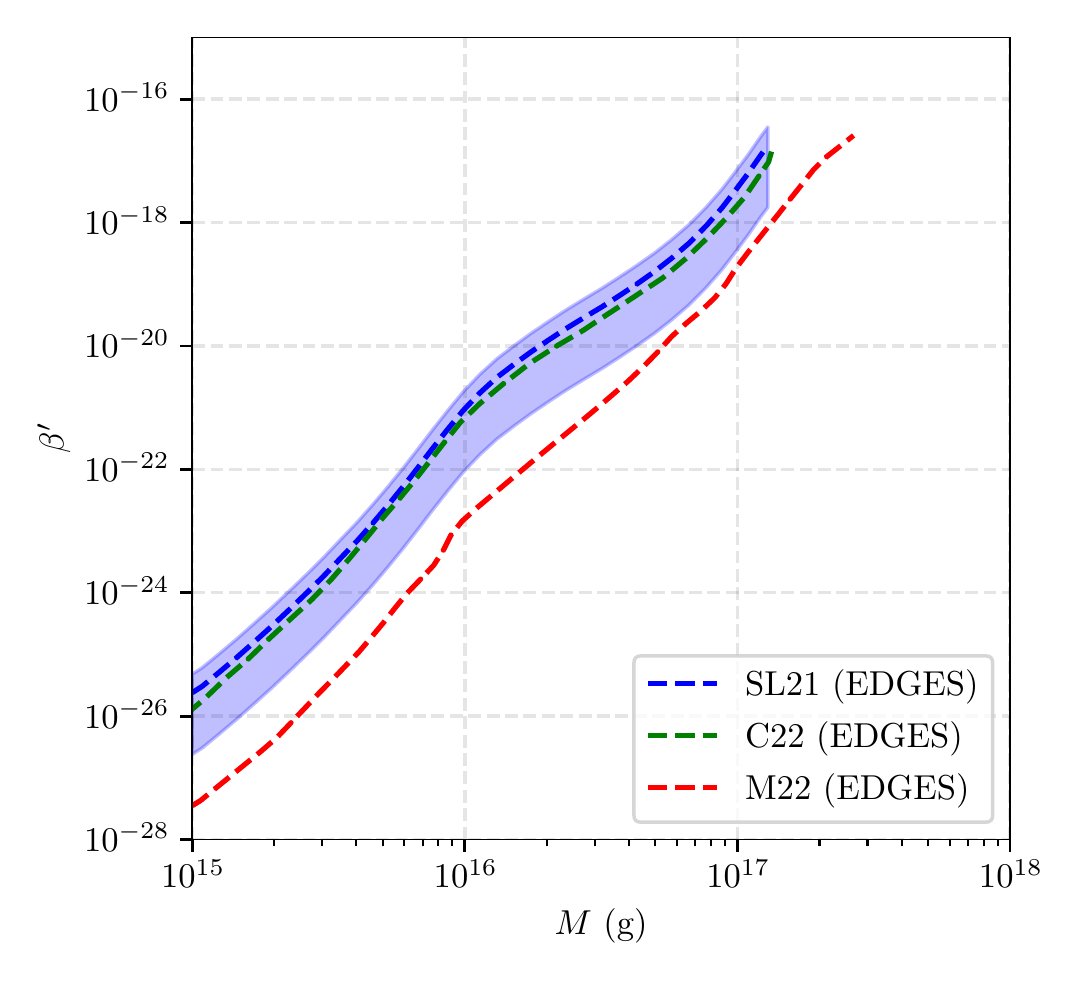}
		\caption{$21\,$cm constraints on PBHs. All constraints are shown as dashed lines as the EDGES results have not yet been confirmed by other instruments. The limit of Saha \& Laha~\cite{2021arXiv211210794S} is shown as a blue dashed line (SL21), with the EDGES error bars echoed as a blue shaded area. The limits of Cang \etal~\cite{2022JCAP...03..012C} (C22) and Mittal \etal~\cite{2022JCAP...03..030M} (M22) are represented as resp.~green and red dashed lines.}
		\label{fig:21cm}
	\end{figure}

	The $21\,$cm constraints of Refs.~\cite{2021arXiv211210794S,2022JCAP...03..012C,2022JCAP...03..030M} are shown in Fig.~\ref{fig:21cm}. It is difficult to assess which limit is the more robust as different models and external effects are considered in those papers. Furthermore, since the $21\,$cm data have not yet been confirmed by other surveys, we leave these limits as prospective but note that they would constitute the strongest to date in the $M\sim 10^{14}-10^{18}\,$g mass range. The model of Cang \etal~\cite{2022JCAP...03..012C} seems to embed all the relevant PBH ingredients.
	
	\section{Cosmological backgrounds}
	\label{sec:backgrounds}
	
	The cosmological backgrounds are conceptually the simplest limits on the PBH abundance from HR; hence they were the first ones to be computed. Assume that there exists a cosmological (or present) density of PBHs, then they evaporate by generating a continuous flux of (B)SM particles. If those particles are cosmologically stable (photons, neutrinos, gravitons, DM particles), they travel towards Earth in straight lines. If not, they decay into stable particles that contribute as secondary components to the stable backgrounds. The background is then obtained by stacking and redshifting the instantaneous emission with formula \eqref{eq:HR_integrated} with correct $t_1$ and $t_2$. The limiting times are obtained simply as follows. $t_1$ is the minimum time, it should start from the PBH formation, but interaction of the particles introduces an optical depth factor $e^{-\tau(t,E)}$ that effectively cuts off the integral at $t\sim 1\,$s for neutrinos and $t\sim t\mrm{CMB}\sim 380\,000\,$yr for photons. The limiting time for DM particles depends on their (self-)interactions while gravitons propagate freely in any era of the universe. $t_2$ is the maximum time, it is simply the minimum between the age of the universe $t_0$ and the lifetime of PBHs $\tau$. These backgrounds are described in Sections~\ref{sec:bckg_photon}, \ref{sec:bckg_neutrino}, \ref{sec:bckg_graviton} and \ref{sec:DM_production}.
	
	The quasi-instantaneous contribution from particles currently produced by PBHs in the Galaxy or in defined targets (M31, dSphs, \textit{etc.}) may be added to describe more accurately the flux along a particular line of sight (see Section~\ref{sec:present_backgrounds}).
	
	\subsection{Photons}
	\label{sec:bckg_photon}
	
	The photon constraint was the first ever constraint on PBHs linked to their HR. Hereafter, we denote as ``EGXB'' the extragalactic $\gamma$/X-ray background. Chapline~\cite{1975Natur.253..251C} depicted the PBH evaporation contribution to the photon background as:
	\begin{quote}
		\textit{[\dots] a continuation of the 3$\,$K blackbody spectrum and} [rising] \textit{slowly to peak at an energy determined by the smallest black hole mass now existing.} \cite{1975Natur.253..251C}
	\end{quote}
	The strongest PHL limit would come from PBHs evaporating today with $M\sim M_*$ initially, as they dominate the integral \eqref{eq:HR_integrated}. This corresponds to a BH temperature $T\sim 100\,$MeV, an energy range that was poorly explored.
	
	This background is the simplest as the photon detection techniques have been mastered for a long time. Furthermore, photons are reasonably products of any interacting (particle) DM candidate, unless they interact only gravitationally. The characteristics of the diffuse photon background could then discriminate between the different DM models~\cite{1992VA.....35..439O,2004PhR...402..267O}, and the refinement of the ``Grand Unified photon spectrum''~\cite{1990ComAp..14..323R} over the years~\cite{2018ApSpe..72..663H} has progressively tightened the constraints.
	
	Following the first numerical calculation of the HR rate by Page~\cite{1976PhRvD..13..198P,1976PhRvD..14.3260P,1977PhRvD..16.2402P}, Page \& Hawking~\cite{1976ApJ...206....1P} estimated precisely the constraint that the $100\,$MeV measurement of the diffuse background by Fichtel \etal~\cite{1975ApJ...198..163F} imposes on the cosmologically stacked abundance of PBHs. They obtained that the stacked flux follows a power-law $\d F/\d E \sim E^{-3}$, which when compared to $\gamma$-ray measurements raised a constraint $\Omega\mrm{PBH} \lesssim 10^{-8}$ around $M_*$. This computation could be applied to other fields, with somewhat different optical depths, but:
	\begin{quote}
		\textit{It would be very difficult to detect the gravitons or neutrinos because they have such small interaction cross-section.} \cite{1976ApJ...206....1P}
	\end{quote}
	(see however below). Concerning CRs, one further difficulty is that:
	\begin{quote}
		\textit{The charged particles would be deflected by magnetic fields and so would not propagate freely to Earth.} \cite{1976ApJ...206....1P}
	\end{quote}
	thus requiring sophisticated propagation models. This first constraint was so tight that it immediately implied that PBHs could not be very abundant in the universe, and the resulting limit on their (original) power-law spectrum resulted in strong constraints on the amplitude of primordial fluctuations. It was denoted as the ``Page Hawking limit'' (PHL). No observations of the diffuse photon backgrounds were performed until the 1990's. Hence, the PHL was not re-assessed until then.
	
	Halzen \etal~\cite{1991Natur.353..807H} returned to the diffuse photon bound in their 1991 review. This review is very interesting in that it is the first to use the newly published MG\&W model~\cite{1990PhRvD..41.3052M,1991PhRvD..44..376M}. The low-energy tail of the photon emission spectrum is greatly enhanced by taking into account the decay of QCD jets, but this also corresponds to regions of more abundant background. Hence, the DM fraction into PBHs is not constrained above $M\sim 10^{17}\,$g. A more precise constraint $\Omega\mrm{PBH} = 7.6\times10^{-9}$ was obtained using the same old data from Ref.~\cite{1975ApJ...198..163F}. MacGibbon \& Carr~\cite{1991ApJ...371..447M} obtained the same order of magnitude with the lower energy~\cite{1977ApJ...217..306S} data. The interactions between the PBH photon flux and the IGM are discussed, with the conclusion that only photons emitted by PBHs with $M\gtrsim 10^{13}\,$g after the CMB can reach detectors.
	
	Great improvement came from the COMPTEL~\cite{1996A&AS..120C.619K} (1996) and EGRET~\cite{1998ApJ...494..523S} (1998) measurements which observed the MeV$-$GeV energy range, first used by Kim \etal~\cite{1999PhRvD..59f3004K} in the PBH context. Following the discovery of the ``critical collapse'' mechanism for PBH formation, Kribs \etal~\cite{1999PhRvD..60j3510K} refined the PHL for this extended mass function numerically, which can be compared to the analytical extrapolation of Yokoyama~\cite{1998PhRvD..58j7502Y}. The first estimation of the limits for ``bumpy'' mass functions was given in \cite{2009PhRvD..79j3511B}. Early accretion impact was considered in~\cite{2006JCAP...06..003S}.
	
	The cosmological PBH flux slope ($E^{-3}$) is not the same as the diffuse background one, thus PBHs cannot be the sole source of this background~\cite{1998PhR...307..141C}.\footnote{See however Refs.~\cite{1999PhRvD..59f3009C,2006astro.ph.12659B} in which models challenging the MG\&W one are used to comply more to the data.} Barrau \etal~\cite{2003astro.ph..4528B} were the first, to our knowledge, to include a background component to the PHL. They argue that the contribution of blazars and active galactic nuclei (AGNs) is ``guaranteed'', translating in a reduced parameter space for PBHs $\Omega\mrm{PBH} \lesssim 3.3\times 10^{-9}$. 
	
	In their 2010 review, Carr \etal~\cite{2010PhRvD..81j4019C} used for the first time the Fermi-LAT data~\cite{2010PhRvL.104j1101A} above the GeV energy scale to obtain the limit on PBHs taking the secondary spectra into account (see also~\cite{2021RPPh...84k6902C} with more recent Fermi-LAT data~\cite{2015ApJ...799...86A}). The secondary spectra calculation relies on \texttt{PYTHIA}, but the transition from primary only to primary and secondary photon spectrum is abrupt because of the computation method. The strongest limit $\Omega\mrm{PBH} \sim 10^{-10}$ is obtained at $M_*$.
	
	Arbey \etal~were the first to compute the PHL limit with numerical GFs thanks to \texttt{BlackHawk}~\cite{2020PhRvD.101b3010A}. This was in fact the first application of the code to PBH constraints, with a complete study of the impact of extended mass distributions and non-zero PBH spin (see also~\cite{2021PhRvD.104b3516R,2021arXiv211003333G}). The latest Fermi-LAT data were used~\cite{2015ApJ...799...86A}. The enhanced HR rates of KBHs results in more stringent constraints, and extended mass functions broadens the excluded region towards higher masses (up to $10^{18}\,$g for broad functions of near-extremal PBHs). This opened a new period of intense work on the PHL. A low energy X-ray background was assumed in Ref.~\cite{2020PhLB..80835624B} to obtain more stringent constraints. Ref.~\cite{2022PhRvD.105f3008C} in particular considered the latest AGN and star forming galaxy models as well as PBH electron-positron contribution to the photon flux. Low-energy e$^\pm$ annihilation was also studied in Refs.~\cite{2021PhRvD.103j3025I} and multiple extended PBH distributions were constrained.
	
	Despite repetition of the EGRET survey~\cite{2004ApJ...613..956S}, the lower MeV range is, to this date, still unexplored with a high resolution instrument (better than COMPTEL). Several propositions have been made, including AdEPT, AMEGO, ASTROGRAM, GECCO, GRAMS, MAST, PANGU and XGIS-THESEUS. On the one hand, the possibility to reach very low energies with high accuracy and effective detection area makes the prospective limits from XGIS-THESEUS among the most stringent ones with the $21\,$cm signal~\cite{2021arXiv211003333G}. On the other hand, the necessity to account precisely for the MeV PBH physics requires dedicated sub-GeV particle physics codes, such as \texttt{Hazma}. The latter was proposed as an amelioration to the original low-energy extrapolation of the \texttt{PYTHIA} results inside \texttt{BlackHawk v1}~\cite{2021PhRvL.126q1101C}, and was implemented recently in \texttt{BlackHawk v2}~\cite{2021EPJC...81..910A}.
	
	However, low-energy hadronization suffers from large uncertainties, inner Bremsstrahlung effects may dominate at keV energies~\cite{2008PhRvD..78f4044P}, the design of prospective instruments is not definitive, statistical treatments and background models vary from one study to the other. All these aspects make it very difficult to compare the PBH constraints from one study to the other. This was clearly demonstrated by the author in Fig.~7 of~\cite{2022EPJC...82..384A}: letting aside the instrumental design, the previous uncertainties result in $10^{5}$ relative error bars on the PHL. Ref.~\cite{2022EPJC...82..384A} advocates for clear assessment of the assumptions in order to compare limits with unified method, using \textit{e.g.}~the \texttt{BlackHawk} tool \texttt{Isatis}. Furthermore, even with high-resolution surveys down to eV energies, the overwhelming background and the dilution of the PBH density at $M \gtrsim 10^{20}\,$g cause the PHL to stop being a competitive bound on the PBH abundance (see also~\cite{2022MNRAS.510.4992M}).
	
\begin{figure}[!t]
		\centering
		\includegraphics{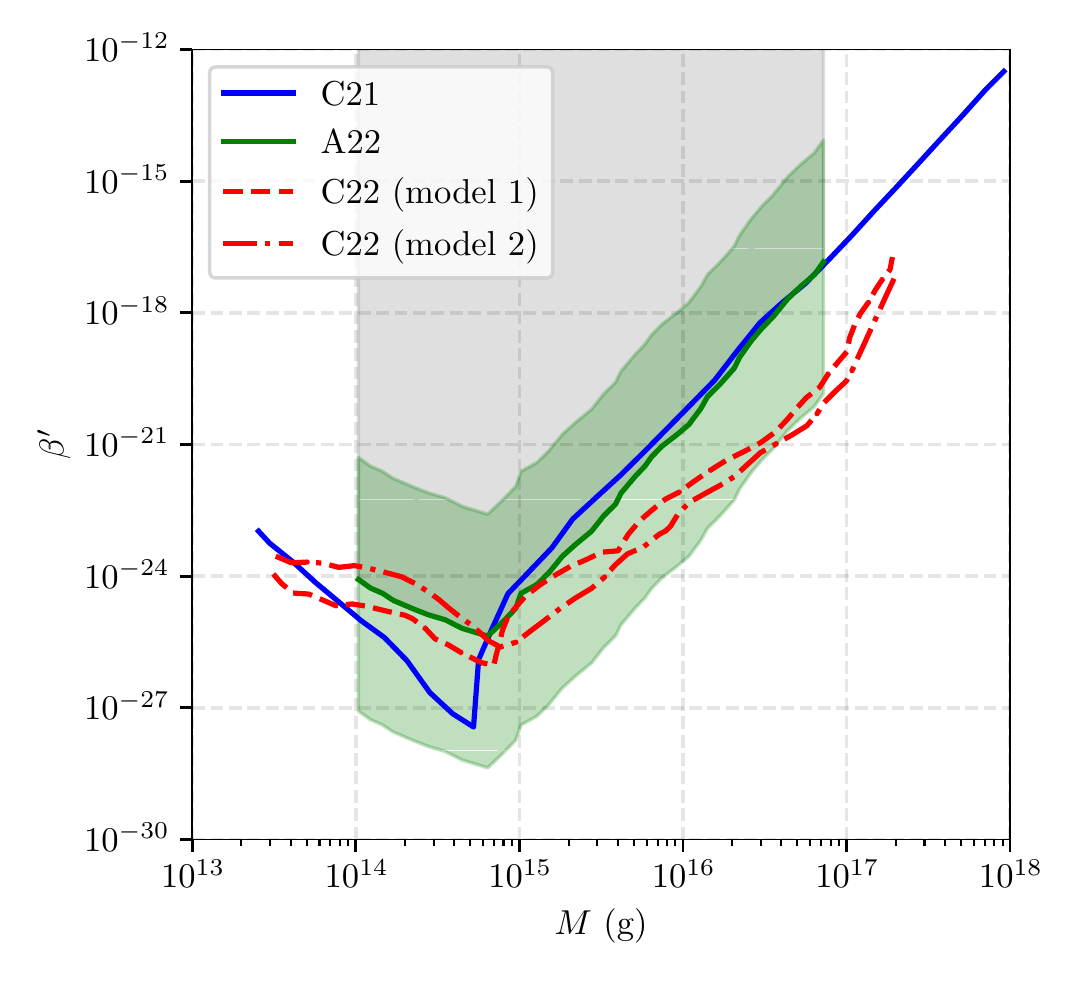}
		\caption{EGXB constraints on PBHs. The solid lines are the limits set by Refs.~\cite{2021RPPh...84k6902C} (C21, blue) and \cite{2022EPJC...82..384A} (A22, green). The (dot-)dashed lines are the ones obtained in Ref.~\cite{2022PhRvD.105f3008C} (C22) with background models of~\cite{2021Natur.597..341R,2019ApJ...880...40I} (model 1) and~\cite{2021Natur.597..341R,2021NatCo..12.5615K} (model 2). The grey shaded area is robustly excluded by Ref.~\cite{2022EPJC...82..384A} while the green hatched area shows the $\sim 10^5$ uncertainty region of the PHL.}
		\label{fig:bckg_photons}
	\end{figure}

	Finally, even if PBHs contribute to some extent to the diffuse background, there is a degeneracy with unresolved sources and other unknown background components. Multimessenger correlations with \textit{e.g.}~stochastic GWs from PBH formation could remove the degeneracy and provide positive PBH detection~\cite{2021arXiv211214588M,2022arXiv220204653A}. We show the PHLs from Refs.~\cite{2021RPPh...84k6902C,2022PhRvD.105f3008C,2022EPJC...82..384A} in Fig.~\ref{fig:bckg_photons}, taking Ref.~\cite{2022EPJC...82..384A} as the most conservative one.
	
	\subsection{Neutrinos}
	\label{sec:bckg_neutrino}
	
	As stated above, neutrinos were at first too elusive to be detected directly inside detectors~\cite{1968PhRvL..20.1205D}. But the neutrino background is particularly interesting as it would give a privileged access to the universe conditions \textit{before} the CMB and BBN epochs, as neutrinos decoupled from the plasma as early as $t\sim 1\,$s. For a long time, only upper limits on the PBH neutrino background existed~\cite{1976ApJ...206....8C,1978Ap.....14..185V}.
	
	The expected isotropic neutrino background was computed in the MG\&W model by Refs.~\cite{1991ApJ...371..447M,1995PhRvD..52.3239H}, with mere prospects for future detection. Halzen \etal~\cite{1995PhRvD..52.3239H} pointed at the differences in the flux from the different neutrino flavors, originating in the secondary PBH spectrum calculations: $\tau$ neutrinos for example are expected to have a lower flux because their production involves heavier primary particles. Halzen \etal~compared the PBH neutrino flux to the known astrophysical backgrounds: while the Solar background dominates by far at lower energies, the PBH intensity could be above the atmospheric and supernova contributions at $E \gtrsim 10\,$MeV~\cite{1988PhRvD..38...85G}. Furthermore, the Solar background is composed of electronic (anti)neutrinos whereas PBHs produce them in all flavors.
	
	Bugaev \& Konishchev~\cite{2002PhRvD..65l3005B,2002PhRvD..66h4004B} were the first to set a prospective constraint on the PBH abundance with the expected Super-Kamiokande sensitivity. They considered a scale-invariant PBH mass function with the critical collapse mechanism and secondary particles from analytical fragmentation functions. They also took into account precise optical depth of neutrinos.\footnote{Just like for photons, a photosphere model of PBH evaporation gives very different constraints~\cite{2006astro.ph.12659B}.} The first effective constraints were finally set with LSD and KamLAND data~\cite{2004astro.ph.12640B}. A more robust neutrino constraint was computed by Carr \etal~\cite{2010PhRvD..81j4019C}, where they estimated the secondary neutrino fluxes from PBHs and used the first Super-Kamiokande upper limits on the electronic antineutrino~\cite{2003PhRvL..90f1101M}.
	
	Lunardini \& Perez-Gonzalez~\cite{2020JCAP...08..014L} were the first to compute the neutrino constraints from PBHs in the light of their Dirac/Majorana nature, ``under the hypothesis that black holes emit neutrino mass eigenstates'' rather than flavor ones~\cite{2005CQGra..22.4247B}. Indeed, the oscillation mechanism~\cite{1958JETP....6..429P,1968JETP...26..984P} proves that neutrinos are massive: BHs ``might be the only emitters of neutrino mass eigenstates''. They predicted that only the flux of RHNs from early evaporated PBHs, once redshifted, could lie above the Solar background at very low energy, in the range of the PTOLEMY detector.
	
	Dasgupta \etal~\cite{2020PhRvL.125j1101D} recomputed the constraints from $\overline{\nu}\mrm{e}$ from Super-Kamiokande using \texttt{BlackHawk}, hence numerical GFs for the first time (with only tiny difference relative to~\cite{2010PhRvD..81j4019C}). They represented the HR of neutrinos as mass eigenstates, so that the  they don't oscillate during propagation. However, the secondary neutrinos oscillate as they are generated as flavor eigenstates by EW interactions, but taking the oscillation matrix into account results only in $\sim 2\%$ modification of the constraints~\cite{2021PhRvD.103d3010W}. They also considered log-normal PBH mass functions.
	
\begin{figure}[!t]
		\centering
		\includegraphics{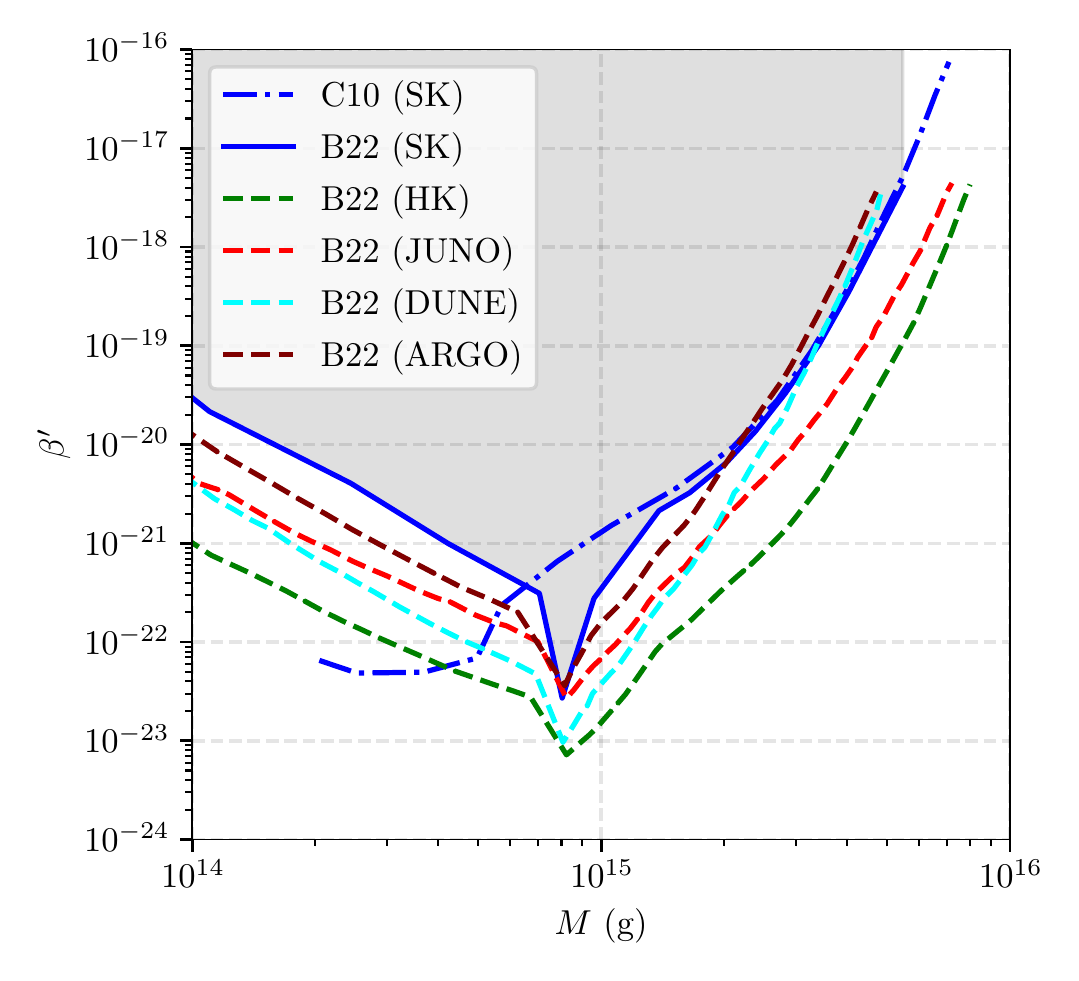}
		\caption{Neutrino constraints on PBHs. The blue lines correspond to the Super-Kamiokande limit from Ref.~\cite{2010PhRvD..81j4019C} (C10, dot-dashed) and~\cite{2022arXiv220314979B} (B22, solid). The dashed lines are derived from the prospective instruments Hyper-Kamiokande (green), JUNO (red), DUNE (cyan) and ARGO (purple)~\cite{2022arXiv220314979B}. The grey shaded area is robustly excluded by Super-Kamiokande~\cite{2022arXiv220314979B}.}
		\label{fig:bckg_neutrinos}
	\end{figure}

	Until now, no direct detection of a neutrino background has been reported at $E > 10\,$MeV, with two consequences: first, the neutrino background from astrophysical sources is still subject to debate; second, the PBH constraints are only upper bounds. Wang \etal~\cite{2021PhRvD.103d3010W} computed the background neutrino constraints at the prospective JUNO observatory, Calabrese \etal~\cite{Calabrese} considered the XENONnT and DARWIN designs and de Romeri \etal~\cite{2021JCAP...10..051D} computed the prospective DUNE ($\nu\mrm{e}$) and THEIA ($\overline{\nu}\mrm{e}$) constraints. All these prospective constraints have been recomputed recently by Bernal \etal~\cite{2022arXiv220314979B} in a unified framework with the release of the public numerical tool \texttt{nuHawkHunter}. They are also the first to estimate the constraints down to neutrino decoupling epoch from $M\sim 10^{12}\,$g PBHs. In Fig.~\ref{fig:bckg_neutrinos}, we show the old constraint from SK~\cite{2010PhRvD..81j4019C}, with the more robust to date from Ref.~\cite{2022arXiv220314979B} as well as some prospective limits (HK, JUNO, DUNE, ARGO). We see that the most stringent limits will be set by Super-Kamiokande.
	
	(RH)Neutrino emission has other consequences. If emitted before BBN/CMB, they could participate in the effective number of neutrino species $\Delta N\mrm{eff}$~\cite{2020JCAP...08..014L}. Second, they act as an irreducible background in particle DM direct searches, denoted as the ``neutrino floor'', because they interact with the nuclei and electrons composing DM detectors~\cite{Calabrese}. Third, they can scatter on and boost DM particles~\cite{2021arXiv210805608C}; this would modify the constraints in direct detection experiments because boosted DM would interact more strongly with the detectors (the resulting constraints on $f\mrm{PBH}$ completely depend on the interaction cross-section between neutrinos and particle DM).
	
	\subsection{Gravitons}
	\label{sec:bckg_graviton}
	
	If there exists a gauge boson mediating the gravitational interaction, namely the massless spin 2 graviton, then it should be emitted together with the rest of the SM particles by PBH evaporation. In fact, the first PBH HR papers all considered gravitons besides photons and neutrinos, as the massless set of particles emitted by heavy BHs. This particle has in fact never been formally identified and a coherent quantum theory of gravitation is yet to be found. Thus, most of the later studies have deprecated the graviton emission by PBH HR.
	
	Vainer \& Nasel'skii~\cite{1978SvA....22..138V} seem to have been the first to predict a background of gravitons generated by cosmological PBH evaporation. This background would have very short wavelength, so that:
	\begin{quote}
		\textit{[\dots] the discovery of a additional peak in the spectrum of the primordial gravitational radiation} [at $\sim 10\,$\AA] \textit{would serve as an important argument favoring the existence of low-mass PBHs.} \cite{1978SvA....22..138V}
	\end{quote}
	This particular signature, unmodified since the dawn of the universe but redshift effects (the gravitons have no optical depth), would be the only reasonable observable on PBHs evaporated before BBN ($M \lesssim 10^{9}\,$g). As a corollary, the graviton background could be the only signature of PBH scenarios of baryogenesis (see \textit{e.g.}~\cite{2014PhRvD..89j3501F}).
	
	Alas, those gravitons would have had very high frequency $f\propto T(M\sim 10^{4}\,\text{g})$ at emission and the subsequent redshift leaves them with still very high frequency $f\sim 10^{14}-10^{16}\,$Hz today~\cite{1978SvA....22..138V,2004CQGra..21.3347B,2011PhRvD..84b4028D}. Detection of these high-frequency GWs is a technical challenge, and some proposals have been made in that direction~\cite{1996PhRvD..54.6040G}. Most of them rely on the Gertsenshtein effect that converts high-frequency gravitons into photons in the presence of a magnetic field~\cite{2013PhRvD..87j4007D} (GRAPH mixing). The magnetic field could either be of astrophysical origin or monitored in a laboratory. The most promising PBHs are those with $M\sim 10^9\,$g, just at the edge of the BBN constraint, because their radiation is less redshifted than the smaller ones, that evaporate earlier. The spectrum of GWs shows features that depend on the expansion history of the universe.
	
	The graviton background from PBHs was computed by several authors, including Refs.~\cite{2004CQGra..21.3347B,2009PhRvL.103k1303A,2011PhRvD..84b4028D}, with numerical GFs used for the first time in~\cite{2016JCAP...10..034D}. In fact, PBHs generate several GW backgrounds, \textit{e.g.}~from scalar-induced effects, on top of HR gravitons. Inomata \etal~\cite{2020PhRvD.101l3533I} computed all these contributions and point to errors in the previous studies in the redshift calculation. The first experimental upper limits on high-frequency GWs were derived by Ejlli \etal~\cite{2019EPJC...79.1032E} in 2019 using data from photon cavities. While the bounds are far from the expected backgrounds, this gives a proof of principle.
	
	Gravitons injected in the very early universe have another consequence: as relativistic particles, they participate in the effective number of neutrinos $\Delta N\mrm{eff}$ as dark radiation (DR). Hence, their abundance is constrained by BBN and CMB observations and this in turn would constrain the abundance of PBHs in the $M \lesssim 10^9\,$g mass range. Hooper \etal~\cite{2020arXiv200400618H} gave the first calculation of the graviton impact on $\Delta N\mrm{eff}$ and concluded that they would be constrained only by the future CMB-S4 experiments (see also~\cite{2020EPJP..135..552M,2021GrCo...27..315M}). Arbey \etal~\cite{2021PhRvD.103l3549A} showed that precision calculation of the graviton density at the time of matter-radiation equality is necessary as the constraints depend strongly on the assumptions regarding the reheating temperature associated to PBH domination as well as the model for the effective number of relativistic dofs $g_*(T)$ as a function of temperature.
	
	\subsection{Present backgrounds}
	\label{sec:present_backgrounds}
	
	PBHs radiate particles continuously, meaning that on top of the cosmologically integrated background, there is also a contemporary background coming from zero-redshift sources such as the Milky Way, nearby galaxies such as M31 or DM-dominated objects like dSphs (Reticulum II, Leo T). The emission from distant sources is easy to assess as they are observed as a whole, thus the only free parameter is their DM density which enters as a normalization factor in the PBH flux. The emission from the Milky Way is trickier as one does not have a clear account of the DM density and spatial distribution in the Galaxy. This is a pity as the Galactic center (GC) contribution to the global background is two orders of magnitude larger than the ``darkest'' known objects~\cite{2021PhRvL.126q1101C}, and it forms an irreducible background to the isotropic surveys. Many DM halo profiles have been proposed in the past decades, with the most popular being the isothermal and Einasto profiles~\cite{Einasto}, and obviously the Navarro--Frenck--White profile~\cite{1997ApJ...490..493N} (NFW). Some generalized version of this profile is usually used in particle DM/PBH studies to obtain the column density of DM along one particular line of sight (los). The flux of some stable particle $i$ from PBHs in the Galaxy per unit solid angle is given by
	\begin{equation}
		\dfrac{\d \Phi_i^{\rm gal}}{\d E} = \dfrac{1}{A\mrm{gal}}\dfrac{J\mrm{gal}}{4\pi} \dfrac{\d^2 N_i}{\d t\d E}\,,\label{eq:flux_gal}
	\end{equation}
	where
	\begin{equation}
		J\mrm{gal} \equiv \dfrac{1}{\Delta\Omega} \int_{\Delta\Omega} \d\Omega \int_{\rm los} \rho\mrm{gal}(r(l,\Omega))\,\d l\,,
	\end{equation}
	where $\Delta\Omega$ is the field of view and $A\mrm{gal}$ is the normalization of the \textit{local} PBH density. We see that uncertainties in the halo profile propagate to the particle flux, with nearly 2 orders of magnitude spread in the constraints for the $68\%$ CL parameters of the NFW profile obtained with the \texttt{Isatis} tool for \texttt{BlackHawk}~\cite{2022EPJC...82..384A}.
	
	To our knowledge, the first evaluation of the \textit{Galactic} PBH contribution to the $\gamma$-ray background was given by Wright~\cite{1996ApJ...459..487W}, claiming that the Galactic EGRET survey~\cite{1997ApJ...481..205H} highlighted a PBH halo. Wright noted further that the intensity of the EGXB and Galactic contributions were coincidentally of the same order of magnitude in the MeV $-$ GeV range of EGRET, hence the Galactic contribution cannot be neglected. It would appear as an anisotropic feature in the diffuse background~\cite{1998ApJ...501L...1C}. The Galactic contribution is usually taken into account in the latest PBH EGXB studies~\cite{2021PhRvD.104b3516R,2021PhRvD.103j3025I,2021arXiv211003333G,2022PhRvD.105f3008C,2022arXiv220207483B}, and some papers even focused on this particular background~\cite{2008PhLB..670..174B,2009A&A...502...37L,2016PhRvD..94d4029C,2021PhRvL.126q1101C,2021arXiv210110370C}, \textit{e.g.}~with the high-resolution INTEGRAL/SPI data~\cite{2022A&A...660A.130S} that gives access to the spatial distribution of $\gamma$-ray sources in the GC~\cite{2020PhRvD.101l3514L,2021PhRvD.103j3025I,2022arXiv220207483B}. The Galactic limit is usually the strongest one in the high-energy range as the high-energy contribution from cosmological PBHs is diluted by the redshift.
	
	The isotropic neutrino background also embeds a Galactic contribution. As the observation of neutrinos is an all-sky average, only 1\% difference is found when tweaking the galactic profile~\cite{2021PhRvD.103d3010W}. Furthermore, contrarily to photons, the isotropic component dominates because only low energy neutrinos $E \sim 10-50\,$MeV are surveyed~\cite{2022arXiv220314979B}.
	
	Reticulum II is for now the only DM-dominated source that benefited from a dedicated $\gamma$-ray survey in the context of particle DM/PBH studies~\cite{2022MNRAS.511..914S}, while globular clusters could have captured numerous PBHs at the time of their formation, resulting in bright sources inside the Galaxy~\cite{1999A&A...343....1D}.
	
	\section{Cosmic rays}
	\label{sec:CRs}
	
	Cosmic rays (CRs) are energetic particles that interact either in the upper atmosphere, leading to cascades of secondary particles, or that are detected by satellites. Within this review, CRs will be used as a denomination for charged particles such as electrons/positrons, or protons/antiprotons and heavier nuclei. The huge difficulty regarding CRs, as stated above, is that they interact with the magnetic field of the Galaxy and therefore do not propagate in straight lines. A second difficulty comes from the Solar winds that hinder low-energy CRs from reaching the Earth, which acts as an effective energy cut-off at $E \lesssim $ GeV. Both Galactic propagation~\cite{1975SvPhU..18..931G} and Solar winds suffer from large uncertainties: the Galactic magnetic field is poorly known and re-acceleration mechanisms could exist~\cite{2017A&A...605A..17B,2021PhRvD.104h3005G}, while the Solar winds vary periodically following two cycles of 11 and $22\,$yr. Combination of the two effects is described by respectively a diffusion model and a ``force-field'' approximation.\footnote{Online material is available on the \href{http://www.marcocirelli.net/PPPC4DMID.html}{PPPC4DMID website} for propagation calculation.} In the context of antiproton detection, Wells \etal~\cite{1999ApJ...518..570W} claimed that should a probe escape the heliopause, it would be ridden of the Solar modulation and have a direct access to the local ISM flux of charged particles---this does not alleviate the Galactic propagation uncertainties.
	
	The complicated propagation of charged particles in the Galaxy results in a segregation of the origin of the particles. Only CRs emitted \textit{inside} the Galaxy in recent epochs could contribute to the CR background in a sizeable way.\footnote{Carr \etal~\cite{2010PhRvD..81j4019C} propose a different constraint from antiprotons that were produced \textit{before} galaxy formation by evaporated $M\sim10^{13}\,$g PBHs.} The CR limits further apply to \textit{local} PBHs as the continuous energy losses prevent charged particles from travelling efficiently from distant regions of the Galaxy. The factors that can affect the CR spectrum are annihilation on background particles, inverse Compton scattering on CMB photons, Bremsstrahlung effects, e$^\pm$ pair production off nuclei, synchrotron radiation due to ambient magnetic fields and ionization~\cite{1991ApJ...371..447M}.
	
	CRs provide a mean of \textit{direct} detection of local PBHs, and the fact that the CR limits were of the same order of magnitude as the $\gamma$-ray bounds in the 1980's was interpreted as a correlative evidence for some PBH population in the Galactic halo~\cite{1981Natur.293..120K}. This was merely coincidental, and the constraints have evolved a lot since then.
	
	\subsection{Electrons \& positrons}
	
	Electrons and positrons are the simplest CRs as they are fundamental particles. They could be released by atoms ionized in the ISM, or they could originate in the evaporation of PBHs. The fact that CR measurements show that the ratio $n_{{\rm e}^+}/n_{{\rm e}^-} \ll 1$, while PBHs should produce both in equal quantity, advocates for use of the e$^+$ flux as a limiting constraint~\cite{1979SvA....23..402N}. Electrons and positrons can be detected either directly by satellite experiments, or they can produce radiation (\textit{e.g.}~synchrotron) that allows for indirect detection in radio/X-rays. A particularly interesting feature is the $511\,$keV line corresponding to e$^+$-e$^-$ annihilation. Only PBHs with mass $M \lesssim 10^{18}\,$g can produce sizeable amounts of e$^\pm$, thus limiting the scope of such constraints.
	
	\subsubsection{Direct detection}
	
	The first direct detection constraint on electrons and positrons is due to Carr~\cite{1976ApJ...206....8C} using the $100\,$MeV background of charged particles~\cite{1969ApJ...158..771F,Cummings:1973qya}, with the conclusion that this limit was not competitive with the strong $\gamma$-ray one. The positron flux was recomputed within the MG\&W model in Ref.~\cite{1991ApJ...371..447M}, as the PBH scenario was backed by evidence that $n_{{\rm e}^+}/n_{{\rm e}^-} \longrightarrow 0.5$ at low energy. No more direct detection constraints were set until Boudaud \& Cirelli~\cite{2019PhRvL.122d1104B} used the fact that Voyager 1 had crossed the heliopause and measured the ISM electron/positron flux~\cite{2013Sci...341..150S} to place new cleaner constraints on the local density of PBHs, without the complicated Solar modulation, based on the DM study~\cite{2017PhRvL.119b1103B}. The GO limit is used and only primary electrons are considered, with uncertainties coming from the Galactic halo profile and the propagation model (with or without re-acceleration). The constraints were also computed for the first time with a log-normal mass function using the Carr method.
	
	\subsubsection{Indirect detection}
	
	Indirect detection of e$^\pm$ from PBHs can proceed from different mechanisms. Charged particles loose energy when propagating in a magnetic field due to synchrotron radiation, which is typically in the radio wavelengths. They can also scatter off the CMB photons, which results in $\gamma$-rays in the visible to near ultraviolet wavelengths~\cite{2021arXiv210914955M}. Finally, the EM interactions between e$^\pm$ and the ISM may heat up the gas clouds which would then radiate in unexpected wavelengths~\cite{2021MNRAS.504.5475K}. Synchrotron radiation in the Galaxy due to PBH evaporation was first predicted by Nasel'skii \& Pelikhov~\cite{1979SvA....23..402N}, but was completely left aside until recently. Chan \& Lee~\cite{2020MNRAS.497.1212C} used archival radio data of the GC to constrain the rate of injection of charged particles by PBHs, relying on \texttt{BlackHawk} for the secondary electron spectrum; they considered log-normal and power-law mass distributions.
	Dutta \etal~\cite{2021JCAP...03..011D} studied the indirect detection of energetic electrons and positron emission by PBHs through their inverse Compton scattering on the CMB photons, which would produce a radio signal detectable at SKA observation of nearby dSphs (Segue I, Ursa Major II, $\omega$-Centauri) and the Coma cluster. They considered only primary e$^\pm$. The constraints depend on the propagation model but could be competitive for some targets.
	Mukhopadhyay \etal~\cite{2021arXiv210914955M} computed the synchrotron and inverse Compton scattering from e$^\pm$ injected by PBHs in the GC with only the GO limit for HR used. Future radio instruments like SKA would have enough sensitivity to constrain the PBH abundance very tightly by this mean. Lee \& Chan~\cite{2021ApJ...912...24L} extended this analysis to constrain the abundance of PBHs in cool-core galaxy clusters, but this is not competitive.
	
	Kim~\cite{2021MNRAS.504.5475K} were the first to compute the heating rate of the ISM by electrons/positrons injected by PBHs in the DM-dominated object Leo T, using \texttt{BlackHawk}. They advocated for the search for other DM dominated dSphs to constrain further the PBH abundance. The same target was used by Laha \etal in Ref.~\cite{2021PhLB..82036459L} where they extended the study to spinning PBHs and accounted for photon heating. They found very different results compared to Ref.~\cite{2021MNRAS.504.5475K}---constraints are not competitive.
	
	\subsubsection{511 keV line}
	
	A special consequence of PBH radiation of e$^\pm$ in a dense medium is the chance that positrons scatter off astrophysical electrons. This can result in two signals: a peaked signal at $511\,$keV if annihilation is immediate, or a low-energy tail at $E<511\,$keV if positronium is formed. A $511\,$keV feature was indeed detected some 50 years ago in the direction of the GC~\cite{1972ApJ...172L...1J,1973ApJ...184..103J,1975ApJ...201..593H}. A PBH origin was first suggested by~\cite{1978ApJ...225L..11L} and numerically estimated in~\cite{1980A&A....81..263O,1980Ap&SS..71..371O}, while there may be other possible origins~\cite{2011RvMP...83.1001P}.
	The $511\,$keV feature was confirmed with more recent data, in particular by INTEGRAL/SPI measurements~\cite{2005A&A...441..513K,2006A&A...445..579J,2006A&A...450.1013W}. In fact, the updated $511\,$keV line constraint proved competitive with $\gamma$-ray constraints for $M\sim 10^{14}-10^{17}\,$g PBHs, and Ref.~\cite{2008PhLB..670..174B} used a log-normal mass distributions for GC PBHs for the first time.
	
	Modern studies use the spatial shape of the line to constrain the PBH origin more efficiently, but this depends on the DM halo model and on the propagation and rate of annihilation in the GC~\cite{2019PhRvL.123y1101L,2019PhRvL.123y1102D}.
	Dasgupta \etal~\cite{2020PhRvL.125j1101D} updated the INTEGRAL $511\,$keV constraints using \texttt{BlackHawk} to compute the positron emission rate (see also~\cite{2021PhRvD.104f3033K}). Monochromatic and log-normal mass functions were considered, as well as a NFW/isothermal halo profile (resulting in one order of magnitude change in the constraints).
	
	If PBHs as DM are at the origin of the $511\,$keV feature in the GC, one would expect that other DM-dominated objects should present a similar line. No $511\,$keV line is observed in the Reticulum II dedicated search with INTEGRAL/SPI, which made Siegert \etal~\cite{2022MNRAS.511..914S} place (non-competitive) constraints on the PBH abundance in this dSph. The propagation model inside Ret II is very uncertain; thus they advocated for more dSph observations.
	
\begin{figure}[!t]
		\centering
		\includegraphics{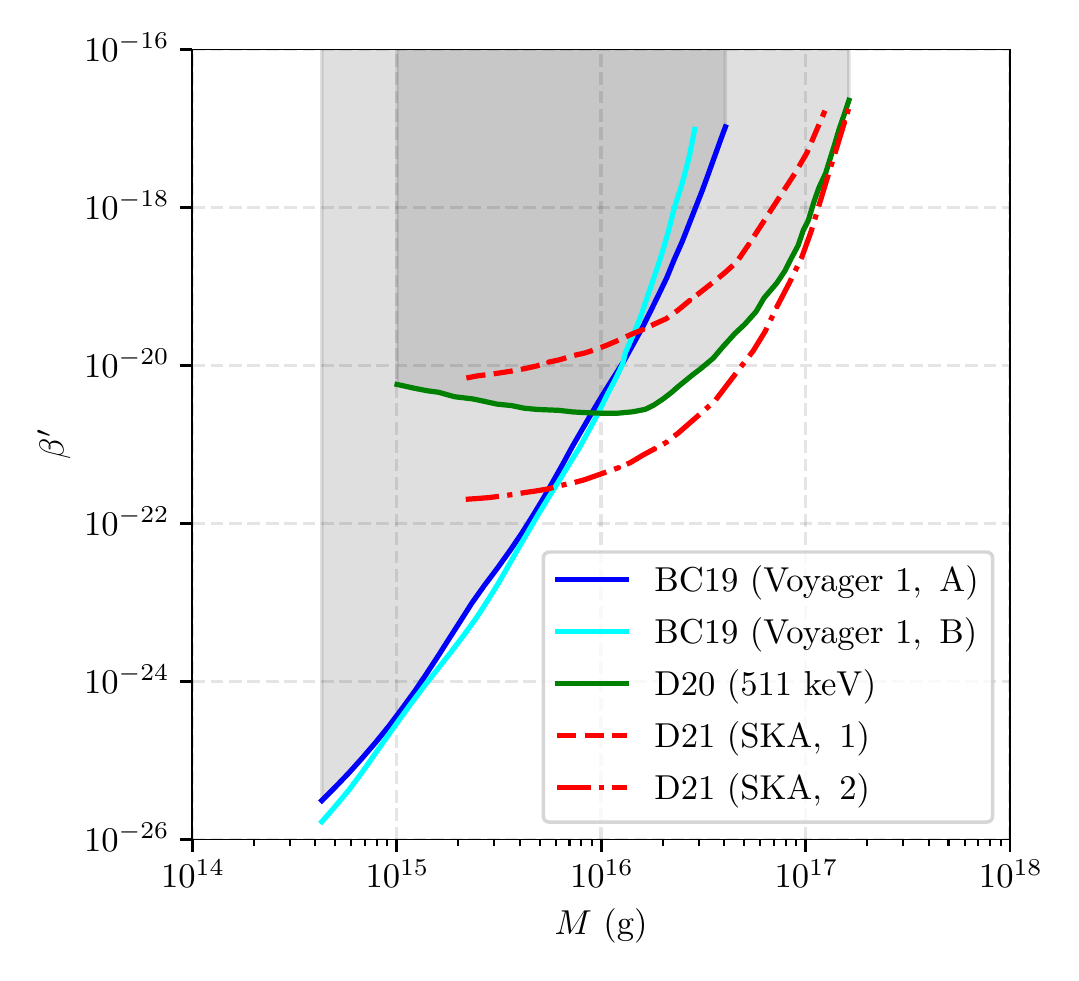}
		\caption{Electron/positron constraints on PBHs. The solid lines represent the Voyager 1 constraints~\cite{2019PhRvL.122d1104B} (BC19, blue and cyan, with different propagation models) and the $511\,$keV line constraint from the GC~\cite{2020PhRvL.125j1101D} (D20, green). Prospective limits from SKA on the e$^\pm$ interactions inside Segue I are shown in red (D21, discontinuous lines with two propagation models).}
		\label{fig:electrons}
	\end{figure}

	The robust e$^\pm$ limits obtained by Refs.~\cite{2019PhRvL.122d1104B,2020PhRvL.125j1101D} as well as the prospective limits of Ref.~\cite{2021JCAP...03..011D} are shown in Fig.~\ref{fig:electrons}.
	
	\subsection{Antimatter}
	
	The situation is more intricate for antiproton CRs. Just like for e$^\pm$, the ratio $\overline{\rm p}/$p $\ll 1$, which would lead to strong PBH constraints. Furthermore, all the recent studies show that \textit{there is no need of a primary source} to explain antiproton observations; that is, classical astrophysical production by spalliation of ISM material is sufficient to explain the observed spectrum~\cite{1999ApJ...518..570W}. The propagation of antiprotons is again strongly hindered by Solar modulation, and no human-made probe with an antinucleus detector has yet left the heliopause. The fact that antiprotons are so heavy ($m\mrm{p} \sim 1\,$GeV) prevents PBHs from $M\gtrsim M_*$ to emit them in sizeable amounts. Thus, antiprotons probe the \textit{local} density of PBHs that are at the beginning of evanescence, that is why some papers have related the $\overline{\rm p}$ constraints to the rate of PBH explosions~\cite{2002A&A...388..676B,2017AdSpR..60..806A} (see Section~\ref{sec:final_burst}) rather than to the DM density fraction.
	
	PBH generation of an antiproton flux was first discussed by Kiraly \etal~\cite{1981Natur.293..120K}. Considering the models of CR propagation in the Milky Way at the time, it was estimated that $\overline{\rm p}$ were 3 times more abundant than what was expected~\cite{1979PhRvL..43.1196G,1981ApJ...248.1179B}. They remarked that $\gamma$-ray, $511\,$keV and $\overline{\rm p}$ constraints were of the same order of magnitude, which suggested a common PBH origin. The constraint was refined in~\cite{1982Natur.297..379T} which found it to be more stringent than the $\gamma$-ray limit due to local PBH clustering.
	
	The $\overline{\rm p}$ flux was recomputed in the MG\&W model by~\cite{1991ApJ...371..447M}, and new data from the BESS experiments~\cite{1995PhRvL..75.3792Y} allowed for precise low-energy antiproton measurements thanks to Solar minimum activity in the 1990's. These data were used by Maki, Mitsui \& Orito~\cite{1996PhRvL..76.3474M,1996PhLB..389..169M} to strengthen the antiproton constraints with their own PBH jet hadronization code based on \texttt{JETSET}/\texttt{PYTHIA}~\cite{1994CoPhC..82...74S}.
	
	Barrau \etal~recomputed the $\overline{\rm p}$ limits from PBHs in~\cite{2002A&A...388..676B}, with an updated propagation model and BESS/CAPRICE/AMS data~\cite{2001AdSpR..27..693Y}. They obtained the flux on Earth with different extended mass functions for the first time. The secondary antiproton spectrum was calculated with the MG\&W model and fragmentation functions from \texttt{JETSET}/\texttt{PYTHIA}. They obtained PBH constraints as stringent as $\gamma$-ray ones, $\Omega\mrm{PBH}(M_*) \lesssim 4\times10^{-9}$.
	
	Abe \etal~\cite{2012PhRvL.108e1102A} (see also~\cite{2013AdSpR..51..227B,2017AdSpR..60..806A}) used the measurement of the 2007/2008 BESS-Polar II experiment performed during the next solar minimum after that of BESS95/BESS98 to re-assess the antiproton constraints. They got a limit translated into a burst rate \rate{1.2\times10^{-3}}.
	Aramaki \etal~\cite{2014APh....59...12A} examined what would be the antiproton constraints on PBHs set by the prospective instrument GAPS, to operate within the next years, with particular focus on the very low antiproton energies $E < 0.25\,$GeV.
	
	The fact that PBHs can emit (anti)nucleons inside dense QCD jets raises the question of their production of (anti)nuclei. Antihelium would be the simplest of them and the very low secondary production of antihelium in standard astrophysical mechanisms would lead to extremely strong constraints with \textit{e.g.} GAPS~\cite{2020JCAP...08..035V}. Barrau and collaborators~\cite{2003PhLB..551..218B,2003A&A...398..403B} (see also~\cite{2017JCAP...02..018H}) computed the antihelium emission rate from PBHs using a ``coalescence model'' that manually sticks together (anti)nucleons radiated with collinear momentum inside \texttt{JETSET}/\texttt{PYTHIA}.
	
	The advantage with antiprotons and antideuterons is that great instrumental progress has been achieved (AMS-02) or is expected (GAPS) in the near future, and particular attention has been given to antihelium regarding its very low astrophysical background~\cite{2020JCAP...08..035V}. However, since 2017 no new PBH constraints have been placed on PBHs. We do not produce a constraint plot from antiprotons because they apply to a very narrow mass range $M\sim M_*$, and are mainly expressed as final burst rate limits (see Section~\ref{sec:final_burst}).
	
	\section{Final bursts}
	\label{sec:final_burst}
	
	The final burst of PBHs is assuredly the type of HR constraint that has received the most attention (in terms of published papers). The competition between the different evaporation models listed in Section~\ref{sec:theory} (``elementary'' and ``composite'' particle models for the primary spectra, photosphere or no photosphere for the secondary spectra) gave rise to various observational signatures, ranging from final bursts with $\mu$s duration and low-energy MeV $\gamma$-ray signature (``composite'' model with a photosphere) to long duration bursts of $1-1000\,$s with increasing $\gamma$ energy (``elementary'' model without photosphere)~\cite{2009arXiv0906.3182B}. This resulted in order of magnitude spacing in the corresponding limits~\cite{1978IrAJ...13..173P,1978MNRAS.183..205P}. The constraints are usually expressed as the number of PBH explosions $R$ per cubic parsec per year.
	
	The MG\&W model settled down the basic features of PBH evaporation, and the subsequent paper~\cite{2008PhRvD..78f4043M} debunked all the competing photosphere models, making it easier to compare the limits on the rate of PBH explosions from different studies~\cite{1991Natur.353..807H,1993PhRvL..71.2524A,1994ApJ...436..254S,2016APh....80...90U}, and the different observational techniques are now well described~\cite{1996SSRv...75...67P,2021arXiv211101198D}. Interestingly, the MG\&W model predicts that PBH explosions could generate CRs up to the Planck energy scale~\cite{2000APh....12..269B}, which could explain the EeV CRs that challenged the expected Greisen--Zatsepin--Kuz'min cutoff~\cite{1966PhRvL..16..748G,1966JETPL...4...78Z}. Ref.~\cite{2000APh....12..269B} extrapolated the fragmentation functions for jet decays up to Planck energies and obtained the rate of emission of photons. Integration on time gives a slope $E^{-2}$ which does not correspond to the CR slope at these energies~\cite{1998PhRvL..81.1163T}. Comparison with those data gives a limit \rate{2.6\times10^6}, with amelioration foreseen from the Pierre Auger observatory.
	
	PBH explosions could be detected if the source is close enough to Earth, $D \lesssim 10\,$pc, and if the PBH lifetime is exactly the age of the universe. The combination of both considerations limits the observable bursts to a narrow mass range around $M_*$, which is already strongly constrained by the diffuse $\gamma$-ray background.
	
	Maybe the most important stake of PBH final burst detection is that it would give a privileged access to ultra-high energy physics, possibly exposing new BSM dofs like supersymmetric particles~\cite{1992ApJ...401L..57C,2021arXiv210510506B}. Page \& Hawking~\cite{1976ApJ...206....1P} stated that:
	\begin{quote}
		\textit{A definite observation of $\gamma$-rays from a primordial black hole would be a tremendous vindication of general relativity and quantum theory and would give us important information about the early universe and strong interactions at high energy, information that probably could not be obtained in another way.} \cite{1976ApJ...206....1P}
	\end{quote}
	and Kapusta~\cite{1999astro.ph.11309K} predicted that:
	\begin{quote}
		\textit{Experimental discovery of exploding black holes will be one of the great challenges of the new millennium.} \cite{1999astro.ph.11309K}
	\end{quote}
	
	\subsection{Photons}
	
	Several models for PBH explosion rates have been proposed across the years. Rees \& Blandford~\cite{1977Natur.266..333R,1977MNRAS.181..489B} argued that if PBHs were embedded in a strong magnetic field, then their energetic emission should form a plasma that would be seen in radio wavelengths, leading to very strong constraints~\cite{1977Natur.267..499J}; no such radio pulse has been robustly attributed to a PBH so far~\cite{2013ApJ...767...40V}. 
	
	Cline and collaborators~\cite{1992ApJ...401L..57C,1999ApJ...527..827C} and Belyanin and collaborators~\cite{1996MNRAS.283..626B,1998R&QE...41...22B,1998AdSpR..22.1111B} proposed a model in-between the Hagedorn ``composite particle'' model and the SM and computed the formation of a photosphere around exploding PBHs~\cite{1999PhRvD..59f3009C}. The signature would be a very short burst peaking at $\Lambda\mrm{QCD}$ energies. In particular, the high-energy slope would have behaviour $E^{-4}$. The search for a new category of short bursts in \textit{e.g.}~the BATSE~\cite{2011NewA...16...33C} and dedicated SGARFACE~\cite{2009APh....31..102S} data has not resulted in PBH detection so far.
	
	Heckler~\cite{1997PhRvD..55..480H,1997PhRvL..78.3430H} then proposed a model of particle interactions around PBHs that would generate a photosphere even in the MG\&W model of PBH evaporation. Efficient number-changing interactions would allow the expanding fireball to reach local thermodynamical equilibrium and a plasma would develop a first time at $T \sim \Lambda\mrm{QCD}$ and a second time at $T\sim \Lambda\mrm{QED}$, with $E^{-4}$ slope at high energies. A QED photosphere is also obtained by Daghigh \& Kapusta~\cite{1999astro.ph.11309K,2002PhRvD..65f4028D}, but they get a $E^{-3}$ slope.
	
	The first SM-inspired experimental search of PBH bursts is due to Porter \& Weekes~\cite{1977ApJ...212..224P} (see also~\cite{1978Natur.271..731F,1979Natur.277..199P,1980Natur.284..433B,1982MNRAS.199.1007B}). Halzen \etal~\cite{1991Natur.353..807H} were the first to use the MG\&W model to obtain more robust constraints. Alexandreas \etal~\cite{1993PhRvL..71.2524A} argued that if PBHs are a component of DM, they must be clustered inside Galaxies, which increases their \textit{local} abundance compared to the cosmological DM density---recall that PBH explosions are detectable only within a $\sim$ pc sphere; they obtained a PBH burst rate \rate{8.5\times10^{5}} with CYGNUS data. Linton \etal~\cite{2006JCAP...01..013L} got \rate{1.08\times10^6} at the Whipple observatory, and were the first to present limits that depend (moderately) on the burst duration.
	
	Petkov \etal~\cite{2008AstL...34..509P} proposed a broken power-law fit to the time-integrated PBH photon spectrum; whose parameters depend on the remaining PBH lifetime or, equivalently, on the energy threshold of the instrument considered. This fitting function is still used as a template in modern PBH burst searches. Surprisingly, the first ever numerical light curve of a PBH explosion was obtained by Ukwatta \etal~\cite{2016APh....80...90U} as late as 2016, in the $50\,$GeV$-100\,$TeV energy range. They advocated for the use of a binned maximum likelihood search that takes into account the time profile of the bursts, such as the characteristic ``spectral lag'' between the first low-energy and the last high-energy photons. The numerical GFs from \texttt{BlackHawk}, interfaced with the high-energy secondary spectra of \texttt{HDMSpectra}, were used for the first time by Capanema \etal~\cite{2021JCAP...12..051C} to constrain the PBH explosion rate.\footnote{Following this paper, \texttt{HDMSpectra} secondaries has been implemented inside \texttt{BlackHawk v2.0}~\cite{2021EPJC...81..910A}.}
	
	Modern limits on PBH bursts in the MG\&W model were obtained with high-resolution TeV IACT instruments (for ``Imaging Atmospheric Cherenkov Telescopes'') such as, by chronological order, MAGIC~\cite{Cassanyes:2015wpr}, VERITAS~\cite{2017ICRC...35..691A},\footnote{The VERITAS limit should be enhanced by a factor of $\sim 2$ in the near future~\cite{2019ICRC...36..719K}.} HESS~\cite{HESS:2021rto} and prospective CTA~\cite{Cassanyes:2015wpr}; and water Cerenkov telescopes such as Milagro~\cite{2015APh....64....4A}, its successor HAWC~\cite{2020JCAP...04..026A} and the prospective SWGO~\cite{2021JCAP...08..040L}.\footnote{The MAGIC and CTA data points were provided in~\cite{2021arXiv211101198D}.} They rely on the detection of respectively astmospheric Cerenkov radiation or secondary particle showers from spalliation with atoms due to high-energy photons and are sensitive to the last seconds of PBH explosions. The Fermi-LAT limits were presented by Ackermann \etal~\cite{2018ApJ...857...49A} and are somewhat of different nature as they are direct satellite observations of GeV $\gamma$-rays; the notion of ``burst'' is quite spurious in that case as durations as long as $10^8\,$s are considered, with a limit \rate{10^4}. However, such long duration ``bursts'' allow to look for proper motion of the source as a confirmation of local PBH origin. All the limits are given with a dependency on the burst duration and templates are used to search for PBH explosions. The rate of PBH explosion constrained by HAWC (the most restrictive) is \rate{10^3}, which is still orders of magnitude above the limit deduced from diffuse $\gamma$-rays. Quantitative conversion depends strongly on the PBH mass function, as shown by the crude estimate
	\begin{equation}
		R \sim \dfrac{\d n}{\d t} \sim \frac{\d n}{\d M}\,\dfrac{\d M}{\d t}\,.
	\end{equation}
	
	\subsection{Other particles}
	
	A way of testing BH quantum properties against theoretical models would be to detect a neutrino counterpart~\cite{1992ApJ...401L..57C,1995PhRvD..52.3239H,1998R&QE...41...22B,2003PhRvD..67d4006D} (or antiprotons~\cite{1996PhRvL..76.3474M}) and to examine the relative fraction of the different particles ($\gamma$, $\nu$'s, gravitons) reaching a detector~\cite{2021JCAP...12..051C}. Multimessenger PBH searches are one of the aims of the AMON program~\cite{2013APh....45...56S,2015ICRC...34..328T,2019arXiv191201317P}.
	
	The first multimessenger constraints were set by Capanema \etal~\cite{2021JCAP...12..051C}. They used for the first time the \texttt{BlackHawk} numerical GFs, convolved with the high-energy particle physics code \texttt{HDMSpectra} to obtain the light curves of PBHs in the photon and neutrino channel. They argued that while neutrinos are more elusive and thus are associated with reduced constraints \rate{10^7} from IceCube (correcting manifest errors in~\cite{2019ICRC...36..863D}), the combined observation of a burst in both $\gamma$-rays and energetic neutrinos would serve as a smoking-gun for PBH identification. Furthermore, the ratio of photon and neutrino energy in the time-integrated burst would give access to the high-energy behaviour of those particles, way above the LHC scope.
	
	As stated above, antiproton constraints can be interpreted as local PBH explosion rates. Indeed, antiprotons are emitted only by PBHs with mass $M\lesssim M_*$, thus PBHs that are completing their evaporation in the present universe. BESS limits translate into \rate{1.2\times10^{-3}} (for very long ``burst'' duration), which would be the strongest limit to date, but it depends on the propagation model~\cite{2012PhRvL.108e1102A}. Within the same paradigm, future GAPS antihelium constraints would place even more stringent constraints~\cite{2017JCAP...02..018H}.
	
\begin{figure}[!t]
		\centering
		\includegraphics{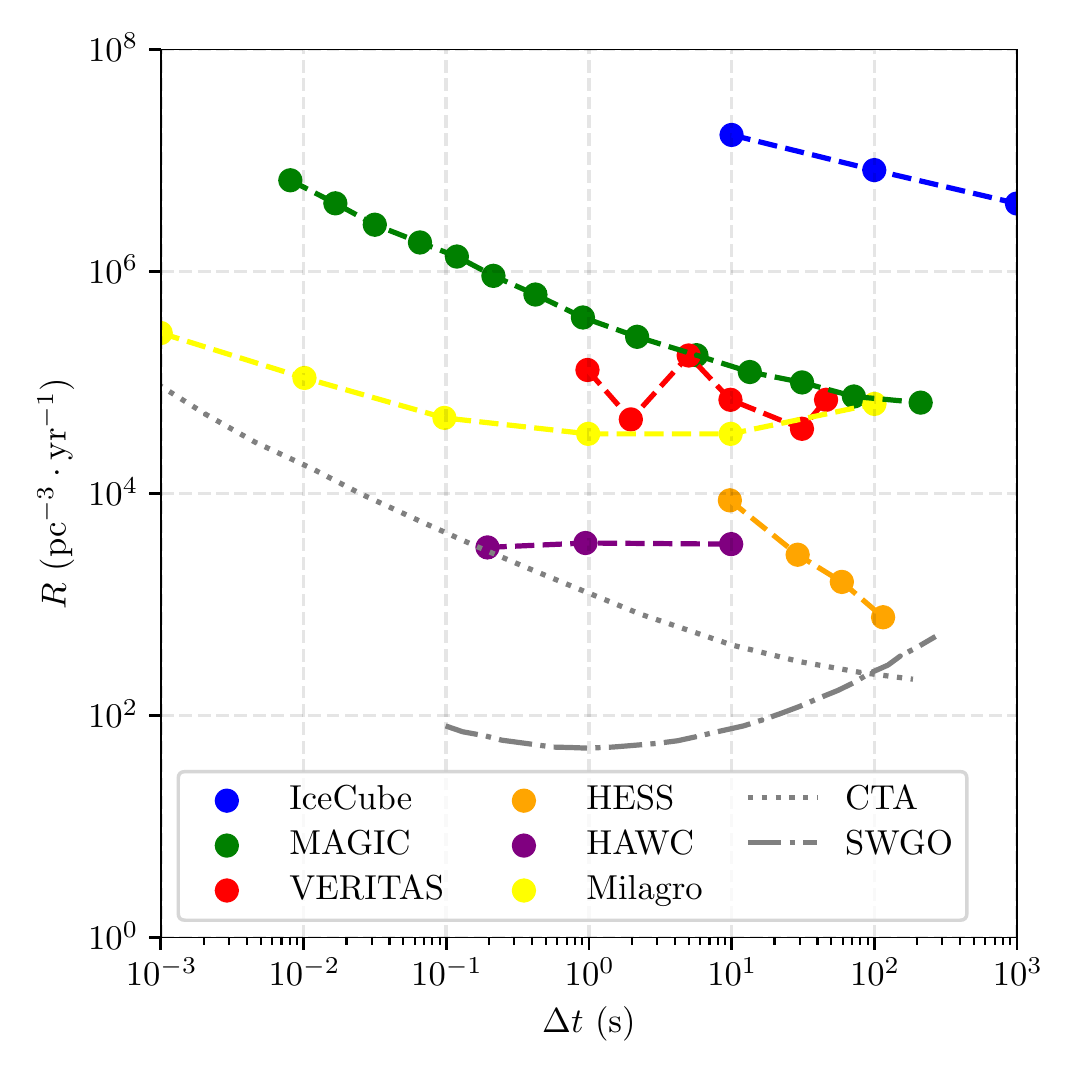}
		\caption{$\gamma$-ray and neutrino constraints on the rate of PBH explosions. The limits set by the existing $\gamma$-ray facilities MAGIC~\cite{2021arXiv211101198D} (green), VERITAS~\cite{2017ICRC...35..691A} (red), HESS~\cite{HESS:2021rto} (orange), HAWC~\cite{2020JCAP...04..026A} (purple), Milagro~\cite{2015APh....64....4A} (yellow) and the neutrino observatory IceCube~\cite{2021JCAP...12..051C} are shown as filled circles. The prospective limits from CTA~\cite{2021arXiv211101198D} and SWGO~\cite{2021JCAP...08..040L} are shown as discontinuous grey lines (dotted and dot-dashed respectively).}
		\label{fig:final_burst}
	\end{figure}

	In Fig.~\ref{fig:final_burst}, we show the PBH burst rate limits from Refs.~\cite{2021arXiv211101198D,2017ICRC...35..691A,HESS:2021rto,2015APh....64....4A,2020JCAP...04..026A,2021JCAP...12..051C}, as well as the prospective limits from Refs.~\cite{2021JCAP...08..040L,2021arXiv211101198D}.
	
	\subsection{Direct observation}
	
	A very exciting prospect would be to observe an evaporating PBH directly. This would give unprecedented access to the steady rate of primary HR emission, which encodes the quantum structure of the PBH horizon, and to the number of (B)SM dofs as well as their high-energy behaviour, which would be reflected in the secondary branching ratios.
	
	Very speculative studies considered PBHs that accumulated inside the Earth due to gravitational friction~\cite{1990Ap&SS.168..277T,1995SSRv...74..467D}; constrained by neutrino observations from the surface~\cite{2021JCAP...04..026A}. More recently, the hypothesis that the putative Solar System Planet 9 is a PBH captured by the Sun's gravitational attraction~\cite{2020PhRvL.125e1103S} has raised the interest of sending a probe towards it; Planet 9 however remains to be located and its HR would be very faint~\cite{2020arXiv200602944A}.
	
	Refs.~\cite{2008PhRvD..78f4044P,2014MNRAS.441.2878S} argued that BHs in the vicinity of the Earth would be detectable with standard optical telescopes due to visible light emission via \textit{e.g.}~inner Bremsstrahlung radiation of emitted charged particles. Ref.~\cite{2022PhRvD.105d5022N} examined thoroughly for the first time what would be the ``picture'' of a BH taken by a detector sensitive to its HR. They calculated the two-point correlation function $\langle X_1,X_2 \rangle$ of the signal received by two independent receptors and reconstruct the BH ``image'' by interferometry---Fourier transform of $\langle X_1,X_2 \rangle$. It is shown in that paper that highly spinning KBHs would exhibit rather different pictures compared to SBHs.
	
	The most promising detection means remains the measure of the proper motion of an unresolved close-by $\gamma$-ray source, that would be identified as a PBH~\cite{2012MNRAS.425..862G,2016APh....80...90U}.
	
	\section{Beyond monochromatic and Schwarzschild constraints}
	\label{sec:prospects}
	
	In the above discussion, we have focused on the simple case of Schwarzschild PBHs with monochromatic mass distribution. Below, we consider extended mass functions, rotating PBHs and more speculative non-standard BH metrics and BSM particles.
	
	\subsection{Extended mass functions}
	
	As stated in Section~\ref{sec:PBH_formation}, realistic PBH formation mechanisms rather imply extended mass functions. These can be broad, like the power-law mass function resulting from the collapse of scale-invariant perturbations, or narrow, like the log-normal mass function that derives from a peak in the perturbation power spectrum. PBHs that formed from perturbations at a single scale may still exhibit an extended mass function if critical collapse mechanisms apply. In any case, subsequent evolution of PBHs such as accretion of nearby material or hierarchical mergers may broaden any initial distribution.
	
	On the other hand, continuous evaporation makes any distribution evolve in a non-trivial way. The PBHs with initial mass $M\gtrsim M_*$ end up as a fine-tuned low-mass tail $M\ll M_*$ today. Hence, the constraints on $M\sim M_*$ PBHs must be treated carefully, in particular in the case of the final burst constraints~\cite{2001PhRvD..65b7301G} or for the very strong present-day $\gamma$-ray constraints, \textit{e.g.}~from the diffuse Galactic background~\cite{2010PhRvD..81j4019C,2016PhRvD..94d4029C}.
	
	Refs.~\cite{2021arXiv211210422C,2022arXiv220305743M} computed numerically the evolution of an initial distribution (\textit{e.g.}~log-normal) and showed that instantaneous constraints can be computed with \texttt{BlackHawk} provided that the \textit{evolved} current distribution is used. This reduces the computational resources needed for a full time-dependent \texttt{BlackHawk} calculation for an extended distribution.
	
	From an opposite point of view, one can consider existing PBH constraints and try to find the best-fitting extended function that would explain the observations (see \textit{e.g.}~\cite{2022arXiv220313008M} for the $21\,$cm signal). This poses the \textit{mathematical} problem of finding the maximum density function that satisfies all the constraints; Ref.~\cite{2018JCAP...04..007L} demonstrated that the solution is a finite linear combination of nearly-monochromatic mass functions.
	
	Two effects of having an extended mass function are in competition~\cite{2020PhRvD.101b3010A}:
	\begin{enumerate}
		\item an extended mass function results in an enhanced \textit{total} density in PBHs;
		\item the high- and low-mass tails may be affected by other constraints.
	\end{enumerate}
	Eventually, if an extended mass function is centered around a peak of constraints (\textit{e.g.}~at $M\mrm{c} \sim M_*$ for EGXB constraints), then an extended mass function results in less stringent constraints because effect 1 is dominant. If it is centered away from the peak (\textit{e.g.}~at $M\mrm{c} \gg M_*$ for EGXB constraints), then the constraints are stronger because the non-zero low-mass tail ($M \sim M_* \ll M\mrm{c}$) is severely constrained, effect 2 dominates. As we are mainly interested in open windows for PBH DM, extended mass functions result globally in a \textit{diminution} of the available parameter space for PBHs. A 2D plot of the constraints obtained with the Carr method shows that in the case of a log-normal mass function, a width $\sigma \gtrsim 2$ closes entirely the PBH window $10^{17}-10^{23}\,$g~\cite{2021RPPh...84k6902C}. The power-law mass function is totally excluded unless very restrictive artificial cut-offs $M\mrm{min/max}$ are set, which effectively removes most of the ``extended'' behaviour~\cite{2017PhRvD..96b3514C}.
	
\begin{figure}[!t]
		\centering
		\includegraphics{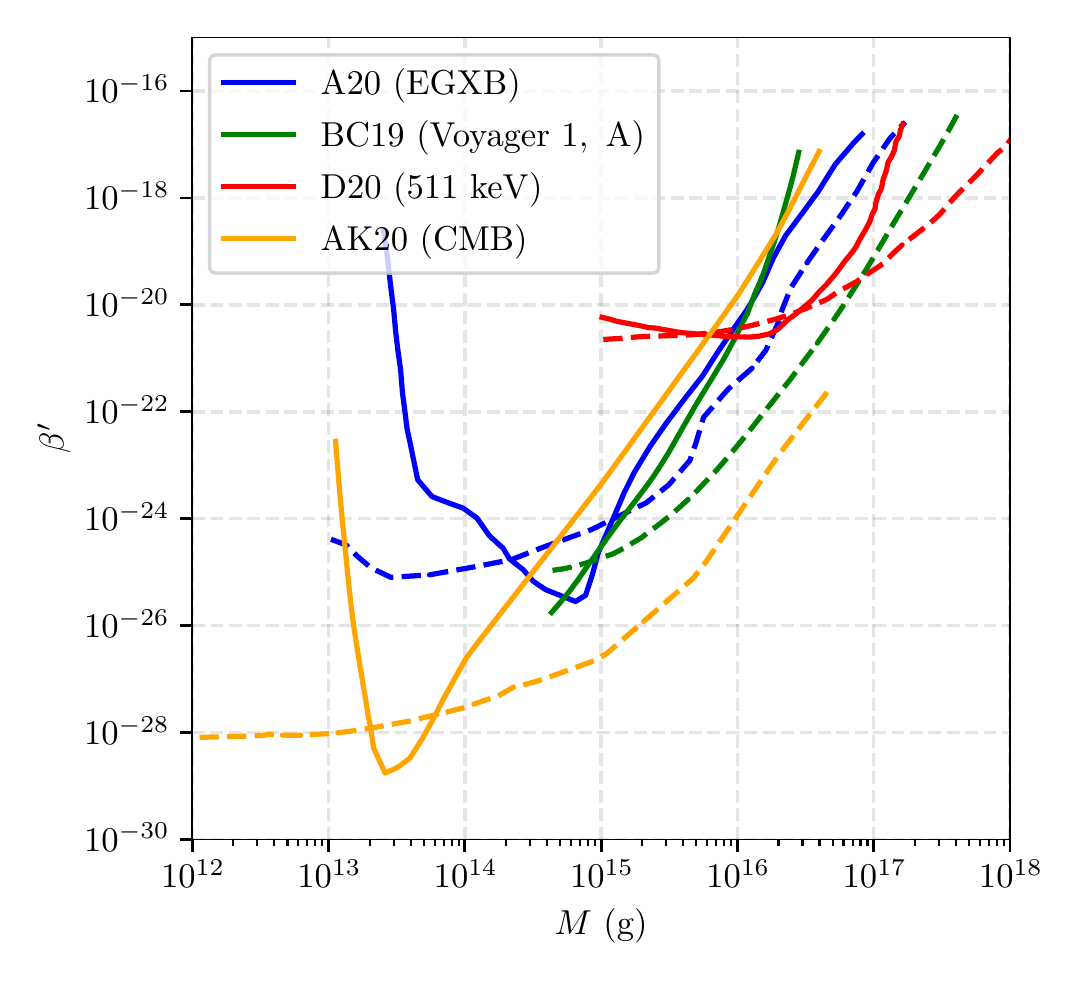}
		\caption{Extended log-normal mass function constraints on PBHs. Solid (dashed) lines are monochromatic (log-normal, with $\sigma = 1$) constraints. The EGXB limits are from~\cite{2020PhRvD.101b3010A} (A20), the Voyager 1 limits from~\cite{2019PhRvL.122d1104B} (BC19), the $511\,$keV limits from~\cite{2020PhRvL.125j1101D} (D20) and the CMB limits from~\cite{2020JCAP...06..018A} (AK20).}
		\label{fig:extended}
	\end{figure}

	As an illustration, we have recollected some HR constraints computed for an extended log-normal mass function with $\sigma = 1$ in Fig.~\ref{fig:extended}, compared to the equivalent figure with $\sigma = 0$ (monochromatic). The two effects detailed above are clearly observable. Equivalent comparison could be made for power-law or critical collapse mass functions.
	
	Log-normal constraints have been computed for the CMB~\cite{2018JCAP...03..018S,2019arXiv190706485P,2020JCAP...06..018A}, the $21\,$cm signal~\cite{2022JCAP...03..012C,2022arXiv220313008M}, the EGXB/GC~\cite{2020PhRvD.101b3010A,2021PhRvD.104b3516R,2022arXiv220305743M}, the diffuse neutrino background~\cite{2020PhRvL.125j1101D,2021JCAP...10..051D,2022arXiv220314979B}, the local positron flux~\cite{2019PhRvL.122d1104B}, the GC radio signal~\cite{2020MNRAS.497.1212C}, the $511\,$keV line~\cite{2008PhLB..670..174B,2019PhRvL.123y1101L,2019PhRvL.123y1102D,2020PhRvL.125j1101D} and gas heating~\cite{2021ApJ...912...24L}. This distribution is by far the most ``popular''.
	Critical collapse constraints have been computed for baryogenesis~\cite{2021PhRvD.103h3504B}, BBN~\cite{2021JCAP...05..042L}, the EGXB~\cite{1999PhRvD..60j3510K,1998PhRvD..58j7502Y,2006astro.ph.12659B,2002PhRvD..66h4004B} and the diffuse neutrino background~\cite{2006astro.ph.12659B}.
	Power-law constraints have been used originally for all constraints, as the PBH mass spectrum was supposed to be scale-invariant. More recently, power-law constraints have been obtained for the $21\,$cm signal~\cite{2022JCAP...03..012C,2022arXiv220313008M}, the CMB~\cite{2019arXiv190706485P}, the GC radio signal~\cite{2020MNRAS.497.1212C} and the $511\,$keV line~\cite{2019PhRvL.123y1101L}.
	
	\subsection{Spinning black holes}
	
	As already stated in Section~\ref{sec:PBH_formation}, PBHs were initially thought to form without sizeable spin. Whenever spin was mentioned, it was assumed that, just like electric charge, it was evaporated more rapidly than mass because of enhanced emission rates for dofs with angular momentum aligned with that of the BH (see Eq.~\eqref{eq:HR_KN}). This idea was challenged as early as 1975~\cite{1976A&A....52..427C} and confirmed numerically in the paper by Page~\cite{1976PhRvD..14.3260P}, the emission rates of spinning PBHs were computed for the first time. Page intuited that enhanced emission rates could make mass $M$ decrease faster than angular momentum $J$, which would result in \textit{increasing} dimensionless spin $a^* \equiv J/M^2$. Depending on the available dofs, the spin could even stabilize around a non-zero value: emission of scalar dofs only results in an asymptotic value $a^* \sim 0.5$, whatever be the initial non-zero spin~\cite{1997PhRvL..78.3249C,1998PhRvD..58d4012T}. This was confirmed recently in the case of an axiverse with a number of axion dofs much larger than the number of SM dofs~\cite{2021arXiv211013602C}.
	
	One sees that no trivial conclusion may be drawn concerning the value of the spin of PBHs today or at the time of their past evaporation. Refs.~\cite{1997PhRvL..78.3249C,1998PhRvD..58d4012T}, and more recently Ref.~\cite{2020MNRAS.494.1257A} (using \texttt{BlackHawk}) proved that even for the SM content only, PBHs with initial mass $M \gtrsim 10^{16}\,$g, if only subject to HR, could still have a spin value today higher than the generalized Thorne limit from BH disk accretion. Such near-extremal BHs would then point towards primordial origin, as mergers of contemporary BHs result only in final spin $a^*\sim0.7$~\cite{2016ApJ...825L..19H}. Hierarchical mergers in the early universe, or formation during an EMDE, respectively produce BH spin distributions with a peak at $a^*\sim 0.7$~\cite{2017ApJ...840L..24F,2021ApJ...914L..18D} or near-extremal PBHs~\cite{2017PhRvD..96h3517H}.
	
	A most interesting consequence of near-extremal spin on HR is that the rate of emission of particles is \textit{globally} enhanced, with a different enhancement factor for different particle spins. Spin $1/2$ fermions and scalars are not very much affected with a factor $\sim 10$ increase, while vectors ($\sim 100$) and tensors ($\sim 10^4$) are preferentially emitted. These modified primary emission rates, with more energy going into tensors and scalars, induces a non-trivial modification of the secondary rates that are a convolution of the primary emission with decay and hadronization. Overall, (secondary) photons and gravitons are the secondary particles that are enhanced the most, while the lifetime of BHs is only reduced by at most $60\%$~\cite{2020MNRAS.494.1257A}.
	
	To our knowledge, Dong \etal~\cite{2016JCAP...10..034D} were the first to consider numerically the effect of enhanced emission rates on the graviton spectrum from PBH evaporation in the early universe. Despite a factor of $10^4$ increase on the graviton spectrum today, its very high frequency still prevents direct observation of this background~\cite{2019EPJC...79.1032E}. Hooper \etal~\cite{2020arXiv200400618H} then considered a more promising detection prospect, namely the constraints on the effective number of neutrinos $\Delta N\mrm{eff}$ at BBN/CMB time. In particular, the CMB-S4 experiment would be sensitive to the hot background of gravitons (or different spin) DR. This background is enhanced in the case of spinning PBHs, with better detection reach. They use the peak value $a^*$ of the extended spin distribution obtained by early hierarchical mergers as well as a benchmark near-extremal value $a^* = 0.99$.
	
	Finally, the code \texttt{BlackHawk} was released which immediately embedded the possibility of computing the spectra from spinning PBHs.\footnote{Note that the newly released \texttt{ULYSSES} package also contains a spinning PBH option~\cite{2022PhRvD.105a5022C,2022PhRvD.105a5023C}.} It was first applied to extended mass function of non-zero spin PBHs in the context of the PHL~\cite{2020PhRvD.101b3010A}, and then to extended spin distributions of monochromatic mass in the context of DR from hot gravitons in the early universe~\cite{2021PhRvD.103l3549A}.
	
	Numerous publications concerning PBHs now present an estimation of the constraints in the spinning case, as required from completeness or from particular formation mechanisms. These include the EGXB/GC $\gamma$-rays~\cite{2020PhRvD.101b3010A,2021PhRvD.104b3516R,2021PhRvD.103j3025I,2021arXiv211003333G,2022MNRAS.510.4992M}, the $511\,$keV line~\cite{2020PhRvL.125j1101D}, the diffuse neutrino background~\cite{2020PhRvL.125j1101D,2021JCAP...10..051D}, the Planet 9 hypothesis~\cite{2020arXiv200602944A}, gas heating by e$^\pm$~\cite{2021PhLB..82036459L}, (W)DM emission~\cite{2021GrCo...27..315M,2022PhRvD.105a5022C,2022PhRvD.105a5023C}, DR~\cite{2021PhRvD.103l3549A}, CMB anisotropies~\cite{2022JCAP...03..012C} and the $21\,$cm signal~\cite{2022MNRAS.510.4236N,2022JCAP...03..012C,2021arXiv211210794S}. In particular, Capanema \etal~\cite{2021JCAP...12..051C} argued that the measure of the neutrino-to-photon ratio energy spectrum in the final burst of PBHs would provide an estimation of the PBH spin prior to evanescence. 
	
\begin{figure}[!t]
		\centering
		\includegraphics{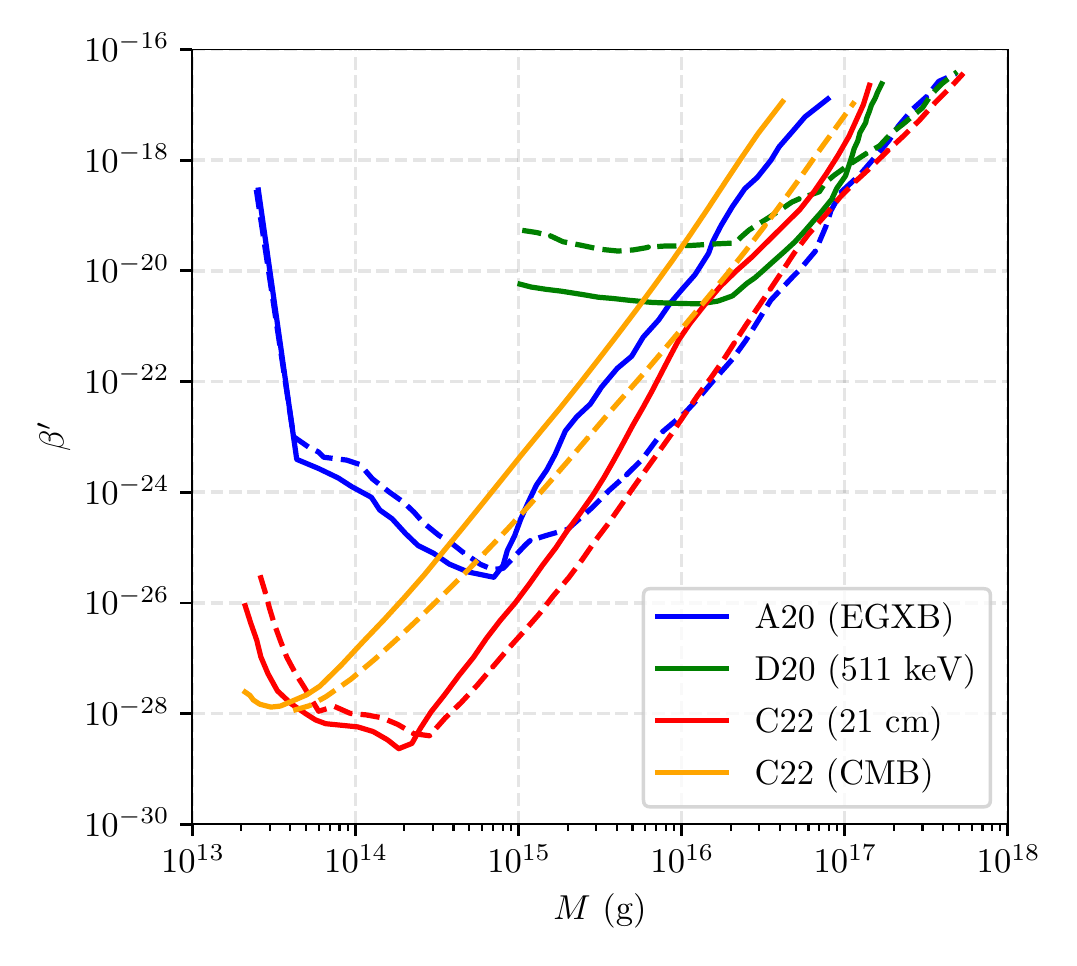}
		\caption{Monochromatic constraints on spinning PBHs. Solid (dashed) lines are for SBHs (KBHs) with $a^* = 0$ ($a^*=0.999(9)$, depending on the references). The EGXB limits are from~\cite{2020PhRvD.101b3010A} (A20), the $511\,$keV line from~\cite{2020PhRvL.125j1101D} (D20) and the CMB anisotropies/$21\,$cm signal from~\cite{2022JCAP...03..012C} (C22).}
		\label{fig:spinning}
	\end{figure}

	Spinning PBHs are overall more tightly constrained at high masses because of their enhanced photon HR, but less tightly at low masses because of complex secondary spectra effects. These effects are shown in Fig.~\ref{fig:spinning} which compares the same constraints for SBHs and near-extremal KBHs from Refs.~\cite{2020PhRvD.101b3010A,2020PhRvL.125j1101D,2022JCAP...03..012C}. There exists no complete study of the constraints in the whole HR mass range in the spinning case (\textit{e.g.}~on BBN).
	
	\subsection{Non-standard black holes}
	\label{sec:non_standard_BHs}
	
	In this Section, we discuss non-standard BHs and the associated PBH constraints. By non-standard, we mean BH solutions of the Einstein equations that do not belong to the Kerr--Newman family.
	
	\subsubsection{Higher-dimensional black holes}
	
	Large extra dimensions were proposed by Randall \& Sundrum~\cite{1999PhRvL..83.3370R,1999PhRvL..83.4690R} (RS) and by Arkani-Hamed, Dimopoulos \& Dvali~\cite{1998PhLB..429..263A,1998PhLB..436..257A,1999PhRvD..59h6004A} (ADD) in 1999 in a series of papers aimed at solving the ``hierarchy problem'', namely the fact that the gravitational coupling constant is so much smaller than that of the other fundamental interactions. In the RS model, there is a single extra dimension, while in the ADD model there can be many. The effect of higher dimensions is to reduce the Planck mass $M\mrm{Pl}$ to an effective value $M\mrm{eff} \ll M\mrm{Pl}$, typically constrained to be larger than $10\,$TeV. The first theoretical derivation of the thermodynamics properties of higher-dimensional BHs (HDBHs) were done by Myers \& Perry~\cite{1986AnPhy.172..304M} in the 1980's. HDBHs of mass $M$ are cooler and larger than SBHs of the same mass, and their GFs as computed in the early 2000's are very different from those of SBHs~\cite{2004IJMPA..19.4899K,2015qabh.book..229K}. Hence, the constraints linked to their evaporation are completely modified.
	
	HDBHs have been the subject of particular interest in the early 2000's because the LHC accelerator would have been capable of producing Planck-size BHs in TeV collisions~\cite{2009LNP...769..387K}. Many work was pursued to identify the BH evaporation signatures in the LHC detectors or in CR experiments, with no positive evidence so far.\footnote{See \textit{e.g.}~the BH event generators listed in Section~\ref{sec:numerical_recipes}.} Higher-dimensional BHs are the only non-standard BHs for which HR PBH constraints have been studied thoroughly.
	
	For the case of RS PBHs, constraints were computed for baryogenesis~\cite{2015PhRvD..91d5002A,2021PhRvD.103b3509A}, BBN~\cite{2003PhRvD..68b3507C}, CMB distortions~\cite{2003PhRvD..68b3507C}, EGXB~\cite{2003PhRvD..68b3507C,2003PhRvD..68j3510S}, antiprotons~\cite{2005PhRvD..71f3512S} and high-energy neutrinos~\cite{2005ICRC....9...33J}. In particular, it was argued that PBHs in the RS model may experience a ``thunderbolt'' explosion at the end of their evaporation, reviving the search for radio pulses~\cite{2008JCAP...11..017K}, while the signal might not be distinguishable from 4D PBHs in $\gamma$-ray telescopes~\cite{2015JETP..120..966A}.
	
	The case of ADD PBHs is more involved as the many supplementary dimensions include Kaluza-Klein modes for the graviton. The PHL was derived by Johnson~\cite{2020JCAP...09..046J}, who modified \texttt{BlackHawk} (prior to \texttt{v2.0} update) to include HDBH emission rates and mass evolution. More recently, Friedlander \etal~\cite{2022arXiv220111761F} released new constraints on PBHs from the whole set of phenomena described above (BBN, CMB, EGXB, \textit{etc.}) in a very complete review were they correctly take into account the graviton emission to compute the PBH lifetime. Their numerical methods are published in the code \texttt{CosmoLED}.
	
	Overall, the most dramatic effect of higher dimensions on the PBH constraints is to shift them from order of magnitude different masses, due to the modified $T-M$ relationship. Hence, the $M_*\sim 5\times 10^{14}\,$g peak in the constraints from the EGXB for 4D PBHs is displaced \textit{e.g.}~at $\sim 10^7\,$g for $n=2$ additional dimensions. In the case of numerous extra dimensions $n \gtrsim 6$, the SBH behaviour would be recovered, within the hypothesis that BHs with size $r\mrm{s} \ll R$, where $R$ is the size of the extra dimensions, do not ``feel'' those.
	
	\subsubsection{Other metrics}
	
	As shown above, PBH constraints must be recomputed whenever a different metric is used for the BH solution to the Einstein equations. Different metrics are associated with different field propagation equations and thus to different GFs. Recently, Refs.~\cite{2021PhRvD.103j4010A,2021PhRvD.104h4016A} obtained the general formulas for the computation of HR for the class of spherically symmetric and static BHs. This class is of particular interest since it contains BH solutions with no true singularity at $r = 0$, denoted as ``regular'' BHs. Ref.~\cite{2021PhRvD.104h4016A} used the GFs computed for a particular example of loop quantum gravity-inspired BH solution to recompute the AMEGO constraints from $\gamma$-rays from the GC. They concluded that due to strong corrections on the GFs, the signal (or constraints) for highly quantum deformed PBHs would be distinguishable from that of classical PBHs. This is a invaluable window opened on the study of quantum BHs.\footnote{Loop quantum gravity, higher-dimensional and electrically charged BHs were implemented in the last version of \texttt{BlackHawk v2.0}~\cite{2021EPJC...81..910A}.}
	
	\subsubsection{Black hole remnants}
	\label{sec:PMRs}
	
	In Section~\ref{sec:HR}, we only dealt with PBH evaporation in the context of the standard theory by Hawking. This theory relies on a semi-classical model of SM fields (quantum) and gravity (classical). It should break down at the Planck scale were quantum corrections should be sizeable, with no certain consequences. On the one hand, Hawking proposed that BHs should disappear in a final flash of energy at the end of the evaporation process~\cite{1975CMaPh..43..199H}. On the other hand, it has been argued that Planck-mass remnants (PMRs) may be stabilized against HR. Those PMRs may be a perfect candidate for DM~\cite{1987Natur.329..308M,1992PhRvD..46..645B}, with the inconvenient that they would interact only gravitationally with baryonic matter, hindering direct detection.
	
	In fact, modification of the HR mechanism may result in remnants of any mass, provided that the semi-classical description breaks down sufficiently early in the BH evolution. Backreaction~\cite{2013PhRvD..87l3514T} or collapse memory~\cite{2020PhRvD.102j3523D} effects may effectively slow down and stop evaporation at $M \gg M\mrm{Pl}$. In the conclusive Fig.~\ref{fig:master}, we show as a dashed line the constraint on the PMR density. Should the remnant mass be $M\mrm{r} = \alpha M\mrm{Pl}$, this constraint would be more stringent by a factor $\alpha$, while applying only for $M > \alpha M\mrm{Pl}$.
	
	\subsection{Non-standard particles}
	\label{sec:DM_production}
	
	As we have seen in the context of baryogenesis, the production of BSM particles by evaporating PBHs is a long-standing idea (GUT bosons for the ``Weinberg model'' or RHNs for the ``Fujita model''). HR is a phenomenon based on \textit{classical} general relativity only, hence any existing dof should be emitted. This includes supersymmetric particles~\cite{1999PhRvD..60f3516G,2003astro.ph..4478S} or mirror matter~\cite{1999PhRvD..59j7301B}, as well as possible numerous axions~\cite{2021arXiv211013602C,2021JCAP...08..063S,2021PhRvD.104g5007B,2021PhRvD.104l3536B}.
	
	In the supersymmetric or axion models, the main impact is the modification of the evaporation rate of PBHs: more available dofs provoke faster evaporation~\cite{1991PhRvD..44..376M}. They also mean decreased branching ratios into SM particles, hence less stringent constraints~\cite{2003astro.ph..4478S}.
	In baryogenesis scenarios, the BSM particles are just an intermediate to introduce asymmetric decay rates and produce the observed baryon asymmetry, and they quickly disappear.
	Already in Refs.~\cite{1999PhRvD..60f3516G,1999PhRvD..59j7301B}, the idea was present that some stable particle could be emitted whose cosmological density would then be compatible with DM, a scenario denoted as ``melanopogenesis''~\cite{2019JCAP...05..005M}. Khlopov \etal~\cite{2006CQGra..23.1875K} considered gravitino emission from PBHs; those gravitinos could either be long-lived and participate in DM, or they could decay and possibly spoil BBN.
	
	The scenario of \textit{direct} DM creation by PBHs was then left aside until Fujita \etal~\cite{2014PhRvD..89j3501F} proposed their unified scenario of DM and baryogenesis via the emission of RHNs by PBHs that would dominate expansion in the early universe (see also~\cite{2014PhRvD..90h3535H,2022PTEP.2022c3B02K}). The need for very high PBH densities in order to obtain the correct DM abundance today implies that their mass must lie in the unconstrained mass range $M \lesssim 10^9\,$g. While the formulas that translate the PBH rate of emission of particles into their abundance today (assuming no other interactions like decay or annihilation) are quite simple, the precise amount of DM created by this means depends critically on the spin of the particle considered, as first computed numerically in~\cite{2021EPJP..136..261A} with \texttt{BlackHawk}. Furthermore, as very light PBHs are considered, the created DM would have high momentum at PBH evaporation, and redshift dilution may still leave DM particles with large free-streaming length at matter-radiation equality. These DM candidates, known as ``warm DM'' (WDM, by opposition with the ``cold'' scenario), are tightly constrained as they would spoil structure formation at small scales~\cite{2014PhRvD..89j3501F,2018JCAP...04..009L,2020EPJP..135..552M}. A way of computing the WDM constraints is to use the code \texttt{CLASS}, which embeds CMB calculations with non-cold DM models~\cite{2020JCAP...08..045B,2021EPJP..136..261A} (see also~\cite{2021GrCo...27..315M} which extends this analysis to spinning PBHs).
	
	If the DM particle is very heavy it would not suffer from the WDM constraints~\cite{2019JCAP...05..005M,2018PhRvD..97e5013A,2021arXiv211204836S}. If it is very light, it would participate in the effective number of neutrinos $\Delta N\mrm{eff}$ just like gravitons~\cite{2018JCAP...04..009L, 2019JHEP...08..001H, 2020JCAP...08..014L, 2020arXiv200400618H,2021PhRvD.103l3549A,2021arXiv211013602C,2021GrCo...27..315M}. In general, any particle DM candidate reduces the parameter space available for PBHs in the $M \lesssim 10^9\,$g mass range, but these constraints are highly model-dependent. The DM mass and the PBH mass and density are further degenerate, with heavier PBHs generating more DM, while lighter DM is emitted in greater amounts as the emission starts only at $T \gtrsim m\mrm{DM}$.
	
	The picture is further complicated when allowing for other DM production/decay/annihilation mechanisms, as recently explored in Refs.~\cite{2020PhRvD.102i5018G,2021JCAP...03..007B,2021JCAP...03..015B,2021PhLB..81536129B,2022JCAP...03..031B}. Particle DM directly evaporated by PBHs in the contemporary universe is not thermalized, and it could also be constantly boosted by scattering with the HR products, both mechanisms increasing its detectability in recoil experiments~\cite{2021arXiv210805608C,2022arXiv220314443L,2022arXiv220317093C,2022PhRvD.105b1302C}.\footnote{A scenario of self-interacting particle DM and PBHs with mass $M \gtrsim 10^{18}\,$g would be very tightly constrained because of the accumulation of DM in mini-halos around PBHs; those halos would turn into bright $\gamma$-ray sources. In fact, (interacting) particle DM and asteroid mass PBHs are totally incompatible, which would place the most stringent constraints on the PBH abundance in this heavier mass range (see \textit{e.g.}~\cite{2021MNRAS.506.3648C} and references therein).}
	
	As it seems that melanopogenesis is a perfectly viable scenario for $M \lesssim 10^9\,$g, the possible constraints will come from indirect observations such as $\Delta N\mrm{eff}$ or the stochastic backgrounds of GWs from PBHs (formation, evaporation). The melanopogenesis scenario attracts increasing attention since the late 2010's, motivating the implementation of BSM dofs inside \texttt{BlackHawk v2.0}~\cite{2021EPJC...81..910A} and the development of the dedicated package \texttt{ULYSSES}~\cite{2022PhRvD.105a5022C,2022PhRvD.105a5023C}, which also embeds spinning PBHs.
	
	\section{Conclusion}
	\label{sec:conclusion}
	
	Light primordial black holes are fascinating objects deeply linked to general relativity, quantum mechanics and thermodynamics. Their abundance depends on the conditions in the early universe, and their evaporation gives a privileged access to very high energy physics, way beyond the scope of terrestrial accelerators and even cosmic rays.
	
	In this review, we have discussed the phenomenon of Hawking radiation, which is the process by which PBHs lose mass and finally evaporate. HR may be the only way of constraining the abundance of PBHs: if there are too many of them, they could participate in baryogenesis, spoil BBN, distort the CMB, generate backgrounds of stable particles and even produce the cosmological DM. On the other hand, PBHs themselves, if they survived evaporation until today, may represent a sizeable fraction of the DM, and even 100\% of it in the as yet unconstrained ``HR window'' $10^{17}-10^{23}\,$g. In the near future, this window will shrink by several orders of magnitude, mainly due to $21\,$cm surveys (see Section~\ref{sec:21cm}), MeV astronomy and microlensing data~\cite{2022arXiv220308967B}.
	
\begin{figure}[!t]
		\centering
		\includegraphics{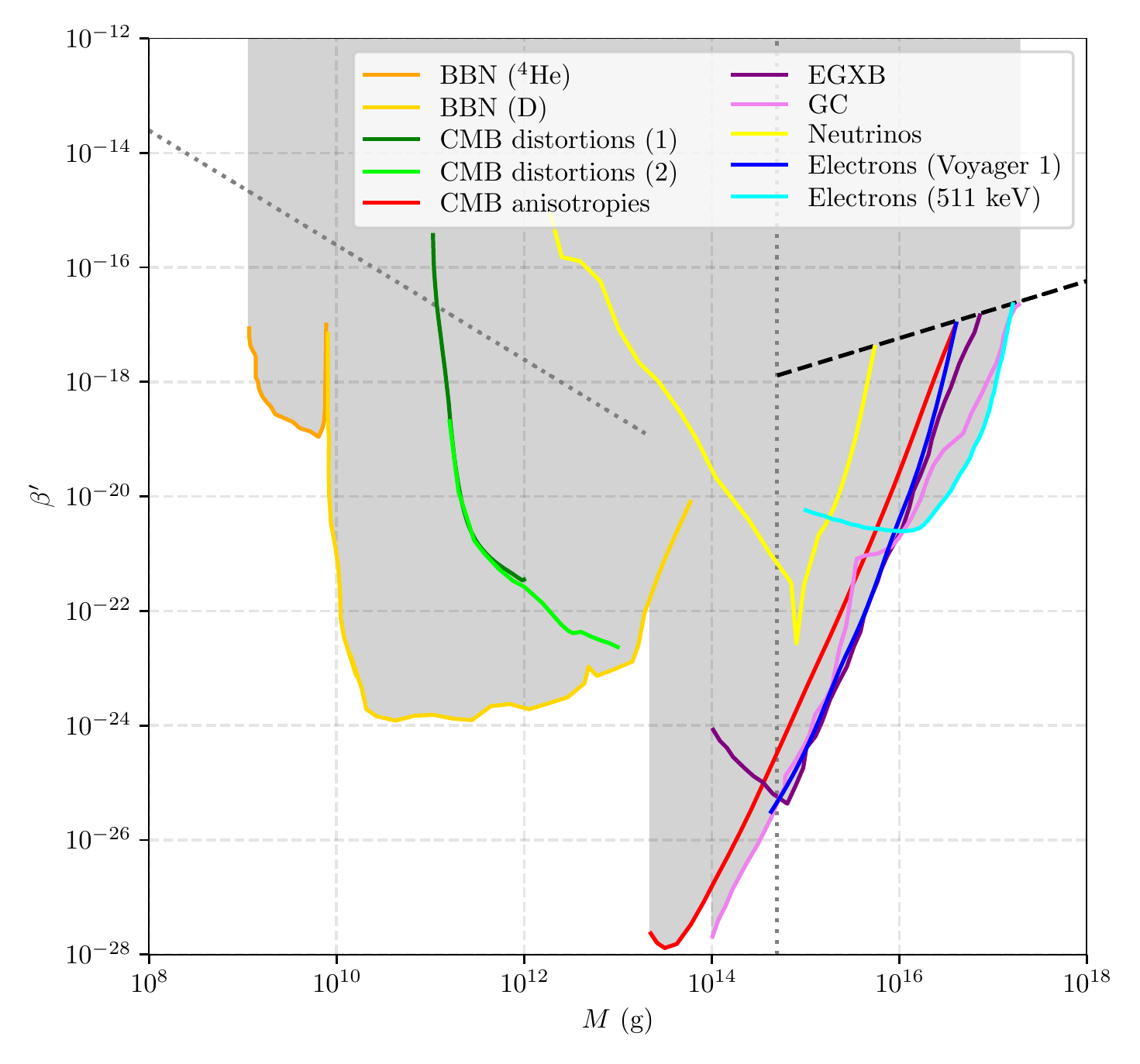}
		\caption{Master constraint plot on PBHs. This plot shows in a single picture the constraints from Figs.~\ref{fig:CMB_distortions} (CMB distortions), \ref{fig:CMB_anisotropies} (CMB anisotropies), \ref{fig:bckg_photons} (diffuse photons), \ref{fig:bckg_neutrinos} (diffuse neutrinos) and \ref{fig:electrons} (electrons), to which we refer the reader for the detailed constraints and references. The critical PBH mass $M_* \sim 5\times 10^{14}\,$g is represented by a vertical dotted grey line. Another dotted grey line separates the regions of RD and transient PBH domination; interrupted at the PBH mass whose lifetime equals the time of matter-radiation equality. The overclosure constraint is shown as a dashed black line, interrupted at $M_*$.
		}
		\label{fig:master}
	\end{figure}

	A summary of the constraints discussed in the present review is given in Fig.~\ref{fig:master}, which represents the conclusion of this work. In this Figure, we see that the mass range $10^9-10^{17}\,$g is tightly constrained by very different and complementary observations : BBN, CMB, the EGXB, neutrinos and electrons. Prospects in all of these domains plus baryogenesis, gravitons, and CRs have been discussed. While it seems that direct observation of the burst of a single PBH may be statistically impossible, cosmological constraints are tightening. Constraints are extended to broad mass distributions, and spinning PBHs are under scrutiny. More speculative work focuses on PBHs with non-standard geometry and particle DM emission by HR is attracting more and more attention.
	
\begin{figure}[!t]
		\centering
		\includegraphics{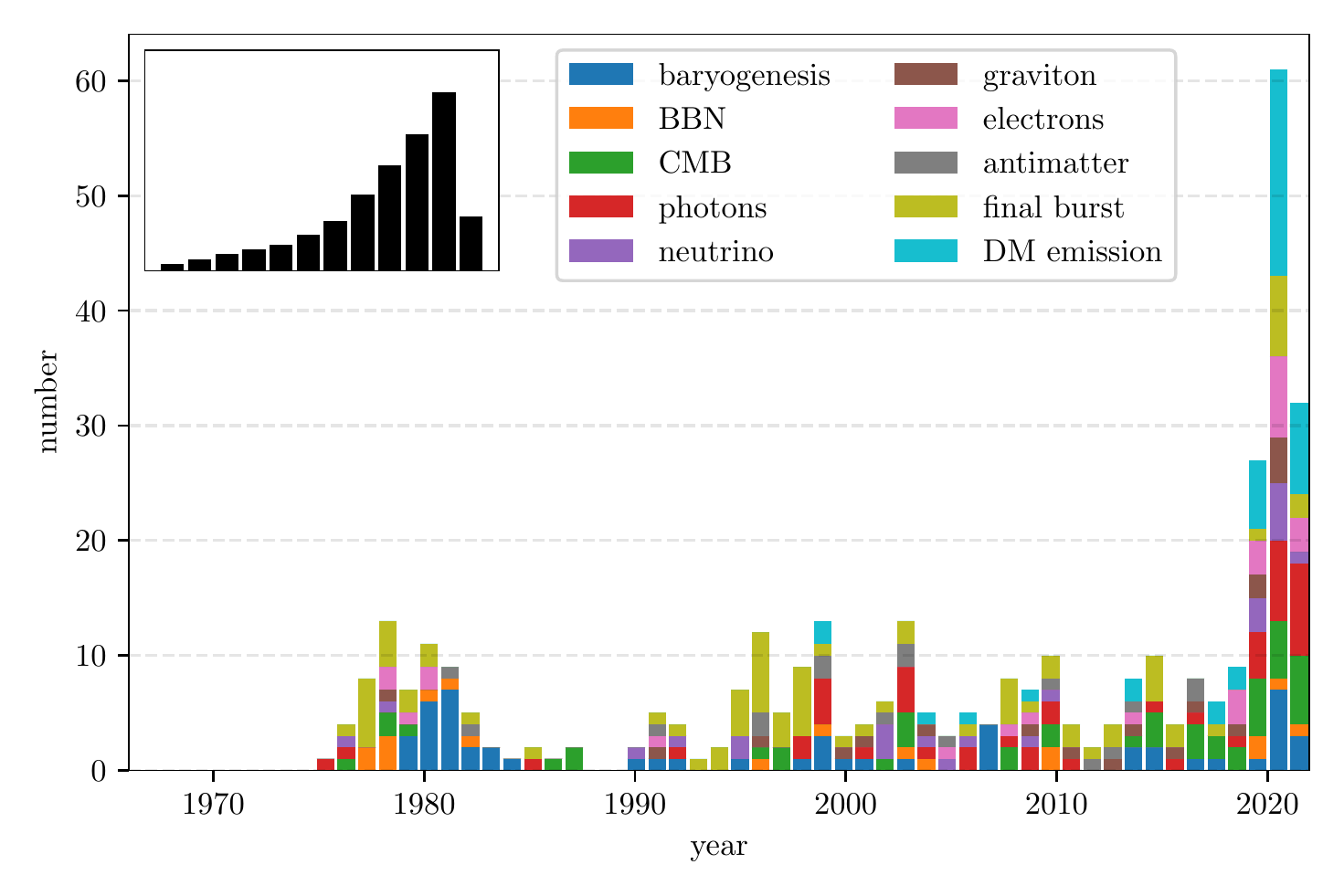}
		\caption{Number of publications concerning PBH constraints, classified by theme. The subplot shows the total number of publications over the same period, with arbitrary units.}
		\label{fig:interest}
	\end{figure}

	In the time of black hole observation, with the latest EHT capture of Sagitarius A* and the increasing catalog of BH merger events from LIGO/VIRGO, PBHs have never received so much attention. In the context of HR constraints, this is shown spectacularly by Fig.~\ref{fig:interest} which presents an histogram of the number of published papers that deal with PBH HR constraints. The subplot shows for comparison the total number of publications during the same period, demonstrating that the two are not correlated. One sees that the general increase in the rate of publication recently hides huge discrepancies between the different ``themes'' used as a classification, and reflects the general history of ideas and observations on top of particular interests of the research teams. The most striking feature is the number of papers that deal with mixed models of particle DM and PBHs since 2019. Historical analysis of this plot would stretch on many (interesting) pages, but we will limit ourselves noting that HR constraints have entered a precision era, with several public numerical tools (\texttt{BlackHawk}, \texttt{CosmoLED}, \texttt{ULYSSES}) and very diverse and complementary observational data (diffuse $\gamma$-rays, neutrino limits, CMB). All proportions kept, it seems that the present epoch represents a golden age of PBH studies, just like the end of the 1970's.\footnote{We may have overlooked papers when building Fig.~\ref{fig:interest}. Some of them are classified into several categories, because they deal with several subjects, and others are not represented (non refeered papers, proceedings, reviews). The general idea is however clear and many details appear on that Figure, such as the ``gap'' of PBH publications in the 1980's-1990's corresponding to the surge in particle DM studies.}
	
	\bibliographystyle{biblio.bst}
	\bibliography{biblio}

\providecommand{\noopsort}[1]{}\providecommand{\singleletter}[1]{#1}%
\begin{thebibliography}{537}%
\makeatletter
\providecommand \@ifxundefined [1]{%
 \@ifx{#1\undefined}
}%
\providecommand \@ifnum [1]{%
 \ifnum #1\expandafter \@firstoftwo
 \else \expandafter \@secondoftwo
 \fi
}%
\providecommand \@ifx [1]{%
 \ifx #1\expandafter \@firstoftwo
 \else \expandafter \@secondoftwo
 \fi
}%
\providecommand \natexlab [1]{#1}%
\providecommand \enquote  [1]{``#1''}%
\providecommand \bibnamefont  [1]{#1}%
\providecommand \bibfnamefont [1]{#1}%
\providecommand \citenamefont [1]{#1}%
\providecommand \href@noop [0]{\@secondoftwo}%
\providecommand \href [0]{\begingroup \@sanitize@url \@href}%
\providecommand \@href[1]{\@@startlink{#1}\@@href}%
\providecommand \@@href[1]{\endgroup#1\@@endlink}%
\providecommand \@sanitize@url [0]{\catcode `\\12\catcode `\$12\catcode
  `\&12\catcode `\#12\catcode `\^12\catcode `\_12\catcode `\%12\relax}%
\providecommand \@@startlink[1]{}%
\providecommand \@@endlink[0]{}%
\providecommand \url  [0]{\begingroup\@sanitize@url \@url }%
\providecommand \@url [1]{\endgroup\@href {#1}{\urlprefix }}%
\providecommand \urlprefix  [0]{URL }%
\providecommand \Eprint [0]{\href }%
\providecommand \doibase [0]{https://doi.org/}%
\providecommand \selectlanguage [0]{\@gobble}%
\providecommand \bibinfo  [0]{\@secondoftwo}%
\providecommand \bibfield  [0]{\@secondoftwo}%
\providecommand \translation [1]{[#1]}%
\providecommand \BibitemOpen [0]{}%
\providecommand \bibitemStop [0]{}%
\providecommand \bibitemNoStop [0]{.\EOS\space}%
\providecommand \EOS [0]{\spacefactor3000\relax}%
\providecommand \BibitemShut  [1]{\csname bibitem#1\endcsname}%
\let\auto@bib@innerbib\@empty
\bibitem [{\citenamefont {{Arbey}}\ and\ \citenamefont
  {{Mahmoudi}}(2021)}]{2021PrPNP.11903865A}%
  \BibitemOpen
  \bibfield  {author} {\bibinfo {author} {\bibfnamefont {A.}~\bibnamefont
  {{Arbey}}}\ and\ \bibinfo {author} {\bibfnamefont {F.}~\bibnamefont
  {{Mahmoudi}}},\ }\href {https://doi.org/10.1016/j.ppnp.2021.103865}
  {\bibfield  {journal} {\bibinfo  {journal} {\PrPNP}\ }\textbf {\bibinfo
  {volume} {119}},\ \bibinfo {eid} {103865} (\bibinfo {year} {2021})},\ \Eprint
  {https://arxiv.org/abs/2104.11488}  { arXiv:2104.11488 [hep-ph]}\BibitemShut
  {NoStop}%
\bibitem [{\citenamefont {{Bertone}}\ and\ \citenamefont
  {{Hooper}}(2018)}]{2018RvMP...90d5002B}%
  \BibitemOpen
  \bibfield  {author} {\bibinfo {author} {\bibfnamefont {G.}~\bibnamefont
  {{Bertone}}}\ and\ \bibinfo {author} {\bibfnamefont {D.}~\bibnamefont
  {{Hooper}}},\ }\href {https://doi.org/10.1103/RevModPhys.90.045002}
  {\bibfield  {journal} {\bibinfo  {journal} {\RMP}\ }\textbf {\bibinfo
  {volume} {90}},\ \bibinfo {eid} {045002} (\bibinfo {year} {2018})},\ \Eprint
  {https://arxiv.org/abs/1605.04909}  { arXiv:1605.04909
  [astro-ph.CO]}\BibitemShut {NoStop}%
\bibitem [{\citenamefont {{Carr}}\ and\ \citenamefont
  {{K{\"u}hnel}}(2020)}]{2020ARNPS..70..355C}%
  \BibitemOpen
  \bibfield  {author} {\bibinfo {author} {\bibfnamefont {B.}~\bibnamefont
  {{Carr}}}\ and\ \bibinfo {author} {\bibfnamefont {F.}~\bibnamefont
  {{K{\"u}hnel}}},\ }\href {https://doi.org/10.1146/annurev-nucl-050520-125911}
  {\bibfield  {journal} {\bibinfo  {journal} {\ARNPS}\ }\textbf {\bibinfo
  {volume} {70}},\ \bibinfo {pages} {355} (\bibinfo {year} {2020})},\ \Eprint
  {https://arxiv.org/abs/2006.02838}  { arXiv:2006.02838
  [astro-ph.CO]}\BibitemShut {NoStop}%
\bibitem [{\citenamefont {{Green}}\ and\ \citenamefont
  {{Kavanagh}}(2021)}]{2021JPhG...48d3001G}%
  \BibitemOpen
  \bibfield  {author} {\bibinfo {author} {\bibfnamefont {A.~M.}\ \bibnamefont
  {{Green}}}\ and\ \bibinfo {author} {\bibfnamefont {B.~J.}\ \bibnamefont
  {{Kavanagh}}},\ }\href {https://doi.org/10.1088/1361-6471/abc534} {\bibfield
  {journal} {\bibinfo  {journal} {\JPhG}\ }\textbf {\bibinfo {volume} {48}},\
  \bibinfo {eid} {043001} (\bibinfo {year} {2021})},\ \Eprint
  {https://arxiv.org/abs/2007.10722}  { arXiv:2007.10722
  [astro-ph.CO]}\BibitemShut {NoStop}%
\bibitem [{\citenamefont {{Bird}}\ \emph {et~al.}(2022)\citenamefont {{Bird}}
  \emph {et~al.}}]{2022arXiv220308967B}%
  \BibitemOpen
  \bibfield  {author} {\bibinfo {author} {\bibfnamefont {S.}~\bibnamefont
  {{Bird}}} \emph {et~al.},\ }\href@noop {} {\bibfield  {journal} {\bibinfo
  {journal} {\arxiv}\ } (\bibinfo {year} {2022})},\ \Eprint
  {https://arxiv.org/abs/2203.08967}  { arXiv:2203.08967 [hep-ph]}\BibitemShut
  {NoStop}%
\bibitem [{\citenamefont {{Sasaki}}\ \emph {et~al.}(2018)\citenamefont
  {{Sasaki}}, \citenamefont {{Suyama}}, \citenamefont {{Tanaka}},\ and\
  \citenamefont {{Yokoyama}}}]{2018CQGra..35f3001S}%
  \BibitemOpen
  \bibfield  {author} {\bibinfo {author} {\bibfnamefont {M.}~\bibnamefont
  {{Sasaki}}}, \bibinfo {author} {\bibfnamefont {T.}~\bibnamefont {{Suyama}}},
  \bibinfo {author} {\bibfnamefont {T.}~\bibnamefont {{Tanaka}}},\ and\
  \bibinfo {author} {\bibfnamefont {S.}~\bibnamefont {{Yokoyama}}},\ }\href
  {https://doi.org/10.1088/1361-6382/aaa7b4} {\bibfield  {journal} {\bibinfo
  {journal} {\CQG}\ }\textbf {\bibinfo {volume} {35}},\ \bibinfo {eid} {063001}
  (\bibinfo {year} {2018})},\ \Eprint {https://arxiv.org/abs/1801.05235}  {
  arXiv:1801.05235 [astro-ph.CO]}\BibitemShut {NoStop}%
\bibitem [{\citenamefont {{Engel}}\ \emph {et~al.}(2022)\citenamefont {{Engel}}
  \emph {et~al.}}]{2022arXiv220307360E}%
  \BibitemOpen
  \bibfield  {author} {\bibinfo {author} {\bibfnamefont {K.}~\bibnamefont
  {{Engel}}} \emph {et~al.},\ }\href@noop {} {\bibfield  {journal} {\bibinfo
  {journal} {\arxiv}\ } (\bibinfo {year} {2022})},\ \Eprint
  {https://arxiv.org/abs/2203.07360}  { arXiv:2203.07360
  [astro-ph.HE]}\BibitemShut {NoStop}%
\bibitem [{\citenamefont {{Capanema}}\ \emph {et~al.}(2021)\citenamefont
  {{Capanema}}, \citenamefont {{Esmaeili}},\ and\ \citenamefont
  {{Esmaili}}}]{2021JCAP...12..051C}%
  \BibitemOpen
  \bibfield  {author} {\bibinfo {author} {\bibfnamefont {A.}~\bibnamefont
  {{Capanema}}}, \bibinfo {author} {\bibfnamefont {A.}~\bibnamefont
  {{Esmaeili}}},\ and\ \bibinfo {author} {\bibfnamefont {A.}~\bibnamefont
  {{Esmaili}}},\ }\href {https://doi.org/10.1088/1475-7516/2021/12/051}
  {\bibfield  {journal} {\bibinfo  {journal} {\jcap}\ }\textbf {\bibinfo
  {volume} {2021}},\ \bibinfo {eid} {051} (\bibinfo {year} {2021})},\ \Eprint
  {https://arxiv.org/abs/2110.05637}  { arXiv:2110.05637 [hep-ph]}\BibitemShut
  {NoStop}%
\bibitem [{\citenamefont {{Carr}}\ \emph
  {et~al.}(2021{\natexlab{a}})\citenamefont {{Carr}}, \citenamefont {{Kohri}},
  \citenamefont {{Sendouda}},\ and\ \citenamefont
  {{Yokoyama}}}]{2021RPPh...84k6902C}%
  \BibitemOpen
  \bibfield  {author} {\bibinfo {author} {\bibfnamefont {B.}~\bibnamefont
  {{Carr}}}, \bibinfo {author} {\bibfnamefont {K.}~\bibnamefont {{Kohri}}},
  \bibinfo {author} {\bibfnamefont {Y.}~\bibnamefont {{Sendouda}}},\ and\
  \bibinfo {author} {\bibfnamefont {J.}~\bibnamefont {{Yokoyama}}},\ }\href
  {https://doi.org/10.1088/1361-6633/ac1e31} {\bibfield  {journal} {\bibinfo
  {journal} {\RPPh}\ }\textbf {\bibinfo {volume} {84}},\ \bibinfo {eid}
  {116902} (\bibinfo {year} {2021}{\natexlab{a}})},\ \Eprint
  {https://arxiv.org/abs/2002.12778}  { arXiv:2002.12778
  [astro-ph.CO]}\BibitemShut {NoStop}%
\bibitem [{\citenamefont {{Arbey}}\ and\ \citenamefont
  {{Auffinger}}(2019)}]{2019EPJC...79..693A}%
  \BibitemOpen
  \bibfield  {author} {\bibinfo {author} {\bibfnamefont {A.}~\bibnamefont
  {{Arbey}}}\ and\ \bibinfo {author} {\bibfnamefont {J.}~\bibnamefont
  {{Auffinger}}},\ }\href {https://doi.org/10.1140/epjc/s10052-019-7161-1}
  {\bibfield  {journal} {\bibinfo  {journal} {\epjc}\ }\textbf {\bibinfo
  {volume} {79}},\ \bibinfo {eid} {693} (\bibinfo {year} {2019})},\ \Eprint
  {https://arxiv.org/abs/1905.04268}  { arXiv:1905.04268 [gr-qc]}\BibitemShut
  {NoStop}%
\bibitem [{\citenamefont {{Friedlander}}\ \emph {et~al.}(2022)\citenamefont
  {{Friedlander}}, \citenamefont {{Mack}}, \citenamefont {{Schon}},
  \citenamefont {{Song}},\ and\ \citenamefont
  {{Vincent}}}]{2022arXiv220111761F}%
  \BibitemOpen
  \bibfield  {author} {\bibinfo {author} {\bibfnamefont {A.}~\bibnamefont
  {{Friedlander}}}, \bibinfo {author} {\bibfnamefont {K.~J.}\ \bibnamefont
  {{Mack}}}, \bibinfo {author} {\bibfnamefont {S.}~\bibnamefont {{Schon}}},
  \bibinfo {author} {\bibfnamefont {N.}~\bibnamefont {{Song}}},\ and\ \bibinfo
  {author} {\bibfnamefont {A.~C.}\ \bibnamefont {{Vincent}}},\ }\href@noop {}
  {\bibfield  {journal} {\bibinfo  {journal} {\arxiv}\ } (\bibinfo {year}
  {2022})},\ \Eprint {https://arxiv.org/abs/2201.11761}  { arXiv:2201.11761
  [hep-ph]}\BibitemShut {NoStop}%
\bibitem [{\citenamefont {{Bernal}}\ \emph
  {et~al.}(2022{\natexlab{a}})\citenamefont {{Bernal}}, \citenamefont
  {{Mu{\~n}oz-Albornoz}}, \citenamefont {{Palomares-Ruiz}},\ and\ \citenamefont
  {{Villanueva-Domingo}}}]{2022arXiv220314979B}%
  \BibitemOpen
  \bibfield  {author} {\bibinfo {author} {\bibfnamefont {N.}~\bibnamefont
  {{Bernal}}}, \bibinfo {author} {\bibfnamefont {V.}~\bibnamefont
  {{Mu{\~n}oz-Albornoz}}}, \bibinfo {author} {\bibfnamefont {S.}~\bibnamefont
  {{Palomares-Ruiz}}},\ and\ \bibinfo {author} {\bibfnamefont {P.}~\bibnamefont
  {{Villanueva-Domingo}}},\ }\href@noop {} {\bibfield  {journal} {\bibinfo
  {journal} {\arxiv}\ } (\bibinfo {year} {2022}{\natexlab{a}})},\ \Eprint
  {https://arxiv.org/abs/2203.14979}  { arXiv:2203.14979 [hep-ph]}\BibitemShut
  {NoStop}%
\bibitem [{\citenamefont {{Cheek}}\ \emph
  {et~al.}(2022{\natexlab{a}})\citenamefont {{Cheek}}, \citenamefont
  {{Heurtier}}, \citenamefont {{Perez-Gonzalez}},\ and\ \citenamefont
  {{Turner}}}]{2022PhRvD.105a5022C}%
  \BibitemOpen
  \bibfield  {author} {\bibinfo {author} {\bibfnamefont {A.}~\bibnamefont
  {{Cheek}}}, \bibinfo {author} {\bibfnamefont {L.}~\bibnamefont {{Heurtier}}},
  \bibinfo {author} {\bibfnamefont {Y.~F.}\ \bibnamefont {{Perez-Gonzalez}}},\
  and\ \bibinfo {author} {\bibfnamefont {J.}~\bibnamefont {{Turner}}},\ }\href
  {https://doi.org/10.1103/PhysRevD.105.015022} {\bibfield  {journal} {\bibinfo
   {journal} {\prd}\ }\textbf {\bibinfo {volume} {105}},\ \bibinfo {eid}
  {015022} (\bibinfo {year} {2022}{\natexlab{a}})},\ \Eprint
  {https://arxiv.org/abs/2107.00013}  { arXiv:2107.00013 [hep-ph]}\BibitemShut
  {NoStop}%
\bibitem [{\citenamefont {{Cheek}}\ \emph
  {et~al.}(2022{\natexlab{b}})\citenamefont {{Cheek}}, \citenamefont
  {{Heurtier}}, \citenamefont {{Perez-Gonzalez}},\ and\ \citenamefont
  {{Turner}}}]{2022PhRvD.105a5023C}%
  \BibitemOpen
  \bibfield  {author} {\bibinfo {author} {\bibfnamefont {A.}~\bibnamefont
  {{Cheek}}}, \bibinfo {author} {\bibfnamefont {L.}~\bibnamefont {{Heurtier}}},
  \bibinfo {author} {\bibfnamefont {Y.~F.}\ \bibnamefont {{Perez-Gonzalez}}},\
  and\ \bibinfo {author} {\bibfnamefont {J.}~\bibnamefont {{Turner}}},\ }\href
  {https://doi.org/10.1103/PhysRevD.105.015023} {\bibfield  {journal} {\bibinfo
   {journal} {\prd}\ }\textbf {\bibinfo {volume} {105}},\ \bibinfo {eid}
  {015023} (\bibinfo {year} {2022}{\natexlab{b}})},\ \Eprint
  {https://arxiv.org/abs/2107.00016}  { arXiv:2107.00016 [hep-ph]}\BibitemShut
  {NoStop}%
\bibitem [{\citenamefont {{Schwarzschild}}(1916)}]{1916SPAW.......189S}%
  \BibitemOpen
  \bibfield  {author} {\bibinfo {author} {\bibfnamefont {K.}~\bibnamefont
  {{Schwarzschild}}},\ }\href@noop {} {\bibfield  {journal} {\bibinfo
  {journal} {Sitzungsberichte der K{\"o}niglich Preu{\ss}ischen Akademie der
  Wissenschaften (Berlin)}\ ,\ \bibinfo {pages} {189}} (\bibinfo {year}
  {1916})},\ \bibinfo {note}
  {[\href{https://arxiv.org/abs/physics/9905030}{arxiv:physics/9905030
  (1999)}]}\BibitemShut {NoStop}%
\bibitem [{\citenamefont {{Landsman}}(2021)}]{2021FoPh...51...42L}%
  \BibitemOpen
  \bibfield  {author} {\bibinfo {author} {\bibfnamefont {K.}~\bibnamefont
  {{Landsman}}},\ }\href {https://doi.org/10.1007/s10701-021-00432-1}
  {\bibfield  {journal} {\bibinfo  {journal} {Found.~Phys.}\ }\textbf {\bibinfo
  {volume} {51}},\ \bibinfo {eid} {42} (\bibinfo {year} {2021})},\ \Eprint
  {https://arxiv.org/abs/2101.02687}  { arXiv:2101.02687 [gr-qc]}\BibitemShut
  {NoStop}%
\bibitem [{\citenamefont {{Van den Bergh}}(1969)}]{1969Natur.224..891V}%
  \BibitemOpen
  \bibfield  {author} {\bibinfo {author} {\bibfnamefont {S.}~\bibnamefont {{Van
  den Bergh}}},\ }\href {https://doi.org/10.1038/224891a0} {\bibfield
  {journal} {\bibinfo  {journal} {\nat}\ }\textbf {\bibinfo {volume} {224}},\
  \bibinfo {pages} {891} (\bibinfo {year} {1969})}\BibitemShut {NoStop}%
\bibitem [{\citenamefont {{Wolfe}}\ and\ \citenamefont
  {{Burbidge}}(1970)}]{1970ApJ...161..419W}%
  \BibitemOpen
  \bibfield  {author} {\bibinfo {author} {\bibfnamefont {A.~M.}\ \bibnamefont
  {{Wolfe}}}\ and\ \bibinfo {author} {\bibfnamefont {G.~R.}\ \bibnamefont
  {{Burbidge}}},\ }\href {https://doi.org/10.1086/150549} {\bibfield  {journal}
  {\bibinfo  {journal} {\apj}\ }\textbf {\bibinfo {volume} {161}},\ \bibinfo
  {pages} {419} (\bibinfo {year} {1970})}\BibitemShut {NoStop}%
\bibitem [{\citenamefont {{Reissner}}(1916)}]{1916AnP...355..106R}%
  \BibitemOpen
  \bibfield  {author} {\bibinfo {author} {\bibfnamefont {H.}~\bibnamefont
  {{Reissner}}},\ }\href {https://doi.org/10.1002/andp.19163550905} {\bibfield
  {journal} {\bibinfo  {journal} {Annalen der Physik}\ }\textbf {\bibinfo
  {volume} {355}},\ \bibinfo {pages} {106} (\bibinfo {year}
  {1916})}\BibitemShut {NoStop}%
\bibitem [{\citenamefont {{Nordstr{\"o}m}}(1918)}]{1918KNAB...20.1238N}%
  \BibitemOpen
  \bibfield  {author} {\bibinfo {author} {\bibfnamefont {G.}~\bibnamefont
  {{Nordstr{\"o}m}}},\ }\href@noop {} {\bibfield  {journal} {\bibinfo
  {journal} {Koninklijke Nederlandse Akademie van Wetenschappen Proceedings
  Series B Physical Sciences}\ }\textbf {\bibinfo {volume} {20}},\ \bibinfo
  {pages} {1238} (\bibinfo {year} {1918})}\BibitemShut {NoStop}%
\bibitem [{\citenamefont {{Kerr}}(1963)}]{1963PhRvL..11..237K}%
  \BibitemOpen
  \bibfield  {author} {\bibinfo {author} {\bibfnamefont {R.~P.}\ \bibnamefont
  {{Kerr}}},\ }\href {https://doi.org/10.1103/PhysRevLett.11.237} {\bibfield
  {journal} {\bibinfo  {journal} {\prl}\ }\textbf {\bibinfo {volume} {11}},\
  \bibinfo {pages} {237} (\bibinfo {year} {1963})}\BibitemShut {NoStop}%
\bibitem [{\citenamefont {{Newman}}\ and\ \citenamefont
  {{Janis}}(1965)}]{1965JMP.....6..915N}%
  \BibitemOpen
  \bibfield  {author} {\bibinfo {author} {\bibfnamefont {E.~T.}\ \bibnamefont
  {{Newman}}}\ and\ \bibinfo {author} {\bibfnamefont {A.~I.}\ \bibnamefont
  {{Janis}}},\ }\href {https://doi.org/10.1063/1.1704350} {\bibfield  {journal}
  {\bibinfo  {journal} {\JMP}\ }\textbf {\bibinfo {volume} {6}},\ \bibinfo
  {pages} {915} (\bibinfo {year} {1965})}\BibitemShut {NoStop}%
\bibitem [{\citenamefont {{Newman}}\ \emph {et~al.}(1965)\citenamefont
  {{Newman}}, \citenamefont {{Couch}}, \citenamefont {{Chinnapared}},
  \citenamefont {{Exton}}, \citenamefont {{Prakash}},\ and\ \citenamefont
  {{Torrence}}}]{1965JMP.....6..918N}%
  \BibitemOpen
  \bibfield  {author} {\bibinfo {author} {\bibfnamefont {E.~T.}\ \bibnamefont
  {{Newman}}}, \bibinfo {author} {\bibfnamefont {E.}~\bibnamefont {{Couch}}},
  \bibinfo {author} {\bibfnamefont {K.}~\bibnamefont {{Chinnapared}}}, \bibinfo
  {author} {\bibfnamefont {A.}~\bibnamefont {{Exton}}}, \bibinfo {author}
  {\bibfnamefont {A.}~\bibnamefont {{Prakash}}},\ and\ \bibinfo {author}
  {\bibfnamefont {R.}~\bibnamefont {{Torrence}}},\ }\href
  {https://doi.org/10.1063/1.1704351} {\bibfield  {journal} {\bibinfo
  {journal} {\JMP}\ }\textbf {\bibinfo {volume} {6}},\ \bibinfo {pages} {918}
  (\bibinfo {year} {1965})}\BibitemShut {NoStop}%
\bibitem [{\citenamefont {{Tolman}}(1939)}]{1939PhRv...55..364T}%
  \BibitemOpen
  \bibfield  {author} {\bibinfo {author} {\bibfnamefont {R.~C.}\ \bibnamefont
  {{Tolman}}},\ }\href {https://doi.org/10.1103/PhysRev.55.364} {\bibfield
  {journal} {\bibinfo  {journal} {\pr}\ }\textbf {\bibinfo {volume} {55}},\
  \bibinfo {pages} {364} (\bibinfo {year} {1939})}\BibitemShut {NoStop}%
\bibitem [{\citenamefont {{Oppenheimer}}\ and\ \citenamefont
  {{Volkoff}}(1939)}]{1939PhRv...55..374O}%
  \BibitemOpen
  \bibfield  {author} {\bibinfo {author} {\bibfnamefont {J.~R.}\ \bibnamefont
  {{Oppenheimer}}}\ and\ \bibinfo {author} {\bibfnamefont {G.~M.}\ \bibnamefont
  {{Volkoff}}},\ }\href {https://doi.org/10.1103/PhysRev.55.374} {\bibfield
  {journal} {\bibinfo  {journal} {\pr}\ }\textbf {\bibinfo {volume} {55}},\
  \bibinfo {pages} {374} (\bibinfo {year} {1939})}\BibitemShut {NoStop}%
\bibitem [{\citenamefont {{Bombaci}}(1996)}]{1996A&A...305..871B}%
  \BibitemOpen
  \bibfield  {author} {\bibinfo {author} {\bibfnamefont {I.}~\bibnamefont
  {{Bombaci}}},\ }\href@noop {} {\bibfield  {journal} {\bibinfo  {journal}
  {\aap}\ }\textbf {\bibinfo {volume} {305}},\ \bibinfo {pages} {871} (\bibinfo
  {year} {1996})}\BibitemShut {NoStop}%
\bibitem [{\citenamefont {{Zel'dovich}}\ and\ \citenamefont
  {{Novikov}}(1967)}]{1967SvA....10..602Z}%
  \BibitemOpen
  \bibfield  {author} {\bibinfo {author} {\bibfnamefont {Y.~B.}\ \bibnamefont
  {{Zel'dovich}}}\ and\ \bibinfo {author} {\bibfnamefont {I.~D.}\ \bibnamefont
  {{Novikov}}},\ }\href@noop {} {\bibfield  {journal} {\bibinfo  {journal}
  {\sovast}\ }\textbf {\bibinfo {volume} {10}},\ \bibinfo {pages} {602}
  (\bibinfo {year} {1967})},\ \bibinfo {note} {[\azh~\textbf{43}, 758
  (1966)]}\BibitemShut {NoStop}%
\bibitem [{\citenamefont {{Novikov}}(1965)}]{1965SvA.....8..857N}%
  \BibitemOpen
  \bibfield  {author} {\bibinfo {author} {\bibfnamefont {I.~D.}\ \bibnamefont
  {{Novikov}}},\ }\href@noop {} {\bibfield  {journal} {\bibinfo  {journal}
  {\sovast}\ }\textbf {\bibinfo {volume} {8}},\ \bibinfo {pages} {857}
  (\bibinfo {year} {1965})},\ \bibinfo {note} {[\azh~\textbf{41}, 1075
  (1964)]}\BibitemShut {NoStop}%
\bibitem [{\citenamefont {{Hawking}}(1971)}]{1971MNRAS.152...75H}%
  \BibitemOpen
  \bibfield  {author} {\bibinfo {author} {\bibfnamefont {S.}~\bibnamefont
  {{Hawking}}},\ }\href {https://doi.org/10.1093/mnras/152.1.75} {\bibfield
  {journal} {\bibinfo  {journal} {\mnras}\ }\textbf {\bibinfo {volume} {152}},\
  \bibinfo {pages} {75} (\bibinfo {year} {1971})}\BibitemShut {NoStop}%
\bibitem [{\citenamefont {{Carr}}\ and\ \citenamefont
  {{Hawking}}(1974)}]{1974MNRAS.168..399C}%
  \BibitemOpen
  \bibfield  {author} {\bibinfo {author} {\bibfnamefont {B.~J.}\ \bibnamefont
  {{Carr}}}\ and\ \bibinfo {author} {\bibfnamefont {S.~W.}\ \bibnamefont
  {{Hawking}}},\ }\href {https://doi.org/10.1093/mnras/168.2.399} {\bibfield
  {journal} {\bibinfo  {journal} {\mnras}\ }\textbf {\bibinfo {volume} {168}},\
  \bibinfo {pages} {399} (\bibinfo {year} {1974})}\BibitemShut {NoStop}%
\bibitem [{\citenamefont {{Meszaros}}(1974)}]{1974A&A....37..225M}%
  \BibitemOpen
  \bibfield  {author} {\bibinfo {author} {\bibfnamefont {P.}~\bibnamefont
  {{Meszaros}}},\ }\href@noop {} {\bibfield  {journal} {\bibinfo  {journal}
  {\aap}\ }\textbf {\bibinfo {volume} {37}},\ \bibinfo {pages} {225} (\bibinfo
  {year} {1974})}\BibitemShut {NoStop}%
\bibitem [{\citenamefont {{Meszaros}}(1975)}]{1975A&A....38....5M}%
  \BibitemOpen
  \bibfield  {author} {\bibinfo {author} {\bibfnamefont {P.}~\bibnamefont
  {{Meszaros}}},\ }\href@noop {} {\bibfield  {journal} {\bibinfo  {journal}
  {\aap}\ }\textbf {\bibinfo {volume} {38}},\ \bibinfo {pages} {5} (\bibinfo
  {year} {1975})}\BibitemShut {NoStop}%
\bibitem [{\citenamefont {{Nadezhin}}\ \emph {et~al.}(1978)\citenamefont
  {{Nadezhin}}, \citenamefont {{Novikov}},\ and\ \citenamefont
  {{Polnarev}}}]{1978SvA....22..129N}%
  \BibitemOpen
  \bibfield  {author} {\bibinfo {author} {\bibfnamefont {D.~K.}\ \bibnamefont
  {{Nadezhin}}}, \bibinfo {author} {\bibfnamefont {I.~D.}\ \bibnamefont
  {{Novikov}}},\ and\ \bibinfo {author} {\bibfnamefont {A.~G.}\ \bibnamefont
  {{Polnarev}}},\ }\href@noop {} {\bibfield  {journal} {\bibinfo  {journal}
  {\sovast}\ }\textbf {\bibinfo {volume} {22}},\ \bibinfo {pages} {129}
  (\bibinfo {year} {1978})},\ \bibinfo {note} {[\azh~\textbf{55}, 216
  (1978)]}\BibitemShut {NoStop}%
\bibitem [{\citenamefont {{Escriv{\`a}}}(2022)}]{2022Univ....8...66E}%
  \BibitemOpen
  \bibfield  {author} {\bibinfo {author} {\bibfnamefont {A.}~\bibnamefont
  {{Escriv{\`a}}}},\ }\href {https://doi.org/10.3390/universe8020066}
  {\bibfield  {journal} {\bibinfo  {journal} {Universe}\ }\textbf {\bibinfo
  {volume} {8}},\ \bibinfo {pages} {66} (\bibinfo {year} {2022})},\ \Eprint
  {https://arxiv.org/abs/2111.12693}  { arXiv:2111.12693 [gr-qc]}\BibitemShut
  {NoStop}%
\bibitem [{\citenamefont {{Akrami}}\ \emph {et~al.}(2020)\citenamefont
  {{Akrami}} \emph {et~al.}}]{2020A&A...641A..10P}%
  \BibitemOpen
  \bibfield  {author} {\bibinfo {author} {\bibfnamefont {Y.}~\bibnamefont
  {{Akrami}}} \emph {et~al.} (\bibinfo {collaboration} {Planck}),\ }\href
  {https://doi.org/10.1051/0004-6361/201833887} {\bibfield  {journal} {\bibinfo
   {journal} {\aap}\ }\textbf {\bibinfo {volume} {641}},\ \bibinfo {eid} {A10}
  (\bibinfo {year} {2020})},\ \Eprint {https://arxiv.org/abs/1807.06211}  {
  arXiv:1807.06211 [astro-ph.CO]}\BibitemShut {NoStop}%
\bibitem [{\citenamefont {{Baumann}}(2011)}]{2011pls..conf..523B}%
  \BibitemOpen
  \bibfield  {author} {\bibinfo {author} {\bibfnamefont {D.}~\bibnamefont
  {{Baumann}}},\ }in\ \href {https://doi.org/10.1142/9789814327183\_0010}
  {\emph {\bibinfo {booktitle} {Physics of the Large and the Small: TASI
  2009}}},\ \bibinfo {editor} {edited by\ \bibinfo {editor} {\bibfnamefont
  {C.}~\bibnamefont {{Csaki}}}\ and\ \bibinfo {editor} {\bibfnamefont
  {S.}~\bibnamefont {{Dodelson}}}}\ (\bibinfo {year} {2011})\ pp.\ \bibinfo
  {pages} {523--686},\ \Eprint {https://arxiv.org/abs/0907.5424}
  {arXiv:0907.5424 [hep-th]} \BibitemShut {NoStop}%
\bibitem [{\citenamefont {{Bedroya}}\ and\ \citenamefont
  {{Vafa}}(2020)}]{2020JHEP...09..123B}%
  \BibitemOpen
  \bibfield  {author} {\bibinfo {author} {\bibfnamefont {A.}~\bibnamefont
  {{Bedroya}}}\ and\ \bibinfo {author} {\bibfnamefont {C.}~\bibnamefont
  {{Vafa}}},\ }\href {https://doi.org/10.1007/JHEP09(2020)123} {\bibfield
  {journal} {\bibinfo  {journal} {\jhep}\ }\textbf {\bibinfo {volume} {2020}},\
  \bibinfo {eid} {123} (\bibinfo {year} {2020})},\ \Eprint
  {https://arxiv.org/abs/1909.11063}  { arXiv:1909.11063 [hep-th]}\BibitemShut
  {NoStop}%
\bibitem [{\citenamefont {{Carr}}\ and\ \citenamefont
  {{Silk}}(2018)}]{2018MNRAS.478.3756C}%
  \BibitemOpen
  \bibfield  {author} {\bibinfo {author} {\bibfnamefont {B.}~\bibnamefont
  {{Carr}}}\ and\ \bibinfo {author} {\bibfnamefont {J.}~\bibnamefont
  {{Silk}}},\ }\href {https://doi.org/10.1093/mnras/sty1204} {\bibfield
  {journal} {\bibinfo  {journal} {\mnras}\ }\textbf {\bibinfo {volume} {478}},\
  \bibinfo {pages} {3756} (\bibinfo {year} {2018})},\ \Eprint
  {https://arxiv.org/abs/1801.00672}  { arXiv:1801.00672
  [astro-ph.CO]}\BibitemShut {NoStop}%
\bibitem [{\citenamefont {{Carr}}\ \emph
  {et~al.}(2021{\natexlab{b}})\citenamefont {{Carr}}, \citenamefont
  {{K{\"u}hnel}},\ and\ \citenamefont {{Visinelli}}}]{2021MNRAS.501.2029C}%
  \BibitemOpen
  \bibfield  {author} {\bibinfo {author} {\bibfnamefont {B.}~\bibnamefont
  {{Carr}}}, \bibinfo {author} {\bibfnamefont {F.}~\bibnamefont
  {{K{\"u}hnel}}},\ and\ \bibinfo {author} {\bibfnamefont {L.}~\bibnamefont
  {{Visinelli}}},\ }\href {https://doi.org/10.1093/mnras/staa3651} {\bibfield
  {journal} {\bibinfo  {journal} {\mnras}\ }\textbf {\bibinfo {volume} {501}},\
  \bibinfo {pages} {2029} (\bibinfo {year} {2021}{\natexlab{b}})},\ \Eprint
  {https://arxiv.org/abs/2008.08077}  { arXiv:2008.08077
  [astro-ph.CO]}\BibitemShut {NoStop}%
\bibitem [{\citenamefont {{Hawking}}\ \emph {et~al.}(1982)\citenamefont
  {{Hawking}}, \citenamefont {{Moss}},\ and\ \citenamefont
  {{Stewart}}}]{1982PhRvD..26.2681H}%
  \BibitemOpen
  \bibfield  {author} {\bibinfo {author} {\bibfnamefont {S.~W.}\ \bibnamefont
  {{Hawking}}}, \bibinfo {author} {\bibfnamefont {I.~G.}\ \bibnamefont
  {{Moss}}},\ and\ \bibinfo {author} {\bibfnamefont {J.~M.}\ \bibnamefont
  {{Stewart}}},\ }\href {https://doi.org/10.1103/PhysRevD.26.2681} {\bibfield
  {journal} {\bibinfo  {journal} {\prd}\ }\textbf {\bibinfo {volume} {26}},\
  \bibinfo {pages} {2681} (\bibinfo {year} {1982})}\BibitemShut {NoStop}%
\bibitem [{\citenamefont {{Liddle}}\ and\ \citenamefont
  {{Green}}(1998)}]{1998PhR...307..125L}%
  \BibitemOpen
  \bibfield  {author} {\bibinfo {author} {\bibfnamefont {A.~R.}\ \bibnamefont
  {{Liddle}}}\ and\ \bibinfo {author} {\bibfnamefont {A.~M.}\ \bibnamefont
  {{Green}}},\ }\href {https://doi.org/10.1016/S0370-1573(98)00069-6}
  {\bibfield  {journal} {\bibinfo  {journal} {\physrep}\ }\textbf {\bibinfo
  {volume} {307}},\ \bibinfo {pages} {125} (\bibinfo {year} {1998})},\ \Eprint
  {https://arxiv.org/abs/gr-qc/9804034}  { arXiv:gr-qc/9804034
  [gr-qc]}\BibitemShut {NoStop}%
\bibitem [{\citenamefont {{Khlopov}}(2010)}]{2010RAA....10..495K}%
  \BibitemOpen
  \bibfield  {author} {\bibinfo {author} {\bibfnamefont {M.~Y.}\ \bibnamefont
  {{Khlopov}}},\ }\href {https://doi.org/10.1088/1674-4527/10/6/001} {\bibfield
   {journal} {\bibinfo  {journal} {Res.~Astron.~Astrophys.}\ }\textbf {\bibinfo
  {volume} {10}},\ \bibinfo {pages} {495} (\bibinfo {year} {2010})},\ \Eprint
  {https://arxiv.org/abs/0801.0116}  { arXiv:0801.0116 [astro-ph]}\BibitemShut
  {NoStop}%
\bibitem [{\citenamefont {{Green}}\ \emph {et~al.}(1997)\citenamefont
  {{Green}}, \citenamefont {{Liddle}},\ and\ \citenamefont
  {{Riotto}}}]{1997PhRvD..56.7559G}%
  \BibitemOpen
  \bibfield  {author} {\bibinfo {author} {\bibfnamefont {A.~M.}\ \bibnamefont
  {{Green}}}, \bibinfo {author} {\bibfnamefont {A.~R.}\ \bibnamefont
  {{Liddle}}},\ and\ \bibinfo {author} {\bibfnamefont {A.}~\bibnamefont
  {{Riotto}}},\ }\href {https://doi.org/10.1103/PhysRevD.56.7559} {\bibfield
  {journal} {\bibinfo  {journal} {\prd}\ }\textbf {\bibinfo {volume} {56}},\
  \bibinfo {pages} {7559} (\bibinfo {year} {1997})},\ \Eprint
  {https://arxiv.org/abs/astro-ph/9705166}  { arXiv:astro-ph/9705166
  [astro-ph]}\BibitemShut {NoStop}%
\bibitem [{\citenamefont {{Kane}}\ \emph {et~al.}(2015)\citenamefont {{Kane}},
  \citenamefont {{Sinha}},\ and\ \citenamefont
  {{Watson}}}]{2015IJMPD..2430022K}%
  \BibitemOpen
  \bibfield  {author} {\bibinfo {author} {\bibfnamefont {G.}~\bibnamefont
  {{Kane}}}, \bibinfo {author} {\bibfnamefont {K.}~\bibnamefont {{Sinha}}},\
  and\ \bibinfo {author} {\bibfnamefont {S.}~\bibnamefont {{Watson}}},\ }\href
  {https://doi.org/10.1142/S0218271815300220} {\bibfield  {journal} {\bibinfo
  {journal} {\IJMPD}\ }\textbf {\bibinfo {volume} {24}},\ \bibinfo {eid}
  {1530022-324} (\bibinfo {year} {2015})},\ \Eprint
  {https://arxiv.org/abs/1502.07746}  { arXiv:1502.07746 [hep-th]}\BibitemShut
  {NoStop}%
\bibitem [{\citenamefont {{Harada}}\ \emph {et~al.}(2016)\citenamefont
  {{Harada}}, \citenamefont {{Yoo}}, \citenamefont {{Kohri}}, \citenamefont
  {{Nakao}},\ and\ \citenamefont {{Jhingan}}}]{2016ApJ...833...61H}%
  \BibitemOpen
  \bibfield  {author} {\bibinfo {author} {\bibfnamefont {T.}~\bibnamefont
  {{Harada}}}, \bibinfo {author} {\bibfnamefont {C.-m.}\ \bibnamefont {{Yoo}}},
  \bibinfo {author} {\bibfnamefont {K.}~\bibnamefont {{Kohri}}}, \bibinfo
  {author} {\bibfnamefont {K.-i.}\ \bibnamefont {{Nakao}}},\ and\ \bibinfo
  {author} {\bibfnamefont {S.}~\bibnamefont {{Jhingan}}},\ }\href
  {https://doi.org/10.3847/1538-4357/833/1/61} {\bibfield  {journal} {\bibinfo
  {journal} {\apj}\ }\textbf {\bibinfo {volume} {833}},\ \bibinfo {eid} {61}
  (\bibinfo {year} {2016})},\ \Eprint {https://arxiv.org/abs/1609.01588}  {
  arXiv:1609.01588 [astro-ph.CO]}\BibitemShut {NoStop}%
\bibitem [{\citenamefont {{Georg}}\ \emph {et~al.}(2016)\citenamefont
  {{Georg}}, \citenamefont {{{\c{S}}eng{\"o}r}},\ and\ \citenamefont
  {{Watson}}}]{2016PhRvD..93l3523G}%
  \BibitemOpen
  \bibfield  {author} {\bibinfo {author} {\bibfnamefont {J.}~\bibnamefont
  {{Georg}}}, \bibinfo {author} {\bibfnamefont {G.}~\bibnamefont
  {{{\c{S}}eng{\"o}r}}},\ and\ \bibinfo {author} {\bibfnamefont
  {S.}~\bibnamefont {{Watson}}},\ }\href
  {https://doi.org/10.1103/PhysRevD.93.123523} {\bibfield  {journal} {\bibinfo
  {journal} {\prd}\ }\textbf {\bibinfo {volume} {93}},\ \bibinfo {eid} {123523}
  (\bibinfo {year} {2016})},\ \Eprint {https://arxiv.org/abs/1603.00023}  {
  arXiv:1603.00023 [hep-ph]}\BibitemShut {NoStop}%
\bibitem [{\citenamefont {{Georg}}\ and\ \citenamefont
  {{Watson}}(2017)}]{2017JHEP...09..138G}%
  \BibitemOpen
  \bibfield  {author} {\bibinfo {author} {\bibfnamefont {J.}~\bibnamefont
  {{Georg}}}\ and\ \bibinfo {author} {\bibfnamefont {S.}~\bibnamefont
  {{Watson}}},\ }\href {https://doi.org/10.1007/JHEP09(2017)138} {\bibfield
  {journal} {\bibinfo  {journal} {\jhep}\ }\textbf {\bibinfo {volume} {2017}},\
  \bibinfo {eid} {138} (\bibinfo {year} {2017})},\ \Eprint
  {https://arxiv.org/abs/1703.04825}  { arXiv:1703.04825
  [astro-ph.CO]}\BibitemShut {NoStop}%
\bibitem [{\citenamefont {{Kokubu}}\ \emph {et~al.}(2018)\citenamefont
  {{Kokubu}}, \citenamefont {{Kyutoku}}, \citenamefont {{Kohri}},\ and\
  \citenamefont {{Harada}}}]{2018PhRvD..98l3024K}%
  \BibitemOpen
  \bibfield  {author} {\bibinfo {author} {\bibfnamefont {T.}~\bibnamefont
  {{Kokubu}}}, \bibinfo {author} {\bibfnamefont {K.}~\bibnamefont {{Kyutoku}}},
  \bibinfo {author} {\bibfnamefont {K.}~\bibnamefont {{Kohri}}},\ and\ \bibinfo
  {author} {\bibfnamefont {T.}~\bibnamefont {{Harada}}},\ }\href
  {https://doi.org/10.1103/PhysRevD.98.123024} {\bibfield  {journal} {\bibinfo
  {journal} {\prd}\ }\textbf {\bibinfo {volume} {98}},\ \bibinfo {eid} {123024}
  (\bibinfo {year} {2018})},\ \Eprint {https://arxiv.org/abs/1810.03490}  {
  arXiv:1810.03490 [astro-ph.CO]}\BibitemShut {NoStop}%
\bibitem [{\citenamefont {{Georg}}\ \emph {et~al.}(2019)\citenamefont
  {{Georg}}, \citenamefont {{Melcher}},\ and\ \citenamefont
  {{Watson}}}]{2019JCAP...11..014G}%
  \BibitemOpen
  \bibfield  {author} {\bibinfo {author} {\bibfnamefont {J.}~\bibnamefont
  {{Georg}}}, \bibinfo {author} {\bibfnamefont {B.}~\bibnamefont {{Melcher}}},\
  and\ \bibinfo {author} {\bibfnamefont {S.}~\bibnamefont {{Watson}}},\ }\href
  {https://doi.org/10.1088/1475-7516/2019/11/014} {\bibfield  {journal}
  {\bibinfo  {journal} {\jcap}\ }\textbf {\bibinfo {volume} {2019}},\ \bibinfo
  {eid} {014} (\bibinfo {year} {2019})},\ \Eprint
  {https://arxiv.org/abs/1902.04082}  { arXiv:1902.04082
  [astro-ph.CO]}\BibitemShut {NoStop}%
\bibitem [{\citenamefont {{Matsubara}}\ \emph {et~al.}(2019)\citenamefont
  {{Matsubara}}, \citenamefont {{Terada}}, \citenamefont {{Kohri}},\ and\
  \citenamefont {{Yokoyama}}}]{2019PhRvD.100l3544M}%
  \BibitemOpen
  \bibfield  {author} {\bibinfo {author} {\bibfnamefont {T.}~\bibnamefont
  {{Matsubara}}}, \bibinfo {author} {\bibfnamefont {T.}~\bibnamefont
  {{Terada}}}, \bibinfo {author} {\bibfnamefont {K.}~\bibnamefont {{Kohri}}},\
  and\ \bibinfo {author} {\bibfnamefont {S.}~\bibnamefont {{Yokoyama}}},\
  }\href {https://doi.org/10.1103/PhysRevD.100.123544} {\bibfield  {journal}
  {\bibinfo  {journal} {\prd}\ }\textbf {\bibinfo {volume} {100}},\ \bibinfo
  {eid} {123544} (\bibinfo {year} {2019})},\ \Eprint
  {https://arxiv.org/abs/1909.04053}  { arXiv:1909.04053
  [astro-ph.CO]}\BibitemShut {NoStop}%
\bibitem [{\citenamefont {{Carr}}(1975)}]{1975ApJ...201....1C}%
  \BibitemOpen
  \bibfield  {author} {\bibinfo {author} {\bibfnamefont {B.~J.}\ \bibnamefont
  {{Carr}}},\ }\href {https://doi.org/10.1086/153853} {\bibfield  {journal}
  {\bibinfo  {journal} {\apj}\ }\textbf {\bibinfo {volume} {201}},\ \bibinfo
  {pages} {1} (\bibinfo {year} {1975})}\BibitemShut {NoStop}%
\bibitem [{\citenamefont {{Yokoyama}}(1998)}]{1998PhRvD..58j7502Y}%
  \BibitemOpen
  \bibfield  {author} {\bibinfo {author} {\bibfnamefont {J.}~\bibnamefont
  {{Yokoyama}}},\ }\href {https://doi.org/10.1103/PhysRevD.58.107502}
  {\bibfield  {journal} {\bibinfo  {journal} {\prd}\ }\textbf {\bibinfo
  {volume} {58}},\ \bibinfo {eid} {107502} (\bibinfo {year} {1998})},\ \Eprint
  {https://arxiv.org/abs/gr-qc/9804041}  { arXiv:gr-qc/9804041
  [gr-qc]}\BibitemShut {NoStop}%
\bibitem [{\citenamefont {{Niemeyer}}\ and\ \citenamefont
  {{Jedamzik}}(1998)}]{1998PhRvL..80.5481N}%
  \BibitemOpen
  \bibfield  {author} {\bibinfo {author} {\bibfnamefont {J.~C.}\ \bibnamefont
  {{Niemeyer}}}\ and\ \bibinfo {author} {\bibfnamefont {K.}~\bibnamefont
  {{Jedamzik}}},\ }\href {https://doi.org/10.1103/PhysRevLett.80.5481}
  {\bibfield  {journal} {\bibinfo  {journal} {\prl}\ }\textbf {\bibinfo
  {volume} {80}},\ \bibinfo {pages} {5481} (\bibinfo {year} {1998})},\ \Eprint
  {https://arxiv.org/abs/astro-ph/9709072}  { arXiv:astro-ph/9709072
  [astro-ph]}\BibitemShut {NoStop}%
\bibitem [{\citenamefont {{Niemeyer}}\ and\ \citenamefont
  {{Jedamzik}}(1999)}]{1999PhRvD..59l4013N}%
  \BibitemOpen
  \bibfield  {author} {\bibinfo {author} {\bibfnamefont {J.~C.}\ \bibnamefont
  {{Niemeyer}}}\ and\ \bibinfo {author} {\bibfnamefont {K.}~\bibnamefont
  {{Jedamzik}}},\ }\href {https://doi.org/10.1103/PhysRevD.59.124013}
  {\bibfield  {journal} {\bibinfo  {journal} {\prd}\ }\textbf {\bibinfo
  {volume} {59}},\ \bibinfo {eid} {124013} (\bibinfo {year} {1999})},\ \Eprint
  {https://arxiv.org/abs/astro-ph/9901292}  { arXiv:astro-ph/9901292
  [astro-ph]}\BibitemShut {NoStop}%
\bibitem [{\citenamefont {{Dolgov}}\ and\ \citenamefont
  {{Silk}}(1993)}]{1993PhRvD..47.4244D}%
  \BibitemOpen
  \bibfield  {author} {\bibinfo {author} {\bibfnamefont {A.}~\bibnamefont
  {{Dolgov}}}\ and\ \bibinfo {author} {\bibfnamefont {J.}~\bibnamefont
  {{Silk}}},\ }\href {https://doi.org/10.1103/PhysRevD.47.4244} {\bibfield
  {journal} {\bibinfo  {journal} {\prd}\ }\textbf {\bibinfo {volume} {47}},\
  \bibinfo {pages} {4244} (\bibinfo {year} {1993})}\BibitemShut {NoStop}%
\bibitem [{\citenamefont {{Tashiro}}\ and\ \citenamefont
  {{Sugiyama}}(2008)}]{2008PhRvD..78b3004T}%
  \BibitemOpen
  \bibfield  {author} {\bibinfo {author} {\bibfnamefont {H.}~\bibnamefont
  {{Tashiro}}}\ and\ \bibinfo {author} {\bibfnamefont {N.}~\bibnamefont
  {{Sugiyama}}},\ }\href {https://doi.org/10.1103/PhysRevD.78.023004}
  {\bibfield  {journal} {\bibinfo  {journal} {\prd}\ }\textbf {\bibinfo
  {volume} {78}},\ \bibinfo {eid} {023004} (\bibinfo {year} {2008})},\ \Eprint
  {https://arxiv.org/abs/0801.3172}  { arXiv:0801.3172 [astro-ph]}\BibitemShut
  {NoStop}%
\bibitem [{\citenamefont {{Germani}}\ and\ \citenamefont
  {{Musco}}(2019)}]{2019PhRvL.122n1302G}%
  \BibitemOpen
  \bibfield  {author} {\bibinfo {author} {\bibfnamefont {C.}~\bibnamefont
  {{Germani}}}\ and\ \bibinfo {author} {\bibfnamefont {I.}~\bibnamefont
  {{Musco}}},\ }\href {https://doi.org/10.1103/PhysRevLett.122.141302}
  {\bibfield  {journal} {\bibinfo  {journal} {\prl}\ }\textbf {\bibinfo
  {volume} {122}},\ \bibinfo {eid} {141302} (\bibinfo {year} {2019})},\ \Eprint
  {https://arxiv.org/abs/1805.04087}  { arXiv:1805.04087
  [astro-ph.CO]}\BibitemShut {NoStop}%
\bibitem [{\citenamefont {{Green}}(2016)}]{2016PhRvD..94f3530G}%
  \BibitemOpen
  \bibfield  {author} {\bibinfo {author} {\bibfnamefont {A.~M.}\ \bibnamefont
  {{Green}}},\ }\href {https://doi.org/10.1103/PhysRevD.94.063530} {\bibfield
  {journal} {\bibinfo  {journal} {\prd}\ }\textbf {\bibinfo {volume} {94}},\
  \bibinfo {eid} {063530} (\bibinfo {year} {2016})},\ \Eprint
  {https://arxiv.org/abs/1609.01143}  { arXiv:1609.01143
  [astro-ph.CO]}\BibitemShut {NoStop}%
\bibitem [{\citenamefont {{Carr}}\ \emph {et~al.}(2017)\citenamefont {{Carr}},
  \citenamefont {{Raidal}}, \citenamefont {{Tenkanen}}, \citenamefont
  {{Vaskonen}},\ and\ \citenamefont {{Veerm{\"a}e}}}]{2017PhRvD..96b3514C}%
  \BibitemOpen
  \bibfield  {author} {\bibinfo {author} {\bibfnamefont {B.}~\bibnamefont
  {{Carr}}}, \bibinfo {author} {\bibfnamefont {M.}~\bibnamefont {{Raidal}}},
  \bibinfo {author} {\bibfnamefont {T.}~\bibnamefont {{Tenkanen}}}, \bibinfo
  {author} {\bibfnamefont {V.}~\bibnamefont {{Vaskonen}}},\ and\ \bibinfo
  {author} {\bibfnamefont {H.}~\bibnamefont {{Veerm{\"a}e}}},\ }\href
  {https://doi.org/10.1103/PhysRevD.96.023514} {\bibfield  {journal} {\bibinfo
  {journal} {\prd}\ }\textbf {\bibinfo {volume} {96}},\ \bibinfo {eid} {023514}
  (\bibinfo {year} {2017})},\ \Eprint {https://arxiv.org/abs/1705.05567}  {
  arXiv:1705.05567 [astro-ph.CO]}\BibitemShut {NoStop}%
\bibitem [{\citenamefont {{Inomata}}\ \emph {et~al.}(2017)\citenamefont
  {{Inomata}}, \citenamefont {{Kawasaki}}, \citenamefont {{Mukaida}},
  \citenamefont {{Tada}},\ and\ \citenamefont
  {{Yanagida}}}]{2017PhRvD..96d3504I}%
  \BibitemOpen
  \bibfield  {author} {\bibinfo {author} {\bibfnamefont {K.}~\bibnamefont
  {{Inomata}}}, \bibinfo {author} {\bibfnamefont {M.}~\bibnamefont
  {{Kawasaki}}}, \bibinfo {author} {\bibfnamefont {K.}~\bibnamefont
  {{Mukaida}}}, \bibinfo {author} {\bibfnamefont {Y.}~\bibnamefont {{Tada}}},\
  and\ \bibinfo {author} {\bibfnamefont {T.~T.}\ \bibnamefont {{Yanagida}}},\
  }\href {https://doi.org/10.1103/PhysRevD.96.043504} {\bibfield  {journal}
  {\bibinfo  {journal} {\prd}\ }\textbf {\bibinfo {volume} {96}},\ \bibinfo
  {eid} {043504} (\bibinfo {year} {2017})},\ \Eprint
  {https://arxiv.org/abs/1701.02544}  { arXiv:1701.02544
  [astro-ph.CO]}\BibitemShut {NoStop}%
\bibitem [{\citenamefont {{K{\"u}hnel}}\ and\ \citenamefont
  {{Freese}}(2017)}]{2017PhRvD..95h3508K}%
  \BibitemOpen
  \bibfield  {author} {\bibinfo {author} {\bibfnamefont {F.}~\bibnamefont
  {{K{\"u}hnel}}}\ and\ \bibinfo {author} {\bibfnamefont {K.}~\bibnamefont
  {{Freese}}},\ }\href {https://doi.org/10.1103/PhysRevD.95.083508} {\bibfield
  {journal} {\bibinfo  {journal} {\prd}\ }\textbf {\bibinfo {volume} {95}},\
  \bibinfo {eid} {083508} (\bibinfo {year} {2017})},\ \Eprint
  {https://arxiv.org/abs/1701.07223}  { arXiv:1701.07223
  [astro-ph.CO]}\BibitemShut {NoStop}%
\bibitem [{\citenamefont {{Bellomo}}\ \emph {et~al.}(2018)\citenamefont
  {{Bellomo}}, \citenamefont {{Bernal}}, \citenamefont {{Raccanelli}},\ and\
  \citenamefont {{Verde}}}]{2018JCAP...01..004B}%
  \BibitemOpen
  \bibfield  {author} {\bibinfo {author} {\bibfnamefont {N.}~\bibnamefont
  {{Bellomo}}}, \bibinfo {author} {\bibfnamefont {J.~L.}\ \bibnamefont
  {{Bernal}}}, \bibinfo {author} {\bibfnamefont {A.}~\bibnamefont
  {{Raccanelli}}},\ and\ \bibinfo {author} {\bibfnamefont {L.}~\bibnamefont
  {{Verde}}},\ }\href {https://doi.org/10.1088/1475-7516/2018/01/004}
  {\bibfield  {journal} {\bibinfo  {journal} {\jcap}\ }\textbf {\bibinfo
  {volume} {2018}},\ \bibinfo {eid} {004} (\bibinfo {year} {2018})},\ \Eprint
  {https://arxiv.org/abs/1709.07467}  { arXiv:1709.07467
  [astro-ph.CO]}\BibitemShut {NoStop}%
\bibitem [{\citenamefont {{Poulter}}\ \emph {et~al.}(2019)\citenamefont
  {{Poulter}}, \citenamefont {{Ali-Ha{\"\i}moud}}, \citenamefont {{Hamann}},
  \citenamefont {{White}},\ and\ \citenamefont
  {{Williams}}}]{2019arXiv190706485P}%
  \BibitemOpen
  \bibfield  {author} {\bibinfo {author} {\bibfnamefont {H.}~\bibnamefont
  {{Poulter}}}, \bibinfo {author} {\bibfnamefont {Y.}~\bibnamefont
  {{Ali-Ha{\"\i}moud}}}, \bibinfo {author} {\bibfnamefont {J.}~\bibnamefont
  {{Hamann}}}, \bibinfo {author} {\bibfnamefont {M.}~\bibnamefont {{White}}},\
  and\ \bibinfo {author} {\bibfnamefont {A.~G.}\ \bibnamefont {{Williams}}},\
  }\href@noop {} {\bibfield  {journal} {\bibinfo  {journal} {\arxiv}\ }
  (\bibinfo {year} {2019})},\ \Eprint {https://arxiv.org/abs/1907.06485}  {
  arXiv:1907.06485 [astro-ph.CO]}\BibitemShut {NoStop}%
\bibitem [{\citenamefont {{De Luca}}\ \emph {et~al.}(2019)\citenamefont {{De
  Luca}}, \citenamefont {{Desjacques}}, \citenamefont {{Franciolini}},
  \citenamefont {{Malhotra}},\ and\ \citenamefont
  {{Riotto}}}]{2019JCAP...05..018D}%
  \BibitemOpen
  \bibfield  {author} {\bibinfo {author} {\bibfnamefont {V.}~\bibnamefont {{De
  Luca}}}, \bibinfo {author} {\bibfnamefont {V.}~\bibnamefont {{Desjacques}}},
  \bibinfo {author} {\bibfnamefont {G.}~\bibnamefont {{Franciolini}}}, \bibinfo
  {author} {\bibfnamefont {A.}~\bibnamefont {{Malhotra}}},\ and\ \bibinfo
  {author} {\bibfnamefont {A.}~\bibnamefont {{Riotto}}},\ }\href
  {https://doi.org/10.1088/1475-7516/2019/05/018} {\bibfield  {journal}
  {\bibinfo  {journal} {\jcap}\ }\textbf {\bibinfo {volume} {2019}},\ \bibinfo
  {eid} {018} (\bibinfo {year} {2019})},\ \Eprint
  {https://arxiv.org/abs/1903.01179}  { arXiv:1903.01179
  [astro-ph.CO]}\BibitemShut {NoStop}%
\bibitem [{\citenamefont {{Chiba}}\ and\ \citenamefont
  {{Yokoyama}}(2017)}]{2017PTEP.2017h3E01C}%
  \BibitemOpen
  \bibfield  {author} {\bibinfo {author} {\bibfnamefont {T.}~\bibnamefont
  {{Chiba}}}\ and\ \bibinfo {author} {\bibfnamefont {S.}~\bibnamefont
  {{Yokoyama}}},\ }\href {https://doi.org/10.1093/ptep/ptx087} {\bibfield
  {journal} {\bibinfo  {journal} {\PTEP}\ }\textbf {\bibinfo {volume} {2017}},\
  \bibinfo {eid} {083E01} (\bibinfo {year} {2017})},\ \Eprint
  {https://arxiv.org/abs/1704.06573}  { arXiv:1704.06573 [gr-qc]}\BibitemShut
  {NoStop}%
\bibitem [{\citenamefont {{Harada}}\ \emph {et~al.}(2017)\citenamefont
  {{Harada}}, \citenamefont {{Yoo}}, \citenamefont {{Kohri}},\ and\
  \citenamefont {{Nakao}}}]{2017PhRvD..96h3517H}%
  \BibitemOpen
  \bibfield  {author} {\bibinfo {author} {\bibfnamefont {T.}~\bibnamefont
  {{Harada}}}, \bibinfo {author} {\bibfnamefont {C.-M.}\ \bibnamefont {{Yoo}}},
  \bibinfo {author} {\bibfnamefont {K.}~\bibnamefont {{Kohri}}},\ and\ \bibinfo
  {author} {\bibfnamefont {K.-I.}\ \bibnamefont {{Nakao}}},\ }\href
  {https://doi.org/10.1103/PhysRevD.96.083517} {\bibfield  {journal} {\bibinfo
  {journal} {\prd}\ }\textbf {\bibinfo {volume} {96}},\ \bibinfo {eid} {083517}
  (\bibinfo {year} {2017})},\ \Eprint {https://arxiv.org/abs/1707.03595}  {
  arXiv:1707.03595 [gr-qc]}\BibitemShut {NoStop}%
\bibitem [{\citenamefont {{Cotner}}\ and\ \citenamefont
  {{Kusenko}}(2017)}]{2017PhRvD..96j3002C}%
  \BibitemOpen
  \bibfield  {author} {\bibinfo {author} {\bibfnamefont {E.}~\bibnamefont
  {{Cotner}}}\ and\ \bibinfo {author} {\bibfnamefont {A.}~\bibnamefont
  {{Kusenko}}},\ }\href {https://doi.org/10.1103/PhysRevD.96.103002} {\bibfield
   {journal} {\bibinfo  {journal} {\prd}\ }\textbf {\bibinfo {volume} {96}},\
  \bibinfo {eid} {103002} (\bibinfo {year} {2017})},\ \Eprint
  {https://arxiv.org/abs/1706.09003}  { arXiv:1706.09003
  [astro-ph.CO]}\BibitemShut {NoStop}%
\bibitem [{\citenamefont {{Allahverdi}}\ \emph {et~al.}(2021)\citenamefont
  {{Allahverdi}} \emph {et~al.}}]{2021OJAp....4E...1A}%
  \BibitemOpen
  \bibfield  {author} {\bibinfo {author} {\bibfnamefont {R.}~\bibnamefont
  {{Allahverdi}}} \emph {et~al.},\ }\href
  {https://doi.org/10.21105/astro.2006.16182} {\bibfield  {journal} {\bibinfo
  {journal} {Open J.~Astrophys.}\ }\textbf {\bibinfo {volume} {4}},\ \bibinfo
  {eid} {1} (\bibinfo {year} {2021})},\ \Eprint
  {https://arxiv.org/abs/2006.16182}  { arXiv:2006.16182
  [astro-ph.CO]}\BibitemShut {NoStop}%
\bibitem [{\citenamefont {{Chongchitnan}}\ and\ \citenamefont
  {{Silk}}(2021)}]{2021PhRvD.104h3018C}%
  \BibitemOpen
  \bibfield  {author} {\bibinfo {author} {\bibfnamefont {S.}~\bibnamefont
  {{Chongchitnan}}}\ and\ \bibinfo {author} {\bibfnamefont {J.}~\bibnamefont
  {{Silk}}},\ }\href {https://doi.org/10.1103/PhysRevD.104.083018} {\bibfield
  {journal} {\bibinfo  {journal} {\prd}\ }\textbf {\bibinfo {volume} {104}},\
  \bibinfo {eid} {083018} (\bibinfo {year} {2021})},\ \Eprint
  {https://arxiv.org/abs/2109.12268}  { arXiv:2109.12268
  [astro-ph.CO]}\BibitemShut {NoStop}%
\bibitem [{\citenamefont {{De Luca}}\ \emph {et~al.}(2020)\citenamefont {{De
  Luca}}, \citenamefont {{Franciolini}}, \citenamefont {{Pani}},\ and\
  \citenamefont {{Riotto}}}]{2020JCAP...04..052D}%
  \BibitemOpen
  \bibfield  {author} {\bibinfo {author} {\bibfnamefont {V.}~\bibnamefont {{De
  Luca}}}, \bibinfo {author} {\bibfnamefont {G.}~\bibnamefont {{Franciolini}}},
  \bibinfo {author} {\bibfnamefont {P.}~\bibnamefont {{Pani}}},\ and\ \bibinfo
  {author} {\bibfnamefont {A.}~\bibnamefont {{Riotto}}},\ }\href
  {https://doi.org/10.1088/1475-7516/2020/04/052} {\bibfield  {journal}
  {\bibinfo  {journal} {\jcap}\ }\textbf {\bibinfo {volume} {2020}},\ \bibinfo
  {eid} {052} (\bibinfo {year} {2020})},\ \Eprint
  {https://arxiv.org/abs/2003.02778}  { arXiv:2003.02778
  [astro-ph.CO]}\BibitemShut {NoStop}%
\bibitem [{\citenamefont {{Fishbach}}\ \emph {et~al.}(2017)\citenamefont
  {{Fishbach}}, \citenamefont {{Holz}},\ and\ \citenamefont
  {{Farr}}}]{2017ApJ...840L..24F}%
  \BibitemOpen
  \bibfield  {author} {\bibinfo {author} {\bibfnamefont {M.}~\bibnamefont
  {{Fishbach}}}, \bibinfo {author} {\bibfnamefont {D.~E.}\ \bibnamefont
  {{Holz}}},\ and\ \bibinfo {author} {\bibfnamefont {B.}~\bibnamefont
  {{Farr}}},\ }\href {https://doi.org/10.3847/2041-8213/aa7045} {\bibfield
  {journal} {\bibinfo  {journal} {\apjl}\ }\textbf {\bibinfo {volume} {840}},\
  \bibinfo {eid} {L24} (\bibinfo {year} {2017})},\ \Eprint
  {https://arxiv.org/abs/1703.06869}  { arXiv:1703.06869
  [astro-ph.HE]}\BibitemShut {NoStop}%
\bibitem [{\citenamefont {{Hooper}}\ \emph {et~al.}(2020)\citenamefont
  {{Hooper}}, \citenamefont {{Krnjaic}}, \citenamefont {{March-Russell}},
  \citenamefont {{McDermott}},\ and\ \citenamefont
  {{Petrossian-Byrne}}}]{2020arXiv200400618H}%
  \BibitemOpen
  \bibfield  {author} {\bibinfo {author} {\bibfnamefont {D.}~\bibnamefont
  {{Hooper}}}, \bibinfo {author} {\bibfnamefont {G.}~\bibnamefont {{Krnjaic}}},
  \bibinfo {author} {\bibfnamefont {J.}~\bibnamefont {{March-Russell}}},
  \bibinfo {author} {\bibfnamefont {S.~D.}\ \bibnamefont {{McDermott}}},\ and\
  \bibinfo {author} {\bibfnamefont {R.}~\bibnamefont {{Petrossian-Byrne}}},\
  }\href@noop {} {\bibfield  {journal} {\bibinfo  {journal} {\arxiv}\ }
  (\bibinfo {year} {2020})},\ \Eprint {https://arxiv.org/abs/2004.00618}  {
  arXiv:2004.00618 [astro-ph.CO]}\BibitemShut {NoStop}%
\bibitem [{\citenamefont {{Arbey}}\ \emph
  {et~al.}(2021{\natexlab{a}})\citenamefont {{Arbey}}, \citenamefont
  {{Auffinger}}, \citenamefont {{Sandick}}, \citenamefont {{Shams Es Haghi}},\
  and\ \citenamefont {{Sinha}}}]{2021PhRvD.103l3549A}%
  \BibitemOpen
  \bibfield  {author} {\bibinfo {author} {\bibfnamefont {A.}~\bibnamefont
  {{Arbey}}}, \bibinfo {author} {\bibfnamefont {J.}~\bibnamefont
  {{Auffinger}}}, \bibinfo {author} {\bibfnamefont {P.}~\bibnamefont
  {{Sandick}}}, \bibinfo {author} {\bibfnamefont {B.}~\bibnamefont {{Shams Es
  Haghi}}},\ and\ \bibinfo {author} {\bibfnamefont {K.}~\bibnamefont
  {{Sinha}}},\ }\href {https://doi.org/10.1103/PhysRevD.103.123549} {\bibfield
  {journal} {\bibinfo  {journal} {\prd}\ }\textbf {\bibinfo {volume} {103}},\
  \bibinfo {eid} {123549} (\bibinfo {year} {2021}{\natexlab{a}})},\ \Eprint
  {https://arxiv.org/abs/2104.04051}  { arXiv:2104.04051
  [astro-ph.CO]}\BibitemShut {NoStop}%
\bibitem [{\citenamefont {{Israel}}(1967)}]{1967PhRv..164.1776I}%
  \BibitemOpen
  \bibfield  {author} {\bibinfo {author} {\bibfnamefont {W.}~\bibnamefont
  {{Israel}}},\ }\href {https://doi.org/10.1103/PhysRev.164.1776} {\bibfield
  {journal} {\bibinfo  {journal} {\pr}\ }\textbf {\bibinfo {volume} {164}},\
  \bibinfo {pages} {1776} (\bibinfo {year} {1967})}\BibitemShut {NoStop}%
\bibitem [{\citenamefont {{Israel}}(1968)}]{1968CMaPh...8..245I}%
  \BibitemOpen
  \bibfield  {author} {\bibinfo {author} {\bibfnamefont {W.}~\bibnamefont
  {{Israel}}},\ }\href {https://doi.org/10.1007/BF01645859} {\bibfield
  {journal} {\bibinfo  {journal} {\CMaPh}\ }\textbf {\bibinfo {volume} {8}},\
  \bibinfo {pages} {245} (\bibinfo {year} {1968})}\BibitemShut {NoStop}%
\bibitem [{\citenamefont {{Carter}}(1971)}]{1971PhRvL..26..331C}%
  \BibitemOpen
  \bibfield  {author} {\bibinfo {author} {\bibfnamefont {B.}~\bibnamefont
  {{Carter}}},\ }\href {https://doi.org/10.1103/PhysRevLett.26.331} {\bibfield
  {journal} {\bibinfo  {journal} {\prl}\ }\textbf {\bibinfo {volume} {26}},\
  \bibinfo {pages} {331} (\bibinfo {year} {1971})}\BibitemShut {NoStop}%
\bibitem [{\citenamefont {{Page}}(1993)}]{1993PhRvL..71.3743P}%
  \BibitemOpen
  \bibfield  {author} {\bibinfo {author} {\bibfnamefont {D.~N.}\ \bibnamefont
  {{Page}}},\ }\href {https://doi.org/10.1103/PhysRevLett.71.3743} {\bibfield
  {journal} {\bibinfo  {journal} {\prl}\ }\textbf {\bibinfo {volume} {71}},\
  \bibinfo {pages} {3743} (\bibinfo {year} {1993})},\ \Eprint
  {https://arxiv.org/abs/hep-th/9306083}  { arXiv:hep-th/9306083
  [hep-th]}\BibitemShut {NoStop}%
\bibitem [{\citenamefont {{Bekenstein}}(1973)}]{1973PhRvD...7.2333B}%
  \BibitemOpen
  \bibfield  {author} {\bibinfo {author} {\bibfnamefont {J.~D.}\ \bibnamefont
  {{Bekenstein}}},\ }\href {https://doi.org/10.1103/PhysRevD.7.2333} {\bibfield
   {journal} {\bibinfo  {journal} {\prd}\ }\textbf {\bibinfo {volume} {7}},\
  \bibinfo {pages} {2333} (\bibinfo {year} {1973})}\BibitemShut {NoStop}%
\bibitem [{\citenamefont {{Bardeen}}\ \emph {et~al.}(1973)\citenamefont
  {{Bardeen}}, \citenamefont {{Carter}},\ and\ \citenamefont
  {{Hawking}}}]{1973CMaPh..31..161B}%
  \BibitemOpen
  \bibfield  {author} {\bibinfo {author} {\bibfnamefont {J.~M.}\ \bibnamefont
  {{Bardeen}}}, \bibinfo {author} {\bibfnamefont {B.}~\bibnamefont
  {{Carter}}},\ and\ \bibinfo {author} {\bibfnamefont {S.~W.}\ \bibnamefont
  {{Hawking}}},\ }\href {https://doi.org/10.1007/BF01645742} {\bibfield
  {journal} {\bibinfo  {journal} {\CMaPh}\ }\textbf {\bibinfo {volume} {31}},\
  \bibinfo {pages} {161} (\bibinfo {year} {1973})}\BibitemShut {NoStop}%
\bibitem [{\citenamefont {{Wallace}}(2018)}]{2018SHPMP..64...52W}%
  \BibitemOpen
  \bibfield  {author} {\bibinfo {author} {\bibfnamefont {D.}~\bibnamefont
  {{Wallace}}},\ }\href {https://doi.org/10.1016/j.shpsb.2018.05.002}
  {\bibfield  {journal} {\bibinfo  {journal} {\SHPSB}\ }\textbf {\bibinfo
  {volume} {64}},\ \bibinfo {pages} {52} (\bibinfo {year} {2018})},\ \Eprint
  {https://arxiv.org/abs/1710.02724}  { arXiv:1710.02724 [gr-qc]}\BibitemShut
  {NoStop}%
\bibitem [{\citenamefont {{Wallace}}(2017)}]{2017arXiv171002725W}%
  \BibitemOpen
  \bibfield  {author} {\bibinfo {author} {\bibfnamefont {D.}~\bibnamefont
  {{Wallace}}},\ }\href@noop {} {\bibfield  {journal} {\bibinfo  {journal}
  {\arxiv}\ } (\bibinfo {year} {2017})},\ \Eprint
  {https://arxiv.org/abs/1710.02725}  { arXiv:1710.02725 [gr-qc]}\BibitemShut
  {NoStop}%
\bibitem [{\citenamefont {{Misner}}(1972)}]{1972PhRvL..28..994M}%
  \BibitemOpen
  \bibfield  {author} {\bibinfo {author} {\bibfnamefont {C.~W.}\ \bibnamefont
  {{Misner}}},\ }\href {https://doi.org/10.1103/PhysRevLett.28.994} {\bibfield
  {journal} {\bibinfo  {journal} {\prl}\ }\textbf {\bibinfo {volume} {28}},\
  \bibinfo {pages} {994} (\bibinfo {year} {1972})}\BibitemShut {NoStop}%
\bibitem [{\citenamefont {{Press}}\ and\ \citenamefont
  {{Teukolsky}}(1972)}]{1972Natur.238..211P}%
  \BibitemOpen
  \bibfield  {author} {\bibinfo {author} {\bibfnamefont {W.~H.}\ \bibnamefont
  {{Press}}}\ and\ \bibinfo {author} {\bibfnamefont {S.~A.}\ \bibnamefont
  {{Teukolsky}}},\ }\href {https://doi.org/10.1038/238211a0} {\bibfield
  {journal} {\bibinfo  {journal} {\nat}\ }\textbf {\bibinfo {volume} {238}},\
  \bibinfo {pages} {211} (\bibinfo {year} {1972})}\BibitemShut {NoStop}%
\bibitem [{\citenamefont {{Brito}}\ \emph {et~al.}(2020)\citenamefont
  {{Brito}}, \citenamefont {{Cardoso}},\ and\ \citenamefont
  {{Pani}}}]{2015LNP...906.....B}%
  \BibitemOpen
  \bibfield  {author} {\bibinfo {author} {\bibfnamefont {R.}~\bibnamefont
  {{Brito}}}, \bibinfo {author} {\bibfnamefont {V.}~\bibnamefont {{Cardoso}}},\
  and\ \bibinfo {author} {\bibfnamefont {P.}~\bibnamefont {{Pani}}},\
  }\bibfield  {title} {\bibinfo {title} {{Superradiance}},\ }in\ \href
  {https://doi.org/10.1007/978-3-030-46622-0} {\emph {\bibinfo {booktitle} {New
  Frontiers in Black Hole Physics}}},\ Vol.\ \bibinfo {volume} {971}\ (\bibinfo
  {year} {2020})\ pp.\ \bibinfo {pages} {1--293},\ \Eprint
  {https://arxiv.org/abs/1501.06570} {arXiv:1501.06570 [gr-qc]} \BibitemShut
  {NoStop}%
\bibitem [{\citenamefont {{Teukolsky}}(1973)}]{1973ApJ...185..635T}%
  \BibitemOpen
  \bibfield  {author} {\bibinfo {author} {\bibfnamefont {S.~A.}\ \bibnamefont
  {{Teukolsky}}},\ }\href {https://doi.org/10.1086/152444} {\bibfield
  {journal} {\bibinfo  {journal} {\apj}\ }\textbf {\bibinfo {volume} {185}},\
  \bibinfo {pages} {635} (\bibinfo {year} {1973})}\BibitemShut {NoStop}%
\bibitem [{\citenamefont {{Press}}\ and\ \citenamefont
  {{Teukolsky}}(1973)}]{1973ApJ...185..649P}%
  \BibitemOpen
  \bibfield  {author} {\bibinfo {author} {\bibfnamefont {W.~H.}\ \bibnamefont
  {{Press}}}\ and\ \bibinfo {author} {\bibfnamefont {S.~A.}\ \bibnamefont
  {{Teukolsky}}},\ }\href {https://doi.org/10.1086/152445} {\bibfield
  {journal} {\bibinfo  {journal} {\apj}\ }\textbf {\bibinfo {volume} {185}},\
  \bibinfo {pages} {649} (\bibinfo {year} {1973})}\BibitemShut {NoStop}%
\bibitem [{\citenamefont {{Teukolsky}}\ and\ \citenamefont
  {{Press}}(1974)}]{1974ApJ...193..443T}%
  \BibitemOpen
  \bibfield  {author} {\bibinfo {author} {\bibfnamefont {S.~A.}\ \bibnamefont
  {{Teukolsky}}}\ and\ \bibinfo {author} {\bibfnamefont {W.~H.}\ \bibnamefont
  {{Press}}},\ }\href {https://doi.org/10.1086/153180} {\bibfield  {journal}
  {\bibinfo  {journal} {\apj}\ }\textbf {\bibinfo {volume} {193}},\ \bibinfo
  {pages} {443} (\bibinfo {year} {1974})}\BibitemShut {NoStop}%
\bibitem [{\citenamefont {{Carter}}(1968)}]{1968CMaPh..10..280C}%
  \BibitemOpen
  \bibfield  {author} {\bibinfo {author} {\bibfnamefont {B.}~\bibnamefont
  {{Carter}}},\ }\href {https://doi.org/10.1007/BF03399503} {\bibfield
  {journal} {\bibinfo  {journal} {\CMaPh}\ }\textbf {\bibinfo {volume} {10}},\
  \bibinfo {pages} {280} (\bibinfo {year} {1968})}\BibitemShut {NoStop}%
\bibitem [{\citenamefont {{Starobinski{\v{i}}}}(1973)}]{1973JETP...37...28S}%
  \BibitemOpen
  \bibfield  {author} {\bibinfo {author} {\bibfnamefont {A.~A.}\ \bibnamefont
  {{Starobinski{\v{i}}}}},\ }\href@noop {} {\bibfield  {journal} {\bibinfo
  {journal} {\JETP}\ }\textbf {\bibinfo {volume} {37}},\ \bibinfo {pages} {28}
  (\bibinfo {year} {1973})},\ \bibinfo {note} {[\ZhETF~\textbf{64}, 48
  (1973)]}\BibitemShut {NoStop}%
\bibitem [{\citenamefont {{Starobinski{\v{i}}}}\ and\ \citenamefont
  {{Churilov}}(1974)}]{1974JETP...38....1S}%
  \BibitemOpen
  \bibfield  {author} {\bibinfo {author} {\bibfnamefont {A.~A.}\ \bibnamefont
  {{Starobinski{\v{i}}}}}\ and\ \bibinfo {author} {\bibfnamefont {S.~M.}\
  \bibnamefont {{Churilov}}},\ }\href@noop {} {\bibfield  {journal} {\bibinfo
  {journal} {\JETP}\ }\textbf {\bibinfo {volume} {38}},\ \bibinfo {pages} {1}
  (\bibinfo {year} {1974})},\ \bibinfo {note} {[\ZhETF~\textbf{65}, 3
  (1973)]}\BibitemShut {NoStop}%
\bibitem [{\citenamefont {{Unruh}}(1973)}]{1973PhRvL..31.1265U}%
  \BibitemOpen
  \bibfield  {author} {\bibinfo {author} {\bibfnamefont {W.}~\bibnamefont
  {{Unruh}}},\ }\href {https://doi.org/10.1103/PhysRevLett.31.1265} {\bibfield
  {journal} {\bibinfo  {journal} {\prl}\ }\textbf {\bibinfo {volume} {31}},\
  \bibinfo {pages} {1265} (\bibinfo {year} {1973})}\BibitemShut {NoStop}%
\bibitem [{\citenamefont {{Fackerell}}\ and\ \citenamefont
  {{Crossman}}(1977)}]{1977JMP....18.1849F}%
  \BibitemOpen
  \bibfield  {author} {\bibinfo {author} {\bibfnamefont {E.~D.}\ \bibnamefont
  {{Fackerell}}}\ and\ \bibinfo {author} {\bibfnamefont {R.~G.}\ \bibnamefont
  {{Crossman}}},\ }\href {https://doi.org/10.1063/1.523499} {\bibfield
  {journal} {\bibinfo  {journal} {\JMP}\ }\textbf {\bibinfo {volume} {18}},\
  \bibinfo {pages} {1849} (\bibinfo {year} {1977})}\BibitemShut {NoStop}%
\bibitem [{\citenamefont {{Hawking}}(1974)}]{1974Natur.248...30H}%
  \BibitemOpen
  \bibfield  {author} {\bibinfo {author} {\bibfnamefont {S.~W.}\ \bibnamefont
  {{Hawking}}},\ }\href {https://doi.org/10.1038/248030a0} {\bibfield
  {journal} {\bibinfo  {journal} {\nat}\ }\textbf {\bibinfo {volume} {248}},\
  \bibinfo {pages} {30} (\bibinfo {year} {1974})}\BibitemShut {NoStop}%
\bibitem [{\citenamefont {{Bekenstein}}(1975)}]{1975PhRvD..12.3077B}%
  \BibitemOpen
  \bibfield  {author} {\bibinfo {author} {\bibfnamefont {J.~D.}\ \bibnamefont
  {{Bekenstein}}},\ }\href {https://doi.org/10.1103/PhysRevD.12.3077}
  {\bibfield  {journal} {\bibinfo  {journal} {\prd}\ }\textbf {\bibinfo
  {volume} {12}},\ \bibinfo {pages} {3077} (\bibinfo {year}
  {1975})}\BibitemShut {NoStop}%
\bibitem [{\citenamefont {{Hawking}}(1976)}]{1976PhRvD..13..191H}%
  \BibitemOpen
  \bibfield  {author} {\bibinfo {author} {\bibfnamefont {S.~W.}\ \bibnamefont
  {{Hawking}}},\ }\href {https://doi.org/10.1103/PhysRevD.13.191} {\bibfield
  {journal} {\bibinfo  {journal} {\prd}\ }\textbf {\bibinfo {volume} {13}},\
  \bibinfo {pages} {191} (\bibinfo {year} {1976})}\BibitemShut {NoStop}%
\bibitem [{\citenamefont {{Hawking}}(1975)}]{1975CMaPh..43..199H}%
  \BibitemOpen
  \bibfield  {author} {\bibinfo {author} {\bibfnamefont {S.~W.}\ \bibnamefont
  {{Hawking}}},\ }\href {https://doi.org/10.1007/BF02345020} {\bibfield
  {journal} {\bibinfo  {journal} {\CMaPh}\ }\textbf {\bibinfo {volume} {43}},\
  \bibinfo {pages} {199} (\bibinfo {year} {1975})}\BibitemShut {NoStop}%
\bibitem [{\citenamefont {{Traschen}}(2000)}]{2000mmp..conf..180T}%
  \BibitemOpen
  \bibfield  {author} {\bibinfo {author} {\bibfnamefont {J.}~\bibnamefont
  {{Traschen}}},\ }in\ \href@noop {} {\emph {\bibinfo {booktitle} {Mathematical
  Methods in Physics}}},\ \bibinfo {editor} {edited by\ \bibinfo {editor}
  {\bibfnamefont {A.~A.}\ \bibnamefont {{Bytsenko}}}\ and\ \bibinfo {editor}
  {\bibfnamefont {F.~L.}\ \bibnamefont {{Williams}}}}\ (\bibinfo {year}
  {2000})\ p.\ \bibinfo {pages} {180},\ \Eprint
  {https://arxiv.org/abs/gr-qc/0010055} {arXiv:gr-qc/0010055 [gr-qc]}
  \BibitemShut {NoStop}%
\bibitem [{\citenamefont {{Page}}(1976{\natexlab{a}})}]{1976PhRvD..13..198P}%
  \BibitemOpen
  \bibfield  {author} {\bibinfo {author} {\bibfnamefont {D.~N.}\ \bibnamefont
  {{Page}}},\ }\href {https://doi.org/10.1103/PhysRevD.13.198} {\bibfield
  {journal} {\bibinfo  {journal} {\prd}\ }\textbf {\bibinfo {volume} {13}},\
  \bibinfo {pages} {198} (\bibinfo {year} {1976}{\natexlab{a}})}\BibitemShut
  {NoStop}%
\bibitem [{\citenamefont {{Hartle}}\ and\ \citenamefont
  {{Hawking}}(1976)}]{1976PhRvD..13.2188H}%
  \BibitemOpen
  \bibfield  {author} {\bibinfo {author} {\bibfnamefont {J.~B.}\ \bibnamefont
  {{Hartle}}}\ and\ \bibinfo {author} {\bibfnamefont {S.~W.}\ \bibnamefont
  {{Hawking}}},\ }\href {https://doi.org/10.1103/PhysRevD.13.2188} {\bibfield
  {journal} {\bibinfo  {journal} {\prd}\ }\textbf {\bibinfo {volume} {13}},\
  \bibinfo {pages} {2188} (\bibinfo {year} {1976})}\BibitemShut {NoStop}%
\bibitem [{\citenamefont {{Parikh}}\ and\ \citenamefont
  {{Wilczek}}(2000)}]{2000PhRvL..85.5042P}%
  \BibitemOpen
  \bibfield  {author} {\bibinfo {author} {\bibfnamefont {M.~K.}\ \bibnamefont
  {{Parikh}}}\ and\ \bibinfo {author} {\bibfnamefont {F.}~\bibnamefont
  {{Wilczek}}},\ }\href {https://doi.org/10.1103/PhysRevLett.85.5042}
  {\bibfield  {journal} {\bibinfo  {journal} {\prl}\ }\textbf {\bibinfo
  {volume} {85}},\ \bibinfo {pages} {5042} (\bibinfo {year} {2000})},\ \Eprint
  {https://arxiv.org/abs/hep-th/9907001}  { arXiv:hep-th/9907001
  [hep-th]}\BibitemShut {NoStop}%
\bibitem [{\citenamefont {{Frolov}}(1976)}]{1976SvPhU..19..244F}%
  \BibitemOpen
  \bibfield  {author} {\bibinfo {author} {\bibfnamefont {V.~P.}\ \bibnamefont
  {{Frolov}}},\ }\href {https://doi.org/10.1070/PU1976v019n03ABEH005141}
  {\bibfield  {journal} {\bibinfo  {journal} {\sovphysusp}\ }\textbf {\bibinfo
  {volume} {19}},\ \bibinfo {pages} {244} (\bibinfo {year} {1976})},\ \bibinfo
  {note} {[\uspfiznauk~\textbf{118}, 473 (1976)]}\BibitemShut {NoStop}%
\bibitem [{\citenamefont {{Carter}}(1974)}]{1974PhRvL..33..558C}%
  \BibitemOpen
  \bibfield  {author} {\bibinfo {author} {\bibfnamefont {B.}~\bibnamefont
  {{Carter}}},\ }\href {https://doi.org/10.1103/PhysRevLett.33.558} {\bibfield
  {journal} {\bibinfo  {journal} {\prl}\ }\textbf {\bibinfo {volume} {33}},\
  \bibinfo {pages} {558} (\bibinfo {year} {1974})}\BibitemShut {NoStop}%
\bibitem [{\citenamefont {{Gibbons}}(1975)}]{1975CMaPh..44..245G}%
  \BibitemOpen
  \bibfield  {author} {\bibinfo {author} {\bibfnamefont {G.~W.}\ \bibnamefont
  {{Gibbons}}},\ }\href {https://doi.org/10.1007/BF01609829} {\bibfield
  {journal} {\bibinfo  {journal} {\CMaPh}\ }\textbf {\bibinfo {volume} {44}},\
  \bibinfo {pages} {245} (\bibinfo {year} {1975})}\BibitemShut {NoStop}%
\bibitem [{\citenamefont {{Schwinger}}(1951)}]{1951PhRv...82..664S}%
  \BibitemOpen
  \bibfield  {author} {\bibinfo {author} {\bibfnamefont {J.}~\bibnamefont
  {{Schwinger}}},\ }\href {https://doi.org/10.1103/PhysRev.82.664} {\bibfield
  {journal} {\bibinfo  {journal} {\pr}\ }\textbf {\bibinfo {volume} {82}},\
  \bibinfo {pages} {664} (\bibinfo {year} {1951})}\BibitemShut {NoStop}%
\bibitem [{\citenamefont {{Zaumen}}(1974)}]{1974Natur.247..530Z}%
  \BibitemOpen
  \bibfield  {author} {\bibinfo {author} {\bibfnamefont {W.~T.}\ \bibnamefont
  {{Zaumen}}},\ }\href {https://doi.org/10.1038/247530a0} {\bibfield  {journal}
  {\bibinfo  {journal} {\nat}\ }\textbf {\bibinfo {volume} {247}},\ \bibinfo
  {pages} {530} (\bibinfo {year} {1974})}\BibitemShut {NoStop}%
\bibitem [{\citenamefont {{Page}}(1976{\natexlab{b}})}]{1976PhRvD..14.3260P}%
  \BibitemOpen
  \bibfield  {author} {\bibinfo {author} {\bibfnamefont {D.~N.}\ \bibnamefont
  {{Page}}},\ }\href {https://doi.org/10.1103/PhysRevD.14.3260} {\bibfield
  {journal} {\bibinfo  {journal} {\prd}\ }\textbf {\bibinfo {volume} {14}},\
  \bibinfo {pages} {3260} (\bibinfo {year} {1976}{\natexlab{b}})}\BibitemShut
  {NoStop}%
\bibitem [{\citenamefont {{Page}}(1977)}]{1977PhRvD..16.2402P}%
  \BibitemOpen
  \bibfield  {author} {\bibinfo {author} {\bibfnamefont {D.~N.}\ \bibnamefont
  {{Page}}},\ }\href {https://doi.org/10.1103/PhysRevD.16.2402} {\bibfield
  {journal} {\bibinfo  {journal} {\prd}\ }\textbf {\bibinfo {volume} {16}},\
  \bibinfo {pages} {2402} (\bibinfo {year} {1977})}\BibitemShut {NoStop}%
\bibitem [{\citenamefont {{Carr}}(1976)}]{1976ApJ...206....8C}%
  \BibitemOpen
  \bibfield  {author} {\bibinfo {author} {\bibfnamefont {B.~J.}\ \bibnamefont
  {{Carr}}},\ }\href {https://doi.org/10.1086/154351} {\bibfield  {journal}
  {\bibinfo  {journal} {\apj}\ }\textbf {\bibinfo {volume} {206}},\ \bibinfo
  {pages} {8} (\bibinfo {year} {1976})}\BibitemShut {NoStop}%
\bibitem [{\citenamefont {{Perry}}(1977)}]{1977PhLB...71..234P}%
  \BibitemOpen
  \bibfield  {author} {\bibinfo {author} {\bibfnamefont {M.~J.}\ \bibnamefont
  {{Perry}}},\ }\href {https://doi.org/10.1016/0370-2693(77)90786-9} {\bibfield
   {journal} {\bibinfo  {journal} {\plb}\ }\textbf {\bibinfo {volume} {71}},\
  \bibinfo {pages} {234} (\bibinfo {year} {1977})}\BibitemShut {NoStop}%
\bibitem [{\citenamefont {{MacGibbon}}\ and\ \citenamefont
  {{Webber}}(1990)}]{1990PhRvD..41.3052M}%
  \BibitemOpen
  \bibfield  {author} {\bibinfo {author} {\bibfnamefont {J.~H.}\ \bibnamefont
  {{MacGibbon}}}\ and\ \bibinfo {author} {\bibfnamefont {B.~R.}\ \bibnamefont
  {{Webber}}},\ }\href {https://doi.org/10.1103/PhysRevD.41.3052} {\bibfield
  {journal} {\bibinfo  {journal} {\prd}\ }\textbf {\bibinfo {volume} {41}},\
  \bibinfo {pages} {3052} (\bibinfo {year} {1990})}\BibitemShut {NoStop}%
\bibitem [{\citenamefont {{MacGibbon}}(1991)}]{1991PhRvD..44..376M}%
  \BibitemOpen
  \bibfield  {author} {\bibinfo {author} {\bibfnamefont {J.~H.}\ \bibnamefont
  {{MacGibbon}}},\ }\href {https://doi.org/10.1103/PhysRevD.44.376} {\bibfield
  {journal} {\bibinfo  {journal} {\prd}\ }\textbf {\bibinfo {volume} {44}},\
  \bibinfo {pages} {376} (\bibinfo {year} {1991})}\BibitemShut {NoStop}%
\bibitem [{\citenamefont {{Carter}}\ \emph {et~al.}(1976)\citenamefont
  {{Carter}}, \citenamefont {{Gibbons}}, \citenamefont {{Lin}},\ and\
  \citenamefont {{Perry}}}]{1976A&A....52..427C}%
  \BibitemOpen
  \bibfield  {author} {\bibinfo {author} {\bibfnamefont {B.}~\bibnamefont
  {{Carter}}}, \bibinfo {author} {\bibfnamefont {G.~W.}\ \bibnamefont
  {{Gibbons}}}, \bibinfo {author} {\bibfnamefont {D.~N.~C.}\ \bibnamefont
  {{Lin}}},\ and\ \bibinfo {author} {\bibfnamefont {M.~J.}\ \bibnamefont
  {{Perry}}},\ }\href@noop {} {\bibfield  {journal} {\bibinfo  {journal}
  {\aap}\ }\textbf {\bibinfo {volume} {52}},\ \bibinfo {pages} {427} (\bibinfo
  {year} {1976})}\BibitemShut {NoStop}%
\bibitem [{\citenamefont {{Oliensis}}\ and\ \citenamefont
  {{Hill}}(1984)}]{1984PhLB..143...92O}%
  \BibitemOpen
  \bibfield  {author} {\bibinfo {author} {\bibfnamefont {J.}~\bibnamefont
  {{Oliensis}}}\ and\ \bibinfo {author} {\bibfnamefont {C.~T.}\ \bibnamefont
  {{Hill}}},\ }\href {https://doi.org/10.1016/0370-2693(84)90811-6} {\bibfield
  {journal} {\bibinfo  {journal} {\plb}\ }\textbf {\bibinfo {volume} {143}},\
  \bibinfo {pages} {92} (\bibinfo {year} {1984})}\BibitemShut {NoStop}%
\bibitem [{\citenamefont {{Marchesini}}\ and\ \citenamefont
  {{Webber}}(1988)}]{1988NuPhB.310..461M}%
  \BibitemOpen
  \bibfield  {author} {\bibinfo {author} {\bibfnamefont {G.}~\bibnamefont
  {{Marchesini}}}\ and\ \bibinfo {author} {\bibfnamefont {B.~R.}\ \bibnamefont
  {{Webber}}},\ }\href {https://doi.org/10.1016/0550-3213(88)90089-2}
  {\bibfield  {journal} {\bibinfo  {journal} {\nphysb}\ }\textbf {\bibinfo
  {volume} {310}},\ \bibinfo {pages} {461} (\bibinfo {year}
  {1988})}\BibitemShut {NoStop}%
\bibitem [{\citenamefont {{Cline}}\ and\ \citenamefont
  {{Hong}}(1992)}]{1992ApJ...401L..57C}%
  \BibitemOpen
  \bibfield  {author} {\bibinfo {author} {\bibfnamefont {D.~B.}\ \bibnamefont
  {{Cline}}}\ and\ \bibinfo {author} {\bibfnamefont {W.}~\bibnamefont
  {{Hong}}},\ }\href {https://doi.org/10.1086/186670} {\bibfield  {journal}
  {\bibinfo  {journal} {\apjl}\ }\textbf {\bibinfo {volume} {401}},\ \bibinfo
  {pages} {L57} (\bibinfo {year} {1992})}\BibitemShut {NoStop}%
\bibitem [{\citenamefont
  {{Heckler}}(1997{\natexlab{a}})}]{1997PhRvD..55..480H}%
  \BibitemOpen
  \bibfield  {author} {\bibinfo {author} {\bibfnamefont {A.~F.}\ \bibnamefont
  {{Heckler}}},\ }\href {https://doi.org/10.1103/PhysRevD.55.480} {\bibfield
  {journal} {\bibinfo  {journal} {\prd}\ }\textbf {\bibinfo {volume} {55}},\
  \bibinfo {pages} {480} (\bibinfo {year} {1997}{\natexlab{a}})},\ \Eprint
  {https://arxiv.org/abs/astro-ph/9601029}  { arXiv:astro-ph/9601029
  [astro-ph]}\BibitemShut {NoStop}%
\bibitem [{\citenamefont
  {{Heckler}}(1997{\natexlab{b}})}]{1997PhRvL..78.3430H}%
  \BibitemOpen
  \bibfield  {author} {\bibinfo {author} {\bibfnamefont {A.~F.}\ \bibnamefont
  {{Heckler}}},\ }\href {https://doi.org/10.1103/PhysRevLett.78.3430}
  {\bibfield  {journal} {\bibinfo  {journal} {\prl}\ }\textbf {\bibinfo
  {volume} {78}},\ \bibinfo {pages} {3430} (\bibinfo {year}
  {1997}{\natexlab{b}})},\ \Eprint {https://arxiv.org/abs/astro-ph/9702027}  {
  arXiv:astro-ph/9702027 [astro-ph]}\BibitemShut {NoStop}%
\bibitem [{\citenamefont {{Kapusta}}(1999)}]{1999astro.ph.11309K}%
  \BibitemOpen
  \bibfield  {author} {\bibinfo {author} {\bibfnamefont {J.}~\bibnamefont
  {{Kapusta}}},\ }\href@noop {} {\bibfield  {journal} {\bibinfo  {journal}
  {\arxiv}\ } (\bibinfo {year} {1999})},\ \Eprint
  {https://arxiv.org/abs/astro-ph/9911309}  { arXiv:astro-ph/9911309
  [astro-ph]}\BibitemShut {NoStop}%
\bibitem [{\citenamefont {{Daghigh}}\ and\ \citenamefont
  {{Kapusta}}(2003)}]{2003PhRvD..67d4006D}%
  \BibitemOpen
  \bibfield  {author} {\bibinfo {author} {\bibfnamefont {R.~G.}\ \bibnamefont
  {{Daghigh}}}\ and\ \bibinfo {author} {\bibfnamefont {J.~I.}\ \bibnamefont
  {{Kapusta}}},\ }\href {https://doi.org/10.1103/PhysRevD.67.044006} {\bibfield
   {journal} {\bibinfo  {journal} {\prd}\ }\textbf {\bibinfo {volume} {67}},\
  \bibinfo {eid} {044006} (\bibinfo {year} {2003})},\ \Eprint
  {https://arxiv.org/abs/astro-ph/0211579}  { arXiv:astro-ph/0211579
  [astro-ph]}\BibitemShut {NoStop}%
\bibitem [{\citenamefont {{Belyanin}}\ \emph {et~al.}(1996)\citenamefont
  {{Belyanin}}, \citenamefont {{Kocharovsky}},\ and\ \citenamefont
  {{Kocharovsky}}}]{1996MNRAS.283..626B}%
  \BibitemOpen
  \bibfield  {author} {\bibinfo {author} {\bibfnamefont {A.~A.}\ \bibnamefont
  {{Belyanin}}}, \bibinfo {author} {\bibfnamefont {V.~V.}\ \bibnamefont
  {{Kocharovsky}}},\ and\ \bibinfo {author} {\bibfnamefont {V.~V.}\
  \bibnamefont {{Kocharovsky}}},\ }\href
  {https://doi.org/10.1093/mnras/283.2.626} {\bibfield  {journal} {\bibinfo
  {journal} {\mnras}\ }\textbf {\bibinfo {volume} {283}},\ \bibinfo {pages}
  {626} (\bibinfo {year} {1996})}\BibitemShut {NoStop}%
\bibitem [{\citenamefont {{Rees}}(1977)}]{1977Natur.266..333R}%
  \BibitemOpen
  \bibfield  {author} {\bibinfo {author} {\bibfnamefont {M.~J.}\ \bibnamefont
  {{Rees}}},\ }\href {https://doi.org/10.1038/266333a0} {\bibfield  {journal}
  {\bibinfo  {journal} {\nat}\ }\textbf {\bibinfo {volume} {266}},\ \bibinfo
  {pages} {333} (\bibinfo {year} {1977})}\BibitemShut {NoStop}%
\bibitem [{\citenamefont {{Blandford}}(1977)}]{1977MNRAS.181..489B}%
  \BibitemOpen
  \bibfield  {author} {\bibinfo {author} {\bibfnamefont {R.~D.}\ \bibnamefont
  {{Blandford}}},\ }\href {https://doi.org/10.1093/mnras/181.3.489} {\bibfield
  {journal} {\bibinfo  {journal} {\mnras}\ }\textbf {\bibinfo {volume} {181}},\
  \bibinfo {pages} {489} (\bibinfo {year} {1977})}\BibitemShut {NoStop}%
\bibitem [{\citenamefont {{Nagatani}}(1999)}]{1999PhRvD..59d1301N}%
  \BibitemOpen
  \bibfield  {author} {\bibinfo {author} {\bibfnamefont {Y.}~\bibnamefont
  {{Nagatani}}},\ }\href {https://doi.org/10.1103/PhysRevD.59.041301}
  {\bibfield  {journal} {\bibinfo  {journal} {\prd}\ }\textbf {\bibinfo
  {volume} {59}},\ \bibinfo {eid} {041301} (\bibinfo {year} {1999})},\ \Eprint
  {https://arxiv.org/abs/hep-ph/9811485}  { arXiv:hep-ph/9811485
  [hep-ph]}\BibitemShut {NoStop}%
\bibitem [{\citenamefont {{Hawking}}(1981)}]{1981CMaPh..80..421H}%
  \BibitemOpen
  \bibfield  {author} {\bibinfo {author} {\bibfnamefont {S.~W.}\ \bibnamefont
  {{Hawking}}},\ }\href {https://doi.org/10.1007/BF01208279} {\bibfield
  {journal} {\bibinfo  {journal} {\CMaPh}\ }\textbf {\bibinfo {volume} {80}},\
  \bibinfo {pages} {421} (\bibinfo {year} {1981})}\BibitemShut {NoStop}%
\bibitem [{\citenamefont {{Cline}}\ \emph
  {et~al.}(1999{\natexlab{a}})\citenamefont {{Cline}}, \citenamefont
  {{Mostoslavsky}},\ and\ \citenamefont {{Servant}}}]{1999PhRvD..59f3009C}%
  \BibitemOpen
  \bibfield  {author} {\bibinfo {author} {\bibfnamefont {J.~M.}\ \bibnamefont
  {{Cline}}}, \bibinfo {author} {\bibfnamefont {M.}~\bibnamefont
  {{Mostoslavsky}}},\ and\ \bibinfo {author} {\bibfnamefont {G.}~\bibnamefont
  {{Servant}}},\ }\href {https://doi.org/10.1103/PhysRevD.59.063009} {\bibfield
   {journal} {\bibinfo  {journal} {\prd}\ }\textbf {\bibinfo {volume} {59}},\
  \bibinfo {eid} {063009} (\bibinfo {year} {1999}{\natexlab{a}})},\ \Eprint
  {https://arxiv.org/abs/hep-ph/9810439}  { arXiv:hep-ph/9810439
  [hep-ph]}\BibitemShut {NoStop}%
\bibitem [{\citenamefont {{MacGibbon}}\ \emph {et~al.}(2008)\citenamefont
  {{MacGibbon}}, \citenamefont {{Carr}},\ and\ \citenamefont
  {{Page}}}]{2008PhRvD..78f4043M}%
  \BibitemOpen
  \bibfield  {author} {\bibinfo {author} {\bibfnamefont {J.~H.}\ \bibnamefont
  {{MacGibbon}}}, \bibinfo {author} {\bibfnamefont {B.~J.}\ \bibnamefont
  {{Carr}}},\ and\ \bibinfo {author} {\bibfnamefont {D.~N.}\ \bibnamefont
  {{Page}}},\ }\href {https://doi.org/10.1103/PhysRevD.78.064043} {\bibfield
  {journal} {\bibinfo  {journal} {\prd}\ }\textbf {\bibinfo {volume} {78}},\
  \bibinfo {eid} {064043} (\bibinfo {year} {2008})},\ \Eprint
  {https://arxiv.org/abs/0709.2380}  { arXiv:0709.2380 [astro-ph]}\BibitemShut
  {NoStop}%
\bibitem [{\citenamefont {{Arbey}}\ \emph
  {et~al.}(2021{\natexlab{b}})\citenamefont {{Arbey}}, \citenamefont
  {{Auffinger}}, \citenamefont {{Geiller}}, \citenamefont {{Livine}},\ and\
  \citenamefont {{Sartini}}}]{2021PhRvD.104h4016A}%
  \BibitemOpen
  \bibfield  {author} {\bibinfo {author} {\bibfnamefont {A.}~\bibnamefont
  {{Arbey}}}, \bibinfo {author} {\bibfnamefont {J.}~\bibnamefont
  {{Auffinger}}}, \bibinfo {author} {\bibfnamefont {M.}~\bibnamefont
  {{Geiller}}}, \bibinfo {author} {\bibfnamefont {E.~R.}\ \bibnamefont
  {{Livine}}},\ and\ \bibinfo {author} {\bibfnamefont {F.}~\bibnamefont
  {{Sartini}}},\ }\href {https://doi.org/10.1103/PhysRevD.104.084016}
  {\bibfield  {journal} {\bibinfo  {journal} {\prd}\ }\textbf {\bibinfo
  {volume} {104}},\ \bibinfo {eid} {084016} (\bibinfo {year}
  {2021}{\natexlab{b}})},\ \Eprint {https://arxiv.org/abs/2107.03293}  {
  arXiv:2107.03293 [gr-qc]}\BibitemShut {NoStop}%
\bibitem [{\citenamefont {{Arbey}}\ \emph
  {et~al.}(2021{\natexlab{c}})\citenamefont {{Arbey}}, \citenamefont
  {{Auffinger}}, \citenamefont {{Geiller}}, \citenamefont {{Livine}},\ and\
  \citenamefont {{Sartini}}}]{2021PhRvD.103j4010A}%
  \BibitemOpen
  \bibfield  {author} {\bibinfo {author} {\bibfnamefont {A.}~\bibnamefont
  {{Arbey}}}, \bibinfo {author} {\bibfnamefont {J.}~\bibnamefont
  {{Auffinger}}}, \bibinfo {author} {\bibfnamefont {M.}~\bibnamefont
  {{Geiller}}}, \bibinfo {author} {\bibfnamefont {E.~R.}\ \bibnamefont
  {{Livine}}},\ and\ \bibinfo {author} {\bibfnamefont {F.}~\bibnamefont
  {{Sartini}}},\ }\href {https://doi.org/10.1103/PhysRevD.103.104010}
  {\bibfield  {journal} {\bibinfo  {journal} {\prd}\ }\textbf {\bibinfo
  {volume} {103}},\ \bibinfo {eid} {104010} (\bibinfo {year}
  {2021}{\natexlab{c}})},\ \Eprint {https://arxiv.org/abs/2101.02951}  {
  arXiv:2101.02951 [gr-qc]}\BibitemShut {NoStop}%
\bibitem [{\citenamefont {{Auffinger}}\ \emph {et~al.}(2021)\citenamefont
  {{Auffinger}}, \citenamefont {{Masina}},\ and\ \citenamefont
  {{Orlando}}}]{2021EPJP..136..261A}%
  \BibitemOpen
  \bibfield  {author} {\bibinfo {author} {\bibfnamefont {J.}~\bibnamefont
  {{Auffinger}}}, \bibinfo {author} {\bibfnamefont {I.}~\bibnamefont
  {{Masina}}},\ and\ \bibinfo {author} {\bibfnamefont {G.}~\bibnamefont
  {{Orlando}}},\ }\href {https://doi.org/10.1140/epjp/s13360-021-01247-9}
  {\bibfield  {journal} {\bibinfo  {journal} {\epjp}\ }\textbf {\bibinfo
  {volume} {136}},\ \bibinfo {eid} {261} (\bibinfo {year} {2021})},\ \Eprint
  {https://arxiv.org/abs/2012.09867}  { arXiv:2012.09867 [hep-ph]}\BibitemShut
  {NoStop}%
\bibitem [{\citenamefont {{Chandrasekhar}}\ and\ \citenamefont
  {{Detweiler}}(1975)}]{1975RSPSA.345..145C}%
  \BibitemOpen
  \bibfield  {author} {\bibinfo {author} {\bibfnamefont {S.}~\bibnamefont
  {{Chandrasekhar}}}\ and\ \bibinfo {author} {\bibfnamefont {S.}~\bibnamefont
  {{Detweiler}}},\ }\href {https://doi.org/10.1098/rspa.1975.0130} {\bibfield
  {journal} {\bibinfo  {journal} {\RSPSA}\ }\textbf {\bibinfo {volume} {345}},\
  \bibinfo {pages} {145} (\bibinfo {year} {1975})}\BibitemShut {NoStop}%
\bibitem [{\citenamefont {{Detweiler}}(1976)}]{1976RSPSA.349..217D}%
  \BibitemOpen
  \bibfield  {author} {\bibinfo {author} {\bibfnamefont {S.}~\bibnamefont
  {{Detweiler}}},\ }\href {https://doi.org/10.1098/rspa.1976.0069} {\bibfield
  {journal} {\bibinfo  {journal} {\RSPSA}\ }\textbf {\bibinfo {volume} {349}},\
  \bibinfo {pages} {217} (\bibinfo {year} {1976})}\BibitemShut {NoStop}%
\bibitem [{\citenamefont {{Chandrasekhar}}\ and\ \citenamefont
  {{Detweiler}}(1976)}]{1976RSPSA.350..165C}%
  \BibitemOpen
  \bibfield  {author} {\bibinfo {author} {\bibfnamefont {S.}~\bibnamefont
  {{Chandrasekhar}}}\ and\ \bibinfo {author} {\bibfnamefont {S.}~\bibnamefont
  {{Detweiler}}},\ }\href {https://doi.org/10.1098/rspa.1976.0101} {\bibfield
  {journal} {\bibinfo  {journal} {\RSPSA}\ }\textbf {\bibinfo {volume} {350}},\
  \bibinfo {pages} {165} (\bibinfo {year} {1976})}\BibitemShut {NoStop}%
\bibitem [{\citenamefont {{Chandrasekhar}}\ and\ \citenamefont
  {{Detweiler}}(1977)}]{1977RSPSA.352..325C}%
  \BibitemOpen
  \bibfield  {author} {\bibinfo {author} {\bibfnamefont {S.}~\bibnamefont
  {{Chandrasekhar}}}\ and\ \bibinfo {author} {\bibfnamefont {S.}~\bibnamefont
  {{Detweiler}}},\ }\href {https://doi.org/10.1098/rspa.1977.0002} {\bibfield
  {journal} {\bibinfo  {journal} {\RSPSA}\ }\textbf {\bibinfo {volume} {352}},\
  \bibinfo {pages} {325} (\bibinfo {year} {1977})}\BibitemShut {NoStop}%
\bibitem [{\citenamefont {{Gray}}\ and\ \citenamefont
  {{Visser}}(2015)}]{2015arXiv151205018G}%
  \BibitemOpen
  \bibfield  {author} {\bibinfo {author} {\bibfnamefont {F.}~\bibnamefont
  {{Gray}}}\ and\ \bibinfo {author} {\bibfnamefont {M.}~\bibnamefont
  {{Visser}}},\ }\href@noop {} {\bibfield  {journal} {\bibinfo  {journal}
  {\arxiv}\ } (\bibinfo {year} {2015})},\ \Eprint
  {https://arxiv.org/abs/1512.05018}  { arXiv:1512.05018 [gr-qc]}\BibitemShut
  {NoStop}%
\bibitem [{\citenamefont {{Harris}}\ \emph {et~al.}(2003)\citenamefont
  {{Harris}}, \citenamefont {{Richardson}},\ and\ \citenamefont
  {{Webber}}}]{2003JHEP...08..033H}%
  \BibitemOpen
  \bibfield  {author} {\bibinfo {author} {\bibfnamefont {C.~M.}\ \bibnamefont
  {{Harris}}}, \bibinfo {author} {\bibfnamefont {P.}~\bibnamefont
  {{Richardson}}},\ and\ \bibinfo {author} {\bibfnamefont {B.~R.}\ \bibnamefont
  {{Webber}}},\ }\href {https://doi.org/10.1088/1126-6708/2003/08/033}
  {\bibfield  {journal} {\bibinfo  {journal} {\jhep}\ }\textbf {\bibinfo
  {volume} {2003}},\ \bibinfo {eid} {033} (\bibinfo {year} {2003})},\ \Eprint
  {https://arxiv.org/abs/hep-ph/0307305}  { arXiv:hep-ph/0307305
  [hep-ph]}\BibitemShut {NoStop}%
\bibitem [{\citenamefont {{Frost}}\ \emph {et~al.}(2009)\citenamefont {{Frost}}
  \emph {et~al.}}]{2009JHEP...10..014F}%
  \BibitemOpen
  \bibfield  {author} {\bibinfo {author} {\bibfnamefont {J.~A.}\ \bibnamefont
  {{Frost}}} \emph {et~al.},\ }\href
  {https://doi.org/10.1088/1126-6708/2009/10/014} {\bibfield  {journal}
  {\bibinfo  {journal} {\jhep}\ }\textbf {\bibinfo {volume} {2009}},\ \bibinfo
  {eid} {014} (\bibinfo {year} {2009})},\ \Eprint
  {https://arxiv.org/abs/0904.0979}  { arXiv:0904.0979 [hep-ph]}\BibitemShut
  {NoStop}%
\bibitem [{\citenamefont {{Cavagli{\`a}}}\ \emph {et~al.}(2007)\citenamefont
  {{Cavagli{\`a}}}, \citenamefont {{Godang}}, \citenamefont {{Cremaldi}},\ and\
  \citenamefont {{Summers}}}]{2007CoPhC.177..506C}%
  \BibitemOpen
  \bibfield  {author} {\bibinfo {author} {\bibfnamefont {M.}~\bibnamefont
  {{Cavagli{\`a}}}}, \bibinfo {author} {\bibfnamefont {R.}~\bibnamefont
  {{Godang}}}, \bibinfo {author} {\bibfnamefont {L.}~\bibnamefont
  {{Cremaldi}}},\ and\ \bibinfo {author} {\bibfnamefont {D.}~\bibnamefont
  {{Summers}}},\ }\href {https://doi.org/10.1016/j.cpc.2007.05.011} {\bibfield
  {journal} {\bibinfo  {journal} {\CPC}\ }\textbf {\bibinfo {volume} {177}},\
  \bibinfo {pages} {506} (\bibinfo {year} {2007})},\ \Eprint
  {https://arxiv.org/abs/hep-ph/0609001}  { arXiv:hep-ph/0609001
  [hep-ph]}\BibitemShut {NoStop}%
\bibitem [{\citenamefont {{Dai}}\ \emph {et~al.}(2008)\citenamefont {{Dai}},
  \citenamefont {{Starkman}}, \citenamefont {{Stojkovic}}, \citenamefont
  {{Issever}}, \citenamefont {{Rizvi}},\ and\ \citenamefont
  {{Tseng}}}]{2008PhRvD..77g6007D}%
  \BibitemOpen
  \bibfield  {author} {\bibinfo {author} {\bibfnamefont {D.-C.}\ \bibnamefont
  {{Dai}}}, \bibinfo {author} {\bibfnamefont {G.}~\bibnamefont {{Starkman}}},
  \bibinfo {author} {\bibfnamefont {D.}~\bibnamefont {{Stojkovic}}}, \bibinfo
  {author} {\bibfnamefont {C.}~\bibnamefont {{Issever}}}, \bibinfo {author}
  {\bibfnamefont {E.}~\bibnamefont {{Rizvi}}},\ and\ \bibinfo {author}
  {\bibfnamefont {J.}~\bibnamefont {{Tseng}}},\ }\href
  {https://doi.org/10.1103/PhysRevD.77.076007} {\bibfield  {journal} {\bibinfo
  {journal} {\prd}\ }\textbf {\bibinfo {volume} {77}},\ \bibinfo {eid} {076007}
  (\bibinfo {year} {2008})},\ \Eprint {https://arxiv.org/abs/0711.3012}  {
  arXiv:0711.3012 [hep-ph]}\BibitemShut {NoStop}%
\bibitem [{\citenamefont {{Dai}}\ \emph {et~al.}(2009)\citenamefont {{Dai}},
  \citenamefont {{Issever}}, \citenamefont {{Rizvi}}, \citenamefont
  {{Starkman}}, \citenamefont {{Stojkovic}},\ and\ \citenamefont
  {{Tseng}}}]{2009arXiv0902.3577D}%
  \BibitemOpen
  \bibfield  {author} {\bibinfo {author} {\bibfnamefont {D.-C.}\ \bibnamefont
  {{Dai}}}, \bibinfo {author} {\bibfnamefont {C.}~\bibnamefont {{Issever}}},
  \bibinfo {author} {\bibfnamefont {E.}~\bibnamefont {{Rizvi}}}, \bibinfo
  {author} {\bibfnamefont {G.}~\bibnamefont {{Starkman}}}, \bibinfo {author}
  {\bibfnamefont {D.}~\bibnamefont {{Stojkovic}}},\ and\ \bibinfo {author}
  {\bibfnamefont {J.}~\bibnamefont {{Tseng}}},\ }\href@noop {} {\bibfield
  {journal} {\bibinfo  {journal} {\arxiv}\ } (\bibinfo {year} {2009})},\
  \Eprint {https://arxiv.org/abs/0902.3577}  { arXiv:0902.3577
  [hep-ph]}\BibitemShut {NoStop}%
\bibitem [{\citenamefont {{Dai}}\ \emph {et~al.}(2019)\citenamefont {{Dai}},
  \citenamefont {{Issever}}, \citenamefont {{Rizvi}}, \citenamefont
  {{Starkman}}, \citenamefont {{Stojkovic}},\ and\ \citenamefont
  {{Tseng}}}]{2019CoPhC.236..285D}%
  \BibitemOpen
  \bibfield  {author} {\bibinfo {author} {\bibfnamefont {D.-C.}\ \bibnamefont
  {{Dai}}}, \bibinfo {author} {\bibfnamefont {C.}~\bibnamefont {{Issever}}},
  \bibinfo {author} {\bibfnamefont {E.}~\bibnamefont {{Rizvi}}}, \bibinfo
  {author} {\bibfnamefont {G.}~\bibnamefont {{Starkman}}}, \bibinfo {author}
  {\bibfnamefont {D.}~\bibnamefont {{Stojkovic}}},\ and\ \bibinfo {author}
  {\bibfnamefont {J.}~\bibnamefont {{Tseng}}},\ }\href
  {https://doi.org/10.1016/j.cpc.2018.10.021} {\bibfield  {journal} {\bibinfo
  {journal} {\CPC}\ }\textbf {\bibinfo {volume} {236}},\ \bibinfo {pages} {285}
  (\bibinfo {year} {2019})}\BibitemShut {NoStop}%
\bibitem [{\citenamefont {{Gingrich}}(2010)}]{2010CoPhC.181.1917G}%
  \BibitemOpen
  \bibfield  {author} {\bibinfo {author} {\bibfnamefont {D.~M.}\ \bibnamefont
  {{Gingrich}}},\ }\href {https://doi.org/10.1016/j.cpc.2010.07.027} {\bibfield
   {journal} {\bibinfo  {journal} {\CPC}\ }\textbf {\bibinfo {volume} {181}},\
  \bibinfo {pages} {1917} (\bibinfo {year} {2010})},\ \Eprint
  {https://arxiv.org/abs/0911.5370}  { arXiv:0911.5370 [hep-ph]}\BibitemShut
  {NoStop}%
\bibitem [{\citenamefont {{Arbey}}\ and\ \citenamefont
  {{Auffinger}}(2021)}]{2021EPJC...81..910A}%
  \BibitemOpen
  \bibfield  {author} {\bibinfo {author} {\bibfnamefont {A.}~\bibnamefont
  {{Arbey}}}\ and\ \bibinfo {author} {\bibfnamefont {J.}~\bibnamefont
  {{Auffinger}}},\ }\href {https://doi.org/10.1140/epjc/s10052-021-09702-8}
  {\bibfield  {journal} {\bibinfo  {journal} {\epjc}\ }\textbf {\bibinfo
  {volume} {81}},\ \bibinfo {eid} {910} (\bibinfo {year} {2021})},\ \Eprint
  {https://arxiv.org/abs/2108.02737}  { arXiv:2108.02737 [gr-qc]}\BibitemShut
  {NoStop}%
\bibitem [{\citenamefont {{Coogan}}\ \emph {et~al.}(2020)\citenamefont
  {{Coogan}}, \citenamefont {{Morrison}},\ and\ \citenamefont
  {{Profumo}}}]{2020JCAP...01..056C}%
  \BibitemOpen
  \bibfield  {author} {\bibinfo {author} {\bibfnamefont {A.}~\bibnamefont
  {{Coogan}}}, \bibinfo {author} {\bibfnamefont {L.}~\bibnamefont
  {{Morrison}}},\ and\ \bibinfo {author} {\bibfnamefont {S.}~\bibnamefont
  {{Profumo}}},\ }\href {https://doi.org/10.1088/1475-7516/2020/01/056}
  {\bibfield  {journal} {\bibinfo  {journal} {\jcap}\ }\textbf {\bibinfo
  {volume} {2020}},\ \bibinfo {eid} {056} (\bibinfo {year} {2020})},\ \Eprint
  {https://arxiv.org/abs/1907.11846}  { arXiv:1907.11846 [hep-ph]}\BibitemShut
  {NoStop}%
\bibitem [{\citenamefont {{Bierlich}}\ \emph {et~al.}(2022)\citenamefont
  {{Bierlich}} \emph {et~al.}}]{2022arXiv220311601B}%
  \BibitemOpen
  \bibfield  {author} {\bibinfo {author} {\bibfnamefont {C.}~\bibnamefont
  {{Bierlich}}} \emph {et~al.},\ }\href@noop {} {\bibfield  {journal} {\bibinfo
   {journal} {\arxiv}\ } (\bibinfo {year} {2022})},\ \Eprint
  {https://arxiv.org/abs/2203.11601}  { arXiv:2203.11601 [hep-ph]}\BibitemShut
  {NoStop}%
\bibitem [{\citenamefont {{Bellm}}\ \emph {et~al.}(2020)\citenamefont {{Bellm}}
  \emph {et~al.}}]{2020EPJC...80..452B}%
  \BibitemOpen
  \bibfield  {author} {\bibinfo {author} {\bibfnamefont {J.}~\bibnamefont
  {{Bellm}}} \emph {et~al.},\ }\href
  {https://doi.org/10.1140/epjc/s10052-020-8011-x} {\bibfield  {journal}
  {\bibinfo  {journal} {\epjc}\ }\textbf {\bibinfo {volume} {80}},\ \bibinfo
  {eid} {452} (\bibinfo {year} {2020})},\ \Eprint
  {https://arxiv.org/abs/1912.06509}  { arXiv:1912.06509 [hep-ph]}\BibitemShut
  {NoStop}%
\bibitem [{\citenamefont {{Bauer}}\ \emph {et~al.}(2021)\citenamefont
  {{Bauer}}, \citenamefont {{Rodd}},\ and\ \citenamefont
  {{Webber}}}]{2021JHEP...06..121B}%
  \BibitemOpen
  \bibfield  {author} {\bibinfo {author} {\bibfnamefont {C.~W.}\ \bibnamefont
  {{Bauer}}}, \bibinfo {author} {\bibfnamefont {N.~L.}\ \bibnamefont
  {{Rodd}}},\ and\ \bibinfo {author} {\bibfnamefont {B.~R.}\ \bibnamefont
  {{Webber}}},\ }\href {https://doi.org/10.1007/JHEP06(2021)121} {\bibfield
  {journal} {\bibinfo  {journal} {\jhep}\ }\textbf {\bibinfo {volume} {2021}},\
  \bibinfo {eid} {121} (\bibinfo {year} {2021})},\ \Eprint
  {https://arxiv.org/abs/2007.15001}  { arXiv:2007.15001 [hep-ph]}\BibitemShut
  {NoStop}%
\bibitem [{\citenamefont {{Cirelli}}\ \emph {et~al.}(2011)\citenamefont
  {{Cirelli}} \emph {et~al.}}]{2011JCAP...03..051C}%
  \BibitemOpen
  \bibfield  {author} {\bibinfo {author} {\bibfnamefont {M.}~\bibnamefont
  {{Cirelli}}} \emph {et~al.},\ }\href
  {https://doi.org/10.1088/1475-7516/2011/03/051} {\bibfield  {journal}
  {\bibinfo  {journal} {\jcap}\ }\textbf {\bibinfo {volume} {2011}},\ \bibinfo
  {eid} {051} (\bibinfo {year} {2011})},\ \Eprint
  {https://arxiv.org/abs/1012.4515}  { arXiv:1012.4515 [hep-ph]}\BibitemShut
  {NoStop}%
\bibitem [{\citenamefont {{Ciafaloni}}\ \emph {et~al.}(2011)\citenamefont
  {{Ciafaloni}}, \citenamefont {{Comelli}}, \citenamefont {{Riotto}},
  \citenamefont {{Sala}}, \citenamefont {{Strumia}},\ and\ \citenamefont
  {{Urbano}}}]{2011JCAP...03..019C}%
  \BibitemOpen
  \bibfield  {author} {\bibinfo {author} {\bibfnamefont {P.}~\bibnamefont
  {{Ciafaloni}}}, \bibinfo {author} {\bibfnamefont {D.}~\bibnamefont
  {{Comelli}}}, \bibinfo {author} {\bibfnamefont {A.}~\bibnamefont {{Riotto}}},
  \bibinfo {author} {\bibfnamefont {F.}~\bibnamefont {{Sala}}}, \bibinfo
  {author} {\bibfnamefont {A.}~\bibnamefont {{Strumia}}},\ and\ \bibinfo
  {author} {\bibfnamefont {A.}~\bibnamefont {{Urbano}}},\ }\href
  {https://doi.org/10.1088/1475-7516/2011/03/019} {\bibfield  {journal}
  {\bibinfo  {journal} {\jcap}\ }\textbf {\bibinfo {volume} {2011}},\ \bibinfo
  {eid} {019} (\bibinfo {year} {2011})},\ \Eprint
  {https://arxiv.org/abs/1009.0224}  { arXiv:1009.0224 [hep-ph]}\BibitemShut
  {NoStop}%
\bibitem [{\citenamefont {{Bartelmann}}(2010)}]{2010CQGra..27w3001B}%
  \BibitemOpen
  \bibfield  {author} {\bibinfo {author} {\bibfnamefont {M.}~\bibnamefont
  {{Bartelmann}}},\ }\href {https://doi.org/10.1088/0264-9381/27/23/233001}
  {\bibfield  {journal} {\bibinfo  {journal} {\CQG}\ }\textbf {\bibinfo
  {volume} {27}},\ \bibinfo {eid} {233001} (\bibinfo {year} {2010})},\ \Eprint
  {https://arxiv.org/abs/1010.3829}  { arXiv:1010.3829
  [astro-ph.CO]}\BibitemShut {NoStop}%
\bibitem [{\citenamefont {{Alcock}}\ \emph {et~al.}(2000)\citenamefont
  {{Alcock}} \emph {et~al.}}]{2000ApJ...542..281A}%
  \BibitemOpen
  \bibfield  {author} {\bibinfo {author} {\bibfnamefont {C.}~\bibnamefont
  {{Alcock}}} \emph {et~al.} (\bibinfo {collaboration} {MACHO}),\ }\href
  {https://doi.org/10.1086/309512} {\bibfield  {journal} {\bibinfo  {journal}
  {\apj}\ }\textbf {\bibinfo {volume} {542}},\ \bibinfo {pages} {281} (\bibinfo
  {year} {2000})},\ \Eprint {https://arxiv.org/abs/astro-ph/0001272}  {
  arXiv:astro-ph/0001272 [astro-ph]}\BibitemShut {NoStop}%
\bibitem [{\citenamefont {{Tisserand}}\ \emph {et~al.}(2007)\citenamefont
  {{Tisserand}} \emph {et~al.}}]{2007A&A...469..387T}%
  \BibitemOpen
  \bibfield  {author} {\bibinfo {author} {\bibfnamefont {P.}~\bibnamefont
  {{Tisserand}}} \emph {et~al.} (\bibinfo {collaboration} {EROS-2}),\ }\href
  {https://doi.org/10.1051/0004-6361:20066017} {\bibfield  {journal} {\bibinfo
  {journal} {\aap}\ }\textbf {\bibinfo {volume} {469}},\ \bibinfo {pages} {387}
  (\bibinfo {year} {2007})},\ \Eprint {https://arxiv.org/abs/astro-ph/0607207}
  { arXiv:astro-ph/0607207 [astro-ph]}\BibitemShut {NoStop}%
\bibitem [{\citenamefont {{Niikura}}\ \emph
  {et~al.}(2019{\natexlab{a}})\citenamefont {{Niikura}}, \citenamefont
  {{Takada}}, \citenamefont {{Yokoyama}}, \citenamefont {{Sumi}},\ and\
  \citenamefont {{Masaki}}}]{2019PhRvD..99h3503N}%
  \BibitemOpen
  \bibfield  {author} {\bibinfo {author} {\bibfnamefont {H.}~\bibnamefont
  {{Niikura}}}, \bibinfo {author} {\bibfnamefont {M.}~\bibnamefont {{Takada}}},
  \bibinfo {author} {\bibfnamefont {S.}~\bibnamefont {{Yokoyama}}}, \bibinfo
  {author} {\bibfnamefont {T.}~\bibnamefont {{Sumi}}},\ and\ \bibinfo {author}
  {\bibfnamefont {S.}~\bibnamefont {{Masaki}}},\ }\href
  {https://doi.org/10.1103/PhysRevD.99.083503} {\bibfield  {journal} {\bibinfo
  {journal} {\prd}\ }\textbf {\bibinfo {volume} {99}},\ \bibinfo {eid} {083503}
  (\bibinfo {year} {2019}{\natexlab{a}})},\ \Eprint
  {https://arxiv.org/abs/1901.07120}  { arXiv:1901.07120
  [astro-ph.CO]}\BibitemShut {NoStop}%
\bibitem [{\citenamefont {{Mr{\'o}z}}\ \emph {et~al.}(2017)\citenamefont
  {{Mr{\'o}z}} \emph {et~al.}}]{2017Natur.548..183M}%
  \BibitemOpen
  \bibfield  {author} {\bibinfo {author} {\bibfnamefont {P.}~\bibnamefont
  {{Mr{\'o}z}}} \emph {et~al.},\ }\href {https://doi.org/10.1038/nature23276}
  {\bibfield  {journal} {\bibinfo  {journal} {\nat}\ }\textbf {\bibinfo
  {volume} {548}},\ \bibinfo {pages} {183} (\bibinfo {year} {2017})},\ \Eprint
  {https://arxiv.org/abs/1707.07634}  { arXiv:1707.07634
  [astro-ph.EP]}\BibitemShut {NoStop}%
\bibitem [{\citenamefont {{Smyth}}\ \emph {et~al.}(2020)\citenamefont {{Smyth}}
  \emph {et~al.}}]{2020PhRvD.101f3005S}%
  \BibitemOpen
  \bibfield  {author} {\bibinfo {author} {\bibfnamefont {N.}~\bibnamefont
  {{Smyth}}} \emph {et~al.},\ }\href
  {https://doi.org/10.1103/PhysRevD.101.063005} {\bibfield  {journal} {\bibinfo
   {journal} {\prd}\ }\textbf {\bibinfo {volume} {101}},\ \bibinfo {eid}
  {063005} (\bibinfo {year} {2020})},\ \Eprint
  {https://arxiv.org/abs/1910.01285}  { arXiv:1910.01285
  [astro-ph.CO]}\BibitemShut {NoStop}%
\bibitem [{\citenamefont {{Niikura}}\ \emph
  {et~al.}(2019{\natexlab{b}})\citenamefont {{Niikura}} \emph
  {et~al.}}]{2019NatAs...3..524N}%
  \BibitemOpen
  \bibfield  {author} {\bibinfo {author} {\bibfnamefont {H.}~\bibnamefont
  {{Niikura}}} \emph {et~al.},\ }\href
  {https://doi.org/10.1038/s41550-019-0723-1} {\bibfield  {journal} {\bibinfo
  {journal} {\natast}\ }\textbf {\bibinfo {volume} {3}},\ \bibinfo {pages}
  {524} (\bibinfo {year} {2019}{\natexlab{b}})},\ \Eprint
  {https://arxiv.org/abs/1701.02151}  { arXiv:1701.02151
  [astro-ph.CO]}\BibitemShut {NoStop}%
\bibitem [{\citenamefont {{Barnacka}}\ \emph {et~al.}(2012)\citenamefont
  {{Barnacka}}, \citenamefont {{Glicenstein}},\ and\ \citenamefont
  {{Moderski}}}]{2012PhRvD..86d3001B}%
  \BibitemOpen
  \bibfield  {author} {\bibinfo {author} {\bibfnamefont {A.}~\bibnamefont
  {{Barnacka}}}, \bibinfo {author} {\bibfnamefont {J.~F.}\ \bibnamefont
  {{Glicenstein}}},\ and\ \bibinfo {author} {\bibfnamefont {R.}~\bibnamefont
  {{Moderski}}},\ }\href {https://doi.org/10.1103/PhysRevD.86.043001}
  {\bibfield  {journal} {\bibinfo  {journal} {\prd}\ }\textbf {\bibinfo
  {volume} {86}},\ \bibinfo {eid} {043001} (\bibinfo {year} {2012})},\ \Eprint
  {https://arxiv.org/abs/1204.2056}  { arXiv:1204.2056
  [astro-ph.CO]}\BibitemShut {NoStop}%
\bibitem [{\citenamefont {{Katz}}\ \emph {et~al.}(2018)\citenamefont {{Katz}},
  \citenamefont {{Kopp}}, \citenamefont {{Sibiryakov}},\ and\ \citenamefont
  {{Xue}}}]{2018JCAP...12..005K}%
  \BibitemOpen
  \bibfield  {author} {\bibinfo {author} {\bibfnamefont {A.}~\bibnamefont
  {{Katz}}}, \bibinfo {author} {\bibfnamefont {J.}~\bibnamefont {{Kopp}}},
  \bibinfo {author} {\bibfnamefont {S.}~\bibnamefont {{Sibiryakov}}},\ and\
  \bibinfo {author} {\bibfnamefont {W.}~\bibnamefont {{Xue}}},\ }\href
  {https://doi.org/10.1088/1475-7516/2018/12/005} {\bibfield  {journal}
  {\bibinfo  {journal} {\jcap}\ }\textbf {\bibinfo {volume} {2018}},\ \bibinfo
  {eid} {005} (\bibinfo {year} {2018})},\ \Eprint
  {https://arxiv.org/abs/1807.11495}  { arXiv:1807.11495
  [astro-ph.CO]}\BibitemShut {NoStop}%
\bibitem [{\citenamefont {{Katz}}\ \emph {et~al.}(2020)\citenamefont {{Katz}},
  \citenamefont {{Kopp}}, \citenamefont {{Sibiryakov}},\ and\ \citenamefont
  {{Xue}}}]{2020MNRAS.496..564K}%
  \BibitemOpen
  \bibfield  {author} {\bibinfo {author} {\bibfnamefont {A.}~\bibnamefont
  {{Katz}}}, \bibinfo {author} {\bibfnamefont {J.}~\bibnamefont {{Kopp}}},
  \bibinfo {author} {\bibfnamefont {S.}~\bibnamefont {{Sibiryakov}}},\ and\
  \bibinfo {author} {\bibfnamefont {W.}~\bibnamefont {{Xue}}},\ }\href
  {https://doi.org/10.1093/mnras/staa1497} {\bibfield  {journal} {\bibinfo
  {journal} {\mnras}\ }\textbf {\bibinfo {volume} {496}},\ \bibinfo {pages}
  {564} (\bibinfo {year} {2020})},\ \Eprint {https://arxiv.org/abs/1912.07620}
  { arXiv:1912.07620 [astro-ph.CO]}\BibitemShut {NoStop}%
\bibitem [{\citenamefont {{Carr}}\ and\ \citenamefont
  {{Sakellariadou}}(1999)}]{1999ApJ...516..195C}%
  \BibitemOpen
  \bibfield  {author} {\bibinfo {author} {\bibfnamefont {B.~J.}\ \bibnamefont
  {{Carr}}}\ and\ \bibinfo {author} {\bibfnamefont {M.}~\bibnamefont
  {{Sakellariadou}}},\ }\href {https://doi.org/10.1086/307071} {\bibfield
  {journal} {\bibinfo  {journal} {\apj}\ }\textbf {\bibinfo {volume} {516}},\
  \bibinfo {pages} {195} (\bibinfo {year} {1999})}\BibitemShut {NoStop}%
\bibitem [{\citenamefont {{Monroy-Rodr{\'\i}guez}}\ and\ \citenamefont
  {{Allen}}(2014)}]{2014ApJ...790..159M}%
  \BibitemOpen
  \bibfield  {author} {\bibinfo {author} {\bibfnamefont {M.~A.}\ \bibnamefont
  {{Monroy-Rodr{\'\i}guez}}}\ and\ \bibinfo {author} {\bibfnamefont
  {C.}~\bibnamefont {{Allen}}},\ }\href
  {https://doi.org/10.1088/0004-637X/790/2/159} {\bibfield  {journal} {\bibinfo
   {journal} {\apj}\ }\textbf {\bibinfo {volume} {790}},\ \bibinfo {eid} {159}
  (\bibinfo {year} {2014})},\ \Eprint {https://arxiv.org/abs/1406.5169}  {
  arXiv:1406.5169 [astro-ph.GA]}\BibitemShut {NoStop}%
\bibitem [{\citenamefont {{Capela}}\ \emph
  {et~al.}(2013{\natexlab{a}})\citenamefont {{Capela}}, \citenamefont
  {{Pshirkov}},\ and\ \citenamefont {{Tinyakov}}}]{2013PhRvD..87b3507C}%
  \BibitemOpen
  \bibfield  {author} {\bibinfo {author} {\bibfnamefont {F.}~\bibnamefont
  {{Capela}}}, \bibinfo {author} {\bibfnamefont {M.}~\bibnamefont
  {{Pshirkov}}},\ and\ \bibinfo {author} {\bibfnamefont {P.}~\bibnamefont
  {{Tinyakov}}},\ }\href {https://doi.org/10.1103/PhysRevD.87.023507}
  {\bibfield  {journal} {\bibinfo  {journal} {\prd}\ }\textbf {\bibinfo
  {volume} {87}},\ \bibinfo {eid} {023507} (\bibinfo {year}
  {2013}{\natexlab{a}})},\ \Eprint {https://arxiv.org/abs/1209.6021}  {
  arXiv:1209.6021 [astro-ph.CO]}\BibitemShut {NoStop}%
\bibitem [{\citenamefont {{Capela}}\ \emph
  {et~al.}(2013{\natexlab{b}})\citenamefont {{Capela}}, \citenamefont
  {{Pshirkov}},\ and\ \citenamefont {{Tinyakov}}}]{2013PhRvD..87l3524C}%
  \BibitemOpen
  \bibfield  {author} {\bibinfo {author} {\bibfnamefont {F.}~\bibnamefont
  {{Capela}}}, \bibinfo {author} {\bibfnamefont {M.}~\bibnamefont
  {{Pshirkov}}},\ and\ \bibinfo {author} {\bibfnamefont {P.}~\bibnamefont
  {{Tinyakov}}},\ }\href {https://doi.org/10.1103/PhysRevD.87.123524}
  {\bibfield  {journal} {\bibinfo  {journal} {\prd}\ }\textbf {\bibinfo
  {volume} {87}},\ \bibinfo {eid} {123524} (\bibinfo {year}
  {2013}{\natexlab{b}})},\ \Eprint {https://arxiv.org/abs/1301.4984}  {
  arXiv:1301.4984 [astro-ph.CO]}\BibitemShut {NoStop}%
\bibitem [{\citenamefont {{Capela}}\ \emph {et~al.}(2014)\citenamefont
  {{Capela}}, \citenamefont {{Pshirkov}},\ and\ \citenamefont
  {{Tinyakov}}}]{2014arXiv1402.4671C}%
  \BibitemOpen
  \bibfield  {author} {\bibinfo {author} {\bibfnamefont {F.}~\bibnamefont
  {{Capela}}}, \bibinfo {author} {\bibfnamefont {M.}~\bibnamefont
  {{Pshirkov}}},\ and\ \bibinfo {author} {\bibfnamefont {P.}~\bibnamefont
  {{Tinyakov}}},\ }\href@noop {} {\bibfield  {journal} {\bibinfo  {journal}
  {\arxiv}\ } (\bibinfo {year} {2014})},\ \Eprint
  {https://arxiv.org/abs/1402.4671}  { arXiv:1402.4671
  [astro-ph.CO]}\BibitemShut {NoStop}%
\bibitem [{\citenamefont {{Defillon}}\ \emph {et~al.}(2014)\citenamefont
  {{Defillon}}, \citenamefont {{Granet}}, \citenamefont {{Tinyakov}},\ and\
  \citenamefont {{Tytgat}}}]{2014PhRvD..90j3522D}%
  \BibitemOpen
  \bibfield  {author} {\bibinfo {author} {\bibfnamefont {G.}~\bibnamefont
  {{Defillon}}}, \bibinfo {author} {\bibfnamefont {E.}~\bibnamefont
  {{Granet}}}, \bibinfo {author} {\bibfnamefont {P.}~\bibnamefont
  {{Tinyakov}}},\ and\ \bibinfo {author} {\bibfnamefont {M.~H.~G.}\
  \bibnamefont {{Tytgat}}},\ }\href
  {https://doi.org/10.1103/PhysRevD.90.103522} {\bibfield  {journal} {\bibinfo
  {journal} {\prd}\ }\textbf {\bibinfo {volume} {90}},\ \bibinfo {eid} {103522}
  (\bibinfo {year} {2014})},\ \Eprint {https://arxiv.org/abs/1409.0469}  {
  arXiv:1409.0469 [gr-qc]}\BibitemShut {NoStop}%
\bibitem [{\citenamefont {{Montero-Camacho}}\ \emph {et~al.}(2019)\citenamefont
  {{Montero-Camacho}}, \citenamefont {{Fang}}, \citenamefont {{Vasquez}},
  \citenamefont {{Silva}},\ and\ \citenamefont
  {{Hirata}}}]{2019JCAP...08..031M}%
  \BibitemOpen
  \bibfield  {author} {\bibinfo {author} {\bibfnamefont {P.}~\bibnamefont
  {{Montero-Camacho}}}, \bibinfo {author} {\bibfnamefont {X.}~\bibnamefont
  {{Fang}}}, \bibinfo {author} {\bibfnamefont {G.}~\bibnamefont {{Vasquez}}},
  \bibinfo {author} {\bibfnamefont {M.}~\bibnamefont {{Silva}}},\ and\ \bibinfo
  {author} {\bibfnamefont {C.~M.}\ \bibnamefont {{Hirata}}},\ }\href
  {https://doi.org/10.1088/1475-7516/2019/08/031} {\bibfield  {journal}
  {\bibinfo  {journal} {\jcap}\ }\textbf {\bibinfo {volume} {2019}},\ \bibinfo
  {eid} {031} (\bibinfo {year} {2019})},\ \Eprint
  {https://arxiv.org/abs/1906.05950}  { arXiv:1906.05950
  [astro-ph.CO]}\BibitemShut {NoStop}%
\bibitem [{\citenamefont {{Inoue}}\ and\ \citenamefont
  {{Kusenko}}(2017)}]{2017JCAP...10..034I}%
  \BibitemOpen
  \bibfield  {author} {\bibinfo {author} {\bibfnamefont {Y.}~\bibnamefont
  {{Inoue}}}\ and\ \bibinfo {author} {\bibfnamefont {A.}~\bibnamefont
  {{Kusenko}}},\ }\href {https://doi.org/10.1088/1475-7516/2017/10/034}
  {\bibfield  {journal} {\bibinfo  {journal} {\jcap}\ }\textbf {\bibinfo
  {volume} {2017}},\ \bibinfo {eid} {034} (\bibinfo {year} {2017})},\ \Eprint
  {https://arxiv.org/abs/1705.00791}  { arXiv:1705.00791
  [astro-ph.CO]}\BibitemShut {NoStop}%
\bibitem [{\citenamefont {{Serpico}}\ \emph {et~al.}(2020)\citenamefont
  {{Serpico}}, \citenamefont {{Poulin}}, \citenamefont {{Inman}},\ and\
  \citenamefont {{Kohri}}}]{2020PhRvR...2b3204S}%
  \BibitemOpen
  \bibfield  {author} {\bibinfo {author} {\bibfnamefont {P.~D.}\ \bibnamefont
  {{Serpico}}}, \bibinfo {author} {\bibfnamefont {V.}~\bibnamefont {{Poulin}}},
  \bibinfo {author} {\bibfnamefont {D.}~\bibnamefont {{Inman}}},\ and\ \bibinfo
  {author} {\bibfnamefont {K.}~\bibnamefont {{Kohri}}},\ }\href
  {https://doi.org/10.1103/PhysRevResearch.2.023204} {\bibfield  {journal}
  {\bibinfo  {journal} {\PhRvR}\ }\textbf {\bibinfo {volume} {2}},\ \bibinfo
  {eid} {023204} (\bibinfo {year} {2020})},\ \Eprint
  {https://arxiv.org/abs/2002.10771}  { arXiv:2002.10771
  [astro-ph.CO]}\BibitemShut {NoStop}%
\bibitem [{\citenamefont {{Nakama}}\ \emph {et~al.}(2018)\citenamefont
  {{Nakama}}, \citenamefont {{Carr}},\ and\ \citenamefont
  {{Silk}}}]{2018PhRvD..97d3525N}%
  \BibitemOpen
  \bibfield  {author} {\bibinfo {author} {\bibfnamefont {T.}~\bibnamefont
  {{Nakama}}}, \bibinfo {author} {\bibfnamefont {B.}~\bibnamefont {{Carr}}},\
  and\ \bibinfo {author} {\bibfnamefont {J.}~\bibnamefont {{Silk}}},\ }\href
  {https://doi.org/10.1103/PhysRevD.97.043525} {\bibfield  {journal} {\bibinfo
  {journal} {\prd}\ }\textbf {\bibinfo {volume} {97}},\ \bibinfo {eid} {043525}
  (\bibinfo {year} {2018})},\ \Eprint {https://arxiv.org/abs/1710.06945}  {
  arXiv:1710.06945 [astro-ph.CO]}\BibitemShut {NoStop}%
\bibitem [{\citenamefont {{Abbott}}\ \emph
  {et~al.}(2016{\natexlab{a}})\citenamefont {{Abbott}} \emph
  {et~al.}}]{2016PhRvL.116f1102A}%
  \BibitemOpen
  \bibfield  {author} {\bibinfo {author} {\bibfnamefont {B.~P.}\ \bibnamefont
  {{Abbott}}} \emph {et~al.} (\bibinfo {collaboration} {LIGO, VIRGO}),\ }\href
  {https://doi.org/10.1103/PhysRevLett.116.061102} {\bibfield  {journal}
  {\bibinfo  {journal} {\prl}\ }\textbf {\bibinfo {volume} {116}},\ \bibinfo
  {eid} {061102} (\bibinfo {year} {2016}{\natexlab{a}})},\ \Eprint
  {https://arxiv.org/abs/1602.03837}  { arXiv:1602.03837 [gr-qc]}\BibitemShut
  {NoStop}%
\bibitem [{\citenamefont {{Abbott}}\ \emph
  {et~al.}(2016{\natexlab{b}})\citenamefont {{Abbott}} \emph
  {et~al.}}]{2016ApJ...833L...1A}%
  \BibitemOpen
  \bibfield  {author} {\bibinfo {author} {\bibfnamefont {B.~P.}\ \bibnamefont
  {{Abbott}}} \emph {et~al.} (\bibinfo {collaboration} {LIGO, VIRGO}),\ }\href
  {https://doi.org/10.3847/2041-8205/833/1/L1} {\bibfield  {journal} {\bibinfo
  {journal} {\apjl}\ }\textbf {\bibinfo {volume} {833}},\ \bibinfo {eid} {L1}
  (\bibinfo {year} {2016}{\natexlab{b}})},\ \Eprint
  {https://arxiv.org/abs/1602.03842}  { arXiv:1602.03842
  [astro-ph.HE]}\BibitemShut {NoStop}%
\bibitem [{\citenamefont {{Bird}}\ \emph {et~al.}(2016)\citenamefont {{Bird}}
  \emph {et~al.}}]{2016PhRvL.116t1301B}%
  \BibitemOpen
  \bibfield  {author} {\bibinfo {author} {\bibfnamefont {S.}~\bibnamefont
  {{Bird}}} \emph {et~al.},\ }\href
  {https://doi.org/10.1103/PhysRevLett.116.201301} {\bibfield  {journal}
  {\bibinfo  {journal} {\prl}\ }\textbf {\bibinfo {volume} {116}},\ \bibinfo
  {eid} {201301} (\bibinfo {year} {2016})},\ \Eprint
  {https://arxiv.org/abs/1603.00464}  { arXiv:1603.00464
  [astro-ph.CO]}\BibitemShut {NoStop}%
\bibitem [{\citenamefont {{Sasaki}}\ \emph {et~al.}(2016)\citenamefont
  {{Sasaki}}, \citenamefont {{Suyama}}, \citenamefont {{Tanaka}},\ and\
  \citenamefont {{Yokoyama}}}]{2016PhRvL.117f1101S}%
  \BibitemOpen
  \bibfield  {author} {\bibinfo {author} {\bibfnamefont {M.}~\bibnamefont
  {{Sasaki}}}, \bibinfo {author} {\bibfnamefont {T.}~\bibnamefont {{Suyama}}},
  \bibinfo {author} {\bibfnamefont {T.}~\bibnamefont {{Tanaka}}},\ and\
  \bibinfo {author} {\bibfnamefont {S.}~\bibnamefont {{Yokoyama}}},\ }\href
  {https://doi.org/10.1103/PhysRevLett.117.061101} {\bibfield  {journal}
  {\bibinfo  {journal} {\prl}\ }\textbf {\bibinfo {volume} {117}},\ \bibinfo
  {eid} {061101} (\bibinfo {year} {2016})},\ \bibinfo {note}
  {[\prl~\textbf{121}, 059901 (2018)]},\ \Eprint
  {https://arxiv.org/abs/1603.08338}  { arXiv:1603.08338
  [astro-ph.CO]}\BibitemShut {NoStop}%
\bibitem [{\citenamefont {{Raidal}}\ \emph {et~al.}(2017)\citenamefont
  {{Raidal}}, \citenamefont {{Vaskonen}},\ and\ \citenamefont
  {{Veerm{\"a}e}}}]{2017JCAP...09..037R}%
  \BibitemOpen
  \bibfield  {author} {\bibinfo {author} {\bibfnamefont {M.}~\bibnamefont
  {{Raidal}}}, \bibinfo {author} {\bibfnamefont {V.}~\bibnamefont
  {{Vaskonen}}},\ and\ \bibinfo {author} {\bibfnamefont {H.}~\bibnamefont
  {{Veerm{\"a}e}}},\ }\href {https://doi.org/10.1088/1475-7516/2017/09/037}
  {\bibfield  {journal} {\bibinfo  {journal} {\jcap}\ }\textbf {\bibinfo
  {volume} {2017}},\ \bibinfo {eid} {037} (\bibinfo {year} {2017})},\ \Eprint
  {https://arxiv.org/abs/1707.01480}  { arXiv:1707.01480
  [astro-ph.CO]}\BibitemShut {NoStop}%
\bibitem [{\citenamefont {{Abbott}}\ \emph {et~al.}(2019)\citenamefont
  {{Abbott}} \emph {et~al.}}]{2019PhRvD.100b4017A}%
  \BibitemOpen
  \bibfield  {author} {\bibinfo {author} {\bibfnamefont {B.~P.}\ \bibnamefont
  {{Abbott}}} \emph {et~al.} (\bibinfo {collaboration} {LIGO, VIRGO}),\ }\href
  {https://doi.org/10.1103/PhysRevD.100.024017} {\bibfield  {journal} {\bibinfo
   {journal} {\prd}\ }\textbf {\bibinfo {volume} {100}},\ \bibinfo {eid}
  {024017} (\bibinfo {year} {2019})},\ \Eprint
  {https://arxiv.org/abs/1904.08976}  { arXiv:1904.08976
  [astro-ph.CO]}\BibitemShut {NoStop}%
\bibitem [{\citenamefont {{Chen}}\ \emph {et~al.}(2020)\citenamefont {{Chen}},
  \citenamefont {{Yuan}},\ and\ \citenamefont {{Huang}}}]{2020PhRvL.124y1101C}%
  \BibitemOpen
  \bibfield  {author} {\bibinfo {author} {\bibfnamefont {Z.-C.}\ \bibnamefont
  {{Chen}}}, \bibinfo {author} {\bibfnamefont {C.}~\bibnamefont {{Yuan}}},\
  and\ \bibinfo {author} {\bibfnamefont {Q.-G.}\ \bibnamefont {{Huang}}},\
  }\href {https://doi.org/10.1103/PhysRevLett.124.251101} {\bibfield  {journal}
  {\bibinfo  {journal} {\prl}\ }\textbf {\bibinfo {volume} {124}},\ \bibinfo
  {eid} {251101} (\bibinfo {year} {2020})},\ \Eprint
  {https://arxiv.org/abs/1910.12239}  { arXiv:1910.12239
  [astro-ph.CO]}\BibitemShut {NoStop}%
\bibitem [{\citenamefont {{Caldwell}}\ \emph {et~al.}(2022)\citenamefont
  {{Caldwell}} \emph {et~al.}}]{2022arXiv220307972C}%
  \BibitemOpen
  \bibfield  {author} {\bibinfo {author} {\bibfnamefont {R.}~\bibnamefont
  {{Caldwell}}} \emph {et~al.},\ }\href@noop {} {\bibfield  {journal} {\bibinfo
   {journal} {\arxiv}\ } (\bibinfo {year} {2022})},\ \Eprint
  {https://arxiv.org/abs/2203.07972}  { arXiv:2203.07972 [gr-qc]}\BibitemShut
  {NoStop}%
\bibitem [{\citenamefont {{Franciolini}}\ \emph {et~al.}(2021)\citenamefont
  {{Franciolini}} \emph {et~al.}}]{2021arXiv210503349F}%
  \BibitemOpen
  \bibfield  {author} {\bibinfo {author} {\bibfnamefont {G.}~\bibnamefont
  {{Franciolini}}} \emph {et~al.},\ }\href@noop {} {\bibfield  {journal}
  {\bibinfo  {journal} {\arxiv}\ } (\bibinfo {year} {2021})},\ \Eprint
  {https://arxiv.org/abs/2105.03349}  { arXiv:2105.03349 [gr-qc]}\BibitemShut
  {NoStop}%
\bibitem [{\citenamefont {{Chen}}\ and\ \citenamefont
  {{Huang}}(2020)}]{2020JCAP...08..039C}%
  \BibitemOpen
  \bibfield  {author} {\bibinfo {author} {\bibfnamefont {Z.-C.}\ \bibnamefont
  {{Chen}}}\ and\ \bibinfo {author} {\bibfnamefont {Q.-G.}\ \bibnamefont
  {{Huang}}},\ }\href {https://doi.org/10.1088/1475-7516/2020/08/039}
  {\bibfield  {journal} {\bibinfo  {journal} {\jcap}\ }\textbf {\bibinfo
  {volume} {2020}},\ \bibinfo {eid} {039} (\bibinfo {year} {2020})},\ \Eprint
  {https://arxiv.org/abs/1904.02396}  { arXiv:1904.02396
  [astro-ph.CO]}\BibitemShut {NoStop}%
\bibitem [{\citenamefont {{Coupechoux}}\ \emph {et~al.}(2022)\citenamefont
  {{Coupechoux}} \emph {et~al.}}]{2022PhRvD.105f4063C}%
  \BibitemOpen
  \bibfield  {author} {\bibinfo {author} {\bibfnamefont {J.~F.}\ \bibnamefont
  {{Coupechoux}}} \emph {et~al.},\ }\href
  {https://doi.org/10.1103/PhysRevD.105.064063} {\bibfield  {journal} {\bibinfo
   {journal} {\prd}\ }\textbf {\bibinfo {volume} {105}},\ \bibinfo {eid}
  {064063} (\bibinfo {year} {2022})},\ \Eprint
  {https://arxiv.org/abs/2106.05805}  { arXiv:2106.05805 [gr-qc]}\BibitemShut
  {NoStop}%
\bibitem [{\citenamefont {{Dasgupta}}\ \emph {et~al.}(2021)\citenamefont
  {{Dasgupta}}, \citenamefont {{Laha}},\ and\ \citenamefont
  {{Ray}}}]{2021PhRvL.126n1105D}%
  \BibitemOpen
  \bibfield  {author} {\bibinfo {author} {\bibfnamefont {B.}~\bibnamefont
  {{Dasgupta}}}, \bibinfo {author} {\bibfnamefont {R.}~\bibnamefont {{Laha}}},\
  and\ \bibinfo {author} {\bibfnamefont {A.}~\bibnamefont {{Ray}}},\ }\href
  {https://doi.org/10.1103/PhysRevLett.126.141105} {\bibfield  {journal}
  {\bibinfo  {journal} {\prl}\ }\textbf {\bibinfo {volume} {126}},\ \bibinfo
  {eid} {141105} (\bibinfo {year} {2021})},\ \Eprint
  {https://arxiv.org/abs/2009.01825}  { arXiv:2009.01825
  [astro-ph.HE]}\BibitemShut {NoStop}%
\bibitem [{\citenamefont {{Inomata}}\ \emph {et~al.}(2020)\citenamefont
  {{Inomata}}, \citenamefont {{Kawasaki}}, \citenamefont {{Mukaida}},
  \citenamefont {{Terada}},\ and\ \citenamefont
  {{Yanagida}}}]{2020PhRvD.101l3533I}%
  \BibitemOpen
  \bibfield  {author} {\bibinfo {author} {\bibfnamefont {K.}~\bibnamefont
  {{Inomata}}}, \bibinfo {author} {\bibfnamefont {M.}~\bibnamefont
  {{Kawasaki}}}, \bibinfo {author} {\bibfnamefont {K.}~\bibnamefont
  {{Mukaida}}}, \bibinfo {author} {\bibfnamefont {T.}~\bibnamefont
  {{Terada}}},\ and\ \bibinfo {author} {\bibfnamefont {T.~T.}\ \bibnamefont
  {{Yanagida}}},\ }\href {https://doi.org/10.1103/PhysRevD.101.123533}
  {\bibfield  {journal} {\bibinfo  {journal} {\prd}\ }\textbf {\bibinfo
  {volume} {101}},\ \bibinfo {eid} {123533} (\bibinfo {year} {2020})},\ \Eprint
  {https://arxiv.org/abs/2003.10455}  { arXiv:2003.10455
  [astro-ph.CO]}\BibitemShut {NoStop}%
\bibitem [{\citenamefont {{Dom{\`e}nech}}\ \emph {et~al.}(2021)\citenamefont
  {{Dom{\`e}nech}}, \citenamefont {{Takhistov}},\ and\ \citenamefont
  {{Sasaki}}}]{2021PhLB..82336722D}%
  \BibitemOpen
  \bibfield  {author} {\bibinfo {author} {\bibfnamefont {G.}~\bibnamefont
  {{Dom{\`e}nech}}}, \bibinfo {author} {\bibfnamefont {V.}~\bibnamefont
  {{Takhistov}}},\ and\ \bibinfo {author} {\bibfnamefont {M.}~\bibnamefont
  {{Sasaki}}},\ }\href {https://doi.org/10.1016/j.physletb.2021.136722}
  {\bibfield  {journal} {\bibinfo  {journal} {\plb}\ }\textbf {\bibinfo
  {volume} {823}},\ \bibinfo {eid} {136722} (\bibinfo {year} {2021})},\ \Eprint
  {https://arxiv.org/abs/2105.06816}  { arXiv:2105.06816
  [astro-ph.CO]}\BibitemShut {NoStop}%
\bibitem [{\citenamefont {{Vaskonen}}\ and\ \citenamefont
  {{Veerm{\"a}e}}(2021)}]{2021PhRvL.126e1303V}%
  \BibitemOpen
  \bibfield  {author} {\bibinfo {author} {\bibfnamefont {V.}~\bibnamefont
  {{Vaskonen}}}\ and\ \bibinfo {author} {\bibfnamefont {H.}~\bibnamefont
  {{Veerm{\"a}e}}},\ }\href {https://doi.org/10.1103/PhysRevLett.126.051303}
  {\bibfield  {journal} {\bibinfo  {journal} {\prl}\ }\textbf {\bibinfo
  {volume} {126}},\ \bibinfo {eid} {051303} (\bibinfo {year} {2021})},\ \Eprint
  {https://arxiv.org/abs/2009.07832}  { arXiv:2009.07832
  [astro-ph.CO]}\BibitemShut {NoStop}%
\bibitem [{\citenamefont {{K{\"u}hnel}}(2020)}]{2020EPJC...80..243K}%
  \BibitemOpen
  \bibfield  {author} {\bibinfo {author} {\bibfnamefont {F.}~\bibnamefont
  {{K{\"u}hnel}}},\ }\href {https://doi.org/10.1140/epjc/s10052-020-7807-z}
  {\bibfield  {journal} {\bibinfo  {journal} {\epjc}\ }\textbf {\bibinfo
  {volume} {80}},\ \bibinfo {eid} {243} (\bibinfo {year} {2020})},\ \Eprint
  {https://arxiv.org/abs/1909.04742}  { arXiv:1909.04742
  [astro-ph.CO]}\BibitemShut {NoStop}%
\bibitem [{\citenamefont {{Auclair}}\ \emph {et~al.}(2022)\citenamefont
  {{Auclair}} \emph {et~al.}}]{2022arXiv220405434A}%
  \BibitemOpen
  \bibfield  {author} {\bibinfo {author} {\bibfnamefont {P.}~\bibnamefont
  {{Auclair}}} \emph {et~al.} (\bibinfo {collaboration} {LISA}),\ }\href@noop
  {} {\bibfield  {journal} {\bibinfo  {journal} {\arxiv}\ } (\bibinfo {year}
  {2022})},\ \Eprint {https://arxiv.org/abs/2204.05434}  { arXiv:2204.05434
  [astro-ph.CO]}\BibitemShut {NoStop}%
\bibitem [{\citenamefont {{Fields}}\ \emph {et~al.}(2020)\citenamefont
  {{Fields}}, \citenamefont {{Olive}}, \citenamefont {{Yeh}},\ and\
  \citenamefont {{Young}}}]{2020JCAP...03..010F}%
  \BibitemOpen
  \bibfield  {author} {\bibinfo {author} {\bibfnamefont {B.~D.}\ \bibnamefont
  {{Fields}}}, \bibinfo {author} {\bibfnamefont {K.~A.}\ \bibnamefont
  {{Olive}}}, \bibinfo {author} {\bibfnamefont {T.-H.}\ \bibnamefont {{Yeh}}},\
  and\ \bibinfo {author} {\bibfnamefont {C.}~\bibnamefont {{Young}}},\ }\href
  {https://doi.org/10.1088/1475-7516/2020/03/010} {\bibfield  {journal}
  {\bibinfo  {journal} {\jcap}\ }\textbf {\bibinfo {volume} {2020}},\ \bibinfo
  {eid} {010} (\bibinfo {year} {2020})},\ \Eprint
  {https://arxiv.org/abs/1912.01132}  { arXiv:1912.01132
  [astro-ph.CO]}\BibitemShut {NoStop}%
\bibitem [{\citenamefont {{Sakharov}}(1967)}]{1967JETPL...5...24S}%
  \BibitemOpen
  \bibfield  {author} {\bibinfo {author} {\bibfnamefont {A.~D.}\ \bibnamefont
  {{Sakharov}}},\ }\href@noop {} {\bibfield  {journal} {\bibinfo  {journal}
  {\JETPL}\ }\textbf {\bibinfo {volume} {5}},\ \bibinfo {pages} {24} (\bibinfo
  {year} {1967})},\ \bibinfo {note} {[\ZhETF~\textbf{5}, 32
  (1967)]}\BibitemShut {NoStop}%
\bibitem [{\citenamefont {{Tanabashi}}\ \emph {et~al.}(2018)\citenamefont
  {{Tanabashi}} \emph {et~al.}}]{2018PhRvD..98c0001T}%
  \BibitemOpen
  \bibfield  {author} {\bibinfo {author} {\bibfnamefont {M.}~\bibnamefont
  {{Tanabashi}}} \emph {et~al.} (\bibinfo {collaboration} {Particle Data
  Group}),\ }\href {https://doi.org/10.1103/PhysRevD.98.030001} {\bibfield
  {journal} {\bibinfo  {journal} {\prd}\ }\textbf {\bibinfo {volume} {98}},\
  \bibinfo {pages} {030001} (\bibinfo {year} {2018})}\BibitemShut {NoStop}%
\bibitem [{\citenamefont {{Zel'dovich}}\ and\ \citenamefont
  {{Starobinski{\v{i}}}}(1976)}]{1976JETPL..24..571Z}%
  \BibitemOpen
  \bibfield  {author} {\bibinfo {author} {\bibfnamefont {Y.~B.}\ \bibnamefont
  {{Zel'dovich}}}\ and\ \bibinfo {author} {\bibfnamefont {A.~A.}\ \bibnamefont
  {{Starobinski{\v{i}}}}},\ }\href@noop {} {\bibfield  {journal} {\bibinfo
  {journal} {\JETPL}\ }\textbf {\bibinfo {volume} {24}},\ \bibinfo {pages}
  {571} (\bibinfo {year} {1976})},\ \bibinfo {note} {[\ZhETF~\textbf{24}, 616
  (1976)]}\BibitemShut {NoStop}%
\bibitem [{\citenamefont {{Chaudhuri}}\ and\ \citenamefont
  {{Dolgov}}(2021)}]{2021JETP..133..552C}%
  \BibitemOpen
  \bibfield  {author} {\bibinfo {author} {\bibfnamefont {A.}~\bibnamefont
  {{Chaudhuri}}}\ and\ \bibinfo {author} {\bibfnamefont {A.}~\bibnamefont
  {{Dolgov}}},\ }\href {https://doi.org/10.1134/S1063776121110078} {\bibfield
  {journal} {\bibinfo  {journal} {\JETP}\ }\textbf {\bibinfo {volume} {133}},\
  \bibinfo {pages} {552} (\bibinfo {year} {2021})},\ \Eprint
  {https://arxiv.org/abs/2001.11219}  { arXiv:2001.11219
  [astro-ph.CO]}\BibitemShut {NoStop}%
\bibitem [{\citenamefont {{Toussaint}}\ \emph {et~al.}(1979)\citenamefont
  {{Toussaint}}, \citenamefont {{Treiman}}, \citenamefont {{Wilczek}},\ and\
  \citenamefont {{Zee}}}]{1979PhRvD..19.1036T}%
  \BibitemOpen
  \bibfield  {author} {\bibinfo {author} {\bibfnamefont {D.}~\bibnamefont
  {{Toussaint}}}, \bibinfo {author} {\bibfnamefont {S.~B.}\ \bibnamefont
  {{Treiman}}}, \bibinfo {author} {\bibfnamefont {F.}~\bibnamefont
  {{Wilczek}}},\ and\ \bibinfo {author} {\bibfnamefont {A.}~\bibnamefont
  {{Zee}}},\ }\href {https://doi.org/10.1103/PhysRevD.19.1036} {\bibfield
  {journal} {\bibinfo  {journal} {\prd}\ }\textbf {\bibinfo {volume} {19}},\
  \bibinfo {pages} {1036} (\bibinfo {year} {1979})}\BibitemShut {NoStop}%
\bibitem [{\citenamefont {{Grillo}}(1980)}]{1980PhLB...94..364G}%
  \BibitemOpen
  \bibfield  {author} {\bibinfo {author} {\bibfnamefont {A.~F.}\ \bibnamefont
  {{Grillo}}},\ }\href {https://doi.org/10.1016/0370-2693(80)90897-7}
  {\bibfield  {journal} {\bibinfo  {journal} {\plb}\ }\textbf {\bibinfo
  {volume} {94}},\ \bibinfo {pages} {364} (\bibinfo {year} {1980})}\BibitemShut
  {NoStop}%
\bibitem [{\citenamefont {{Barrow}}(1980{\natexlab{a}})}]{1980SHEP....1..183B}%
  \BibitemOpen
  \bibfield  {author} {\bibinfo {author} {\bibfnamefont {J.~D.}\ \bibnamefont
  {{Barrow}}},\ }\href {https://doi.org/10.1080/01422418008225252} {\bibfield
  {journal} {\bibinfo  {journal} {\SHEP}\ }\textbf {\bibinfo {volume} {1}},\
  \bibinfo {pages} {183} (\bibinfo {year} {1980}{\natexlab{a}})}\BibitemShut
  {NoStop}%
\bibitem [{\citenamefont {{Barrow}}\ and\ \citenamefont
  {{Ross}}(1981)}]{1981NuPhB.181..461B}%
  \BibitemOpen
  \bibfield  {author} {\bibinfo {author} {\bibfnamefont {J.~D.}\ \bibnamefont
  {{Barrow}}}\ and\ \bibinfo {author} {\bibfnamefont {G.~G.}\ \bibnamefont
  {{Ross}}},\ }\href {https://doi.org/10.1016/0550-3213(81)90536-8} {\bibfield
  {journal} {\bibinfo  {journal} {\nphysb}\ }\textbf {\bibinfo {volume}
  {181}},\ \bibinfo {pages} {461} (\bibinfo {year} {1981})}\BibitemShut
  {NoStop}%
\bibitem [{\citenamefont {{Dolgov}}(1992)}]{1992PhR...222..309D}%
  \BibitemOpen
  \bibfield  {author} {\bibinfo {author} {\bibfnamefont {A.~D.}\ \bibnamefont
  {{Dolgov}}},\ }\href {https://doi.org/10.1016/0370-1573(92)90107-B}
  {\bibfield  {journal} {\bibinfo  {journal} {\physrep}\ }\textbf {\bibinfo
  {volume} {222}},\ \bibinfo {pages} {309} (\bibinfo {year}
  {1992})}\BibitemShut {NoStop}%
\bibitem [{\citenamefont {{Barrow}}(1980{\natexlab{b}})}]{1980MNRAS.192..427B}%
  \BibitemOpen
  \bibfield  {author} {\bibinfo {author} {\bibfnamefont {J.~D.}\ \bibnamefont
  {{Barrow}}},\ }\href {https://doi.org/10.1093/mnras/192.3.427} {\bibfield
  {journal} {\bibinfo  {journal} {\mnras}\ }\textbf {\bibinfo {volume} {192}},\
  \bibinfo {pages} {427} (\bibinfo {year} {1980}{\natexlab{b}})}\BibitemShut
  {NoStop}%
\bibitem [{\citenamefont {{Hook}}(2014)}]{2014PhRvD..90h3535H}%
  \BibitemOpen
  \bibfield  {author} {\bibinfo {author} {\bibfnamefont {A.}~\bibnamefont
  {{Hook}}},\ }\href {https://doi.org/10.1103/PhysRevD.90.083535} {\bibfield
  {journal} {\bibinfo  {journal} {\prd}\ }\textbf {\bibinfo {volume} {90}},\
  \bibinfo {eid} {083535} (\bibinfo {year} {2014})},\ \Eprint
  {https://arxiv.org/abs/1404.0113}  { arXiv:1404.0113 [hep-ph]}\BibitemShut
  {NoStop}%
\bibitem [{\citenamefont {{Fischler}}\ and\ \citenamefont
  {{Kundu}}(2015)}]{2015PhRvD..92d6008F}%
  \BibitemOpen
  \bibfield  {author} {\bibinfo {author} {\bibfnamefont {W.}~\bibnamefont
  {{Fischler}}}\ and\ \bibinfo {author} {\bibfnamefont {S.}~\bibnamefont
  {{Kundu}}},\ }\href {https://doi.org/10.1103/PhysRevD.92.046008} {\bibfield
  {journal} {\bibinfo  {journal} {\prd}\ }\textbf {\bibinfo {volume} {92}},\
  \bibinfo {eid} {046008} (\bibinfo {year} {2015})},\ \Eprint
  {https://arxiv.org/abs/1501.01316}  { arXiv:1501.01316 [hep-th]}\BibitemShut
  {NoStop}%
\bibitem [{\citenamefont {{Banks}}\ and\ \citenamefont
  {{Fischler}}(2015)}]{2015arXiv150500472B}%
  \BibitemOpen
  \bibfield  {author} {\bibinfo {author} {\bibfnamefont {T.}~\bibnamefont
  {{Banks}}}\ and\ \bibinfo {author} {\bibfnamefont {W.}~\bibnamefont
  {{Fischler}}},\ }\href@noop {} {\bibfield  {journal} {\bibinfo  {journal}
  {\arxiv}\ } (\bibinfo {year} {2015})},\ \Eprint
  {https://arxiv.org/abs/1505.00472}  { 1505.00472 [hep-th]}\BibitemShut
  {NoStop}%
\bibitem [{\citenamefont {{Hamada}}\ and\ \citenamefont
  {{Iso}}(2017)}]{2017PTEP.2017c3B02H}%
  \BibitemOpen
  \bibfield  {author} {\bibinfo {author} {\bibfnamefont {Y.}~\bibnamefont
  {{Hamada}}}\ and\ \bibinfo {author} {\bibfnamefont {S.}~\bibnamefont
  {{Iso}}},\ }\href {https://doi.org/10.1093/ptep/ptx011} {\bibfield  {journal}
  {\bibinfo  {journal} {\PTEP}\ }\textbf {\bibinfo {volume} {2017}},\ \bibinfo
  {eid} {033B02} (\bibinfo {year} {2017})},\ \Eprint
  {https://arxiv.org/abs/1610.02586}  { arXiv:1610.02586 [hep-ph]}\BibitemShut
  {NoStop}%
\bibitem [{\citenamefont {{Boudon}}\ \emph {et~al.}(2021)\citenamefont
  {{Boudon}}, \citenamefont {{Bose}}, \citenamefont {{Huang}},\ and\
  \citenamefont {{Lombriser}}}]{2021PhRvD.103h3504B}%
  \BibitemOpen
  \bibfield  {author} {\bibinfo {author} {\bibfnamefont {A.}~\bibnamefont
  {{Boudon}}}, \bibinfo {author} {\bibfnamefont {B.}~\bibnamefont {{Bose}}},
  \bibinfo {author} {\bibfnamefont {H.}~\bibnamefont {{Huang}}},\ and\ \bibinfo
  {author} {\bibfnamefont {L.}~\bibnamefont {{Lombriser}}},\ }\href
  {https://doi.org/10.1103/PhysRevD.103.083504} {\bibfield  {journal} {\bibinfo
   {journal} {\prd}\ }\textbf {\bibinfo {volume} {103}},\ \bibinfo {eid}
  {083504} (\bibinfo {year} {2021})},\ \Eprint
  {https://arxiv.org/abs/2010.14426}  { arXiv:2010.14426
  [astro-ph.CO]}\BibitemShut {NoStop}%
\bibitem [{\citenamefont {{Smyth}}\ \emph {et~al.}(2022)\citenamefont
  {{Smyth}}, \citenamefont {{Santos-Olmsted}},\ and\ \citenamefont
  {{Profumo}}}]{2022JCAP...03..013S}%
  \BibitemOpen
  \bibfield  {author} {\bibinfo {author} {\bibfnamefont {N.}~\bibnamefont
  {{Smyth}}}, \bibinfo {author} {\bibfnamefont {L.}~\bibnamefont
  {{Santos-Olmsted}}},\ and\ \bibinfo {author} {\bibfnamefont {S.}~\bibnamefont
  {{Profumo}}},\ }\href {https://doi.org/10.1088/1475-7516/2022/03/013}
  {\bibfield  {journal} {\bibinfo  {journal} {\jcap}\ }\textbf {\bibinfo
  {volume} {2022}},\ \bibinfo {eid} {013} (\bibinfo {year} {2022})},\ \Eprint
  {https://arxiv.org/abs/2110.14660}  { arXiv:2110.14660 [hep-ph]}\BibitemShut
  {NoStop}%
\bibitem [{\citenamefont {{Turner}}\ and\ \citenamefont
  {{Schramm}}(1979)}]{1979Natur.279..303T}%
  \BibitemOpen
  \bibfield  {author} {\bibinfo {author} {\bibfnamefont {M.~S.}\ \bibnamefont
  {{Turner}}}\ and\ \bibinfo {author} {\bibfnamefont {D.~N.}\ \bibnamefont
  {{Schramm}}},\ }\href {https://doi.org/10.1038/279303a0} {\bibfield
  {journal} {\bibinfo  {journal} {\nat}\ }\textbf {\bibinfo {volume} {279}},\
  \bibinfo {pages} {303} (\bibinfo {year} {1979})}\BibitemShut {NoStop}%
\bibitem [{\citenamefont {{Weinberg}}(1979)}]{1979PhRvL..42..850W}%
  \BibitemOpen
  \bibfield  {author} {\bibinfo {author} {\bibfnamefont {S.}~\bibnamefont
  {{Weinberg}}},\ }\href {https://doi.org/10.1103/PhysRevLett.42.850}
  {\bibfield  {journal} {\bibinfo  {journal} {\prl}\ }\textbf {\bibinfo
  {volume} {42}},\ \bibinfo {pages} {850} (\bibinfo {year} {1979})}\BibitemShut
  {NoStop}%
\bibitem [{\citenamefont {{Nanopoulos}}\ and\ \citenamefont
  {{Weinberg}}(1979)}]{1979PhRvD..20.2484N}%
  \BibitemOpen
  \bibfield  {author} {\bibinfo {author} {\bibfnamefont {D.~V.}\ \bibnamefont
  {{Nanopoulos}}}\ and\ \bibinfo {author} {\bibfnamefont {S.}~\bibnamefont
  {{Weinberg}}},\ }\href {https://doi.org/10.1103/PhysRevD.20.2484} {\bibfield
  {journal} {\bibinfo  {journal} {\prd}\ }\textbf {\bibinfo {volume} {20}},\
  \bibinfo {pages} {2484} (\bibinfo {year} {1979})}\BibitemShut {NoStop}%
\bibitem [{\citenamefont {{Turner}}(1979)}]{1979PhLB...89..155T}%
  \BibitemOpen
  \bibfield  {author} {\bibinfo {author} {\bibfnamefont {M.~S.}\ \bibnamefont
  {{Turner}}},\ }\href {https://doi.org/10.1016/0370-2693(79)90095-9}
  {\bibfield  {journal} {\bibinfo  {journal} {\plb}\ }\textbf {\bibinfo
  {volume} {89}},\ \bibinfo {pages} {155} (\bibinfo {year} {1979})}\BibitemShut
  {NoStop}%
\bibitem [{\citenamefont {{Barrow}}(1980{\natexlab{c}})}]{1980MNRAS.192P..19B}%
  \BibitemOpen
  \bibfield  {author} {\bibinfo {author} {\bibfnamefont {J.~D.}\ \bibnamefont
  {{Barrow}}},\ }\href {https://doi.org/10.1093/mnras/192.1.19P} {\bibfield
  {journal} {\bibinfo  {journal} {\mnras}\ }\textbf {\bibinfo {volume} {192}},\
  \bibinfo {pages} {19P} (\bibinfo {year} {1980}{\natexlab{c}})}\BibitemShut
  {NoStop}%
\bibitem [{\citenamefont {{Barrow}}\ \emph
  {et~al.}(1991{\natexlab{a}})\citenamefont {{Barrow}}, \citenamefont
  {{Copeland}}, \citenamefont {{Kolb}},\ and\ \citenamefont
  {{Liddle}}}]{1991PhRvD..43..977B}%
  \BibitemOpen
  \bibfield  {author} {\bibinfo {author} {\bibfnamefont {J.~D.}\ \bibnamefont
  {{Barrow}}}, \bibinfo {author} {\bibfnamefont {E.~J.}\ \bibnamefont
  {{Copeland}}}, \bibinfo {author} {\bibfnamefont {E.~W.}\ \bibnamefont
  {{Kolb}}},\ and\ \bibinfo {author} {\bibfnamefont {A.~R.}\ \bibnamefont
  {{Liddle}}},\ }\href {https://doi.org/10.1103/PhysRevD.43.977} {\bibfield
  {journal} {\bibinfo  {journal} {\prd}\ }\textbf {\bibinfo {volume} {43}},\
  \bibinfo {pages} {977} (\bibinfo {year} {1991}{\natexlab{a}})}\BibitemShut
  {NoStop}%
\bibitem [{\citenamefont {{Barrow}}\ \emph
  {et~al.}(1991{\natexlab{b}})\citenamefont {{Barrow}}, \citenamefont
  {{Copeland}}, \citenamefont {{Kolb}},\ and\ \citenamefont
  {{Liddle}}}]{1991PhRvD..43..984B}%
  \BibitemOpen
  \bibfield  {author} {\bibinfo {author} {\bibfnamefont {J.~D.}\ \bibnamefont
  {{Barrow}}}, \bibinfo {author} {\bibfnamefont {E.~J.}\ \bibnamefont
  {{Copeland}}}, \bibinfo {author} {\bibfnamefont {E.~W.}\ \bibnamefont
  {{Kolb}}},\ and\ \bibinfo {author} {\bibfnamefont {A.~R.}\ \bibnamefont
  {{Liddle}}},\ }\href {https://doi.org/10.1103/PhysRevD.43.984} {\bibfield
  {journal} {\bibinfo  {journal} {\prd}\ }\textbf {\bibinfo {volume} {43}},\
  \bibinfo {pages} {984} (\bibinfo {year} {1991}{\natexlab{b}})}\BibitemShut
  {NoStop}%
\bibitem [{\citenamefont {{Baumann}}\ \emph {et~al.}(2007)\citenamefont
  {{Baumann}}, \citenamefont {{Steinhardt}},\ and\ \citenamefont
  {{Turok}}}]{2007hep.th....3250B}%
  \BibitemOpen
  \bibfield  {author} {\bibinfo {author} {\bibfnamefont {D.}~\bibnamefont
  {{Baumann}}}, \bibinfo {author} {\bibfnamefont {P.~J.}\ \bibnamefont
  {{Steinhardt}}},\ and\ \bibinfo {author} {\bibfnamefont {N.}~\bibnamefont
  {{Turok}}},\ }\href@noop {} {\bibfield  {journal} {\bibinfo  {journal}
  {\arxiv}\ } (\bibinfo {year} {2007})},\ \Eprint
  {https://arxiv.org/abs/hep-th/0703250}  { arXiv:hep-th/0703250
  [hep-th]}\BibitemShut {NoStop}%
\bibitem [{\citenamefont {{Hooper}}\ and\ \citenamefont
  {{Krnjaic}}(2021)}]{2021PhRvD.103d3504H}%
  \BibitemOpen
  \bibfield  {author} {\bibinfo {author} {\bibfnamefont {D.}~\bibnamefont
  {{Hooper}}}\ and\ \bibinfo {author} {\bibfnamefont {G.}~\bibnamefont
  {{Krnjaic}}},\ }\href {https://doi.org/10.1103/PhysRevD.103.043504}
  {\bibfield  {journal} {\bibinfo  {journal} {\prd}\ }\textbf {\bibinfo
  {volume} {103}},\ \bibinfo {eid} {043504} (\bibinfo {year} {2021})},\ \Eprint
  {https://arxiv.org/abs/2010.01134}  { arXiv:2010.01134 [hep-ph]}\BibitemShut
  {NoStop}%
\bibitem [{\citenamefont {{Dolgov}}(1981)}]{1981PhRvD..24.1042D}%
  \BibitemOpen
  \bibfield  {author} {\bibinfo {author} {\bibfnamefont {A.~D.}\ \bibnamefont
  {{Dolgov}}},\ }\href {https://doi.org/10.1103/PhysRevD.24.1042} {\bibfield
  {journal} {\bibinfo  {journal} {\prd}\ }\textbf {\bibinfo {volume} {24}},\
  \bibinfo {pages} {1042} (\bibinfo {year} {1981})}\BibitemShut {NoStop}%
\bibitem [{\citenamefont {{Dolgov}}(1980)}]{1980JETP...52..169D}%
  \BibitemOpen
  \bibfield  {author} {\bibinfo {author} {\bibfnamefont {A.~D.}\ \bibnamefont
  {{Dolgov}}},\ }\href@noop {} {\bibfield  {journal} {\bibinfo  {journal}
  {\JETP}\ }\textbf {\bibinfo {volume} {52}},\ \bibinfo {pages} {169} (\bibinfo
  {year} {1980})},\ \bibinfo {note} {[\ZhETF~\textbf{79}, 337
  (1980)]}\BibitemShut {NoStop}%
\bibitem [{\citenamefont {{Ambrosone}}\ \emph {et~al.}(2022)\citenamefont
  {{Ambrosone}}, \citenamefont {{Calabrese}}, \citenamefont {{Fiorillo}},
  \citenamefont {{Miele}},\ and\ \citenamefont
  {{Morisi}}}]{2022PhRvD.105d5001A}%
  \BibitemOpen
  \bibfield  {author} {\bibinfo {author} {\bibfnamefont {A.}~\bibnamefont
  {{Ambrosone}}}, \bibinfo {author} {\bibfnamefont {R.}~\bibnamefont
  {{Calabrese}}}, \bibinfo {author} {\bibfnamefont {D.~F.~G.}\ \bibnamefont
  {{Fiorillo}}}, \bibinfo {author} {\bibfnamefont {G.}~\bibnamefont
  {{Miele}}},\ and\ \bibinfo {author} {\bibfnamefont {S.}~\bibnamefont
  {{Morisi}}},\ }\href {https://doi.org/10.1103/PhysRevD.105.045001} {\bibfield
   {journal} {\bibinfo  {journal} {\prd}\ }\textbf {\bibinfo {volume} {105}},\
  \bibinfo {eid} {045001} (\bibinfo {year} {2022})},\ \Eprint
  {https://arxiv.org/abs/2106.11980}  { arXiv:2106.11980 [hep-ph]}\BibitemShut
  {NoStop}%
\bibitem [{\citenamefont {{Kuzmin}}\ \emph {et~al.}(1985)\citenamefont
  {{Kuzmin}}, \citenamefont {{Rubakov}},\ and\ \citenamefont
  {{Shaposhnikov}}}]{1985PhLB..155...36K}%
  \BibitemOpen
  \bibfield  {author} {\bibinfo {author} {\bibfnamefont {V.~A.}\ \bibnamefont
  {{Kuzmin}}}, \bibinfo {author} {\bibfnamefont {V.~A.}\ \bibnamefont
  {{Rubakov}}},\ and\ \bibinfo {author} {\bibfnamefont {M.~E.}\ \bibnamefont
  {{Shaposhnikov}}},\ }\href {https://doi.org/10.1016/0370-2693(85)91028-7}
  {\bibfield  {journal} {\bibinfo  {journal} {\plb}\ }\textbf {\bibinfo
  {volume} {155}},\ \bibinfo {pages} {36} (\bibinfo {year} {1985})}\BibitemShut
  {NoStop}%
\bibitem [{\citenamefont {{Alexander}}\ and\ \citenamefont
  {{M{\'e}sz{\'a}ros}}(2007)}]{2007hep.th....3070A}%
  \BibitemOpen
  \bibfield  {author} {\bibinfo {author} {\bibfnamefont {S.}~\bibnamefont
  {{Alexander}}}\ and\ \bibinfo {author} {\bibfnamefont {P.}~\bibnamefont
  {{M{\'e}sz{\'a}ros}}},\ }\href@noop {} {\bibfield  {journal} {\bibinfo
  {journal} {arXiv e-prints}\ } (\bibinfo {year} {2007})},\ \Eprint
  {https://arxiv.org/abs/hep-th/0703070}  { hep-th/0703070
  [hep-th]}\BibitemShut {NoStop}%
\bibitem [{\citenamefont {{Fujita}}\ \emph {et~al.}(2014)\citenamefont
  {{Fujita}}, \citenamefont {{Harigaya}}, \citenamefont {{Kawasaki}},\ and\
  \citenamefont {{Matsuda}}}]{2014PhRvD..89j3501F}%
  \BibitemOpen
  \bibfield  {author} {\bibinfo {author} {\bibfnamefont {T.}~\bibnamefont
  {{Fujita}}}, \bibinfo {author} {\bibfnamefont {K.}~\bibnamefont
  {{Harigaya}}}, \bibinfo {author} {\bibfnamefont {M.}~\bibnamefont
  {{Kawasaki}}},\ and\ \bibinfo {author} {\bibfnamefont {R.}~\bibnamefont
  {{Matsuda}}},\ }\href {https://doi.org/10.1103/PhysRevD.89.103501} {\bibfield
   {journal} {\bibinfo  {journal} {\prd}\ }\textbf {\bibinfo {volume} {89}},\
  \bibinfo {eid} {103501} (\bibinfo {year} {2014})},\ \Eprint
  {https://arxiv.org/abs/1401.1909}  { arXiv:1401.1909
  [astro-ph.CO]}\BibitemShut {NoStop}%
\bibitem [{\citenamefont {{Fukugita}}\ and\ \citenamefont
  {{Yanagida}}(1986)}]{1986PhLB..174...45F}%
  \BibitemOpen
  \bibfield  {author} {\bibinfo {author} {\bibfnamefont {M.}~\bibnamefont
  {{Fukugita}}}\ and\ \bibinfo {author} {\bibfnamefont {T.}~\bibnamefont
  {{Yanagida}}},\ }\href {https://doi.org/10.1016/0370-2693(86)91126-3}
  {\bibfield  {journal} {\bibinfo  {journal} {\plb}\ }\textbf {\bibinfo
  {volume} {174}},\ \bibinfo {pages} {45} (\bibinfo {year} {1986})}\BibitemShut
  {NoStop}%
\bibitem [{\citenamefont {{Baldes}}\ \emph {et~al.}(2020)\citenamefont
  {{Baldes}}, \citenamefont {{Decant}}, \citenamefont {{Hooper}},\ and\
  \citenamefont {{Lopez-Honorez}}}]{2020JCAP...08..045B}%
  \BibitemOpen
  \bibfield  {author} {\bibinfo {author} {\bibfnamefont {I.}~\bibnamefont
  {{Baldes}}}, \bibinfo {author} {\bibfnamefont {Q.}~\bibnamefont {{Decant}}},
  \bibinfo {author} {\bibfnamefont {D.~C.}\ \bibnamefont {{Hooper}}},\ and\
  \bibinfo {author} {\bibfnamefont {L.}~\bibnamefont {{Lopez-Honorez}}},\
  }\href {https://doi.org/10.1088/1475-7516/2020/08/045} {\bibfield  {journal}
  {\bibinfo  {journal} {\jcap}\ }\textbf {\bibinfo {volume} {2020}},\ \bibinfo
  {eid} {045} (\bibinfo {year} {2020})},\ \Eprint
  {https://arxiv.org/abs/2004.14773}  { arXiv:2004.14773
  [astro-ph.CO]}\BibitemShut {NoStop}%
\bibitem [{\citenamefont {{Perez-Gonzalez}}\ and\ \citenamefont
  {{Turner}}(2021)}]{2021PhRvD.104j3021P}%
  \BibitemOpen
  \bibfield  {author} {\bibinfo {author} {\bibfnamefont {Y.~F.}\ \bibnamefont
  {{Perez-Gonzalez}}}\ and\ \bibinfo {author} {\bibfnamefont {J.}~\bibnamefont
  {{Turner}}},\ }\href {https://doi.org/10.1103/PhysRevD.104.103021} {\bibfield
   {journal} {\bibinfo  {journal} {\prd}\ }\textbf {\bibinfo {volume} {104}},\
  \bibinfo {eid} {103021} (\bibinfo {year} {2021})},\ \Eprint
  {https://arxiv.org/abs/2010.03565}  { arXiv:2010.03565 [hep-ph]}\BibitemShut
  {NoStop}%
\bibitem [{\citenamefont {{Das}}\ \emph {et~al.}(2021)\citenamefont {{Das}},
  \citenamefont {{Mahanta}},\ and\ \citenamefont
  {{Borah}}}]{2021JCAP...11..019D}%
  \BibitemOpen
  \bibfield  {author} {\bibinfo {author} {\bibfnamefont {S.~J.}\ \bibnamefont
  {{Das}}}, \bibinfo {author} {\bibfnamefont {D.}~\bibnamefont {{Mahanta}}},\
  and\ \bibinfo {author} {\bibfnamefont {D.}~\bibnamefont {{Borah}}},\ }\href
  {https://doi.org/10.1088/1475-7516/2021/11/019} {\bibfield  {journal}
  {\bibinfo  {journal} {\jcap}\ }\textbf {\bibinfo {volume} {2021}},\ \bibinfo
  {eid} {019} (\bibinfo {year} {2021})}\BibitemShut {NoStop}%
\bibitem [{\citenamefont {{Datta}}\ \emph {et~al.}(2021)\citenamefont
  {{Datta}}, \citenamefont {{Ghosal}},\ and\ \citenamefont
  {{Samanta}}}]{2021JCAP...08..021D}%
  \BibitemOpen
  \bibfield  {author} {\bibinfo {author} {\bibfnamefont {S.}~\bibnamefont
  {{Datta}}}, \bibinfo {author} {\bibfnamefont {A.}~\bibnamefont {{Ghosal}}},\
  and\ \bibinfo {author} {\bibfnamefont {R.}~\bibnamefont {{Samanta}}},\ }\href
  {https://doi.org/10.1088/1475-7516/2021/08/021} {\bibfield  {journal}
  {\bibinfo  {journal} {\jcap}\ }\textbf {\bibinfo {volume} {2021}},\ \bibinfo
  {eid} {021} (\bibinfo {year} {2021})},\ \Eprint
  {https://arxiv.org/abs/2012.14981}  { arXiv:2012.14981 [hep-ph]}\BibitemShut
  {NoStop}%
\bibitem [{\citenamefont {{Bernal}}\ \emph
  {et~al.}(2022{\natexlab{b}})\citenamefont {{Bernal}}, \citenamefont {{Fong}},
  \citenamefont {{Perez-Gonzalez}},\ and\ \citenamefont
  {{Turner}}}]{2022arXiv220308823B}%
  \BibitemOpen
  \bibfield  {author} {\bibinfo {author} {\bibfnamefont {N.}~\bibnamefont
  {{Bernal}}}, \bibinfo {author} {\bibfnamefont {C.~S.}\ \bibnamefont
  {{Fong}}}, \bibinfo {author} {\bibfnamefont {Y.~F.}\ \bibnamefont
  {{Perez-Gonzalez}}},\ and\ \bibinfo {author} {\bibfnamefont {J.}~\bibnamefont
  {{Turner}}},\ }\href@noop {} {\bibfield  {journal} {\bibinfo  {journal}
  {\arxiv}\ } (\bibinfo {year} {2022}{\natexlab{b}})},\ \Eprint
  {https://arxiv.org/abs/2203.08823}  { arXiv:2203.08823 [hep-ph]}\BibitemShut
  {NoStop}%
\bibitem [{\citenamefont {{Barman}}\ \emph {et~al.}(2022)\citenamefont
  {{Barman}}, \citenamefont {{Borah}}, \citenamefont {{Das}},\ and\
  \citenamefont {{Roshan}}}]{2022JCAP...03..031B}%
  \BibitemOpen
  \bibfield  {author} {\bibinfo {author} {\bibfnamefont {B.}~\bibnamefont
  {{Barman}}}, \bibinfo {author} {\bibfnamefont {D.}~\bibnamefont {{Borah}}},
  \bibinfo {author} {\bibfnamefont {S.~J.}\ \bibnamefont {{Das}}},\ and\
  \bibinfo {author} {\bibfnamefont {R.}~\bibnamefont {{Roshan}}},\ }\href
  {https://doi.org/10.1088/1475-7516/2022/03/031} {\bibfield  {journal}
  {\bibinfo  {journal} {\jcap}\ }\textbf {\bibinfo {volume} {2022}},\ \bibinfo
  {eid} {031} (\bibinfo {year} {2022})},\ \Eprint
  {https://arxiv.org/abs/2111.08034}  { arXiv:2111.08034 [hep-ph]}\BibitemShut
  {NoStop}%
\bibitem [{\citenamefont {{Lindley}}(1981)}]{1981MNRAS.196..317L}%
  \BibitemOpen
  \bibfield  {author} {\bibinfo {author} {\bibfnamefont {D.}~\bibnamefont
  {{Lindley}}},\ }\href {https://doi.org/10.1093/mnras/196.2.317} {\bibfield
  {journal} {\bibinfo  {journal} {\mnras}\ }\textbf {\bibinfo {volume} {196}},\
  \bibinfo {pages} {317} (\bibinfo {year} {1981})}\BibitemShut {NoStop}%
\bibitem [{\citenamefont {{Lindely}}(1982)}]{1982MNRAS.199..775L}%
  \BibitemOpen
  \bibfield  {author} {\bibinfo {author} {\bibfnamefont {D.}~\bibnamefont
  {{Lindely}}},\ }\href {https://doi.org/10.1093/mnras/199.3.775} {\bibfield
  {journal} {\bibinfo  {journal} {\mnras}\ }\textbf {\bibinfo {volume} {199}},\
  \bibinfo {pages} {775} (\bibinfo {year} {1982})}\BibitemShut {NoStop}%
\bibitem [{\citenamefont {{Majumdar}}\ \emph {et~al.}(1995)\citenamefont
  {{Majumdar}}, \citenamefont {{Gupta}},\ and\ \citenamefont
  {{Saxena}}}]{1995IJMPD...4..517M}%
  \BibitemOpen
  \bibfield  {author} {\bibinfo {author} {\bibfnamefont {A.~S.}\ \bibnamefont
  {{Majumdar}}}, \bibinfo {author} {\bibfnamefont {P.~D.}\ \bibnamefont
  {{Gupta}}},\ and\ \bibinfo {author} {\bibfnamefont {R.~P.}\ \bibnamefont
  {{Saxena}}},\ }\href {https://doi.org/10.1142/S0218271895000363} {\bibfield
  {journal} {\bibinfo  {journal} {International Journal of Modern Physics D}\
  }\textbf {\bibinfo {volume} {4}},\ \bibinfo {pages} {517} (\bibinfo {year}
  {1995})}\BibitemShut {NoStop}%
\bibitem [{\citenamefont {{Upadhyay}}\ \emph {et~al.}(1999)\citenamefont
  {{Upadhyay}}, \citenamefont {{Das Gupta}},\ and\ \citenamefont
  {{Saxena}}}]{1999PhRvD..60f3513U}%
  \BibitemOpen
  \bibfield  {author} {\bibinfo {author} {\bibfnamefont {N.}~\bibnamefont
  {{Upadhyay}}}, \bibinfo {author} {\bibfnamefont {P.}~\bibnamefont {{Das
  Gupta}}},\ and\ \bibinfo {author} {\bibfnamefont {R.~P.}\ \bibnamefont
  {{Saxena}}},\ }\href {https://doi.org/10.1103/PhysRevD.60.063513} {\bibfield
  {journal} {\bibinfo  {journal} {\prd}\ }\textbf {\bibinfo {volume} {60}},\
  \bibinfo {eid} {063513} (\bibinfo {year} {1999})},\ \Eprint
  {https://arxiv.org/abs/astro-ph/9903253}  { arXiv:astro-ph/9903253
  [astro-ph]}\BibitemShut {NoStop}%
\bibitem [{\citenamefont {{Bugaev}}\ \emph {et~al.}(2003)\citenamefont
  {{Bugaev}}, \citenamefont {{Elbakidze}},\ and\ \citenamefont
  {{Konishche}}}]{2003PAN....66..476B}%
  \BibitemOpen
  \bibfield  {author} {\bibinfo {author} {\bibfnamefont {E.~V.}\ \bibnamefont
  {{Bugaev}}}, \bibinfo {author} {\bibfnamefont {M.~G.}\ \bibnamefont
  {{Elbakidze}}},\ and\ \bibinfo {author} {\bibfnamefont {K.~V.}\ \bibnamefont
  {{Konishche}}},\ }\href {https://doi.org/10.1134/1.1563709} {\bibfield
  {journal} {\bibinfo  {journal} {\PAN}\ }\textbf {\bibinfo {volume} {66}},\
  \bibinfo {pages} {476} (\bibinfo {year} {2003})},\ \Eprint
  {https://arxiv.org/abs/astro-ph/0110660}  { arXiv:astro-ph/0110660
  [astro-ph]}\BibitemShut {NoStop}%
\bibitem [{\citenamefont {{Arbey}}(2012)}]{2012CoPhC.183.1822A}%
  \BibitemOpen
  \bibfield  {author} {\bibinfo {author} {\bibfnamefont {A.}~\bibnamefont
  {{Arbey}}},\ }\href {https://doi.org/10.1016/j.cpc.2012.03.018} {\bibfield
  {journal} {\bibinfo  {journal} {\CPC}\ }\textbf {\bibinfo {volume} {183}},\
  \bibinfo {pages} {1822} (\bibinfo {year} {2012})},\ \Eprint
  {https://arxiv.org/abs/1106.1363}  { arXiv:1106.1363
  [astro-ph.CO]}\BibitemShut {NoStop}%
\bibitem [{\citenamefont {{Arbey}}\ \emph
  {et~al.}(2020{\natexlab{a}})\citenamefont {{Arbey}}, \citenamefont
  {{Auffinger}}, \citenamefont {{Hickerson}},\ and\ \citenamefont
  {{Jenssen}}}]{2020CoPhC.24806982A}%
  \BibitemOpen
  \bibfield  {author} {\bibinfo {author} {\bibfnamefont {A.}~\bibnamefont
  {{Arbey}}}, \bibinfo {author} {\bibfnamefont {J.}~\bibnamefont
  {{Auffinger}}}, \bibinfo {author} {\bibfnamefont {K.~P.}\ \bibnamefont
  {{Hickerson}}},\ and\ \bibinfo {author} {\bibfnamefont {E.~S.}\ \bibnamefont
  {{Jenssen}}},\ }\href {https://doi.org/10.1016/j.cpc.2019.106982} {\bibfield
  {journal} {\bibinfo  {journal} {\CPC}\ }\textbf {\bibinfo {volume} {248}},\
  \bibinfo {eid} {106982} (\bibinfo {year} {2020}{\natexlab{a}})}\BibitemShut
  {NoStop}%
\bibitem [{\citenamefont {{Carr}}\ \emph {et~al.}(2010)\citenamefont {{Carr}},
  \citenamefont {{Kohri}}, \citenamefont {{Sendouda}},\ and\ \citenamefont
  {{Yokoyama}}}]{2010PhRvD..81j4019C}%
  \BibitemOpen
  \bibfield  {author} {\bibinfo {author} {\bibfnamefont {B.~J.}\ \bibnamefont
  {{Carr}}}, \bibinfo {author} {\bibfnamefont {K.}~\bibnamefont {{Kohri}}},
  \bibinfo {author} {\bibfnamefont {Y.}~\bibnamefont {{Sendouda}}},\ and\
  \bibinfo {author} {\bibfnamefont {J.}~\bibnamefont {{Yokoyama}}},\ }\href
  {https://doi.org/10.1103/PhysRevD.81.104019} {\bibfield  {journal} {\bibinfo
  {journal} {\prd}\ }\textbf {\bibinfo {volume} {81}},\ \bibinfo {eid} {104019}
  (\bibinfo {year} {2010})},\ \Eprint {https://arxiv.org/abs/0912.5297}  {
  arXiv:0912.5297 [astro-ph.CO]}\BibitemShut {NoStop}%
\bibitem [{\citenamefont {{Acharya}}\ and\ \citenamefont
  {{Khatri}}(2020{\natexlab{a}})}]{2020JCAP...06..018A}%
  \BibitemOpen
  \bibfield  {author} {\bibinfo {author} {\bibfnamefont {S.~K.}\ \bibnamefont
  {{Acharya}}}\ and\ \bibinfo {author} {\bibfnamefont {R.}~\bibnamefont
  {{Khatri}}},\ }\href {https://doi.org/10.1088/1475-7516/2020/06/018}
  {\bibfield  {journal} {\bibinfo  {journal} {\jcap}\ }\textbf {\bibinfo
  {volume} {2020}},\ \bibinfo {eid} {018} (\bibinfo {year}
  {2020}{\natexlab{a}})},\ \Eprint {https://arxiv.org/abs/2002.00898}  {
  arXiv:2002.00898 [astro-ph.CO]}\BibitemShut {NoStop}%
\bibitem [{\citenamefont {{Luo}}\ \emph {et~al.}(2021)\citenamefont {{Luo}},
  \citenamefont {{Chen}}, \citenamefont {{Kusakabe}},\ and\ \citenamefont
  {{Kajino}}}]{2021JCAP...05..042L}%
  \BibitemOpen
  \bibfield  {author} {\bibinfo {author} {\bibfnamefont {Y.}~\bibnamefont
  {{Luo}}}, \bibinfo {author} {\bibfnamefont {C.}~\bibnamefont {{Chen}}},
  \bibinfo {author} {\bibfnamefont {M.}~\bibnamefont {{Kusakabe}}},\ and\
  \bibinfo {author} {\bibfnamefont {T.}~\bibnamefont {{Kajino}}},\ }\href
  {https://doi.org/10.1088/1475-7516/2021/05/042} {\bibfield  {journal}
  {\bibinfo  {journal} {\jcap}\ }\textbf {\bibinfo {volume} {2021}},\ \bibinfo
  {eid} {042} (\bibinfo {year} {2021})},\ \Eprint
  {https://arxiv.org/abs/2011.10937}  { arXiv:2011.10937
  [astro-ph.CO]}\BibitemShut {NoStop}%
\bibitem [{\citenamefont {{Keith}}\ \emph {et~al.}(2020)\citenamefont
  {{Keith}}, \citenamefont {{Hooper}}, \citenamefont {{Blinov}},\ and\
  \citenamefont {{McDermott}}}]{2020PhRvD.102j3512K}%
  \BibitemOpen
  \bibfield  {author} {\bibinfo {author} {\bibfnamefont {C.}~\bibnamefont
  {{Keith}}}, \bibinfo {author} {\bibfnamefont {D.}~\bibnamefont {{Hooper}}},
  \bibinfo {author} {\bibfnamefont {N.}~\bibnamefont {{Blinov}}},\ and\
  \bibinfo {author} {\bibfnamefont {S.~D.}\ \bibnamefont {{McDermott}}},\
  }\href {https://doi.org/10.1103/PhysRevD.102.103512} {\bibfield  {journal}
  {\bibinfo  {journal} {\prd}\ }\textbf {\bibinfo {volume} {102}},\ \bibinfo
  {eid} {103512} (\bibinfo {year} {2020})},\ \Eprint
  {https://arxiv.org/abs/2006.03608}  { arXiv:2006.03608
  [astro-ph.CO]}\BibitemShut {NoStop}%
\bibitem [{\citenamefont {{Vainer}}\ and\ \citenamefont
  {{Nasel'skii}}(1977)}]{1977SvAL....3...76V}%
  \BibitemOpen
  \bibfield  {author} {\bibinfo {author} {\bibfnamefont {B.~V.}\ \bibnamefont
  {{Vainer}}}\ and\ \bibinfo {author} {\bibfnamefont {P.~D.}\ \bibnamefont
  {{Nasel'skii}}},\ }\href@noop {} {\bibfield  {journal} {\bibinfo  {journal}
  {\sovastl}\ }\textbf {\bibinfo {volume} {3}},\ \bibinfo {pages} {76}
  (\bibinfo {year} {1977})},\ \bibinfo {note} {[\PAZh~\textbf{3}, 147
  (1977)]}\BibitemShut {NoStop}%
\bibitem [{\citenamefont {{Vainer}}\ and\ \citenamefont
  {{Nasel'skii}}(1978)}]{1978SvA....22..138V}%
  \BibitemOpen
  \bibfield  {author} {\bibinfo {author} {\bibfnamefont {B.~V.}\ \bibnamefont
  {{Vainer}}}\ and\ \bibinfo {author} {\bibfnamefont {P.~D.}\ \bibnamefont
  {{Nasel'skii}}},\ }\href@noop {} {\bibfield  {journal} {\bibinfo  {journal}
  {\sovast}\ }\textbf {\bibinfo {volume} {22}},\ \bibinfo {pages} {138}
  (\bibinfo {year} {1978})},\ \bibinfo {note} {[\azh~\textbf{55}, 231
  (1978)]}\BibitemShut {NoStop}%
\bibitem [{\citenamefont {{Zel'dovich}}\ \emph {et~al.}(1977)\citenamefont
  {{Zel'dovich}}, \citenamefont {{Starobinski{\v{i}}}}, \citenamefont
  {{Khlopov}},\ and\ \citenamefont {{Chechetkin}}}]{1977SvAL....3..110Z}%
  \BibitemOpen
  \bibfield  {author} {\bibinfo {author} {\bibfnamefont {I.~B.}\ \bibnamefont
  {{Zel'dovich}}}, \bibinfo {author} {\bibfnamefont {A.~A.}\ \bibnamefont
  {{Starobinski{\v{i}}}}}, \bibinfo {author} {\bibfnamefont {M.~I.}\
  \bibnamefont {{Khlopov}}},\ and\ \bibinfo {author} {\bibfnamefont {V.~M.}\
  \bibnamefont {{Chechetkin}}},\ }\href@noop {} {\bibfield  {journal} {\bibinfo
   {journal} {\sovastl}\ }\textbf {\bibinfo {volume} {3}},\ \bibinfo {pages}
  {110} (\bibinfo {year} {1977})},\ \bibinfo {note} {[\PAZh~\textbf{3}, 208
  (1977)]}\BibitemShut {NoStop}%
\bibitem [{\citenamefont {{Wagoner}}\ \emph {et~al.}(1967)\citenamefont
  {{Wagoner}}, \citenamefont {{Fowler}},\ and\ \citenamefont
  {{Hoyle}}}]{1967ApJ...148....3W}%
  \BibitemOpen
  \bibfield  {author} {\bibinfo {author} {\bibfnamefont {R.~V.}\ \bibnamefont
  {{Wagoner}}}, \bibinfo {author} {\bibfnamefont {W.~A.}\ \bibnamefont
  {{Fowler}}},\ and\ \bibinfo {author} {\bibfnamefont {F.}~\bibnamefont
  {{Hoyle}}},\ }\href {https://doi.org/10.1086/149126} {\bibfield  {journal}
  {\bibinfo  {journal} {\apj}\ }\textbf {\bibinfo {volume} {148}},\ \bibinfo
  {pages} {3} (\bibinfo {year} {1967})}\BibitemShut {NoStop}%
\bibitem [{\citenamefont {{Wagoner}}(1969)}]{1969ApJS...18..247W}%
  \BibitemOpen
  \bibfield  {author} {\bibinfo {author} {\bibfnamefont {R.~V.}\ \bibnamefont
  {{Wagoner}}},\ }\href {https://doi.org/10.1086/190191} {\bibfield  {journal}
  {\bibinfo  {journal} {\apjs}\ }\textbf {\bibinfo {volume} {18}},\ \bibinfo
  {pages} {247} (\bibinfo {year} {1969})}\BibitemShut {NoStop}%
\bibitem [{\citenamefont {{Wagoner}}(1973)}]{1973ApJ...179..343W}%
  \BibitemOpen
  \bibfield  {author} {\bibinfo {author} {\bibfnamefont {R.~V.}\ \bibnamefont
  {{Wagoner}}},\ }\href {https://doi.org/10.1086/151873} {\bibfield  {journal}
  {\bibinfo  {journal} {\apj}\ }\textbf {\bibinfo {volume} {179}},\ \bibinfo
  {pages} {343} (\bibinfo {year} {1973})}\BibitemShut {NoStop}%
\bibitem [{\citenamefont {{Kohri}}\ and\ \citenamefont
  {{Yokoyama}}(1999)}]{1999PhRvD..61b3501K}%
  \BibitemOpen
  \bibfield  {author} {\bibinfo {author} {\bibfnamefont {K.}~\bibnamefont
  {{Kohri}}}\ and\ \bibinfo {author} {\bibfnamefont {J.}~\bibnamefont
  {{Yokoyama}}},\ }\href {https://doi.org/10.1103/PhysRevD.61.023501}
  {\bibfield  {journal} {\bibinfo  {journal} {\prd}\ }\textbf {\bibinfo
  {volume} {61}},\ \bibinfo {eid} {023501} (\bibinfo {year} {1999})},\ \Eprint
  {https://arxiv.org/abs/astro-ph/9908160}  { arXiv:astro-ph/9908160
  [astro-ph]}\BibitemShut {NoStop}%
\bibitem [{\citenamefont {{Dimopoulos}}\ \emph
  {et~al.}(1988{\natexlab{a}})\citenamefont {{Dimopoulos}}, \citenamefont
  {{Esmailzadeh}}, \citenamefont {{Starkman}},\ and\ \citenamefont
  {{Hall}}}]{1988PhRvL..60....7D}%
  \BibitemOpen
  \bibfield  {author} {\bibinfo {author} {\bibfnamefont {S.}~\bibnamefont
  {{Dimopoulos}}}, \bibinfo {author} {\bibfnamefont {R.}~\bibnamefont
  {{Esmailzadeh}}}, \bibinfo {author} {\bibfnamefont {G.~D.}\ \bibnamefont
  {{Starkman}}},\ and\ \bibinfo {author} {\bibfnamefont {L.~J.}\ \bibnamefont
  {{Hall}}},\ }\href {https://doi.org/10.1103/PhysRevLett.60.7} {\bibfield
  {journal} {\bibinfo  {journal} {\prl}\ }\textbf {\bibinfo {volume} {60}},\
  \bibinfo {pages} {7} (\bibinfo {year} {1988}{\natexlab{a}})}\BibitemShut
  {NoStop}%
\bibitem [{\citenamefont {{Dimopoulos}}\ \emph
  {et~al.}(1988{\natexlab{b}})\citenamefont {{Dimopoulos}}, \citenamefont
  {{Esmailzadeh}}, \citenamefont {{Hall}},\ and\ \citenamefont
  {{Starkman}}}]{1988ApJ...330..545D}%
  \BibitemOpen
  \bibfield  {author} {\bibinfo {author} {\bibfnamefont {S.}~\bibnamefont
  {{Dimopoulos}}}, \bibinfo {author} {\bibfnamefont {R.}~\bibnamefont
  {{Esmailzadeh}}}, \bibinfo {author} {\bibfnamefont {L.~J.}\ \bibnamefont
  {{Hall}}},\ and\ \bibinfo {author} {\bibfnamefont {G.~D.}\ \bibnamefont
  {{Starkman}}},\ }\href {https://doi.org/10.1086/166493} {\bibfield  {journal}
  {\bibinfo  {journal} {\apj}\ }\textbf {\bibinfo {volume} {330}},\ \bibinfo
  {pages} {545} (\bibinfo {year} {1988}{\natexlab{b}})}\BibitemShut {NoStop}%
\bibitem [{\citenamefont {{Dimopoulos}}\ \emph {et~al.}(1989)\citenamefont
  {{Dimopoulos}}, \citenamefont {{Esmailzadeh}}, \citenamefont {{Hall}},\ and\
  \citenamefont {{Starkman}}}]{1989NuPhB.311..699D}%
  \BibitemOpen
  \bibfield  {author} {\bibinfo {author} {\bibfnamefont {S.}~\bibnamefont
  {{Dimopoulos}}}, \bibinfo {author} {\bibfnamefont {R.}~\bibnamefont
  {{Esmailzadeh}}}, \bibinfo {author} {\bibfnamefont {L.~J.}\ \bibnamefont
  {{Hall}}},\ and\ \bibinfo {author} {\bibfnamefont {G.~D.}\ \bibnamefont
  {{Starkman}}},\ }\href {https://doi.org/10.1016/0550-3213(89)90173-9}
  {\bibfield  {journal} {\bibinfo  {journal} {\nphysb}\ }\textbf {\bibinfo
  {volume} {311}},\ \bibinfo {pages} {699} (\bibinfo {year}
  {1989})}\BibitemShut {NoStop}%
\bibitem [{\citenamefont {{Miyama}}\ and\ \citenamefont
  {{Sato}}(1978)}]{1978PThPh..59.1012M}%
  \BibitemOpen
  \bibfield  {author} {\bibinfo {author} {\bibfnamefont {S.}~\bibnamefont
  {{Miyama}}}\ and\ \bibinfo {author} {\bibfnamefont {K.}~\bibnamefont
  {{Sato}}},\ }\href {https://doi.org/10.1143/PTP.59.1012} {\bibfield
  {journal} {\bibinfo  {journal} {\PThPh}\ }\textbf {\bibinfo {volume} {59}},\
  \bibinfo {pages} {1012} (\bibinfo {year} {1978})}\BibitemShut {NoStop}%
\bibitem [{\citenamefont {{Vainer}}\ \emph {et~al.}(1978)\citenamefont
  {{Vainer}}, \citenamefont {{Dryzhakova}},\ and\ \citenamefont
  {{Nasel'skii}}}]{1978SvAL....4..185V}%
  \BibitemOpen
  \bibfield  {author} {\bibinfo {author} {\bibfnamefont {B.~V.}\ \bibnamefont
  {{Vainer}}}, \bibinfo {author} {\bibfnamefont {O.~V.}\ \bibnamefont
  {{Dryzhakova}}},\ and\ \bibinfo {author} {\bibfnamefont {P.~D.}\ \bibnamefont
  {{Nasel'skii}}},\ }\href@noop {} {\bibfield  {journal} {\bibinfo  {journal}
  {\sovastl}\ }\textbf {\bibinfo {volume} {4}},\ \bibinfo {pages} {185}
  (\bibinfo {year} {1978})},\ \bibinfo {note} {[\PAZh~\textbf{4}, 344
  (1978)]}\BibitemShut {NoStop}%
\bibitem [{\citenamefont {{Sedel'nikov}}(1996)}]{1996AstL...22..797S}%
  \BibitemOpen
  \bibfield  {author} {\bibinfo {author} {\bibfnamefont {E.~V.}\ \bibnamefont
  {{Sedel'nikov}}},\ }\href@noop {} {\bibfield  {journal} {\bibinfo  {journal}
  {\astl}\ }\textbf {\bibinfo {volume} {22}},\ \bibinfo {pages} {797} (\bibinfo
  {year} {1996})},\ \bibinfo {note} {[\PAZh~\textbf{22}, 889
  (1996)]}\BibitemShut {NoStop}%
\bibitem [{\citenamefont {{Lindley}}(1980)}]{1980MNRAS.193..593L}%
  \BibitemOpen
  \bibfield  {author} {\bibinfo {author} {\bibfnamefont {D.}~\bibnamefont
  {{Lindley}}},\ }\href {https://doi.org/10.1093/mnras/193.3.593} {\bibfield
  {journal} {\bibinfo  {journal} {\mnras}\ }\textbf {\bibinfo {volume} {193}},\
  \bibinfo {pages} {593} (\bibinfo {year} {1980})}\BibitemShut {NoStop}%
\bibitem [{\citenamefont {{Acharya}}\ and\ \citenamefont
  {{Khatri}}(2019{\natexlab{a}})}]{2019JCAP...12..046A}%
  \BibitemOpen
  \bibfield  {author} {\bibinfo {author} {\bibfnamefont {S.~K.}\ \bibnamefont
  {{Acharya}}}\ and\ \bibinfo {author} {\bibfnamefont {R.}~\bibnamefont
  {{Khatri}}},\ }\href {https://doi.org/10.1088/1475-7516/2019/12/046}
  {\bibfield  {journal} {\bibinfo  {journal} {\jcap}\ }\textbf {\bibinfo
  {volume} {2019}},\ \bibinfo {eid} {046} (\bibinfo {year}
  {2019}{\natexlab{a}})},\ \Eprint {https://arxiv.org/abs/1910.06272}  {
  arXiv:1910.06272 [astro-ph.CO]}\BibitemShut {NoStop}%
\bibitem [{\citenamefont {{Poulin}}\ and\ \citenamefont
  {{Serpico}}(2015)}]{2015PhRvD..91j3007P}%
  \BibitemOpen
  \bibfield  {author} {\bibinfo {author} {\bibfnamefont {V.}~\bibnamefont
  {{Poulin}}}\ and\ \bibinfo {author} {\bibfnamefont {P.~D.}\ \bibnamefont
  {{Serpico}}},\ }\href {https://doi.org/10.1103/PhysRevD.91.103007} {\bibfield
   {journal} {\bibinfo  {journal} {\prd}\ }\textbf {\bibinfo {volume} {91}},\
  \bibinfo {eid} {103007} (\bibinfo {year} {2015})},\ \Eprint
  {https://arxiv.org/abs/1503.04852}  { arXiv:1503.04852
  [astro-ph.CO]}\BibitemShut {NoStop}%
\bibitem [{\citenamefont {{Jedamzik}}(2004)}]{2004PhRvD..70f3524J}%
  \BibitemOpen
  \bibfield  {author} {\bibinfo {author} {\bibfnamefont {K.}~\bibnamefont
  {{Jedamzik}}},\ }\href {https://doi.org/10.1103/PhysRevD.70.063524}
  {\bibfield  {journal} {\bibinfo  {journal} {\prd}\ }\textbf {\bibinfo
  {volume} {70}},\ \bibinfo {eid} {063524} (\bibinfo {year} {2004})},\ \Eprint
  {https://arxiv.org/abs/astro-ph/0402344}  { arXiv:astro-ph/0402344
  [astro-ph]}\BibitemShut {NoStop}%
\bibitem [{\citenamefont {{Kawasaki}}\ \emph {et~al.}(2005)\citenamefont
  {{Kawasaki}}, \citenamefont {{Kohri}},\ and\ \citenamefont
  {{Moroi}}}]{2005PhRvD..71h3502K}%
  \BibitemOpen
  \bibfield  {author} {\bibinfo {author} {\bibfnamefont {M.}~\bibnamefont
  {{Kawasaki}}}, \bibinfo {author} {\bibfnamefont {K.}~\bibnamefont
  {{Kohri}}},\ and\ \bibinfo {author} {\bibfnamefont {T.}~\bibnamefont
  {{Moroi}}},\ }\href {https://doi.org/10.1103/PhysRevD.71.083502} {\bibfield
  {journal} {\bibinfo  {journal} {\prd}\ }\textbf {\bibinfo {volume} {71}},\
  \bibinfo {eid} {083502} (\bibinfo {year} {2005})},\ \Eprint
  {https://arxiv.org/abs/astro-ph/0408426}  { arXiv:astro-ph/0408426
  [astro-ph]}\BibitemShut {NoStop}%
\bibitem [{\citenamefont {{Kawasaki}}\ \emph {et~al.}(2018)\citenamefont
  {{Kawasaki}}, \citenamefont {{Kohri}}, \citenamefont {{Moroi}},\ and\
  \citenamefont {{Takaesu}}}]{2018PhRvD..97b3502K}%
  \BibitemOpen
  \bibfield  {author} {\bibinfo {author} {\bibfnamefont {M.}~\bibnamefont
  {{Kawasaki}}}, \bibinfo {author} {\bibfnamefont {K.}~\bibnamefont {{Kohri}}},
  \bibinfo {author} {\bibfnamefont {T.}~\bibnamefont {{Moroi}}},\ and\ \bibinfo
  {author} {\bibfnamefont {Y.}~\bibnamefont {{Takaesu}}},\ }\href
  {https://doi.org/10.1103/PhysRevD.97.023502} {\bibfield  {journal} {\bibinfo
  {journal} {\prd}\ }\textbf {\bibinfo {volume} {97}},\ \bibinfo {eid} {023502}
  (\bibinfo {year} {2018})},\ \Eprint {https://arxiv.org/abs/1709.01211}  {
  arXiv:1709.01211 [hep-ph]}\BibitemShut {NoStop}%
\bibitem [{\citenamefont {{Kawasaki}}\ \emph {et~al.}(2020)\citenamefont
  {{Kawasaki}}, \citenamefont {{Kohri}}, \citenamefont {{Moroi}}, \citenamefont
  {{Murai}},\ and\ \citenamefont {{Murayama}}}]{2020JCAP...12..048K}%
  \BibitemOpen
  \bibfield  {author} {\bibinfo {author} {\bibfnamefont {M.}~\bibnamefont
  {{Kawasaki}}}, \bibinfo {author} {\bibfnamefont {K.}~\bibnamefont {{Kohri}}},
  \bibinfo {author} {\bibfnamefont {T.}~\bibnamefont {{Moroi}}}, \bibinfo
  {author} {\bibfnamefont {K.}~\bibnamefont {{Murai}}},\ and\ \bibinfo {author}
  {\bibfnamefont {H.}~\bibnamefont {{Murayama}}},\ }\href
  {https://doi.org/10.1088/1475-7516/2020/12/048} {\bibfield  {journal}
  {\bibinfo  {journal} {\jcap}\ }\textbf {\bibinfo {volume} {2020}},\ \bibinfo
  {eid} {048} (\bibinfo {year} {2020})},\ \Eprint
  {https://arxiv.org/abs/2006.14803}  { arXiv:2006.14803 [hep-ph]}\BibitemShut
  {NoStop}%
\bibitem [{\citenamefont {{Penzias}}\ and\ \citenamefont
  {{Wilson}}(1965)}]{1965ApJ...142..419P}%
  \BibitemOpen
  \bibfield  {author} {\bibinfo {author} {\bibfnamefont {A.~A.}\ \bibnamefont
  {{Penzias}}}\ and\ \bibinfo {author} {\bibfnamefont {R.~W.}\ \bibnamefont
  {{Wilson}}},\ }\href {https://doi.org/10.1086/148307} {\bibfield  {journal}
  {\bibinfo  {journal} {\apj}\ }\textbf {\bibinfo {volume} {142}},\ \bibinfo
  {pages} {419} (\bibinfo {year} {1965})}\BibitemShut {NoStop}%
\bibitem [{\citenamefont {{Mather}}\ \emph {et~al.}(1990)\citenamefont
  {{Mather}} \emph {et~al.}}]{1990ApJ...354L..37M}%
  \BibitemOpen
  \bibfield  {author} {\bibinfo {author} {\bibfnamefont {J.~C.}\ \bibnamefont
  {{Mather}}} \emph {et~al.},\ }\href {https://doi.org/10.1086/185717}
  {\bibfield  {journal} {\bibinfo  {journal} {\apjl}\ }\textbf {\bibinfo
  {volume} {354}},\ \bibinfo {pages} {L37} (\bibinfo {year}
  {1990})}\BibitemShut {NoStop}%
\bibitem [{\citenamefont {{Bennett}}\ \emph {et~al.}(2003)\citenamefont
  {{Bennett}} \emph {et~al.}}]{2003ApJS..148....1B}%
  \BibitemOpen
  \bibfield  {author} {\bibinfo {author} {\bibfnamefont {C.~L.}\ \bibnamefont
  {{Bennett}}} \emph {et~al.} (\bibinfo {collaboration} {WMAP}),\ }\href
  {https://doi.org/10.1086/377253} {\bibfield  {journal} {\bibinfo  {journal}
  {\apjs}\ }\textbf {\bibinfo {volume} {148}},\ \bibinfo {pages} {1} (\bibinfo
  {year} {2003})},\ \Eprint {https://arxiv.org/abs/astro-ph/0302207}  {
  arXiv:astro-ph/0302207 [astro-ph]}\BibitemShut {NoStop}%
\bibitem [{\citenamefont {{Hinshaw}}\ \emph {et~al.}(2003)\citenamefont
  {{Hinshaw}} \emph {et~al.}}]{2003ApJS..148..135H}%
  \BibitemOpen
  \bibfield  {author} {\bibinfo {author} {\bibfnamefont {G.}~\bibnamefont
  {{Hinshaw}}} \emph {et~al.} (\bibinfo {collaboration} {WMAP}),\ }\href
  {https://doi.org/10.1086/377225} {\bibfield  {journal} {\bibinfo  {journal}
  {\apjs}\ }\textbf {\bibinfo {volume} {148}},\ \bibinfo {pages} {135}
  (\bibinfo {year} {2003})},\ \Eprint {https://arxiv.org/abs/astro-ph/0302217}
  { arXiv:astro-ph/0302217 [astro-ph]}\BibitemShut {NoStop}%
\bibitem [{\citenamefont {{Aghanim}}\ \emph {et~al.}(2016)\citenamefont
  {{Aghanim}} \emph {et~al.}}]{2016A&A...594A..11P}%
  \BibitemOpen
  \bibfield  {author} {\bibinfo {author} {\bibfnamefont {N.}~\bibnamefont
  {{Aghanim}}} \emph {et~al.} (\bibinfo {collaboration} {Planck}),\ }\href
  {https://doi.org/10.1051/0004-6361/201526926} {\bibfield  {journal} {\bibinfo
   {journal} {\aap}\ }\textbf {\bibinfo {volume} {594}},\ \bibinfo {eid} {A11}
  (\bibinfo {year} {2016})},\ \Eprint {https://arxiv.org/abs/1507.02704}  {
  arXiv:1507.02704 [astro-ph.CO]}\BibitemShut {NoStop}%
\bibitem [{\citenamefont {{Aghanim}}\ \emph
  {et~al.}(2020{\natexlab{a}})\citenamefont {{Aghanim}} \emph
  {et~al.}}]{2020A&A...641A...5P}%
  \BibitemOpen
  \bibfield  {author} {\bibinfo {author} {\bibfnamefont {N.}~\bibnamefont
  {{Aghanim}}} \emph {et~al.} (\bibinfo {collaboration} {Planck}),\ }\href
  {https://doi.org/10.1051/0004-6361/201936386} {\bibfield  {journal} {\bibinfo
   {journal} {\aap}\ }\textbf {\bibinfo {volume} {641}},\ \bibinfo {eid} {A5}
  (\bibinfo {year} {2020}{\natexlab{a}})},\ \Eprint
  {https://arxiv.org/abs/1907.12875}  { arXiv:1907.12875
  [astro-ph.CO]}\BibitemShut {NoStop}%
\bibitem [{\citenamefont {{Abazajian}}\ \emph {et~al.}(2016)\citenamefont
  {{Abazajian}} \emph {et~al.}}]{2016arXiv161002743A}%
  \BibitemOpen
  \bibfield  {author} {\bibinfo {author} {\bibfnamefont {K.~N.}\ \bibnamefont
  {{Abazajian}}} \emph {et~al.} (\bibinfo {collaboration} {CMB-S4}),\
  }\href@noop {} {\bibfield  {journal} {\bibinfo  {journal} {\arxiv}\ }
  (\bibinfo {year} {2016})},\ \Eprint {https://arxiv.org/abs/1610.02743}  {
  arXiv:1610.02743 [astro-ph.CO]}\BibitemShut {NoStop}%
\bibitem [{\citenamefont {{Baumann}}\ \emph {et~al.}(2018)\citenamefont
  {{Baumann}}, \citenamefont {{Green}},\ and\ \citenamefont
  {{Wallisch}}}]{2018JCAP...08..029B}%
  \BibitemOpen
  \bibfield  {author} {\bibinfo {author} {\bibfnamefont {D.}~\bibnamefont
  {{Baumann}}}, \bibinfo {author} {\bibfnamefont {D.}~\bibnamefont {{Green}}},\
  and\ \bibinfo {author} {\bibfnamefont {B.}~\bibnamefont {{Wallisch}}},\
  }\href {https://doi.org/10.1088/1475-7516/2018/08/029} {\bibfield  {journal}
  {\bibinfo  {journal} {\jcap}\ }\textbf {\bibinfo {volume} {2018}},\ \bibinfo
  {eid} {029} (\bibinfo {year} {2018})},\ \Eprint
  {https://arxiv.org/abs/1712.08067}  { arXiv:1712.08067
  [astro-ph.CO]}\BibitemShut {NoStop}%
\bibitem [{\citenamefont {{Hanany}}\ \emph {et~al.}(2019)\citenamefont
  {{Hanany}} \emph {et~al.}}]{2019arXiv190210541H}%
  \BibitemOpen
  \bibfield  {author} {\bibinfo {author} {\bibfnamefont {S.}~\bibnamefont
  {{Hanany}}} \emph {et~al.} (\bibinfo {collaboration} {NASA-PICO}),\
  }\href@noop {} {\bibfield  {journal} {\bibinfo  {journal} {\arxiv}\ }
  (\bibinfo {year} {2019})},\ \Eprint {https://arxiv.org/abs/1902.10541}  {
  arXiv:1902.10541 [astro-ph.IM]}\BibitemShut {NoStop}%
\bibitem [{\citenamefont
  {{Lesgourgues}}(2011{\natexlab{a}})}]{2011arXiv1104.2932L}%
  \BibitemOpen
  \bibfield  {author} {\bibinfo {author} {\bibfnamefont {J.}~\bibnamefont
  {{Lesgourgues}}},\ }\href@noop {} {\bibfield  {journal} {\bibinfo  {journal}
  {\arxiv}\ } (\bibinfo {year} {2011}{\natexlab{a}})},\ \Eprint
  {https://arxiv.org/abs/1104.2932}  { arXiv:1104.2932
  [astro-ph.IM]}\BibitemShut {NoStop}%
\bibitem [{\citenamefont {{Blas}}\ \emph {et~al.}(2011)\citenamefont {{Blas}},
  \citenamefont {{Lesgourgues}},\ and\ \citenamefont
  {{Tram}}}]{2011JCAP...07..034B}%
  \BibitemOpen
  \bibfield  {author} {\bibinfo {author} {\bibfnamefont {D.}~\bibnamefont
  {{Blas}}}, \bibinfo {author} {\bibfnamefont {J.}~\bibnamefont
  {{Lesgourgues}}},\ and\ \bibinfo {author} {\bibfnamefont {T.}~\bibnamefont
  {{Tram}}},\ }\href {https://doi.org/10.1088/1475-7516/2011/07/034} {\bibfield
   {journal} {\bibinfo  {journal} {\jcap}\ }\textbf {\bibinfo {volume}
  {2011}},\ \bibinfo {eid} {034} (\bibinfo {year} {2011})},\ \Eprint
  {https://arxiv.org/abs/1104.2933}  { arXiv:1104.2933
  [astro-ph.CO]}\BibitemShut {NoStop}%
\bibitem [{\citenamefont
  {{Lesgourgues}}(2011{\natexlab{b}})}]{2011arXiv1104.2934L}%
  \BibitemOpen
  \bibfield  {author} {\bibinfo {author} {\bibfnamefont {J.}~\bibnamefont
  {{Lesgourgues}}},\ }\href@noop {} {\bibfield  {journal} {\bibinfo  {journal}
  {\arxiv}\ } (\bibinfo {year} {2011}{\natexlab{b}})},\ \Eprint
  {https://arxiv.org/abs/1104.2934}  { arXiv:1104.2934
  [astro-ph.CO]}\BibitemShut {NoStop}%
\bibitem [{\citenamefont {{Lesgourgues}}\ and\ \citenamefont
  {{Tram}}(2011)}]{2011JCAP...09..032L}%
  \BibitemOpen
  \bibfield  {author} {\bibinfo {author} {\bibfnamefont {J.}~\bibnamefont
  {{Lesgourgues}}}\ and\ \bibinfo {author} {\bibfnamefont {T.}~\bibnamefont
  {{Tram}}},\ }\href {https://doi.org/10.1088/1475-7516/2011/09/032} {\bibfield
   {journal} {\bibinfo  {journal} {\jcap}\ }\textbf {\bibinfo {volume}
  {2011}},\ \bibinfo {eid} {032} (\bibinfo {year} {2011})},\ \Eprint
  {https://arxiv.org/abs/1104.2935}  { arXiv:1104.2935
  [astro-ph.CO]}\BibitemShut {NoStop}%
\bibitem [{\citenamefont {{Ali-Ha{\"\i}moud}}\ and\ \citenamefont
  {{Hirata}}(2011)}]{2011PhRvD..83d3513A}%
  \BibitemOpen
  \bibfield  {author} {\bibinfo {author} {\bibfnamefont {Y.}~\bibnamefont
  {{Ali-Ha{\"\i}moud}}}\ and\ \bibinfo {author} {\bibfnamefont {C.~M.}\
  \bibnamefont {{Hirata}}},\ }\href
  {https://doi.org/10.1103/PhysRevD.83.043513} {\bibfield  {journal} {\bibinfo
  {journal} {\prd}\ }\textbf {\bibinfo {volume} {83}},\ \bibinfo {eid} {043513}
  (\bibinfo {year} {2011})},\ \Eprint {https://arxiv.org/abs/1011.3758}  {
  arXiv:1011.3758 [astro-ph.CO]}\BibitemShut {NoStop}%
\bibitem [{\citenamefont {{Lee}}\ and\ \citenamefont
  {{Ali-Ha{\"\i}moud}}(2020)}]{2020PhRvD.102h3517L}%
  \BibitemOpen
  \bibfield  {author} {\bibinfo {author} {\bibfnamefont {N.}~\bibnamefont
  {{Lee}}}\ and\ \bibinfo {author} {\bibfnamefont {Y.}~\bibnamefont
  {{Ali-Ha{\"\i}moud}}},\ }\href {https://doi.org/10.1103/PhysRevD.102.083517}
  {\bibfield  {journal} {\bibinfo  {journal} {\prd}\ }\textbf {\bibinfo
  {volume} {102}},\ \bibinfo {eid} {083517} (\bibinfo {year} {2020})},\ \Eprint
  {https://arxiv.org/abs/2007.14114}  { arXiv:2007.14114
  [astro-ph.CO]}\BibitemShut {NoStop}%
\bibitem [{\citenamefont {{Liu}}\ \emph {et~al.}(2020)\citenamefont {{Liu}},
  \citenamefont {{Ridgway}},\ and\ \citenamefont
  {{Slatyer}}}]{2020PhRvD.101b3530L}%
  \BibitemOpen
  \bibfield  {author} {\bibinfo {author} {\bibfnamefont {H.}~\bibnamefont
  {{Liu}}}, \bibinfo {author} {\bibfnamefont {G.~W.}\ \bibnamefont
  {{Ridgway}}},\ and\ \bibinfo {author} {\bibfnamefont {T.~R.}\ \bibnamefont
  {{Slatyer}}},\ }\href {https://doi.org/10.1103/PhysRevD.101.023530}
  {\bibfield  {journal} {\bibinfo  {journal} {\prd}\ }\textbf {\bibinfo
  {volume} {101}},\ \bibinfo {eid} {023530} (\bibinfo {year} {2020})},\ \Eprint
  {https://arxiv.org/abs/1904.09296}  { arXiv:1904.09296
  [astro-ph.CO]}\BibitemShut {NoStop}%
\bibitem [{\citenamefont {{Narlikar}}\ and\ \citenamefont
  {{Rana}}(1979)}]{1979PhLA...72...75N}%
  \BibitemOpen
  \bibfield  {author} {\bibinfo {author} {\bibfnamefont {J.~V.}\ \bibnamefont
  {{Narlikar}}}\ and\ \bibinfo {author} {\bibfnamefont {N.~C.}\ \bibnamefont
  {{Rana}}},\ }\href {https://doi.org/10.1016/0375-9601(79)90651-0} {\bibfield
  {journal} {\bibinfo  {journal} {\pla}\ }\textbf {\bibinfo {volume} {72}},\
  \bibinfo {pages} {75} (\bibinfo {year} {1979})}\BibitemShut {NoStop}%
\bibitem [{\citenamefont {{Lucca}}\ \emph {et~al.}(2020)\citenamefont
  {{Lucca}}, \citenamefont {{Sch{\"o}neberg}}, \citenamefont {{Hooper}},
  \citenamefont {{Lesgourgues}},\ and\ \citenamefont
  {{Chluba}}}]{2020JCAP...02..026L}%
  \BibitemOpen
  \bibfield  {author} {\bibinfo {author} {\bibfnamefont {M.}~\bibnamefont
  {{Lucca}}}, \bibinfo {author} {\bibfnamefont {N.}~\bibnamefont
  {{Sch{\"o}neberg}}}, \bibinfo {author} {\bibfnamefont {D.~C.}\ \bibnamefont
  {{Hooper}}}, \bibinfo {author} {\bibfnamefont {J.}~\bibnamefont
  {{Lesgourgues}}},\ and\ \bibinfo {author} {\bibfnamefont {J.}~\bibnamefont
  {{Chluba}}},\ }\href {https://doi.org/10.1088/1475-7516/2020/02/026}
  {\bibfield  {journal} {\bibinfo  {journal} {\jcap}\ }\textbf {\bibinfo
  {volume} {2020}},\ \bibinfo {eid} {026} (\bibinfo {year} {2020})},\ \Eprint
  {https://arxiv.org/abs/1910.04619}  { arXiv:1910.04619
  [astro-ph.CO]}\BibitemShut {NoStop}%
\bibitem [{\citenamefont {{Nasel'Skii}}(1978)}]{1978Ap.....14...82N}%
  \BibitemOpen
  \bibfield  {author} {\bibinfo {author} {\bibfnamefont {P.~D.}\ \bibnamefont
  {{Nasel'Skii}}},\ }\href {https://doi.org/10.1007/BF01005367} {\bibfield
  {journal} {\bibinfo  {journal} {Astrophysics}\ }\textbf {\bibinfo {volume}
  {14}},\ \bibinfo {pages} {82} (\bibinfo {year} {1978})},\ \bibinfo {note}
  {[Astrofizika \textbf{14}, 145 (1978)]}\BibitemShut {NoStop}%
\bibitem [{\citenamefont {{Nasel'skii}}(1978)}]{1978SvAL....4..209N}%
  \BibitemOpen
  \bibfield  {author} {\bibinfo {author} {\bibfnamefont {P.~D.}\ \bibnamefont
  {{Nasel'skii}}},\ }\href@noop {} {\bibfield  {journal} {\bibinfo  {journal}
  {\sovastl}\ }\textbf {\bibinfo {volume} {4}},\ \bibinfo {pages} {209}
  (\bibinfo {year} {1978})},\ \bibinfo {note} {[\PAZh~\textbf{4}, 387
  (1978)]}\BibitemShut {NoStop}%
\bibitem [{\citenamefont {{Aghanim}}\ \emph
  {et~al.}(2020{\natexlab{b}})\citenamefont {{Aghanim}} \emph
  {et~al.}}]{2020A&A...641A...6P}%
  \BibitemOpen
  \bibfield  {author} {\bibinfo {author} {\bibfnamefont {N.}~\bibnamefont
  {{Aghanim}}} \emph {et~al.} (\bibinfo {collaboration} {Planck}),\ }\href
  {https://doi.org/10.1051/0004-6361/201833910} {\bibfield  {journal} {\bibinfo
   {journal} {\aap}\ }\textbf {\bibinfo {volume} {641}},\ \bibinfo {eid} {A6}
  (\bibinfo {year} {2020}{\natexlab{b}})},\ \Eprint
  {https://arxiv.org/abs/1807.06209}  { arXiv:1807.06209
  [astro-ph.CO]}\BibitemShut {NoStop}%
\bibitem [{\citenamefont {{Acharya}}\ and\ \citenamefont
  {{Khatri}}(2020{\natexlab{b}})}]{2020JCAP...02..010A}%
  \BibitemOpen
  \bibfield  {author} {\bibinfo {author} {\bibfnamefont {S.~K.}\ \bibnamefont
  {{Acharya}}}\ and\ \bibinfo {author} {\bibfnamefont {R.}~\bibnamefont
  {{Khatri}}},\ }\href {https://doi.org/10.1088/1475-7516/2020/02/010}
  {\bibfield  {journal} {\bibinfo  {journal} {\jcap}\ }\textbf {\bibinfo
  {volume} {2020}},\ \bibinfo {eid} {010} (\bibinfo {year}
  {2020}{\natexlab{b}})},\ \Eprint {https://arxiv.org/abs/1912.10995}  {
  arXiv:1912.10995 [astro-ph.CO]}\BibitemShut {NoStop}%
\bibitem [{\citenamefont {{Nasel'Skii}}\ and\ \citenamefont
  {{Shevelev}}(1978)}]{1978Ap.....14..386N}%
  \BibitemOpen
  \bibfield  {author} {\bibinfo {author} {\bibfnamefont {P.~D.}\ \bibnamefont
  {{Nasel'Skii}}}\ and\ \bibinfo {author} {\bibfnamefont {Y.~G.}\ \bibnamefont
  {{Shevelev}}},\ }\href {https://doi.org/10.1007/BF01007456} {\bibfield
  {journal} {\bibinfo  {journal} {Astrophysics}\ }\textbf {\bibinfo {volume}
  {14}},\ \bibinfo {pages} {386} (\bibinfo {year} {1978})},\ \bibinfo {note}
  {[Astrofizika~\textbf{14}, 679 (1978)]}\BibitemShut {NoStop}%
\bibitem [{\citenamefont {{Acharya}}\ and\ \citenamefont
  {{Khatri}}(2019{\natexlab{b}})}]{2019PhRvD..99l3510A}%
  \BibitemOpen
  \bibfield  {author} {\bibinfo {author} {\bibfnamefont {S.~K.}\ \bibnamefont
  {{Acharya}}}\ and\ \bibinfo {author} {\bibfnamefont {R.}~\bibnamefont
  {{Khatri}}},\ }\href {https://doi.org/10.1103/PhysRevD.99.123510} {\bibfield
  {journal} {\bibinfo  {journal} {\prd}\ }\textbf {\bibinfo {volume} {99}},\
  \bibinfo {eid} {123510} (\bibinfo {year} {2019}{\natexlab{b}})},\ \Eprint
  {https://arxiv.org/abs/1903.04503}  { arXiv:1903.04503
  [astro-ph.CO]}\BibitemShut {NoStop}%
\bibitem [{\citenamefont {{Acharya}}\ and\ \citenamefont
  {{Khatri}}(2019{\natexlab{c}})}]{2019PhRvD..99d3520A}%
  \BibitemOpen
  \bibfield  {author} {\bibinfo {author} {\bibfnamefont {S.~K.}\ \bibnamefont
  {{Acharya}}}\ and\ \bibinfo {author} {\bibfnamefont {R.}~\bibnamefont
  {{Khatri}}},\ }\href {https://doi.org/10.1103/PhysRevD.99.043520} {\bibfield
  {journal} {\bibinfo  {journal} {\prd}\ }\textbf {\bibinfo {volume} {99}},\
  \bibinfo {eid} {043520} (\bibinfo {year} {2019}{\natexlab{c}})},\ \Eprint
  {https://arxiv.org/abs/1808.02897}  { arXiv:1808.02897
  [astro-ph.CO]}\BibitemShut {NoStop}%
\bibitem [{\citenamefont {{Acharya}}\ and\ \citenamefont
  {{Chluba}}(2021)}]{2021arXiv211206699A}%
  \BibitemOpen
  \bibfield  {author} {\bibinfo {author} {\bibfnamefont {S.~K.}\ \bibnamefont
  {{Acharya}}}\ and\ \bibinfo {author} {\bibfnamefont {J.}~\bibnamefont
  {{Chluba}}},\ }\href@noop {} {\bibfield  {journal} {\bibinfo  {journal}
  {\arxiv}\ } (\bibinfo {year} {2021})},\ \Eprint
  {https://arxiv.org/abs/2112.06699}  { 2112.06699 [astro-ph.CO]}\BibitemShut
  {NoStop}%
\bibitem [{\citenamefont {{Chluba}}\ \emph {et~al.}(2020)\citenamefont
  {{Chluba}}, \citenamefont {{Ravenni}},\ and\ \citenamefont
  {{Acharya}}}]{2020MNRAS.498..959C}%
  \BibitemOpen
  \bibfield  {author} {\bibinfo {author} {\bibfnamefont {J.}~\bibnamefont
  {{Chluba}}}, \bibinfo {author} {\bibfnamefont {A.}~\bibnamefont
  {{Ravenni}}},\ and\ \bibinfo {author} {\bibfnamefont {S.~K.}\ \bibnamefont
  {{Acharya}}},\ }\href {https://doi.org/10.1093/mnras/staa2131} {\bibfield
  {journal} {\bibinfo  {journal} {\mnras}\ }\textbf {\bibinfo {volume} {498}},\
  \bibinfo {pages} {959} (\bibinfo {year} {2020})},\ \Eprint
  {https://arxiv.org/abs/2005.11325}  { arXiv:2005.11325
  [astro-ph.CO]}\BibitemShut {NoStop}%
\bibitem [{\citenamefont {{Gibilisco}}(1996)}]{1996IJMPA..11.5541G}%
  \BibitemOpen
  \bibfield  {author} {\bibinfo {author} {\bibfnamefont {M.}~\bibnamefont
  {{Gibilisco}}},\ }\href {https://doi.org/10.1142/S0217751X96002546}
  {\bibfield  {journal} {\bibinfo  {journal} {\IJMPA}\ }\textbf {\bibinfo
  {volume} {11}},\ \bibinfo {pages} {5541} (\bibinfo {year} {1996})},\ \Eprint
  {https://arxiv.org/abs/astro-ph/9611227}  { arXiv:astro-ph/9611227
  [astro-ph]}\BibitemShut {NoStop}%
\bibitem [{\citenamefont
  {{Gibilisco}}(1997{\natexlab{a}})}]{1997IJMPA..12.2855G}%
  \BibitemOpen
  \bibfield  {author} {\bibinfo {author} {\bibfnamefont {M.}~\bibnamefont
  {{Gibilisco}}},\ }\href {https://doi.org/10.1142/S0217751X97001584}
  {\bibfield  {journal} {\bibinfo  {journal} {\IJMPA}\ }\textbf {\bibinfo
  {volume} {12}},\ \bibinfo {pages} {2855} (\bibinfo {year}
  {1997}{\natexlab{a}})},\ \Eprint {https://arxiv.org/abs/astro-ph/9609053}  {
  arXiv:astro-ph/9609053 [astro-ph]}\BibitemShut {NoStop}%
\bibitem [{\citenamefont
  {{Gibilisco}}(1997{\natexlab{b}})}]{1997IJMPA..12.4167G}%
  \BibitemOpen
  \bibfield  {author} {\bibinfo {author} {\bibfnamefont {M.}~\bibnamefont
  {{Gibilisco}}},\ }\href {https://doi.org/10.1142/S0217751X97002280}
  {\bibfield  {journal} {\bibinfo  {journal} {\IJMPA}\ }\textbf {\bibinfo
  {volume} {12}},\ \bibinfo {pages} {4167} (\bibinfo {year}
  {1997}{\natexlab{b}})},\ \Eprint {https://arxiv.org/abs/astro-ph/9604116}  {
  arXiv:astro-ph/9604116 [astro-ph]}\BibitemShut {NoStop}%
\bibitem [{\citenamefont {{Belotsky}}\ and\ \citenamefont
  {{Kirillov}}(2015)}]{2015JCAP...01..041B}%
  \BibitemOpen
  \bibfield  {author} {\bibinfo {author} {\bibfnamefont {K.~M.}\ \bibnamefont
  {{Belotsky}}}\ and\ \bibinfo {author} {\bibfnamefont {A.~A.}\ \bibnamefont
  {{Kirillov}}},\ }\href {https://doi.org/10.1088/1475-7516/2015/01/041}
  {\bibfield  {journal} {\bibinfo  {journal} {\jcap}\ }\textbf {\bibinfo
  {volume} {2015}},\ \bibinfo {eid} {041} (\bibinfo {year} {2015})},\ \Eprint
  {https://arxiv.org/abs/1409.8601}  { arXiv:1409.8601
  [astro-ph.CO]}\BibitemShut {NoStop}%
\bibitem [{\citenamefont {{Belotsky}}\ \emph {et~al.}(2014)\citenamefont
  {{Belotsky}} \emph {et~al.}}]{2014MPLA...2940005B}%
  \BibitemOpen
  \bibfield  {author} {\bibinfo {author} {\bibfnamefont {K.~M.}\ \bibnamefont
  {{Belotsky}}} \emph {et~al.},\ }\href
  {https://doi.org/10.1142/S0217732314400057} {\bibfield  {journal} {\bibinfo
  {journal} {\MPLA}\ }\textbf {\bibinfo {volume} {29}},\ \bibinfo {eid}
  {1440005} (\bibinfo {year} {2014})},\ \Eprint
  {https://arxiv.org/abs/1410.0203}  { arXiv:1410.0203
  [astro-ph.CO]}\BibitemShut {NoStop}%
\bibitem [{\citenamefont {{Belotsky}}\ \emph
  {et~al.}(2015{\natexlab{a}})\citenamefont {{Belotsky}}, \citenamefont
  {{Kirillov}},\ and\ \citenamefont {{Rubin}}}]{2015IJMPD..2445005B}%
  \BibitemOpen
  \bibfield  {author} {\bibinfo {author} {\bibfnamefont {K.~M.}\ \bibnamefont
  {{Belotsky}}}, \bibinfo {author} {\bibfnamefont {A.~A.}\ \bibnamefont
  {{Kirillov}}},\ and\ \bibinfo {author} {\bibfnamefont {S.~G.}\ \bibnamefont
  {{Rubin}}},\ }\href {https://doi.org/10.1142/S0218271815450054} {\bibfield
  {journal} {\bibinfo  {journal} {International Journal of Modern Physics D}\
  }\textbf {\bibinfo {volume} {24}},\ \bibinfo {eid} {1545005-528} (\bibinfo
  {year} {2015}{\natexlab{a}})}\BibitemShut {NoStop}%
\bibitem [{\citenamefont {{Belotsky}}\ \emph
  {et~al.}(2015{\natexlab{b}})\citenamefont {{Belotsky}}, \citenamefont
  {{Kirillov}},\ and\ \citenamefont {{Rubin}}}]{2015PAN....78..387B}%
  \BibitemOpen
  \bibfield  {author} {\bibinfo {author} {\bibfnamefont {K.~M.}\ \bibnamefont
  {{Belotsky}}}, \bibinfo {author} {\bibfnamefont {A.~A.}\ \bibnamefont
  {{Kirillov}}},\ and\ \bibinfo {author} {\bibfnamefont {S.~G.}\ \bibnamefont
  {{Rubin}}},\ }\href {https://doi.org/10.1134/S1063778815020088} {\bibfield
  {journal} {\bibinfo  {journal} {\PAN}\ }\textbf {\bibinfo {volume} {78}},\
  \bibinfo {pages} {387} (\bibinfo {year} {2015}{\natexlab{b}})}\BibitemShut
  {NoStop}%
\bibitem [{\citenamefont {{Belotsky}}\ \emph {et~al.}(2017)\citenamefont
  {{Belotsky}}, \citenamefont {{Kirillov}}, \citenamefont {{Nazarova}},\ and\
  \citenamefont {{Rubin}}}]{2017IJMPD..2650102B}%
  \BibitemOpen
  \bibfield  {author} {\bibinfo {author} {\bibfnamefont {K.~M.}\ \bibnamefont
  {{Belotsky}}}, \bibinfo {author} {\bibfnamefont {A.~A.}\ \bibnamefont
  {{Kirillov}}}, \bibinfo {author} {\bibfnamefont {N.~O.}\ \bibnamefont
  {{Nazarova}}},\ and\ \bibinfo {author} {\bibfnamefont {S.~G.}\ \bibnamefont
  {{Rubin}}},\ }\href {https://doi.org/10.1142/S0218271817501024} {\bibfield
  {journal} {\bibinfo  {journal} {\IJMPD}\ }\textbf {\bibinfo {volume} {26}},\
  \bibinfo {eid} {1750102} (\bibinfo {year} {2017})},\ \Eprint
  {https://arxiv.org/abs/1702.06338}  { arXiv:1702.06338
  [astro-ph.CO]}\BibitemShut {NoStop}%
\bibitem [{\citenamefont {{Schiavone}}\ \emph {et~al.}(2021)\citenamefont
  {{Schiavone}}, \citenamefont {{Montanino}}, \citenamefont {{Mirizzi}},\ and\
  \citenamefont {{Capozzi}}}]{2021JCAP...08..063S}%
  \BibitemOpen
  \bibfield  {author} {\bibinfo {author} {\bibfnamefont {F.}~\bibnamefont
  {{Schiavone}}}, \bibinfo {author} {\bibfnamefont {D.}~\bibnamefont
  {{Montanino}}}, \bibinfo {author} {\bibfnamefont {A.}~\bibnamefont
  {{Mirizzi}}},\ and\ \bibinfo {author} {\bibfnamefont {F.}~\bibnamefont
  {{Capozzi}}},\ }\href {https://doi.org/10.1088/1475-7516/2021/08/063}
  {\bibfield  {journal} {\bibinfo  {journal} {\jcap}\ }\textbf {\bibinfo
  {volume} {2021}},\ \bibinfo {eid} {063} (\bibinfo {year} {2021})},\ \Eprint
  {https://arxiv.org/abs/2107.03420}  { arXiv:2107.03420 [hep-ph]}\BibitemShut
  {NoStop}%
\bibitem [{\citenamefont {{Evoli}}\ \emph {et~al.}(2016)\citenamefont
  {{Evoli}}, \citenamefont {{Leo}}, \citenamefont {{Mirizzi}},\ and\
  \citenamefont {{Montanino}}}]{2016JCAP...05..006E}%
  \BibitemOpen
  \bibfield  {author} {\bibinfo {author} {\bibfnamefont {C.}~\bibnamefont
  {{Evoli}}}, \bibinfo {author} {\bibfnamefont {M.}~\bibnamefont {{Leo}}},
  \bibinfo {author} {\bibfnamefont {A.}~\bibnamefont {{Mirizzi}}},\ and\
  \bibinfo {author} {\bibfnamefont {D.}~\bibnamefont {{Montanino}}},\ }\href
  {https://doi.org/10.1088/1475-7516/2016/05/006} {\bibfield  {journal}
  {\bibinfo  {journal} {\jcap}\ }\textbf {\bibinfo {volume} {2016}},\ \bibinfo
  {eid} {006} (\bibinfo {year} {2016})},\ \Eprint
  {https://arxiv.org/abs/1602.08433}  { arXiv:1602.08433
  [astro-ph.CO]}\BibitemShut {NoStop}%
\bibitem [{\citenamefont {{Poulin}}\ \emph {et~al.}(2017)\citenamefont
  {{Poulin}}, \citenamefont {{Lesgourgues}},\ and\ \citenamefont
  {{Serpico}}}]{2017JCAP...03..043P}%
  \BibitemOpen
  \bibfield  {author} {\bibinfo {author} {\bibfnamefont {V.}~\bibnamefont
  {{Poulin}}}, \bibinfo {author} {\bibfnamefont {J.}~\bibnamefont
  {{Lesgourgues}}},\ and\ \bibinfo {author} {\bibfnamefont {P.~D.}\
  \bibnamefont {{Serpico}}},\ }\href
  {https://doi.org/10.1088/1475-7516/2017/03/043} {\bibfield  {journal}
  {\bibinfo  {journal} {\jcap}\ }\textbf {\bibinfo {volume} {2017}},\ \bibinfo
  {eid} {043} (\bibinfo {year} {2017})},\ \Eprint
  {https://arxiv.org/abs/1610.10051}  { arXiv:1610.10051
  [astro-ph.CO]}\BibitemShut {NoStop}%
\bibitem [{\citenamefont {{Dorosheva}}\ and\ \citenamefont
  {{Nasel'Skii}}(1986)}]{1986Ap.....24..321D}%
  \BibitemOpen
  \bibfield  {author} {\bibinfo {author} {\bibfnamefont {E.~I.}\ \bibnamefont
  {{Dorosheva}}}\ and\ \bibinfo {author} {\bibfnamefont {P.~D.}\ \bibnamefont
  {{Nasel'Skii}}},\ }\href {https://doi.org/10.1007/BF01005728} {\bibfield
  {journal} {\bibinfo  {journal} {Astrophysics}\ }\textbf {\bibinfo {volume}
  {24}},\ \bibinfo {pages} {321} (\bibinfo {year} {1986})},\ \bibinfo {note}
  {[Astrofizika \textbf{24}, 561 (1986)]}\BibitemShut {NoStop}%
\bibitem [{\citenamefont {{Dorosheva}}\ and\ \citenamefont
  {{Nasel'skii}}(1987)}]{1987SvA....31....1D}%
  \BibitemOpen
  \bibfield  {author} {\bibinfo {author} {\bibfnamefont {E.~I.}\ \bibnamefont
  {{Dorosheva}}}\ and\ \bibinfo {author} {\bibfnamefont {P.~D.}\ \bibnamefont
  {{Nasel'skii}}},\ }\href@noop {} {\bibfield  {journal} {\bibinfo  {journal}
  {\sovast}\ }\textbf {\bibinfo {volume} {31}},\ \bibinfo {pages} {1} (\bibinfo
  {year} {1987})},\ \bibinfo {note} {[\azh~\textbf{64}, 1 (1987)]}\BibitemShut
  {NoStop}%
\bibitem [{\citenamefont {{Nasel'skii}}\ and\ \citenamefont
  {{Polnarev}}(1987)}]{1987SvAL...13...67N}%
  \BibitemOpen
  \bibfield  {author} {\bibinfo {author} {\bibfnamefont {P.~D.}\ \bibnamefont
  {{Nasel'skii}}}\ and\ \bibinfo {author} {\bibfnamefont {A.~G.}\ \bibnamefont
  {{Polnarev}}},\ }\href@noop {} {\bibfield  {journal} {\bibinfo  {journal}
  {\sovastl}\ }\textbf {\bibinfo {volume} {13}},\ \bibinfo {pages} {67}
  (\bibinfo {year} {1987})},\ \bibinfo {note} {[\PAZh~\textbf{13}, 167
  (1987)]}\BibitemShut {NoStop}%
\bibitem [{\citenamefont {{Zhang}}\ \emph {et~al.}(2007)\citenamefont
  {{Zhang}}, \citenamefont {{Chen}}, \citenamefont {{Kamionkowski}},
  \citenamefont {{Si}},\ and\ \citenamefont {{Zheng}}}]{2007PhRvD..76f1301Z}%
  \BibitemOpen
  \bibfield  {author} {\bibinfo {author} {\bibfnamefont {L.}~\bibnamefont
  {{Zhang}}}, \bibinfo {author} {\bibfnamefont {X.}~\bibnamefont {{Chen}}},
  \bibinfo {author} {\bibfnamefont {M.}~\bibnamefont {{Kamionkowski}}},
  \bibinfo {author} {\bibfnamefont {Z.-G.}\ \bibnamefont {{Si}}},\ and\
  \bibinfo {author} {\bibfnamefont {Z.}~\bibnamefont {{Zheng}}},\ }\href
  {https://doi.org/10.1103/PhysRevD.76.061301} {\bibfield  {journal} {\bibinfo
  {journal} {\prd}\ }\textbf {\bibinfo {volume} {76}},\ \bibinfo {eid} {061301}
  (\bibinfo {year} {2007})},\ \Eprint {https://arxiv.org/abs/0704.2444}  {
  arXiv:0704.2444 [astro-ph]}\BibitemShut {NoStop}%
\bibitem [{\citenamefont {{Yang}}(2019)}]{2019MNRAS.486.4569Y}%
  \BibitemOpen
  \bibfield  {author} {\bibinfo {author} {\bibfnamefont {Y.}~\bibnamefont
  {{Yang}}},\ }\href {https://doi.org/10.1093/mnras/stz1148} {\bibfield
  {journal} {\bibinfo  {journal} {\mnras}\ }\textbf {\bibinfo {volume} {486}},\
  \bibinfo {pages} {4569} (\bibinfo {year} {2019})},\ \Eprint
  {https://arxiv.org/abs/1904.09104}  { arXiv:1904.09104
  [astro-ph.CO]}\BibitemShut {NoStop}%
\bibitem [{\citenamefont {{Clark}}\ \emph {et~al.}(2017)\citenamefont
  {{Clark}}, \citenamefont {{Dutta}}, \citenamefont {{Gao}}, \citenamefont
  {{Strigari}},\ and\ \citenamefont {{Watson}}}]{2017PhRvD..95h3006C}%
  \BibitemOpen
  \bibfield  {author} {\bibinfo {author} {\bibfnamefont {S.~J.}\ \bibnamefont
  {{Clark}}}, \bibinfo {author} {\bibfnamefont {B.}~\bibnamefont {{Dutta}}},
  \bibinfo {author} {\bibfnamefont {Y.}~\bibnamefont {{Gao}}}, \bibinfo
  {author} {\bibfnamefont {L.~E.}\ \bibnamefont {{Strigari}}},\ and\ \bibinfo
  {author} {\bibfnamefont {S.}~\bibnamefont {{Watson}}},\ }\href
  {https://doi.org/10.1103/PhysRevD.95.083006} {\bibfield  {journal} {\bibinfo
  {journal} {\prd}\ }\textbf {\bibinfo {volume} {95}},\ \bibinfo {eid} {083006}
  (\bibinfo {year} {2017})},\ \Eprint {https://arxiv.org/abs/1612.07738}  {
  arXiv:1612.07738 [astro-ph.CO]}\BibitemShut {NoStop}%
\bibitem [{\citenamefont {{St{\"o}cker}}\ \emph {et~al.}(2018)\citenamefont
  {{St{\"o}cker}}, \citenamefont {{Kr{\"a}mer}}, \citenamefont
  {{Lesgourgues}},\ and\ \citenamefont {{Poulin}}}]{2018JCAP...03..018S}%
  \BibitemOpen
  \bibfield  {author} {\bibinfo {author} {\bibfnamefont {P.}~\bibnamefont
  {{St{\"o}cker}}}, \bibinfo {author} {\bibfnamefont {M.}~\bibnamefont
  {{Kr{\"a}mer}}}, \bibinfo {author} {\bibfnamefont {J.}~\bibnamefont
  {{Lesgourgues}}},\ and\ \bibinfo {author} {\bibfnamefont {V.}~\bibnamefont
  {{Poulin}}},\ }\href {https://doi.org/10.1088/1475-7516/2018/03/018}
  {\bibfield  {journal} {\bibinfo  {journal} {\jcap}\ }\textbf {\bibinfo
  {volume} {2018}},\ \bibinfo {eid} {018} (\bibinfo {year} {2018})},\ \Eprint
  {https://arxiv.org/abs/1801.01871}  { arXiv:1801.01871
  [astro-ph.CO]}\BibitemShut {NoStop}%
\bibitem [{\citenamefont {{Cang}}\ \emph {et~al.}(2021)\citenamefont {{Cang}},
  \citenamefont {{Gao}},\ and\ \citenamefont {{Ma}}}]{2021JCAP...05..051C}%
  \BibitemOpen
  \bibfield  {author} {\bibinfo {author} {\bibfnamefont {J.}~\bibnamefont
  {{Cang}}}, \bibinfo {author} {\bibfnamefont {Y.}~\bibnamefont {{Gao}}},\ and\
  \bibinfo {author} {\bibfnamefont {Y.-Z.}\ \bibnamefont {{Ma}}},\ }\href
  {https://doi.org/10.1088/1475-7516/2021/05/051} {\bibfield  {journal}
  {\bibinfo  {journal} {\jcap}\ }\textbf {\bibinfo {volume} {2021}},\ \bibinfo
  {eid} {051} (\bibinfo {year} {2021})},\ \Eprint
  {https://arxiv.org/abs/2011.12244}  { arXiv:2011.12244
  [astro-ph.CO]}\BibitemShut {NoStop}%
\bibitem [{\citenamefont {{Cang}}\ \emph {et~al.}(2022)\citenamefont {{Cang}},
  \citenamefont {{Gao}},\ and\ \citenamefont {{Ma}}}]{2022JCAP...03..012C}%
  \BibitemOpen
  \bibfield  {author} {\bibinfo {author} {\bibfnamefont {J.}~\bibnamefont
  {{Cang}}}, \bibinfo {author} {\bibfnamefont {Y.}~\bibnamefont {{Gao}}},\ and\
  \bibinfo {author} {\bibfnamefont {Y.-Z.}\ \bibnamefont {{Ma}}},\ }\href
  {https://doi.org/10.1088/1475-7516/2022/03/012} {\bibfield  {journal}
  {\bibinfo  {journal} {\jcap}\ }\textbf {\bibinfo {volume} {2022}},\ \bibinfo
  {eid} {012} (\bibinfo {year} {2022})},\ \Eprint
  {https://arxiv.org/abs/2108.13256}  { arXiv:2108.13256
  [astro-ph.CO]}\BibitemShut {NoStop}%
\bibitem [{\citenamefont {{Mack}}\ and\ \citenamefont
  {{Wesley}}(2008)}]{2008arXiv0805.1531M}%
  \BibitemOpen
  \bibfield  {author} {\bibinfo {author} {\bibfnamefont {K.~J.}\ \bibnamefont
  {{Mack}}}\ and\ \bibinfo {author} {\bibfnamefont {D.~H.}\ \bibnamefont
  {{Wesley}}},\ }\href@noop {} {\bibfield  {journal} {\bibinfo  {journal}
  {\arxiv}\ } (\bibinfo {year} {2008})},\ \Eprint
  {https://arxiv.org/abs/0805.1531}  { arXiv:0805.1531 [astro-ph]}\BibitemShut
  {NoStop}%
\bibitem [{\citenamefont {{Bowman}}\ \emph {et~al.}(2018)\citenamefont
  {{Bowman}}, \citenamefont {{Rogers}}, \citenamefont {{Monsalve}},
  \citenamefont {{Mozdzen}},\ and\ \citenamefont
  {{Mahesh}}}]{2018Natur.555...67B}%
  \BibitemOpen
  \bibfield  {author} {\bibinfo {author} {\bibfnamefont {J.~D.}\ \bibnamefont
  {{Bowman}}}, \bibinfo {author} {\bibfnamefont {A.~E.~E.}\ \bibnamefont
  {{Rogers}}}, \bibinfo {author} {\bibfnamefont {R.~A.}\ \bibnamefont
  {{Monsalve}}}, \bibinfo {author} {\bibfnamefont {T.~J.}\ \bibnamefont
  {{Mozdzen}}},\ and\ \bibinfo {author} {\bibfnamefont {N.}~\bibnamefont
  {{Mahesh}}},\ }\href {https://doi.org/10.1038/nature25792} {\bibfield
  {journal} {\bibinfo  {journal} {\nat}\ }\textbf {\bibinfo {volume} {555}},\
  \bibinfo {pages} {67} (\bibinfo {year} {2018})},\ \Eprint
  {https://arxiv.org/abs/1810.05912}  { arXiv:1810.05912
  [astro-ph.CO]}\BibitemShut {NoStop}%
\bibitem [{\citenamefont {{Clark}}\ \emph {et~al.}(2018)\citenamefont
  {{Clark}}, \citenamefont {{Dutta}}, \citenamefont {{Gao}}, \citenamefont
  {{Ma}},\ and\ \citenamefont {{Strigari}}}]{2018PhRvD..98d3006C}%
  \BibitemOpen
  \bibfield  {author} {\bibinfo {author} {\bibfnamefont {S.~J.}\ \bibnamefont
  {{Clark}}}, \bibinfo {author} {\bibfnamefont {B.}~\bibnamefont {{Dutta}}},
  \bibinfo {author} {\bibfnamefont {Y.}~\bibnamefont {{Gao}}}, \bibinfo
  {author} {\bibfnamefont {Y.-Z.}\ \bibnamefont {{Ma}}},\ and\ \bibinfo
  {author} {\bibfnamefont {L.~E.}\ \bibnamefont {{Strigari}}},\ }\href
  {https://doi.org/10.1103/PhysRevD.98.043006} {\bibfield  {journal} {\bibinfo
  {journal} {\prd}\ }\textbf {\bibinfo {volume} {98}},\ \bibinfo {eid} {043006}
  (\bibinfo {year} {2018})},\ \Eprint {https://arxiv.org/abs/1803.09390}  {
  arXiv:1803.09390 [astro-ph.HE]}\BibitemShut {NoStop}%
\bibitem [{\citenamefont {{Yang}}(2020)}]{2020PhRvD.102h3538Y}%
  \BibitemOpen
  \bibfield  {author} {\bibinfo {author} {\bibfnamefont {Y.}~\bibnamefont
  {{Yang}}},\ }\href {https://doi.org/10.1103/PhysRevD.102.083538} {\bibfield
  {journal} {\bibinfo  {journal} {\prd}\ }\textbf {\bibinfo {volume} {102}},\
  \bibinfo {eid} {083538} (\bibinfo {year} {2020})},\ \Eprint
  {https://arxiv.org/abs/2009.11547}  { arXiv:2009.11547
  [astro-ph.CO]}\BibitemShut {NoStop}%
\bibitem [{\citenamefont {{Halder}}\ \emph {et~al.}(2021)\citenamefont
  {{Halder}}, \citenamefont {{Pandey}}, \citenamefont {{Majumdar}},\ and\
  \citenamefont {{Basu}}}]{2021JCAP...10..033H}%
  \BibitemOpen
  \bibfield  {author} {\bibinfo {author} {\bibfnamefont {A.}~\bibnamefont
  {{Halder}}}, \bibinfo {author} {\bibfnamefont {M.}~\bibnamefont {{Pandey}}},
  \bibinfo {author} {\bibfnamefont {D.}~\bibnamefont {{Majumdar}}},\ and\
  \bibinfo {author} {\bibfnamefont {R.}~\bibnamefont {{Basu}}},\ }\href
  {https://doi.org/10.1088/1475-7516/2021/10/033} {\bibfield  {journal}
  {\bibinfo  {journal} {\jcap}\ }\textbf {\bibinfo {volume} {2021}},\ \bibinfo
  {eid} {033} (\bibinfo {year} {2021})},\ \Eprint
  {https://arxiv.org/abs/2105.14356}  { arXiv:2105.14356
  [astro-ph.CO]}\BibitemShut {NoStop}%
\bibitem [{\citenamefont {{Halder}}\ and\ \citenamefont
  {{Pandey}}(2021)}]{2021MNRAS.508.3446H}%
  \BibitemOpen
  \bibfield  {author} {\bibinfo {author} {\bibfnamefont {A.}~\bibnamefont
  {{Halder}}}\ and\ \bibinfo {author} {\bibfnamefont {M.}~\bibnamefont
  {{Pandey}}},\ }\href {https://doi.org/10.1093/mnras/stab2795} {\bibfield
  {journal} {\bibinfo  {journal} {\mnras}\ }\textbf {\bibinfo {volume} {508}},\
  \bibinfo {pages} {3446} (\bibinfo {year} {2021})},\ \Eprint
  {https://arxiv.org/abs/2101.05228}  { arXiv:2101.05228
  [astro-ph.CO]}\BibitemShut {NoStop}%
\bibitem [{\citenamefont {{Halder}}\ and\ \citenamefont
  {{Banerjee}}(2021)}]{2021PhRvD.103f3044H}%
  \BibitemOpen
  \bibfield  {author} {\bibinfo {author} {\bibfnamefont {A.}~\bibnamefont
  {{Halder}}}\ and\ \bibinfo {author} {\bibfnamefont {S.}~\bibnamefont
  {{Banerjee}}},\ }\href {https://doi.org/10.1103/PhysRevD.103.063044}
  {\bibfield  {journal} {\bibinfo  {journal} {\prd}\ }\textbf {\bibinfo
  {volume} {103}},\ \bibinfo {eid} {063044} (\bibinfo {year} {2021})},\ \Eprint
  {https://arxiv.org/abs/2102.00959}  { arXiv:2102.00959
  [astro-ph.CO]}\BibitemShut {NoStop}%
\bibitem [{\citenamefont {{Mittal}}\ \emph {et~al.}(2022)\citenamefont
  {{Mittal}}, \citenamefont {{Ray}}, \citenamefont {{Kulkarni}},\ and\
  \citenamefont {{Dasgupta}}}]{2022JCAP...03..030M}%
  \BibitemOpen
  \bibfield  {author} {\bibinfo {author} {\bibfnamefont {S.}~\bibnamefont
  {{Mittal}}}, \bibinfo {author} {\bibfnamefont {A.}~\bibnamefont {{Ray}}},
  \bibinfo {author} {\bibfnamefont {G.}~\bibnamefont {{Kulkarni}}},\ and\
  \bibinfo {author} {\bibfnamefont {B.}~\bibnamefont {{Dasgupta}}},\ }\href
  {https://doi.org/10.1088/1475-7516/2022/03/030} {\bibfield  {journal}
  {\bibinfo  {journal} {\jcap}\ }\textbf {\bibinfo {volume} {2022}},\ \bibinfo
  {eid} {030} (\bibinfo {year} {2022})},\ \Eprint
  {https://arxiv.org/abs/2107.02190}  { arXiv:2107.02190
  [astro-ph.CO]}\BibitemShut {NoStop}%
\bibitem [{\citenamefont {{Natwariya}}\ \emph {et~al.}(2022)\citenamefont
  {{Natwariya}}, \citenamefont {{Nayak}},\ and\ \citenamefont
  {{Srivastava}}}]{2022MNRAS.510.4236N}%
  \BibitemOpen
  \bibfield  {author} {\bibinfo {author} {\bibfnamefont {P.~K.}\ \bibnamefont
  {{Natwariya}}}, \bibinfo {author} {\bibfnamefont {A.~C.}\ \bibnamefont
  {{Nayak}}},\ and\ \bibinfo {author} {\bibfnamefont {T.}~\bibnamefont
  {{Srivastava}}},\ }\href {https://doi.org/10.1093/mnras/stab3754} {\bibfield
  {journal} {\bibinfo  {journal} {\mnras}\ }\textbf {\bibinfo {volume} {510}},\
  \bibinfo {pages} {4236} (\bibinfo {year} {2022})},\ \Eprint
  {https://arxiv.org/abs/2107.12358}  { arXiv:2107.12358
  [astro-ph.CO]}\BibitemShut {NoStop}%
\bibitem [{\citenamefont {{Saha}}\ and\ \citenamefont
  {{Laha}}(2021)}]{2021arXiv211210794S}%
  \BibitemOpen
  \bibfield  {author} {\bibinfo {author} {\bibfnamefont {A.~K.}\ \bibnamefont
  {{Saha}}}\ and\ \bibinfo {author} {\bibfnamefont {R.}~\bibnamefont
  {{Laha}}},\ }\href@noop {} {\bibfield  {journal} {\bibinfo  {journal}
  {\arxiv}\ } (\bibinfo {year} {2021})},\ \Eprint
  {https://arxiv.org/abs/2112.10794}  { arXiv:2112.10794
  [astro-ph.CO]}\BibitemShut {NoStop}%
\bibitem [{\citenamefont {{Chapline}}(1975)}]{1975Natur.253..251C}%
  \BibitemOpen
  \bibfield  {author} {\bibinfo {author} {\bibfnamefont {G.~F.}\ \bibnamefont
  {{Chapline}}},\ }\href {https://doi.org/10.1038/253251a0} {\bibfield
  {journal} {\bibinfo  {journal} {\nat}\ }\textbf {\bibinfo {volume} {253}},\
  \bibinfo {pages} {251} (\bibinfo {year} {1975})}\BibitemShut {NoStop}%
\bibitem [{\citenamefont {{Overduin}}\ and\ \citenamefont
  {{Wesson}}(1992)}]{1992VA.....35..439O}%
  \BibitemOpen
  \bibfield  {author} {\bibinfo {author} {\bibfnamefont {J.~M.}\ \bibnamefont
  {{Overduin}}}\ and\ \bibinfo {author} {\bibfnamefont {P.~S.}\ \bibnamefont
  {{Wesson}}},\ }\href {https://doi.org/10.1016/0083-6656(92)90003-O}
  {\bibfield  {journal} {\bibinfo  {journal} {Vistas Astron.}\ }\textbf
  {\bibinfo {volume} {35}},\ \bibinfo {pages} {439} (\bibinfo {year}
  {1992})}\BibitemShut {NoStop}%
\bibitem [{\citenamefont {{Overduin}}\ and\ \citenamefont
  {{Wesson}}(2004)}]{2004PhR...402..267O}%
  \BibitemOpen
  \bibfield  {author} {\bibinfo {author} {\bibfnamefont {J.~M.}\ \bibnamefont
  {{Overduin}}}\ and\ \bibinfo {author} {\bibfnamefont {P.~S.}\ \bibnamefont
  {{Wesson}}},\ }\href {https://doi.org/10.1016/j.physrep.2004.07.006}
  {\bibfield  {journal} {\bibinfo  {journal} {\physrep}\ }\textbf {\bibinfo
  {volume} {402}},\ \bibinfo {pages} {267} (\bibinfo {year} {2004})},\ \Eprint
  {https://arxiv.org/abs/astro-ph/0407207}  { arXiv:astro-ph/0407207
  [astro-ph]}\BibitemShut {NoStop}%
\bibitem [{\citenamefont {{Ressell}}\ and\ \citenamefont
  {{Turner}}(1990)}]{1990ComAp..14..323R}%
  \BibitemOpen
  \bibfield  {author} {\bibinfo {author} {\bibfnamefont {M.~T.}\ \bibnamefont
  {{Ressell}}}\ and\ \bibinfo {author} {\bibfnamefont {M.~S.}\ \bibnamefont
  {{Turner}}},\ }\href@noop {} {\bibfield  {journal} {\bibinfo  {journal}
  {Comments Astrophys.}\ }\textbf {\bibinfo {volume} {14}},\ \bibinfo {pages}
  {323} (\bibinfo {year} {1990})}\BibitemShut {NoStop}%
\bibitem [{\citenamefont {{Hill}}\ \emph {et~al.}(2018)\citenamefont {{Hill}},
  \citenamefont {{Masui}},\ and\ \citenamefont
  {{Scott}}}]{2018ApSpe..72..663H}%
  \BibitemOpen
  \bibfield  {author} {\bibinfo {author} {\bibfnamefont {R.}~\bibnamefont
  {{Hill}}}, \bibinfo {author} {\bibfnamefont {K.~W.}\ \bibnamefont
  {{Masui}}},\ and\ \bibinfo {author} {\bibfnamefont {D.}~\bibnamefont
  {{Scott}}},\ }\href {https://doi.org/10.1177/0003702818767133} {\bibfield
  {journal} {\bibinfo  {journal} {Appl.~Spectrosc.}\ }\textbf {\bibinfo
  {volume} {72}},\ \bibinfo {pages} {663} (\bibinfo {year} {2018})},\ \Eprint
  {https://arxiv.org/abs/1802.03694}  { arXiv:1802.03694
  [astro-ph.CO]}\BibitemShut {NoStop}%
\bibitem [{\citenamefont {{Page}}\ and\ \citenamefont
  {{Hawking}}(1976)}]{1976ApJ...206....1P}%
  \BibitemOpen
  \bibfield  {author} {\bibinfo {author} {\bibfnamefont {D.~N.}\ \bibnamefont
  {{Page}}}\ and\ \bibinfo {author} {\bibfnamefont {S.~W.}\ \bibnamefont
  {{Hawking}}},\ }\href {https://doi.org/10.1086/154350} {\bibfield  {journal}
  {\bibinfo  {journal} {\apj}\ }\textbf {\bibinfo {volume} {206}},\ \bibinfo
  {pages} {1} (\bibinfo {year} {1976})}\BibitemShut {NoStop}%
\bibitem [{\citenamefont {{Fichtel}}\ \emph {et~al.}(1975)\citenamefont
  {{Fichtel}} \emph {et~al.}}]{1975ApJ...198..163F}%
  \BibitemOpen
  \bibfield  {author} {\bibinfo {author} {\bibfnamefont {C.~E.}\ \bibnamefont
  {{Fichtel}}} \emph {et~al.},\ }\href {https://doi.org/10.1086/153590}
  {\bibfield  {journal} {\bibinfo  {journal} {\apj}\ }\textbf {\bibinfo
  {volume} {198}},\ \bibinfo {pages} {163} (\bibinfo {year}
  {1975})}\BibitemShut {NoStop}%
\bibitem [{\citenamefont {{Halzen}}\ \emph {et~al.}(1991)\citenamefont
  {{Halzen}}, \citenamefont {{Zas}}, \citenamefont {{MacGibbon}},\ and\
  \citenamefont {{Weekes}}}]{1991Natur.353..807H}%
  \BibitemOpen
  \bibfield  {author} {\bibinfo {author} {\bibfnamefont {F.}~\bibnamefont
  {{Halzen}}}, \bibinfo {author} {\bibfnamefont {E.}~\bibnamefont {{Zas}}},
  \bibinfo {author} {\bibfnamefont {J.~H.}\ \bibnamefont {{MacGibbon}}},\ and\
  \bibinfo {author} {\bibfnamefont {T.~C.}\ \bibnamefont {{Weekes}}},\ }\href
  {https://doi.org/10.1038/353807a0} {\bibfield  {journal} {\bibinfo  {journal}
  {\nat}\ }\textbf {\bibinfo {volume} {353}},\ \bibinfo {pages} {807} (\bibinfo
  {year} {1991})}\BibitemShut {NoStop}%
\bibitem [{\citenamefont {{Sch\"onfelder}}\ \emph {et~al.}(1977)\citenamefont
  {{Sch\"onfelder}}, \citenamefont {{Graser}},\ and\ \citenamefont
  {{Daugherty}}}]{1977ApJ...217..306S}%
  \BibitemOpen
  \bibfield  {author} {\bibinfo {author} {\bibfnamefont {V.}~\bibnamefont
  {{Sch\"onfelder}}}, \bibinfo {author} {\bibfnamefont {U.}~\bibnamefont
  {{Graser}}},\ and\ \bibinfo {author} {\bibfnamefont {J.}~\bibnamefont
  {{Daugherty}}},\ }\href {https://doi.org/10.1086/155580} {\bibfield
  {journal} {\bibinfo  {journal} {\apj}\ }\textbf {\bibinfo {volume} {217}},\
  \bibinfo {pages} {306} (\bibinfo {year} {1977})}\BibitemShut {NoStop}%
\bibitem [{\citenamefont {{MacGibbon}}\ and\ \citenamefont
  {{Carr}}(1991)}]{1991ApJ...371..447M}%
  \BibitemOpen
  \bibfield  {author} {\bibinfo {author} {\bibfnamefont {J.~H.}\ \bibnamefont
  {{MacGibbon}}}\ and\ \bibinfo {author} {\bibfnamefont {B.~J.}\ \bibnamefont
  {{Carr}}},\ }\href {https://doi.org/10.1086/169909} {\bibfield  {journal}
  {\bibinfo  {journal} {\apj}\ }\textbf {\bibinfo {volume} {371}},\ \bibinfo
  {pages} {447} (\bibinfo {year} {1991})}\BibitemShut {NoStop}%
\bibitem [{\citenamefont {{Kappadath}}\ \emph {et~al.}(1996)\citenamefont
  {{Kappadath}} \emph {et~al.}}]{1996A&AS..120C.619K}%
  \BibitemOpen
  \bibfield  {author} {\bibinfo {author} {\bibfnamefont {S.~C.}\ \bibnamefont
  {{Kappadath}}} \emph {et~al.},\ }\href@noop {} {\bibfield  {journal}
  {\bibinfo  {journal} {\aaps}\ }\textbf {\bibinfo {volume} {120}},\ \bibinfo
  {pages} {619} (\bibinfo {year} {1996})}\BibitemShut {NoStop}%
\bibitem [{\citenamefont {{Sreekumar}}\ \emph {et~al.}(1998)\citenamefont
  {{Sreekumar}} \emph {et~al.}}]{1998ApJ...494..523S}%
  \BibitemOpen
  \bibfield  {author} {\bibinfo {author} {\bibfnamefont {P.}~\bibnamefont
  {{Sreekumar}}} \emph {et~al.},\ }\href {https://doi.org/10.1086/305222}
  {\bibfield  {journal} {\bibinfo  {journal} {\apj}\ }\textbf {\bibinfo
  {volume} {494}},\ \bibinfo {pages} {523} (\bibinfo {year} {1998})},\ \Eprint
  {https://arxiv.org/abs/astro-ph/9709257}  { arXiv:astro-ph/9709257
  [astro-ph]}\BibitemShut {NoStop}%
\bibitem [{\citenamefont {{Kim}}\ \emph {et~al.}(1999)\citenamefont {{Kim}},
  \citenamefont {{Lee}},\ and\ \citenamefont
  {{MacGibbon}}}]{1999PhRvD..59f3004K}%
  \BibitemOpen
  \bibfield  {author} {\bibinfo {author} {\bibfnamefont {H.~I.}\ \bibnamefont
  {{Kim}}}, \bibinfo {author} {\bibfnamefont {C.~H.}\ \bibnamefont {{Lee}}},\
  and\ \bibinfo {author} {\bibfnamefont {J.~H.}\ \bibnamefont {{MacGibbon}}},\
  }\href {https://doi.org/10.1103/PhysRevD.59.063004} {\bibfield  {journal}
  {\bibinfo  {journal} {\prd}\ }\textbf {\bibinfo {volume} {59}},\ \bibinfo
  {eid} {063004} (\bibinfo {year} {1999})},\ \Eprint
  {https://arxiv.org/abs/astro-ph/9901030}  { arXiv:astro-ph/9901030
  [astro-ph]}\BibitemShut {NoStop}%
\bibitem [{\citenamefont {{Kribs}}\ \emph {et~al.}(1999)\citenamefont
  {{Kribs}}, \citenamefont {{Leibovich}},\ and\ \citenamefont
  {{Rothstein}}}]{1999PhRvD..60j3510K}%
  \BibitemOpen
  \bibfield  {author} {\bibinfo {author} {\bibfnamefont {G.~D.}\ \bibnamefont
  {{Kribs}}}, \bibinfo {author} {\bibfnamefont {A.~K.}\ \bibnamefont
  {{Leibovich}}},\ and\ \bibinfo {author} {\bibfnamefont {I.~Z.}\ \bibnamefont
  {{Rothstein}}},\ }\href {https://doi.org/10.1103/PhysRevD.60.103510}
  {\bibfield  {journal} {\bibinfo  {journal} {\prd}\ }\textbf {\bibinfo
  {volume} {60}},\ \bibinfo {eid} {103510} (\bibinfo {year} {1999})},\ \Eprint
  {https://arxiv.org/abs/astro-ph/9904021}  { arXiv:astro-ph/9904021
  [astro-ph]}\BibitemShut {NoStop}%
\bibitem [{\citenamefont {{Bugaev}}\ and\ \citenamefont
  {{Klimai}}(2009)}]{2009PhRvD..79j3511B}%
  \BibitemOpen
  \bibfield  {author} {\bibinfo {author} {\bibfnamefont {E.}~\bibnamefont
  {{Bugaev}}}\ and\ \bibinfo {author} {\bibfnamefont {P.}~\bibnamefont
  {{Klimai}}},\ }\href {https://doi.org/10.1103/PhysRevD.79.103511} {\bibfield
  {journal} {\bibinfo  {journal} {\prd}\ }\textbf {\bibinfo {volume} {79}},\
  \bibinfo {eid} {103511} (\bibinfo {year} {2009})},\ \Eprint
  {https://arxiv.org/abs/0812.4247}  { arXiv:0812.4247 [astro-ph]}\BibitemShut
  {NoStop}%
\bibitem [{\citenamefont {{Sendouda}}\ \emph {et~al.}(2006)\citenamefont
  {{Sendouda}}, \citenamefont {{Nagataki}},\ and\ \citenamefont
  {{Sato}}}]{2006JCAP...06..003S}%
  \BibitemOpen
  \bibfield  {author} {\bibinfo {author} {\bibfnamefont {Y.}~\bibnamefont
  {{Sendouda}}}, \bibinfo {author} {\bibfnamefont {S.}~\bibnamefont
  {{Nagataki}}},\ and\ \bibinfo {author} {\bibfnamefont {K.}~\bibnamefont
  {{Sato}}},\ }\href {https://doi.org/10.1088/1475-7516/2006/06/003} {\bibfield
   {journal} {\bibinfo  {journal} {\jcap}\ }\textbf {\bibinfo {volume}
  {2006}},\ \bibinfo {eid} {003} (\bibinfo {year} {2006})},\ \Eprint
  {https://arxiv.org/abs/astro-ph/0603509}  { arXiv:astro-ph/0603509
  [astro-ph]}\BibitemShut {NoStop}%
\bibitem [{\citenamefont {{Carr}}\ and\ \citenamefont
  {{MacGibbon}}(1998)}]{1998PhR...307..141C}%
  \BibitemOpen
  \bibfield  {author} {\bibinfo {author} {\bibfnamefont {B.~J.}\ \bibnamefont
  {{Carr}}}\ and\ \bibinfo {author} {\bibfnamefont {J.~H.}\ \bibnamefont
  {{MacGibbon}}},\ }\href {https://doi.org/10.1016/S0370-1573(98)00039-8}
  {\bibfield  {journal} {\bibinfo  {journal} {\physrep}\ }\textbf {\bibinfo
  {volume} {307}},\ \bibinfo {pages} {141} (\bibinfo {year}
  {1998})}\BibitemShut {NoStop}%
\bibitem [{\citenamefont {{Bugaev}}\ and\ \citenamefont
  {{Klimai}}(2006)}]{2006astro.ph.12659B}%
  \BibitemOpen
  \bibfield  {author} {\bibinfo {author} {\bibfnamefont {E.}~\bibnamefont
  {{Bugaev}}}\ and\ \bibinfo {author} {\bibfnamefont {P.}~\bibnamefont
  {{Klimai}}},\ }\href@noop {} {\bibfield  {journal} {\bibinfo  {journal}
  {arXiv e-prints}\ } (\bibinfo {year} {2006})},\ \Eprint
  {https://arxiv.org/abs/astro-ph/0612659}  { astro-ph/0612659
  [astro-ph]}\BibitemShut {NoStop}%
\bibitem [{\citenamefont {{Barrau}}\ \emph
  {et~al.}(2003{\natexlab{a}})\citenamefont {{Barrau}}, \citenamefont
  {{Boudoul}},\ and\ \citenamefont {{Derome}}}]{2003astro.ph..4528B}%
  \BibitemOpen
  \bibfield  {author} {\bibinfo {author} {\bibfnamefont {A.}~\bibnamefont
  {{Barrau}}}, \bibinfo {author} {\bibfnamefont {G.}~\bibnamefont
  {{Boudoul}}},\ and\ \bibinfo {author} {\bibfnamefont {L.}~\bibnamefont
  {{Derome}}},\ }\href@noop {} {\bibfield  {journal} {\bibinfo  {journal}
  {\arxiv}\ } (\bibinfo {year} {2003}{\natexlab{a}})},\ \Eprint
  {https://arxiv.org/abs/astro-ph/0304528}  { arXiv:astro-ph/0304528
  [astro-ph]}\BibitemShut {NoStop}%
\bibitem [{\citenamefont {{Abdo}}\ \emph {et~al.}(2010)\citenamefont {{Abdo}}
  \emph {et~al.}}]{2010PhRvL.104j1101A}%
  \BibitemOpen
  \bibfield  {author} {\bibinfo {author} {\bibfnamefont {A.~A.}\ \bibnamefont
  {{Abdo}}} \emph {et~al.} (\bibinfo {collaboration} {Fermi-LAT}),\ }\href
  {https://doi.org/10.1103/PhysRevLett.104.101101} {\bibfield  {journal}
  {\bibinfo  {journal} {\prl}\ }\textbf {\bibinfo {volume} {104}},\ \bibinfo
  {eid} {101101} (\bibinfo {year} {2010})},\ \Eprint
  {https://arxiv.org/abs/1002.3603}  { arXiv:1002.3603
  [astro-ph.HE]}\BibitemShut {NoStop}%
\bibitem [{\citenamefont {{Ackermann}}\ \emph {et~al.}(2015)\citenamefont
  {{Ackermann}} \emph {et~al.}}]{2015ApJ...799...86A}%
  \BibitemOpen
  \bibfield  {author} {\bibinfo {author} {\bibfnamefont {M.}~\bibnamefont
  {{Ackermann}}} \emph {et~al.},\ }\href
  {https://doi.org/10.1088/0004-637X/799/1/86} {\bibfield  {journal} {\bibinfo
  {journal} {\apj}\ }\textbf {\bibinfo {volume} {799}},\ \bibinfo {eid} {86}
  (\bibinfo {year} {2015})},\ \Eprint {https://arxiv.org/abs/1410.3696}  {
  arXiv:1410.3696 [astro-ph.HE]}\BibitemShut {NoStop}%
\bibitem [{\citenamefont {{Arbey}}\ \emph
  {et~al.}(2020{\natexlab{b}})\citenamefont {{Arbey}}, \citenamefont
  {{Auffinger}},\ and\ \citenamefont {{Silk}}}]{2020PhRvD.101b3010A}%
  \BibitemOpen
  \bibfield  {author} {\bibinfo {author} {\bibfnamefont {A.}~\bibnamefont
  {{Arbey}}}, \bibinfo {author} {\bibfnamefont {J.}~\bibnamefont
  {{Auffinger}}},\ and\ \bibinfo {author} {\bibfnamefont {J.}~\bibnamefont
  {{Silk}}},\ }\href {https://doi.org/10.1103/PhysRevD.101.023010} {\bibfield
  {journal} {\bibinfo  {journal} {\prd}\ }\textbf {\bibinfo {volume} {101}},\
  \bibinfo {eid} {023010} (\bibinfo {year} {2020}{\natexlab{b}})},\ \Eprint
  {https://arxiv.org/abs/1906.04750}  { arXiv:1906.04750
  [astro-ph.CO]}\BibitemShut {NoStop}%
\bibitem [{\citenamefont {{Ray}}\ \emph {et~al.}(2021)\citenamefont {{Ray}},
  \citenamefont {{Laha}}, \citenamefont {{Mu{\~n}oz}},\ and\ \citenamefont
  {{Caputo}}}]{2021PhRvD.104b3516R}%
  \BibitemOpen
  \bibfield  {author} {\bibinfo {author} {\bibfnamefont {A.}~\bibnamefont
  {{Ray}}}, \bibinfo {author} {\bibfnamefont {R.}~\bibnamefont {{Laha}}},
  \bibinfo {author} {\bibfnamefont {J.~B.}\ \bibnamefont {{Mu{\~n}oz}}},\ and\
  \bibinfo {author} {\bibfnamefont {R.}~\bibnamefont {{Caputo}}},\ }\href
  {https://doi.org/10.1103/PhysRevD.104.023516} {\bibfield  {journal} {\bibinfo
   {journal} {\prd}\ }\textbf {\bibinfo {volume} {104}},\ \bibinfo {eid}
  {023516} (\bibinfo {year} {2021})},\ \Eprint
  {https://arxiv.org/abs/2102.06714}  { arXiv:2102.06714
  [astro-ph.CO]}\BibitemShut {NoStop}%
\bibitem [{\citenamefont {{Ghosh}}\ \emph {et~al.}(2021)\citenamefont
  {{Ghosh}}, \citenamefont {{Sachdeva}},\ and\ \citenamefont
  {{Singh}}}]{2021arXiv211003333G}%
  \BibitemOpen
  \bibfield  {author} {\bibinfo {author} {\bibfnamefont {D.}~\bibnamefont
  {{Ghosh}}}, \bibinfo {author} {\bibfnamefont {D.}~\bibnamefont
  {{Sachdeva}}},\ and\ \bibinfo {author} {\bibfnamefont {P.}~\bibnamefont
  {{Singh}}},\ }\href@noop {} {\bibfield  {journal} {\bibinfo  {journal}
  {\arxiv}\ } (\bibinfo {year} {2021})},\ \Eprint
  {https://arxiv.org/abs/2110.03333}  { arXiv:2110.03333
  [astro-ph.CO]}\BibitemShut {NoStop}%
\bibitem [{\citenamefont {{Ballesteros}}\ \emph {et~al.}(2020)\citenamefont
  {{Ballesteros}}, \citenamefont {{Coronado-Bl{\'a}zquez}},\ and\ \citenamefont
  {{Gaggero}}}]{2020PhLB..80835624B}%
  \BibitemOpen
  \bibfield  {author} {\bibinfo {author} {\bibfnamefont {G.}~\bibnamefont
  {{Ballesteros}}}, \bibinfo {author} {\bibfnamefont {J.}~\bibnamefont
  {{Coronado-Bl{\'a}zquez}}},\ and\ \bibinfo {author} {\bibfnamefont
  {D.}~\bibnamefont {{Gaggero}}},\ }\href
  {https://doi.org/10.1016/j.physletb.2020.135624} {\bibfield  {journal}
  {\bibinfo  {journal} {\plb}\ }\textbf {\bibinfo {volume} {808}},\ \bibinfo
  {eid} {135624} (\bibinfo {year} {2020})},\ \Eprint
  {https://arxiv.org/abs/1906.10113}  { arXiv:1906.10113
  [astro-ph.CO]}\BibitemShut {NoStop}%
\bibitem [{\citenamefont {{Chen}}\ \emph {et~al.}(2022)\citenamefont {{Chen}},
  \citenamefont {{Zhang}},\ and\ \citenamefont {{Long}}}]{2022PhRvD.105f3008C}%
  \BibitemOpen
  \bibfield  {author} {\bibinfo {author} {\bibfnamefont {S.}~\bibnamefont
  {{Chen}}}, \bibinfo {author} {\bibfnamefont {H.-H.}\ \bibnamefont
  {{Zhang}}},\ and\ \bibinfo {author} {\bibfnamefont {G.}~\bibnamefont
  {{Long}}},\ }\href {https://doi.org/10.1103/PhysRevD.105.063008} {\bibfield
  {journal} {\bibinfo  {journal} {\prd}\ }\textbf {\bibinfo {volume} {105}},\
  \bibinfo {eid} {063008} (\bibinfo {year} {2022})},\ \Eprint
  {https://arxiv.org/abs/2112.15463}  { arXiv:2112.15463
  [astro-ph.CO]}\BibitemShut {NoStop}%
\bibitem [{\citenamefont {{Iguaz}}\ \emph {et~al.}(2021)\citenamefont
  {{Iguaz}}, \citenamefont {{Serpico}},\ and\ \citenamefont
  {{Siegert}}}]{2021PhRvD.103j3025I}%
  \BibitemOpen
  \bibfield  {author} {\bibinfo {author} {\bibfnamefont {J.}~\bibnamefont
  {{Iguaz}}}, \bibinfo {author} {\bibfnamefont {P.~D.}\ \bibnamefont
  {{Serpico}}},\ and\ \bibinfo {author} {\bibfnamefont {T.}~\bibnamefont
  {{Siegert}}},\ }\href {https://doi.org/10.1103/PhysRevD.103.103025}
  {\bibfield  {journal} {\bibinfo  {journal} {\prd}\ }\textbf {\bibinfo
  {volume} {103}},\ \bibinfo {eid} {103025} (\bibinfo {year} {2021})},\ \Eprint
  {https://arxiv.org/abs/2104.03145}  { arXiv:2104.03145
  [astro-ph.CO]}\BibitemShut {NoStop}%
\bibitem [{\citenamefont {{Strong}}\ \emph {et~al.}(2004)\citenamefont
  {{Strong}}, \citenamefont {{Moskalenko}},\ and\ \citenamefont
  {{Reimer}}}]{2004ApJ...613..956S}%
  \BibitemOpen
  \bibfield  {author} {\bibinfo {author} {\bibfnamefont {A.~W.}\ \bibnamefont
  {{Strong}}}, \bibinfo {author} {\bibfnamefont {I.~V.}\ \bibnamefont
  {{Moskalenko}}},\ and\ \bibinfo {author} {\bibfnamefont {O.}~\bibnamefont
  {{Reimer}}},\ }\href {https://doi.org/10.1086/423196} {\bibfield  {journal}
  {\bibinfo  {journal} {\apj}\ }\textbf {\bibinfo {volume} {613}},\ \bibinfo
  {pages} {956} (\bibinfo {year} {2004})},\ \Eprint
  {https://arxiv.org/abs/astro-ph/0405441}  { arXiv:astro-ph/0405441
  [astro-ph]}\BibitemShut {NoStop}%
\bibitem [{\citenamefont {{Coogan}}\ \emph
  {et~al.}(2021{\natexlab{a}})\citenamefont {{Coogan}}, \citenamefont
  {{Morrison}},\ and\ \citenamefont {{Profumo}}}]{2021PhRvL.126q1101C}%
  \BibitemOpen
  \bibfield  {author} {\bibinfo {author} {\bibfnamefont {A.}~\bibnamefont
  {{Coogan}}}, \bibinfo {author} {\bibfnamefont {L.}~\bibnamefont
  {{Morrison}}},\ and\ \bibinfo {author} {\bibfnamefont {S.}~\bibnamefont
  {{Profumo}}},\ }\href {https://doi.org/10.1103/PhysRevLett.126.171101}
  {\bibfield  {journal} {\bibinfo  {journal} {\prl}\ }\textbf {\bibinfo
  {volume} {126}},\ \bibinfo {eid} {171101} (\bibinfo {year}
  {2021}{\natexlab{a}})},\ \Eprint {https://arxiv.org/abs/2010.04797}  {
  arXiv:2010.04797 [astro-ph.CO]}\BibitemShut {NoStop}%
\bibitem [{\citenamefont {{Auffinger}}(2022)}]{2022EPJC...82..384A}%
  \BibitemOpen
  \bibfield  {author} {\bibinfo {author} {\bibfnamefont {J.}~\bibnamefont
  {{Auffinger}}},\ }\href {https://doi.org/10.1140/epjc/s10052-022-10199-y}
  {\bibfield  {journal} {\bibinfo  {journal} {\epjc}\ }\textbf {\bibinfo
  {volume} {82}},\ \bibinfo {eid} {384} (\bibinfo {year} {2022})},\ \Eprint
  {https://arxiv.org/abs/2201.01265}  { arXiv:2201.01265
  [astro-ph.HE]}\BibitemShut {NoStop}%
\bibitem [{\citenamefont {{Mittal}}\ and\ \citenamefont
  {{Kulkarni}}(2022)}]{2022MNRAS.510.4992M}%
  \BibitemOpen
  \bibfield  {author} {\bibinfo {author} {\bibfnamefont {S.}~\bibnamefont
  {{Mittal}}}\ and\ \bibinfo {author} {\bibfnamefont {G.}~\bibnamefont
  {{Kulkarni}}},\ }\href {https://doi.org/10.1093/mnras/stac005} {\bibfield
  {journal} {\bibinfo  {journal} {\mnras}\ }\textbf {\bibinfo {volume} {510}},\
  \bibinfo {pages} {4992} (\bibinfo {year} {2022})},\ \Eprint
  {https://arxiv.org/abs/2110.11975}  { arXiv:2110.11975
  [astro-ph.CO]}\BibitemShut {NoStop}%
\bibitem [{\citenamefont {{Marfatia}}\ and\ \citenamefont
  {{Tseng}}(2021)}]{2021arXiv211214588M}%
  \BibitemOpen
  \bibfield  {author} {\bibinfo {author} {\bibfnamefont {D.}~\bibnamefont
  {{Marfatia}}}\ and\ \bibinfo {author} {\bibfnamefont {P.-Y.}\ \bibnamefont
  {{Tseng}}},\ }\href@noop {} {\bibfield  {journal} {\bibinfo  {journal}
  {\arxiv}\ } (\bibinfo {year} {2021})},\ \Eprint
  {https://arxiv.org/abs/2112.14588}  { arXiv:2112.14588 [hep-ph]}\BibitemShut
  {NoStop}%
\bibitem [{\citenamefont {{Agashe}}\ \emph {et~al.}(2022)\citenamefont
  {{Agashe}} \emph {et~al.}}]{2022arXiv220204653A}%
  \BibitemOpen
  \bibfield  {author} {\bibinfo {author} {\bibfnamefont {K.}~\bibnamefont
  {{Agashe}}} \emph {et~al.},\ }\href@noop {} {\bibfield  {journal} {\bibinfo
  {journal} {\arxiv}\ } (\bibinfo {year} {2022})},\ \Eprint
  {https://arxiv.org/abs/2202.04653}  { arXiv:2202.04653
  [astro-ph.CO]}\BibitemShut {NoStop}%
\bibitem [{\citenamefont {{Roth}}\ \emph {et~al.}(2021)\citenamefont {{Roth}},
  \citenamefont {{Krumholz}}, \citenamefont {{Crocker}},\ and\ \citenamefont
  {{Celli}}}]{2021Natur.597..341R}%
  \BibitemOpen
  \bibfield  {author} {\bibinfo {author} {\bibfnamefont {M.~A.}\ \bibnamefont
  {{Roth}}}, \bibinfo {author} {\bibfnamefont {M.~R.}\ \bibnamefont
  {{Krumholz}}}, \bibinfo {author} {\bibfnamefont {R.~M.}\ \bibnamefont
  {{Crocker}}},\ and\ \bibinfo {author} {\bibfnamefont {S.}~\bibnamefont
  {{Celli}}},\ }\href {https://doi.org/10.1038/s41586-021-03802-x} {\bibfield
  {journal} {\bibinfo  {journal} {\nat}\ }\textbf {\bibinfo {volume} {597}},\
  \bibinfo {pages} {341} (\bibinfo {year} {2021})},\ \Eprint
  {https://arxiv.org/abs/2109.07598}  { arXiv:2109.07598
  [astro-ph.HE]}\BibitemShut {NoStop}%
\bibitem [{\citenamefont {{Inoue}}\ \emph {et~al.}(2019)\citenamefont
  {{Inoue}}, \citenamefont {{Khangulyan}}, \citenamefont {{Inoue}},\ and\
  \citenamefont {{Doi}}}]{2019ApJ...880...40I}%
  \BibitemOpen
  \bibfield  {author} {\bibinfo {author} {\bibfnamefont {Y.}~\bibnamefont
  {{Inoue}}}, \bibinfo {author} {\bibfnamefont {D.}~\bibnamefont
  {{Khangulyan}}}, \bibinfo {author} {\bibfnamefont {S.}~\bibnamefont
  {{Inoue}}},\ and\ \bibinfo {author} {\bibfnamefont {A.}~\bibnamefont
  {{Doi}}},\ }\href {https://doi.org/10.3847/1538-4357/ab2715} {\bibfield
  {journal} {\bibinfo  {journal} {\apj}\ }\textbf {\bibinfo {volume} {880}},\
  \bibinfo {eid} {40} (\bibinfo {year} {2019})},\ \Eprint
  {https://arxiv.org/abs/1904.00554}  { arXiv:1904.00554
  [astro-ph.HE]}\BibitemShut {NoStop}%
\bibitem [{\citenamefont {{Kimura}}\ \emph {et~al.}(2021)\citenamefont
  {{Kimura}}, \citenamefont {{Murase}},\ and\ \citenamefont
  {{M{\'e}sz{\'a}ros}}}]{2021NatCo..12.5615K}%
  \BibitemOpen
  \bibfield  {author} {\bibinfo {author} {\bibfnamefont {S.~S.}\ \bibnamefont
  {{Kimura}}}, \bibinfo {author} {\bibfnamefont {K.}~\bibnamefont {{Murase}}},\
  and\ \bibinfo {author} {\bibfnamefont {P.}~\bibnamefont
  {{M{\'e}sz{\'a}ros}}},\ }\href {https://doi.org/10.1038/s41467-021-25111-7}
  {\bibfield  {journal} {\bibinfo  {journal} {\natcom}\ }\textbf {\bibinfo
  {volume} {12}},\ \bibinfo {eid} {5615} (\bibinfo {year} {2021})},\ \Eprint
  {https://arxiv.org/abs/2005.01934}  { arXiv:2005.01934
  [astro-ph.HE]}\BibitemShut {NoStop}%
\bibitem [{\citenamefont {{Davis}}\ \emph {et~al.}(1968)\citenamefont
  {{Davis}}, \citenamefont {{Harmer}},\ and\ \citenamefont
  {{Hoffman}}}]{1968PhRvL..20.1205D}%
  \BibitemOpen
  \bibfield  {author} {\bibinfo {author} {\bibfnamefont {R.}~\bibnamefont
  {{Davis}}}, \bibinfo {author} {\bibfnamefont {D.~S.}\ \bibnamefont
  {{Harmer}}},\ and\ \bibinfo {author} {\bibfnamefont {K.~C.}\ \bibnamefont
  {{Hoffman}}},\ }\href {https://doi.org/10.1103/PhysRevLett.20.1205}
  {\bibfield  {journal} {\bibinfo  {journal} {\prl}\ }\textbf {\bibinfo
  {volume} {20}},\ \bibinfo {pages} {1205} (\bibinfo {year}
  {1968})}\BibitemShut {NoStop}%
\bibitem [{\citenamefont {{Vainer}}(1978)}]{1978Ap.....14..185V}%
  \BibitemOpen
  \bibfield  {author} {\bibinfo {author} {\bibfnamefont {B.~V.}\ \bibnamefont
  {{Vainer}}},\ }\href {https://doi.org/10.1007/BF01006060} {\bibfield
  {journal} {\bibinfo  {journal} {Astrophysics}\ }\textbf {\bibinfo {volume}
  {14}},\ \bibinfo {pages} {185} (\bibinfo {year} {1978})},\ \bibinfo {note}
  {[Astrofizika \textbf{14}, 325 (1978)]}\BibitemShut {NoStop}%
\bibitem [{\citenamefont {{Halzen}}\ \emph {et~al.}(1995)\citenamefont
  {{Halzen}}, \citenamefont {{Keszthelyi}},\ and\ \citenamefont
  {{Zas}}}]{1995PhRvD..52.3239H}%
  \BibitemOpen
  \bibfield  {author} {\bibinfo {author} {\bibfnamefont {F.}~\bibnamefont
  {{Halzen}}}, \bibinfo {author} {\bibfnamefont {B.}~\bibnamefont
  {{Keszthelyi}}},\ and\ \bibinfo {author} {\bibfnamefont {E.}~\bibnamefont
  {{Zas}}},\ }\href {https://doi.org/10.1103/PhysRevD.52.3239} {\bibfield
  {journal} {\bibinfo  {journal} {\prd}\ }\textbf {\bibinfo {volume} {52}},\
  \bibinfo {pages} {3239} (\bibinfo {year} {1995})},\ \Eprint
  {https://arxiv.org/abs/hep-ph/9502268}  { arXiv:hep-ph/9502268
  [hep-ph]}\BibitemShut {NoStop}%
\bibitem [{\citenamefont {{Gaisser}}\ \emph {et~al.}(1988)\citenamefont
  {{Gaisser}}, \citenamefont {{Stanev}},\ and\ \citenamefont
  {{Barr}}}]{1988PhRvD..38...85G}%
  \BibitemOpen
  \bibfield  {author} {\bibinfo {author} {\bibfnamefont {T.~K.}\ \bibnamefont
  {{Gaisser}}}, \bibinfo {author} {\bibfnamefont {T.}~\bibnamefont
  {{Stanev}}},\ and\ \bibinfo {author} {\bibfnamefont {G.}~\bibnamefont
  {{Barr}}},\ }\href {https://doi.org/10.1103/PhysRevD.38.85} {\bibfield
  {journal} {\bibinfo  {journal} {\prd}\ }\textbf {\bibinfo {volume} {38}},\
  \bibinfo {pages} {85} (\bibinfo {year} {1988})}\BibitemShut {NoStop}%
\bibitem [{\citenamefont {{Bugaev}}\ and\ \citenamefont
  {{Konishchev}}(2002{\natexlab{a}})}]{2002PhRvD..65l3005B}%
  \BibitemOpen
  \bibfield  {author} {\bibinfo {author} {\bibfnamefont {E.~V.}\ \bibnamefont
  {{Bugaev}}}\ and\ \bibinfo {author} {\bibfnamefont {K.~V.}\ \bibnamefont
  {{Konishchev}}},\ }\href {https://doi.org/10.1103/PhysRevD.65.123005}
  {\bibfield  {journal} {\bibinfo  {journal} {\prd}\ }\textbf {\bibinfo
  {volume} {65}},\ \bibinfo {eid} {123005} (\bibinfo {year}
  {2002}{\natexlab{a}})},\ \Eprint {https://arxiv.org/abs/astro-ph/0005295}  {
  arXiv:astro-ph/0005295 [astro-ph]}\BibitemShut {NoStop}%
\bibitem [{\citenamefont {{Bugaev}}\ and\ \citenamefont
  {{Konishchev}}(2002{\natexlab{b}})}]{2002PhRvD..66h4004B}%
  \BibitemOpen
  \bibfield  {author} {\bibinfo {author} {\bibfnamefont {E.~V.}\ \bibnamefont
  {{Bugaev}}}\ and\ \bibinfo {author} {\bibfnamefont {K.~V.}\ \bibnamefont
  {{Konishchev}}},\ }\href {https://doi.org/10.1103/PhysRevD.66.084004}
  {\bibfield  {journal} {\bibinfo  {journal} {\prd}\ }\textbf {\bibinfo
  {volume} {66}},\ \bibinfo {eid} {084004} (\bibinfo {year}
  {2002}{\natexlab{b}})},\ \Eprint {https://arxiv.org/abs/astro-ph/0206082}  {
  arXiv:astro-ph/0206082 [astro-ph]}\BibitemShut {NoStop}%
\bibitem [{\citenamefont {{Bugaev}}\ and\ \citenamefont
  {{Konishchev}}(2004)}]{2004astro.ph.12640B}%
  \BibitemOpen
  \bibfield  {author} {\bibinfo {author} {\bibfnamefont {E.~V.}\ \bibnamefont
  {{Bugaev}}}\ and\ \bibinfo {author} {\bibfnamefont {K.~V.}\ \bibnamefont
  {{Konishchev}}},\ }\href@noop {} {\bibfield  {journal} {\bibinfo  {journal}
  {arXiv e-prints}\ } (\bibinfo {year} {2004})},\ \Eprint
  {https://arxiv.org/abs/astro-ph/0412640}  { astro-ph/0412640
  [astro-ph]}\BibitemShut {NoStop}%
\bibitem [{\citenamefont {{Malek}}(2003)}]{2003PhRvL..90f1101M}%
  \BibitemOpen
  \bibfield  {author} {\bibinfo {author} {\bibfnamefont {M.~o.}\ \bibnamefont
  {{Malek}}} (\bibinfo {collaboration} {Super-Kamiokande}),\ }\href
  {https://doi.org/10.1103/PhysRevLett.90.061101} {\bibfield  {journal}
  {\bibinfo  {journal} {\prl}\ }\textbf {\bibinfo {volume} {90}},\ \bibinfo
  {eid} {061101} (\bibinfo {year} {2003})},\ \Eprint
  {https://arxiv.org/abs/hep-ex/0209028}  { arXiv:hep-ex/0209028
  [hep-ex]}\BibitemShut {NoStop}%
\bibitem [{\citenamefont {{Lunardini}}\ and\ \citenamefont
  {{Perez-Gonzalez}}(2020)}]{2020JCAP...08..014L}%
  \BibitemOpen
  \bibfield  {author} {\bibinfo {author} {\bibfnamefont {C.}~\bibnamefont
  {{Lunardini}}}\ and\ \bibinfo {author} {\bibfnamefont {Y.~F.}\ \bibnamefont
  {{Perez-Gonzalez}}},\ }\href {https://doi.org/10.1088/1475-7516/2020/08/014}
  {\bibfield  {journal} {\bibinfo  {journal} {\jcap}\ }\textbf {\bibinfo
  {volume} {2020}},\ \bibinfo {eid} {014} (\bibinfo {year} {2020})},\ \Eprint
  {https://arxiv.org/abs/1910.07864}  { arXiv:1910.07864 [hep-ph]}\BibitemShut
  {NoStop}%
\bibitem [{\citenamefont {{Bambeck}}\ and\ \citenamefont
  {{Hiscock}}(2005)}]{2005CQGra..22.4247B}%
  \BibitemOpen
  \bibfield  {author} {\bibinfo {author} {\bibfnamefont {D.}~\bibnamefont
  {{Bambeck}}}\ and\ \bibinfo {author} {\bibfnamefont {W.~A.}\ \bibnamefont
  {{Hiscock}}},\ }\href {https://doi.org/10.1088/0264-9381/22/20/006}
  {\bibfield  {journal} {\bibinfo  {journal} {\CQG}\ }\textbf {\bibinfo
  {volume} {22}},\ \bibinfo {pages} {4247} (\bibinfo {year} {2005})},\ \Eprint
  {https://arxiv.org/abs/gr-qc/0506050}  { arXiv:gr-qc/0506050
  [gr-qc]}\BibitemShut {NoStop}%
\bibitem [{\citenamefont {{Pontecorvo}}(1958)}]{1958JETP....6..429P}%
  \BibitemOpen
  \bibfield  {author} {\bibinfo {author} {\bibfnamefont {B.}~\bibnamefont
  {{Pontecorvo}}},\ }\href@noop {} {\bibfield  {journal} {\bibinfo  {journal}
  {\JETP}\ }\textbf {\bibinfo {volume} {6}},\ \bibinfo {pages} {429} (\bibinfo
  {year} {1958})},\ \bibinfo {note} {[\ZhETF~\textbf{33}, 549
  (1957)]}\BibitemShut {NoStop}%
\bibitem [{\citenamefont {{Pontecorvo}}(1968)}]{1968JETP...26..984P}%
  \BibitemOpen
  \bibfield  {author} {\bibinfo {author} {\bibfnamefont {B.}~\bibnamefont
  {{Pontecorvo}}},\ }\href@noop {} {\bibfield  {journal} {\bibinfo  {journal}
  {\JETP}\ }\textbf {\bibinfo {volume} {26}},\ \bibinfo {pages} {984} (\bibinfo
  {year} {1968})},\ \bibinfo {note} {[\ZhETF~\textbf{53}, 1717
  (1967)]}\BibitemShut {NoStop}%
\bibitem [{\citenamefont {{Dasgupta}}\ \emph {et~al.}(2020)\citenamefont
  {{Dasgupta}}, \citenamefont {{Laha}},\ and\ \citenamefont
  {{Ray}}}]{2020PhRvL.125j1101D}%
  \BibitemOpen
  \bibfield  {author} {\bibinfo {author} {\bibfnamefont {B.}~\bibnamefont
  {{Dasgupta}}}, \bibinfo {author} {\bibfnamefont {R.}~\bibnamefont {{Laha}}},\
  and\ \bibinfo {author} {\bibfnamefont {A.}~\bibnamefont {{Ray}}},\ }\href
  {https://doi.org/10.1103/PhysRevLett.125.101101} {\bibfield  {journal}
  {\bibinfo  {journal} {\prl}\ }\textbf {\bibinfo {volume} {125}},\ \bibinfo
  {eid} {101101} (\bibinfo {year} {2020})},\ \Eprint
  {https://arxiv.org/abs/1912.01014}  { arXiv:1912.01014 [hep-ph]}\BibitemShut
  {NoStop}%
\bibitem [{\citenamefont {{Wang}}\ \emph {et~al.}(2021)\citenamefont {{Wang}},
  \citenamefont {{Xia}}, \citenamefont {{Zhang}}, \citenamefont {{Zhou}},\ and\
  \citenamefont {{Chang}}}]{2021PhRvD.103d3010W}%
  \BibitemOpen
  \bibfield  {author} {\bibinfo {author} {\bibfnamefont {S.}~\bibnamefont
  {{Wang}}}, \bibinfo {author} {\bibfnamefont {D.-M.}\ \bibnamefont {{Xia}}},
  \bibinfo {author} {\bibfnamefont {X.}~\bibnamefont {{Zhang}}}, \bibinfo
  {author} {\bibfnamefont {S.}~\bibnamefont {{Zhou}}},\ and\ \bibinfo {author}
  {\bibfnamefont {Z.}~\bibnamefont {{Chang}}},\ }\href
  {https://doi.org/10.1103/PhysRevD.103.043010} {\bibfield  {journal} {\bibinfo
   {journal} {\prd}\ }\textbf {\bibinfo {volume} {103}},\ \bibinfo {eid}
  {043010} (\bibinfo {year} {2021})},\ \Eprint
  {https://arxiv.org/abs/2010.16053}  { arXiv:2010.16053 [hep-ph]}\BibitemShut
  {NoStop}%
\bibitem [{\citenamefont {{Calabrese}}\ \emph
  {et~al.}(2022{\natexlab{a}})\citenamefont {{Calabrese}}, \citenamefont
  {{Fiorillo}}, \citenamefont {{Miele}}, \citenamefont {{Morisi}},\ and\
  \citenamefont {{Palazzo}}}]{Calabrese}%
  \BibitemOpen
  \bibfield  {author} {\bibinfo {author} {\bibfnamefont {R.}~\bibnamefont
  {{Calabrese}}}, \bibinfo {author} {\bibfnamefont {D.~F.~G.}\ \bibnamefont
  {{Fiorillo}}}, \bibinfo {author} {\bibfnamefont {G.}~\bibnamefont {{Miele}}},
  \bibinfo {author} {\bibfnamefont {S.}~\bibnamefont {{Morisi}}},\ and\
  \bibinfo {author} {\bibfnamefont {A.}~\bibnamefont {{Palazzo}}},\ }\href
  {https://doi.org/10.1016/j.physletb.2022.137050} {\bibfield  {journal}
  {\bibinfo  {journal} {\plb}\ }\textbf {\bibinfo {volume} {829}},\ \bibinfo
  {pages} {137050} (\bibinfo {year} {2022}{\natexlab{a}})},\ \Eprint
  {https://arxiv.org/abs/2106.02492}  { arXiv:2106.02492 [hep-ph]}\BibitemShut
  {NoStop}%
\bibitem [{\citenamefont {{De Romeri}}\ \emph {et~al.}(2021)\citenamefont {{De
  Romeri}}, \citenamefont {{Mart{\'\i}nez-Mirav{\'e}}},\ and\ \citenamefont
  {{T{\'o}rtola}}}]{2021JCAP...10..051D}%
  \BibitemOpen
  \bibfield  {author} {\bibinfo {author} {\bibfnamefont {V.}~\bibnamefont {{De
  Romeri}}}, \bibinfo {author} {\bibfnamefont {P.}~\bibnamefont
  {{Mart{\'\i}nez-Mirav{\'e}}}},\ and\ \bibinfo {author} {\bibfnamefont
  {M.}~\bibnamefont {{T{\'o}rtola}}},\ }\href
  {https://doi.org/10.1088/1475-7516/2021/10/051} {\bibfield  {journal}
  {\bibinfo  {journal} {\jcap}\ }\textbf {\bibinfo {volume} {2021}},\ \bibinfo
  {eid} {051} (\bibinfo {year} {2021})},\ \Eprint
  {https://arxiv.org/abs/2106.05013}  { arXiv:2106.05013 [hep-ph]}\BibitemShut
  {NoStop}%
\bibitem [{\citenamefont {{Chao}}\ \emph {et~al.}(2021)\citenamefont {{Chao}},
  \citenamefont {{Li}},\ and\ \citenamefont {{Liao}}}]{2021arXiv210805608C}%
  \BibitemOpen
  \bibfield  {author} {\bibinfo {author} {\bibfnamefont {W.}~\bibnamefont
  {{Chao}}}, \bibinfo {author} {\bibfnamefont {T.}~\bibnamefont {{Li}}},\ and\
  \bibinfo {author} {\bibfnamefont {J.}~\bibnamefont {{Liao}}},\ }\href@noop {}
  {\bibfield  {journal} {\bibinfo  {journal} {\arxiv}\ } (\bibinfo {year}
  {2021})},\ \Eprint {https://arxiv.org/abs/2108.05608}  { arXiv:2108.05608
  [hep-ph]}\BibitemShut {NoStop}%
\bibitem [{\citenamefont {{Bisnovatyi-Kogan}}\ and\ \citenamefont
  {{Rudenko}}(2004)}]{2004CQGra..21.3347B}%
  \BibitemOpen
  \bibfield  {author} {\bibinfo {author} {\bibfnamefont {G.~S.}\ \bibnamefont
  {{Bisnovatyi-Kogan}}}\ and\ \bibinfo {author} {\bibfnamefont {V.~N.}\
  \bibnamefont {{Rudenko}}},\ }\href
  {https://doi.org/10.1088/0264-9381/21/14/001} {\bibfield  {journal} {\bibinfo
   {journal} {\CQG}\ }\textbf {\bibinfo {volume} {21}},\ \bibinfo {pages}
  {3347} (\bibinfo {year} {2004})},\ \Eprint
  {https://arxiv.org/abs/gr-qc/0406089}  { arXiv:gr-qc/0406089
  [gr-qc]}\BibitemShut {NoStop}%
\bibitem [{\citenamefont {{Dolgov}}\ and\ \citenamefont
  {{Ejlli}}(2011)}]{2011PhRvD..84b4028D}%
  \BibitemOpen
  \bibfield  {author} {\bibinfo {author} {\bibfnamefont {A.~D.}\ \bibnamefont
  {{Dolgov}}}\ and\ \bibinfo {author} {\bibfnamefont {D.}~\bibnamefont
  {{Ejlli}}},\ }\href {https://doi.org/10.1103/PhysRevD.84.024028} {\bibfield
  {journal} {\bibinfo  {journal} {\prd}\ }\textbf {\bibinfo {volume} {84}},\
  \bibinfo {eid} {024028} (\bibinfo {year} {2011})},\ \Eprint
  {https://arxiv.org/abs/1105.2303}  { arXiv:1105.2303
  [astro-ph.CO]}\BibitemShut {NoStop}%
\bibitem [{\citenamefont {{Garc{\'\i}a-Bellido}}\ \emph
  {et~al.}(1996)\citenamefont {{Garc{\'\i}a-Bellido}}, \citenamefont
  {{Linde}},\ and\ \citenamefont {{Wands}}}]{1996PhRvD..54.6040G}%
  \BibitemOpen
  \bibfield  {author} {\bibinfo {author} {\bibfnamefont {J.}~\bibnamefont
  {{Garc{\'\i}a-Bellido}}}, \bibinfo {author} {\bibfnamefont {A.}~\bibnamefont
  {{Linde}}},\ and\ \bibinfo {author} {\bibfnamefont {D.}~\bibnamefont
  {{Wands}}},\ }\href {https://doi.org/10.1103/PhysRevD.54.6040} {\bibfield
  {journal} {\bibinfo  {journal} {\prd}\ }\textbf {\bibinfo {volume} {54}},\
  \bibinfo {pages} {6040} (\bibinfo {year} {1996})},\ \Eprint
  {https://arxiv.org/abs/astro-ph/9605094}  { arXiv:astro-ph/9605094
  [astro-ph]}\BibitemShut {NoStop}%
\bibitem [{\citenamefont {{Dolgov}}\ and\ \citenamefont
  {{Ejlli}}(2013)}]{2013PhRvD..87j4007D}%
  \BibitemOpen
  \bibfield  {author} {\bibinfo {author} {\bibfnamefont {A.~D.}\ \bibnamefont
  {{Dolgov}}}\ and\ \bibinfo {author} {\bibfnamefont {D.}~\bibnamefont
  {{Ejlli}}},\ }\href {https://doi.org/10.1103/PhysRevD.87.104007} {\bibfield
  {journal} {\bibinfo  {journal} {\prd}\ }\textbf {\bibinfo {volume} {87}},\
  \bibinfo {eid} {104007} (\bibinfo {year} {2013})},\ \Eprint
  {https://arxiv.org/abs/1303.1556}  { arXiv:1303.1556 [gr-qc]}\BibitemShut
  {NoStop}%
\bibitem [{\citenamefont {{Anantua}}\ \emph {et~al.}(2009)\citenamefont
  {{Anantua}}, \citenamefont {{Easther}},\ and\ \citenamefont
  {{Giblin}}}]{2009PhRvL.103k1303A}%
  \BibitemOpen
  \bibfield  {author} {\bibinfo {author} {\bibfnamefont {R.}~\bibnamefont
  {{Anantua}}}, \bibinfo {author} {\bibfnamefont {R.}~\bibnamefont
  {{Easther}}},\ and\ \bibinfo {author} {\bibfnamefont {J.}~\bibnamefont
  {{Giblin}}, \bibfnamefont {John~T.}},\ }\href
  {https://doi.org/10.1103/PhysRevLett.103.111303} {\bibfield  {journal}
  {\bibinfo  {journal} {\prl}\ }\textbf {\bibinfo {volume} {103}},\ \bibinfo
  {eid} {111303} (\bibinfo {year} {2009})},\ \Eprint
  {https://arxiv.org/abs/0812.0825}  { arXiv:0812.0825 [astro-ph]}\BibitemShut
  {NoStop}%
\bibitem [{\citenamefont {{Dong}}\ \emph {et~al.}(2016)\citenamefont {{Dong}},
  \citenamefont {{Kinney}},\ and\ \citenamefont
  {{Stojkovic}}}]{2016JCAP...10..034D}%
  \BibitemOpen
  \bibfield  {author} {\bibinfo {author} {\bibfnamefont {R.}~\bibnamefont
  {{Dong}}}, \bibinfo {author} {\bibfnamefont {W.~H.}\ \bibnamefont
  {{Kinney}}},\ and\ \bibinfo {author} {\bibfnamefont {D.}~\bibnamefont
  {{Stojkovic}}},\ }\href {https://doi.org/10.1088/1475-7516/2016/10/034}
  {\bibfield  {journal} {\bibinfo  {journal} {\jcap}\ }\textbf {\bibinfo
  {volume} {2016}},\ \bibinfo {eid} {034} (\bibinfo {year} {2016})},\ \Eprint
  {https://arxiv.org/abs/1511.05642}  { arXiv:1511.05642
  [astro-ph.CO]}\BibitemShut {NoStop}%
\bibitem [{\citenamefont {{Ejlli}}\ \emph {et~al.}(2019)\citenamefont
  {{Ejlli}}, \citenamefont {{Ejlli}}, \citenamefont {{Cruise}}, \citenamefont
  {{Pisano}},\ and\ \citenamefont {{Grote}}}]{2019EPJC...79.1032E}%
  \BibitemOpen
  \bibfield  {author} {\bibinfo {author} {\bibfnamefont {A.}~\bibnamefont
  {{Ejlli}}}, \bibinfo {author} {\bibfnamefont {D.}~\bibnamefont {{Ejlli}}},
  \bibinfo {author} {\bibfnamefont {A.~M.}\ \bibnamefont {{Cruise}}}, \bibinfo
  {author} {\bibfnamefont {G.}~\bibnamefont {{Pisano}}},\ and\ \bibinfo
  {author} {\bibfnamefont {H.}~\bibnamefont {{Grote}}},\ }\href
  {https://doi.org/10.1140/epjc/s10052-019-7542-5} {\bibfield  {journal}
  {\bibinfo  {journal} {\epjc}\ }\textbf {\bibinfo {volume} {79}},\ \bibinfo
  {eid} {1032} (\bibinfo {year} {2019})},\ \Eprint
  {https://arxiv.org/abs/1908.00232}  { arXiv:1908.00232 [gr-qc]}\BibitemShut
  {NoStop}%
\bibitem [{\citenamefont {{Masina}}(2020)}]{2020EPJP..135..552M}%
  \BibitemOpen
  \bibfield  {author} {\bibinfo {author} {\bibfnamefont {I.}~\bibnamefont
  {{Masina}}},\ }\href {https://doi.org/10.1140/epjp/s13360-020-00564-9}
  {\bibfield  {journal} {\bibinfo  {journal} {\epjp}\ }\textbf {\bibinfo
  {volume} {135}},\ \bibinfo {eid} {552} (\bibinfo {year} {2020})},\ \Eprint
  {https://arxiv.org/abs/2004.04740}  { arXiv:2004.04740 [hep-ph]}\BibitemShut
  {NoStop}%
\bibitem [{\citenamefont {{Masina}}(2021)}]{2021GrCo...27..315M}%
  \BibitemOpen
  \bibfield  {author} {\bibinfo {author} {\bibfnamefont {I.}~\bibnamefont
  {{Masina}}},\ }\href {https://doi.org/10.1134/S0202289321040101} {\bibfield
  {journal} {\bibinfo  {journal} {Gravit.~Cosmology}\ }\textbf {\bibinfo
  {volume} {27}},\ \bibinfo {pages} {315} (\bibinfo {year} {2021})},\ \Eprint
  {https://arxiv.org/abs/2103.13825}  { arXiv:2103.13825 [gr-qc]}\BibitemShut
  {NoStop}%
\bibitem [{\citenamefont {{Einasto}}(1965)}]{Einasto}%
  \BibitemOpen
  \bibfield  {author} {\bibinfo {author} {\bibfnamefont {J.}~\bibnamefont
  {{Einasto}}},\ }\href@noop {} {\bibfield  {journal} {\bibinfo  {journal}
  {Trudy Inst.~Astrofiz.~Alma-Ata}\ }\textbf {\bibinfo {volume} {5}},\ \bibinfo
  {pages} {87} (\bibinfo {year} {1965})}\BibitemShut {NoStop}%
\bibitem [{\citenamefont {{Navarro}}\ \emph {et~al.}(1997)\citenamefont
  {{Navarro}}, \citenamefont {{Frenk}},\ and\ \citenamefont
  {{White}}}]{1997ApJ...490..493N}%
  \BibitemOpen
  \bibfield  {author} {\bibinfo {author} {\bibfnamefont {J.~F.}\ \bibnamefont
  {{Navarro}}}, \bibinfo {author} {\bibfnamefont {C.~S.}\ \bibnamefont
  {{Frenk}}},\ and\ \bibinfo {author} {\bibfnamefont {S.~D.~M.}\ \bibnamefont
  {{White}}},\ }\href {https://doi.org/10.1086/304888} {\bibfield  {journal}
  {\bibinfo  {journal} {\apj}\ }\textbf {\bibinfo {volume} {490}},\ \bibinfo
  {pages} {493} (\bibinfo {year} {1997})},\ \Eprint
  {https://arxiv.org/abs/astro-ph/9611107}  { arXiv:astro-ph/9611107
  [astro-ph]}\BibitemShut {NoStop}%
\bibitem [{\citenamefont {{Wright}}(1996)}]{1996ApJ...459..487W}%
  \BibitemOpen
  \bibfield  {author} {\bibinfo {author} {\bibfnamefont {E.~L.}\ \bibnamefont
  {{Wright}}},\ }\href {https://doi.org/10.1086/176910} {\bibfield  {journal}
  {\bibinfo  {journal} {\apj}\ }\textbf {\bibinfo {volume} {459}},\ \bibinfo
  {pages} {487} (\bibinfo {year} {1996})},\ \Eprint
  {https://arxiv.org/abs/astro-ph/9509074}  { arXiv:astro-ph/9509074
  [astro-ph]}\BibitemShut {NoStop}%
\bibitem [{\citenamefont {{Hunter}}\ \emph {et~al.}(1997)\citenamefont
  {{Hunter}} \emph {et~al.}}]{1997ApJ...481..205H}%
  \BibitemOpen
  \bibfield  {author} {\bibinfo {author} {\bibfnamefont {S.~D.}\ \bibnamefont
  {{Hunter}}} \emph {et~al.} (\bibinfo {collaboration} {EGRET}),\ }\href
  {https://doi.org/10.1086/304012} {\bibfield  {journal} {\bibinfo  {journal}
  {\apj}\ }\textbf {\bibinfo {volume} {481}},\ \bibinfo {pages} {205} (\bibinfo
  {year} {1997})}\BibitemShut {NoStop}%
\bibitem [{\citenamefont {{Cline}}(1998)}]{1998ApJ...501L...1C}%
  \BibitemOpen
  \bibfield  {author} {\bibinfo {author} {\bibfnamefont {D.~B.}\ \bibnamefont
  {{Cline}}},\ }\href {https://doi.org/10.1086/311433} {\bibfield  {journal}
  {\bibinfo  {journal} {\apjl}\ }\textbf {\bibinfo {volume} {501}},\ \bibinfo
  {pages} {L1} (\bibinfo {year} {1998})}\BibitemShut {NoStop}%
\bibitem [{\citenamefont {{Berteaud}}\ \emph {et~al.}(2022)\citenamefont
  {{Berteaud}}, \citenamefont {{Calore}}, \citenamefont {{Iguaz}},
  \citenamefont {{Serpico}},\ and\ \citenamefont
  {{Siegert}}}]{2022arXiv220207483B}%
  \BibitemOpen
  \bibfield  {author} {\bibinfo {author} {\bibfnamefont {J.}~\bibnamefont
  {{Berteaud}}}, \bibinfo {author} {\bibfnamefont {F.}~\bibnamefont
  {{Calore}}}, \bibinfo {author} {\bibfnamefont {J.}~\bibnamefont {{Iguaz}}},
  \bibinfo {author} {\bibfnamefont {P.~D.}\ \bibnamefont {{Serpico}}},\ and\
  \bibinfo {author} {\bibfnamefont {T.}~\bibnamefont {{Siegert}}},\ }\href@noop
  {} {\bibfield  {journal} {\bibinfo  {journal} {\arxiv}\ } (\bibinfo {year}
  {2022})},\ \Eprint {https://arxiv.org/abs/2202.07483}  { arXiv:2202.07483
  [astro-ph.HE]}\BibitemShut {NoStop}%
\bibitem [{\citenamefont {{Bambi}}\ \emph {et~al.}(2008)\citenamefont
  {{Bambi}}, \citenamefont {{Dolgov}},\ and\ \citenamefont
  {{Petrov}}}]{2008PhLB..670..174B}%
  \BibitemOpen
  \bibfield  {author} {\bibinfo {author} {\bibfnamefont {C.}~\bibnamefont
  {{Bambi}}}, \bibinfo {author} {\bibfnamefont {A.~D.}\ \bibnamefont
  {{Dolgov}}},\ and\ \bibinfo {author} {\bibfnamefont {A.~A.}\ \bibnamefont
  {{Petrov}}},\ }\href {https://doi.org/10.1016/j.physletb.2008.10.057}
  {\bibfield  {journal} {\bibinfo  {journal} {\plb}\ }\textbf {\bibinfo
  {volume} {670}},\ \bibinfo {pages} {174} (\bibinfo {year} {2008})},\ \Eprint
  {https://arxiv.org/abs/0801.2786}  { arXiv:0801.2786 [astro-ph]}\BibitemShut
  {NoStop}%
\bibitem [{\citenamefont {{Lehoucq}}\ \emph {et~al.}(2009)\citenamefont
  {{Lehoucq}}, \citenamefont {{Cass{\'e}}}, \citenamefont {{Casandjian}},\ and\
  \citenamefont {{Grenier}}}]{2009A&A...502...37L}%
  \BibitemOpen
  \bibfield  {author} {\bibinfo {author} {\bibfnamefont {R.}~\bibnamefont
  {{Lehoucq}}}, \bibinfo {author} {\bibfnamefont {M.}~\bibnamefont
  {{Cass{\'e}}}}, \bibinfo {author} {\bibfnamefont {J.~M.}\ \bibnamefont
  {{Casandjian}}},\ and\ \bibinfo {author} {\bibfnamefont {I.}~\bibnamefont
  {{Grenier}}},\ }\href {https://doi.org/10.1051/0004-6361/200911961}
  {\bibfield  {journal} {\bibinfo  {journal} {\aap}\ }\textbf {\bibinfo
  {volume} {502}},\ \bibinfo {pages} {37} (\bibinfo {year} {2009})},\ \Eprint
  {https://arxiv.org/abs/0906.1648}  { arXiv:0906.1648
  [astro-ph.HE]}\BibitemShut {NoStop}%
\bibitem [{\citenamefont {{Carr}}\ \emph {et~al.}(2016)\citenamefont {{Carr}},
  \citenamefont {{Kohri}}, \citenamefont {{Sendouda}},\ and\ \citenamefont
  {{Yokoyama}}}]{2016PhRvD..94d4029C}%
  \BibitemOpen
  \bibfield  {author} {\bibinfo {author} {\bibfnamefont {B.~J.}\ \bibnamefont
  {{Carr}}}, \bibinfo {author} {\bibfnamefont {K.}~\bibnamefont {{Kohri}}},
  \bibinfo {author} {\bibfnamefont {Y.}~\bibnamefont {{Sendouda}}},\ and\
  \bibinfo {author} {\bibfnamefont {J.}~\bibnamefont {{Yokoyama}}},\ }\href
  {https://doi.org/10.1103/PhysRevD.94.044029} {\bibfield  {journal} {\bibinfo
  {journal} {\prd}\ }\textbf {\bibinfo {volume} {94}},\ \bibinfo {eid} {044029}
  (\bibinfo {year} {2016})},\ \Eprint {https://arxiv.org/abs/1604.05349}  {
  arXiv:1604.05349 [astro-ph.CO]}\BibitemShut {NoStop}%
\bibitem [{\citenamefont {{Coogan}}\ \emph
  {et~al.}(2021{\natexlab{b}})\citenamefont {{Coogan}}, \citenamefont
  {{Moiseev}}, \citenamefont {{Morrison}},\ and\ \citenamefont
  {{Profumo}}}]{2021arXiv210110370C}%
  \BibitemOpen
  \bibfield  {author} {\bibinfo {author} {\bibfnamefont {A.}~\bibnamefont
  {{Coogan}}}, \bibinfo {author} {\bibfnamefont {A.}~\bibnamefont {{Moiseev}}},
  \bibinfo {author} {\bibfnamefont {L.}~\bibnamefont {{Morrison}}},\ and\
  \bibinfo {author} {\bibfnamefont {S.}~\bibnamefont {{Profumo}}},\ }\href@noop
  {} {\bibfield  {journal} {\bibinfo  {journal} {\arxiv}\ } (\bibinfo {year}
  {2021}{\natexlab{b}})},\ \Eprint {https://arxiv.org/abs/2101.10370}  {
  arXiv:2101.10370 [astro-ph.HE]}\BibitemShut {NoStop}%
\bibitem [{\citenamefont {{Siegert}}\ \emph
  {et~al.}(2022{\natexlab{a}})\citenamefont {{Siegert}}, \citenamefont
  {{Berteaud}}, \citenamefont {{Calore}}, \citenamefont {{Serpico}},\ and\
  \citenamefont {{Weinberger}}}]{2022A&A...660A.130S}%
  \BibitemOpen
  \bibfield  {author} {\bibinfo {author} {\bibfnamefont {T.}~\bibnamefont
  {{Siegert}}}, \bibinfo {author} {\bibfnamefont {J.}~\bibnamefont
  {{Berteaud}}}, \bibinfo {author} {\bibfnamefont {F.}~\bibnamefont
  {{Calore}}}, \bibinfo {author} {\bibfnamefont {P.~D.}\ \bibnamefont
  {{Serpico}}},\ and\ \bibinfo {author} {\bibfnamefont {C.}~\bibnamefont
  {{Weinberger}}},\ }\href {https://doi.org/10.1051/0004-6361/202142639}
  {\bibfield  {journal} {\bibinfo  {journal} {\aap}\ }\textbf {\bibinfo
  {volume} {660}},\ \bibinfo {eid} {A130} (\bibinfo {year}
  {2022}{\natexlab{a}})},\ \Eprint {https://arxiv.org/abs/2202.04574}  {
  arXiv:2202.04574 [astro-ph.HE]}\BibitemShut {NoStop}%
\bibitem [{\citenamefont {{Laha}}\ \emph {et~al.}(2020)\citenamefont {{Laha}},
  \citenamefont {{Mu{\~n}oz}},\ and\ \citenamefont
  {{Slatyer}}}]{2020PhRvD.101l3514L}%
  \BibitemOpen
  \bibfield  {author} {\bibinfo {author} {\bibfnamefont {R.}~\bibnamefont
  {{Laha}}}, \bibinfo {author} {\bibfnamefont {J.~B.}\ \bibnamefont
  {{Mu{\~n}oz}}},\ and\ \bibinfo {author} {\bibfnamefont {T.~R.}\ \bibnamefont
  {{Slatyer}}},\ }\href {https://doi.org/10.1103/PhysRevD.101.123514}
  {\bibfield  {journal} {\bibinfo  {journal} {\prd}\ }\textbf {\bibinfo
  {volume} {101}},\ \bibinfo {eid} {123514} (\bibinfo {year} {2020})},\ \Eprint
  {https://arxiv.org/abs/2004.00627}  { arXiv:2004.00627
  [astro-ph.CO]}\BibitemShut {NoStop}%
\bibitem [{\citenamefont {{Siegert}}\ \emph
  {et~al.}(2022{\natexlab{b}})\citenamefont {{Siegert}} \emph
  {et~al.}}]{2022MNRAS.511..914S}%
  \BibitemOpen
  \bibfield  {author} {\bibinfo {author} {\bibfnamefont {T.}~\bibnamefont
  {{Siegert}}} \emph {et~al.},\ }\href {https://doi.org/10.1093/mnras/stac008}
  {\bibfield  {journal} {\bibinfo  {journal} {\mnras}\ }\textbf {\bibinfo
  {volume} {511}},\ \bibinfo {pages} {914} (\bibinfo {year}
  {2022}{\natexlab{b}})},\ \Eprint {https://arxiv.org/abs/2109.03791}  {
  arXiv:2109.03791 [astro-ph.HE]}\BibitemShut {NoStop}%
\bibitem [{\citenamefont {{Derishev}}\ and\ \citenamefont
  {{Belyanin}}(1999)}]{1999A&A...343....1D}%
  \BibitemOpen
  \bibfield  {author} {\bibinfo {author} {\bibfnamefont {E.~V.}\ \bibnamefont
  {{Derishev}}}\ and\ \bibinfo {author} {\bibfnamefont {A.~A.}\ \bibnamefont
  {{Belyanin}}},\ }\href@noop {} {\bibfield  {journal} {\bibinfo  {journal}
  {\aap}\ }\textbf {\bibinfo {volume} {343}},\ \bibinfo {pages} {1} (\bibinfo
  {year} {1999})}\BibitemShut {NoStop}%
\bibitem [{\citenamefont {{Ginzburg}}\ and\ \citenamefont
  {{Ptuskin}}(1975)}]{1975SvPhU..18..931G}%
  \BibitemOpen
  \bibfield  {author} {\bibinfo {author} {\bibfnamefont {V.~L.}\ \bibnamefont
  {{Ginzburg}}}\ and\ \bibinfo {author} {\bibfnamefont {V.~S.}\ \bibnamefont
  {{Ptuskin}}},\ }\href {https://doi.org/10.1070/PU1975v018n12ABEH005243}
  {\bibfield  {journal} {\bibinfo  {journal} {\sovphysusp}\ }\textbf {\bibinfo
  {volume} {18}},\ \bibinfo {pages} {931} (\bibinfo {year} {1975})},\ \bibinfo
  {note} {[\uspfiznauk~\textbf{117}, 585 (1975)]}\BibitemShut {NoStop}%
\bibitem [{\citenamefont {{Boudaud}}\ \emph
  {et~al.}(2017{\natexlab{a}})\citenamefont {{Boudaud}} \emph
  {et~al.}}]{2017A&A...605A..17B}%
  \BibitemOpen
  \bibfield  {author} {\bibinfo {author} {\bibfnamefont {M.}~\bibnamefont
  {{Boudaud}}} \emph {et~al.},\ }\href
  {https://doi.org/10.1051/0004-6361/201630321} {\bibfield  {journal} {\bibinfo
   {journal} {\aap}\ }\textbf {\bibinfo {volume} {605}},\ \bibinfo {eid} {A17}
  (\bibinfo {year} {2017}{\natexlab{a}})},\ \Eprint
  {https://arxiv.org/abs/1612.03924}  { arXiv:1612.03924
  [astro-ph.HE]}\BibitemShut {NoStop}%
\bibitem [{\citenamefont {{G{\'e}nolini}}\ \emph {et~al.}(2021)\citenamefont
  {{G{\'e}nolini}} \emph {et~al.}}]{2021PhRvD.104h3005G}%
  \BibitemOpen
  \bibfield  {author} {\bibinfo {author} {\bibfnamefont {Y.}~\bibnamefont
  {{G{\'e}nolini}}} \emph {et~al.},\ }\href
  {https://doi.org/10.1103/PhysRevD.104.083005} {\bibfield  {journal} {\bibinfo
   {journal} {\prd}\ }\textbf {\bibinfo {volume} {104}},\ \bibinfo {eid}
  {083005} (\bibinfo {year} {2021})},\ \Eprint
  {https://arxiv.org/abs/2103.04108}  { arXiv:2103.04108
  [astro-ph.HE]}\BibitemShut {NoStop}%
\bibitem [{\citenamefont {{Wells}}\ \emph {et~al.}(1999)\citenamefont
  {{Wells}}, \citenamefont {{Moiseev}},\ and\ \citenamefont
  {{Ormes}}}]{1999ApJ...518..570W}%
  \BibitemOpen
  \bibfield  {author} {\bibinfo {author} {\bibfnamefont {J.~D.}\ \bibnamefont
  {{Wells}}}, \bibinfo {author} {\bibfnamefont {A.}~\bibnamefont {{Moiseev}}},\
  and\ \bibinfo {author} {\bibfnamefont {J.~F.}\ \bibnamefont {{Ormes}}},\
  }\href {https://doi.org/10.1086/307325} {\bibfield  {journal} {\bibinfo
  {journal} {\apj}\ }\textbf {\bibinfo {volume} {518}},\ \bibinfo {pages} {570}
  (\bibinfo {year} {1999})},\ \Eprint {https://arxiv.org/abs/hep-ph/9811325}  {
  arXiv:hep-ph/9811325 [hep-ph]}\BibitemShut {NoStop}%
\bibitem [{\citenamefont {{Kiraly}}\ \emph {et~al.}(1981)\citenamefont
  {{Kiraly}}, \citenamefont {{Szabelski}}, \citenamefont {{Wdowczyk}},\ and\
  \citenamefont {{Wolfendale}}}]{1981Natur.293..120K}%
  \BibitemOpen
  \bibfield  {author} {\bibinfo {author} {\bibfnamefont {P.}~\bibnamefont
  {{Kiraly}}}, \bibinfo {author} {\bibfnamefont {J.}~\bibnamefont
  {{Szabelski}}}, \bibinfo {author} {\bibfnamefont {J.}~\bibnamefont
  {{Wdowczyk}}},\ and\ \bibinfo {author} {\bibfnamefont {A.~W.}\ \bibnamefont
  {{Wolfendale}}},\ }\href {https://doi.org/10.1038/293120a0} {\bibfield
  {journal} {\bibinfo  {journal} {\nat}\ }\textbf {\bibinfo {volume} {293}},\
  \bibinfo {pages} {120} (\bibinfo {year} {1981})}\BibitemShut {NoStop}%
\bibitem [{\citenamefont {{Nasel'skii}}\ and\ \citenamefont
  {{Pelikhov}}(1979)}]{1979SvA....23..402N}%
  \BibitemOpen
  \bibfield  {author} {\bibinfo {author} {\bibfnamefont {P.~D.}\ \bibnamefont
  {{Nasel'skii}}}\ and\ \bibinfo {author} {\bibfnamefont {N.~V.}\ \bibnamefont
  {{Pelikhov}}},\ }\href@noop {} {\bibfield  {journal} {\bibinfo  {journal}
  {\sovast}\ }\textbf {\bibinfo {volume} {23}},\ \bibinfo {pages} {402}
  (\bibinfo {year} {1979})},\ \bibinfo {note} {[\azh~\textbf{56}, 714
  (1979)]}\BibitemShut {NoStop}%
\bibitem [{\citenamefont {{Fanselow}}\ \emph {et~al.}(1969)\citenamefont
  {{Fanselow}}, \citenamefont {{Hartman}}, \citenamefont {{Hildebrad}},\ and\
  \citenamefont {{Meyer}}}]{1969ApJ...158..771F}%
  \BibitemOpen
  \bibfield  {author} {\bibinfo {author} {\bibfnamefont {J.~L.}\ \bibnamefont
  {{Fanselow}}}, \bibinfo {author} {\bibfnamefont {R.~C.}\ \bibnamefont
  {{Hartman}}}, \bibinfo {author} {\bibfnamefont {R.~H.}\ \bibnamefont
  {{Hildebrad}}},\ and\ \bibinfo {author} {\bibfnamefont {P.}~\bibnamefont
  {{Meyer}}},\ }\href {https://doi.org/10.1086/150236} {\bibfield  {journal}
  {\bibinfo  {journal} {\apj}\ }\textbf {\bibinfo {volume} {158}},\ \bibinfo
  {pages} {771} (\bibinfo {year} {1969})}\BibitemShut {NoStop}%
\bibitem [{\citenamefont {{Cummings}}(1973)}]{Cummings:1973qya}%
  \BibitemOpen
  \bibfield  {author} {\ }\bibinfo {author} {\bibfnamefont {A.~C.}\
  \bibnamefont {{Cummings}}},\ \href@noop {} {Ph.D. thesis},\ \bibinfo
  {school} {Caltech, USA} (\bibinfo {year} {1973})\BibitemShut {NoStop}%
\bibitem [{\citenamefont {{Boudaud}}\ and\ \citenamefont
  {{Cirelli}}(2019)}]{2019PhRvL.122d1104B}%
  \BibitemOpen
  \bibfield  {author} {\bibinfo {author} {\bibfnamefont {M.}~\bibnamefont
  {{Boudaud}}}\ and\ \bibinfo {author} {\bibfnamefont {M.}~\bibnamefont
  {{Cirelli}}},\ }\href {https://doi.org/10.1103/PhysRevLett.122.041104}
  {\bibfield  {journal} {\bibinfo  {journal} {\prl}\ }\textbf {\bibinfo
  {volume} {122}},\ \bibinfo {eid} {041104} (\bibinfo {year} {2019})},\ \Eprint
  {https://arxiv.org/abs/1807.03075}  { arXiv:1807.03075
  [astro-ph.HE]}\BibitemShut {NoStop}%
\bibitem [{\citenamefont {{Stone}}\ \emph {et~al.}(2013)\citenamefont {{Stone}}
  \emph {et~al.}}]{2013Sci...341..150S}%
  \BibitemOpen
  \bibfield  {author} {\bibinfo {author} {\bibfnamefont {E.~C.}\ \bibnamefont
  {{Stone}}} \emph {et~al.},\ }\href {https://doi.org/10.1126/science.1236408}
  {\bibfield  {journal} {\bibinfo  {journal} {Science}\ }\textbf {\bibinfo
  {volume} {341}},\ \bibinfo {pages} {150} (\bibinfo {year}
  {2013})}\BibitemShut {NoStop}%
\bibitem [{\citenamefont {{Boudaud}}\ \emph
  {et~al.}(2017{\natexlab{b}})\citenamefont {{Boudaud}}, \citenamefont
  {{Lavalle}},\ and\ \citenamefont {{Salati}}}]{2017PhRvL.119b1103B}%
  \BibitemOpen
  \bibfield  {author} {\bibinfo {author} {\bibfnamefont {M.}~\bibnamefont
  {{Boudaud}}}, \bibinfo {author} {\bibfnamefont {J.}~\bibnamefont
  {{Lavalle}}},\ and\ \bibinfo {author} {\bibfnamefont {P.}~\bibnamefont
  {{Salati}}},\ }\href {https://doi.org/10.1103/PhysRevLett.119.021103}
  {\bibfield  {journal} {\bibinfo  {journal} {\prl}\ }\textbf {\bibinfo
  {volume} {119}},\ \bibinfo {eid} {021103} (\bibinfo {year}
  {2017}{\natexlab{b}})},\ \Eprint {https://arxiv.org/abs/1612.07698}  {
  arXiv:1612.07698 [astro-ph.HE]}\BibitemShut {NoStop}%
\bibitem [{\citenamefont {{Mukhopadhyay}}\ \emph {et~al.}(2021)\citenamefont
  {{Mukhopadhyay}}, \citenamefont {{Majumdar}},\ and\ \citenamefont
  {{Paul}}}]{2021arXiv210914955M}%
  \BibitemOpen
  \bibfield  {author} {\bibinfo {author} {\bibfnamefont {U.}~\bibnamefont
  {{Mukhopadhyay}}}, \bibinfo {author} {\bibfnamefont {D.}~\bibnamefont
  {{Majumdar}}},\ and\ \bibinfo {author} {\bibfnamefont {A.}~\bibnamefont
  {{Paul}}},\ }\href@noop {} {\bibfield  {journal} {\bibinfo  {journal}
  {\arxiv}\ } (\bibinfo {year} {2021})},\ \Eprint
  {https://arxiv.org/abs/2109.14955}  { 2109.14955 [astro-ph.CO]}\BibitemShut
  {NoStop}%
\bibitem [{\citenamefont {{Kim}}(2021)}]{2021MNRAS.504.5475K}%
  \BibitemOpen
  \bibfield  {author} {\bibinfo {author} {\bibfnamefont {H.}~\bibnamefont
  {{Kim}}},\ }\href {https://doi.org/10.1093/mnras/stab1222} {\bibfield
  {journal} {\bibinfo  {journal} {\mnras}\ }\textbf {\bibinfo {volume} {504}},\
  \bibinfo {pages} {5475} (\bibinfo {year} {2021})},\ \Eprint
  {https://arxiv.org/abs/2007.07739}  { arXiv:2007.07739 [hep-ph]}\BibitemShut
  {NoStop}%
\bibitem [{\citenamefont {{Chan}}\ and\ \citenamefont
  {{Lee}}(2020)}]{2020MNRAS.497.1212C}%
  \BibitemOpen
  \bibfield  {author} {\bibinfo {author} {\bibfnamefont {M.~H.}\ \bibnamefont
  {{Chan}}}\ and\ \bibinfo {author} {\bibfnamefont {C.~M.}\ \bibnamefont
  {{Lee}}},\ }\href {https://doi.org/10.1093/mnras/staa1966} {\bibfield
  {journal} {\bibinfo  {journal} {\mnras}\ }\textbf {\bibinfo {volume} {497}},\
  \bibinfo {pages} {1212} (\bibinfo {year} {2020})},\ \Eprint
  {https://arxiv.org/abs/2007.05677}  { arXiv:2007.05677
  [astro-ph.HE]}\BibitemShut {NoStop}%
\bibitem [{\citenamefont {{Dutta}}\ \emph {et~al.}(2021)\citenamefont
  {{Dutta}}, \citenamefont {{Kar}},\ and\ \citenamefont
  {{Strigari}}}]{2021JCAP...03..011D}%
  \BibitemOpen
  \bibfield  {author} {\bibinfo {author} {\bibfnamefont {B.}~\bibnamefont
  {{Dutta}}}, \bibinfo {author} {\bibfnamefont {A.}~\bibnamefont {{Kar}}},\
  and\ \bibinfo {author} {\bibfnamefont {L.~E.}\ \bibnamefont {{Strigari}}},\
  }\href {https://doi.org/10.1088/1475-7516/2021/03/011} {\bibfield  {journal}
  {\bibinfo  {journal} {\jcap}\ }\textbf {\bibinfo {volume} {2021}},\ \bibinfo
  {eid} {011} (\bibinfo {year} {2021})},\ \Eprint
  {https://arxiv.org/abs/2010.05977}  { arXiv:2010.05977
  [astro-ph.HE]}\BibitemShut {NoStop}%
\bibitem [{\citenamefont {{Lee}}\ and\ \citenamefont {{Ho
  Chan}}(2021)}]{2021ApJ...912...24L}%
  \BibitemOpen
  \bibfield  {author} {\bibinfo {author} {\bibfnamefont {C.}~\bibnamefont
  {{Lee}}}\ and\ \bibinfo {author} {\bibfnamefont {M.}~\bibnamefont {{Ho
  Chan}}},\ }\href {https://doi.org/10.3847/1538-4357/abee72} {\bibfield
  {journal} {\bibinfo  {journal} {\apj}\ }\textbf {\bibinfo {volume} {912}},\
  \bibinfo {eid} {24} (\bibinfo {year} {2021})},\ \Eprint
  {https://arxiv.org/abs/2103.12354}  { arXiv:2103.12354
  [astro-ph.HE]}\BibitemShut {NoStop}%
\bibitem [{\citenamefont {{Laha}}\ \emph {et~al.}(2021)\citenamefont {{Laha}},
  \citenamefont {{Lu}},\ and\ \citenamefont
  {{Takhistov}}}]{2021PhLB..82036459L}%
  \BibitemOpen
  \bibfield  {author} {\bibinfo {author} {\bibfnamefont {R.}~\bibnamefont
  {{Laha}}}, \bibinfo {author} {\bibfnamefont {P.}~\bibnamefont {{Lu}}},\ and\
  \bibinfo {author} {\bibfnamefont {V.}~\bibnamefont {{Takhistov}}},\ }\href
  {https://doi.org/10.1016/j.physletb.2021.136459} {\bibfield  {journal}
  {\bibinfo  {journal} {\plb}\ }\textbf {\bibinfo {volume} {820}},\ \bibinfo
  {eid} {136459} (\bibinfo {year} {2021})},\ \Eprint
  {https://arxiv.org/abs/2009.11837}  { arXiv:2009.11837
  [astro-ph.CO]}\BibitemShut {NoStop}%
\bibitem [{\citenamefont {{Johnson}}\ \emph {et~al.}(1972)\citenamefont
  {{Johnson}}, \citenamefont {{Harnden}},\ and\ \citenamefont
  {{Haymes}}}]{1972ApJ...172L...1J}%
  \BibitemOpen
  \bibfield  {author} {\bibinfo {author} {\bibfnamefont {I.}~\bibnamefont
  {{Johnson}}, \bibfnamefont {W.~N.}}, \bibinfo {author} {\bibfnamefont
  {J.}~\bibnamefont {{Harnden}}, \bibfnamefont {F.~R.}},\ and\ \bibinfo
  {author} {\bibfnamefont {R.~C.}\ \bibnamefont {{Haymes}}},\ }\href
  {https://doi.org/10.1086/180878} {\bibfield  {journal} {\bibinfo  {journal}
  {\apjl}\ }\textbf {\bibinfo {volume} {172}},\ \bibinfo {pages} {L1} (\bibinfo
  {year} {1972})}\BibitemShut {NoStop}%
\bibitem [{\citenamefont {{Johnson}}\ and\ \citenamefont
  {{Haymes}}(1973)}]{1973ApJ...184..103J}%
  \BibitemOpen
  \bibfield  {author} {\bibinfo {author} {\bibfnamefont {I.}~\bibnamefont
  {{Johnson}}, \bibfnamefont {W.~N.}}\ and\ \bibinfo {author} {\bibfnamefont
  {R.~C.}\ \bibnamefont {{Haymes}}},\ }\href {https://doi.org/10.1086/152309}
  {\bibfield  {journal} {\bibinfo  {journal} {\apj}\ }\textbf {\bibinfo
  {volume} {184}},\ \bibinfo {pages} {103} (\bibinfo {year}
  {1973})}\BibitemShut {NoStop}%
\bibitem [{\citenamefont {{Haymes}}\ \emph {et~al.}(1975)\citenamefont
  {{Haymes}} \emph {et~al.}}]{1975ApJ...201..593H}%
  \BibitemOpen
  \bibfield  {author} {\bibinfo {author} {\bibfnamefont {R.~C.}\ \bibnamefont
  {{Haymes}}} \emph {et~al.},\ }\href {https://doi.org/10.1086/153925}
  {\bibfield  {journal} {\bibinfo  {journal} {\apj}\ }\textbf {\bibinfo
  {volume} {201}},\ \bibinfo {pages} {593} (\bibinfo {year}
  {1975})}\BibitemShut {NoStop}%
\bibitem [{\citenamefont {{Leventhal}}\ \emph {et~al.}(1978)\citenamefont
  {{Leventhal}}, \citenamefont {{MacCallum}},\ and\ \citenamefont
  {{Stang}}}]{1978ApJ...225L..11L}%
  \BibitemOpen
  \bibfield  {author} {\bibinfo {author} {\bibfnamefont {M.}~\bibnamefont
  {{Leventhal}}}, \bibinfo {author} {\bibfnamefont {C.~J.}\ \bibnamefont
  {{MacCallum}}},\ and\ \bibinfo {author} {\bibfnamefont {P.~D.}\ \bibnamefont
  {{Stang}}},\ }\href {https://doi.org/10.1086/182782} {\bibfield  {journal}
  {\bibinfo  {journal} {\apjl}\ }\textbf {\bibinfo {volume} {225}},\ \bibinfo
  {pages} {L11} (\bibinfo {year} {1978})}\BibitemShut {NoStop}%
\bibitem [{\citenamefont {{Okeke}}\ and\ \citenamefont
  {{Rees}}(1980)}]{1980A&A....81..263O}%
  \BibitemOpen
  \bibfield  {author} {\bibinfo {author} {\bibfnamefont {P.~N.}\ \bibnamefont
  {{Okeke}}}\ and\ \bibinfo {author} {\bibfnamefont {M.~J.}\ \bibnamefont
  {{Rees}}},\ }\href@noop {} {\bibfield  {journal} {\bibinfo  {journal} {\aap}\
  }\textbf {\bibinfo {volume} {81}},\ \bibinfo {pages} {263} (\bibinfo {year}
  {1980})}\BibitemShut {NoStop}%
\bibitem [{\citenamefont {{Okeke}}(1980)}]{1980Ap&SS..71..371O}%
  \BibitemOpen
  \bibfield  {author} {\bibinfo {author} {\bibfnamefont {P.~N.}\ \bibnamefont
  {{Okeke}}},\ }\href {https://doi.org/10.1007/BF00639396} {\bibfield
  {journal} {\bibinfo  {journal} {\apss}\ }\textbf {\bibinfo {volume} {71}},\
  \bibinfo {pages} {371} (\bibinfo {year} {1980})}\BibitemShut {NoStop}%
\bibitem [{\citenamefont {{Prantzos}}\ \emph {et~al.}(2011)\citenamefont
  {{Prantzos}} \emph {et~al.}}]{2011RvMP...83.1001P}%
  \BibitemOpen
  \bibfield  {author} {\bibinfo {author} {\bibfnamefont {N.}~\bibnamefont
  {{Prantzos}}} \emph {et~al.},\ }\href
  {https://doi.org/10.1103/RevModPhys.83.1001} {\bibfield  {journal} {\bibinfo
  {journal} {\RMP}\ }\textbf {\bibinfo {volume} {83}},\ \bibinfo {pages} {1001}
  (\bibinfo {year} {2011})},\ \Eprint {https://arxiv.org/abs/1009.4620}  {
  arXiv:1009.4620 [astro-ph.HE]}\BibitemShut {NoStop}%
\bibitem [{\citenamefont {{Kn{\"o}dlseder}}\ \emph {et~al.}(2005)\citenamefont
  {{Kn{\"o}dlseder}} \emph {et~al.}}]{2005A&A...441..513K}%
  \BibitemOpen
  \bibfield  {author} {\bibinfo {author} {\bibfnamefont {J.}~\bibnamefont
  {{Kn{\"o}dlseder}}} \emph {et~al.},\ }\href
  {https://doi.org/10.1051/0004-6361:20042063} {\bibfield  {journal} {\bibinfo
  {journal} {\aap}\ }\textbf {\bibinfo {volume} {441}},\ \bibinfo {pages} {513}
  (\bibinfo {year} {2005})},\ \Eprint {https://arxiv.org/abs/astro-ph/0506026}
  { arXiv:astro-ph/0506026 [astro-ph]}\BibitemShut {NoStop}%
\bibitem [{\citenamefont {{Jean}}\ \emph {et~al.}(2006)\citenamefont {{Jean}}
  \emph {et~al.}}]{2006A&A...445..579J}%
  \BibitemOpen
  \bibfield  {author} {\bibinfo {author} {\bibfnamefont {P.}~\bibnamefont
  {{Jean}}} \emph {et~al.},\ }\href
  {https://doi.org/10.1051/0004-6361:20053765} {\bibfield  {journal} {\bibinfo
  {journal} {\aap}\ }\textbf {\bibinfo {volume} {445}},\ \bibinfo {pages} {579}
  (\bibinfo {year} {2006})},\ \Eprint {https://arxiv.org/abs/astro-ph/0509298}
  { arXiv:astro-ph/0509298 [astro-ph]}\BibitemShut {NoStop}%
\bibitem [{\citenamefont {{Weidenspointner}}\ \emph {et~al.}(2006)\citenamefont
  {{Weidenspointner}} \emph {et~al.}}]{2006A&A...450.1013W}%
  \BibitemOpen
  \bibfield  {author} {\bibinfo {author} {\bibfnamefont {G.}~\bibnamefont
  {{Weidenspointner}}} \emph {et~al.} (\bibinfo {collaboration} {INTEGRAL}),\
  }\href {https://doi.org/10.1051/0004-6361:20054046} {\bibfield  {journal}
  {\bibinfo  {journal} {\aap}\ }\textbf {\bibinfo {volume} {450}},\ \bibinfo
  {pages} {1013} (\bibinfo {year} {2006})},\ \Eprint
  {https://arxiv.org/abs/astro-ph/0601673}  { arXiv:astro-ph/0601673
  [astro-ph]}\BibitemShut {NoStop}%
\bibitem [{\citenamefont {{Laha}}(2019)}]{2019PhRvL.123y1101L}%
  \BibitemOpen
  \bibfield  {author} {\bibinfo {author} {\bibfnamefont {R.}~\bibnamefont
  {{Laha}}},\ }\href {https://doi.org/10.1103/PhysRevLett.123.251101}
  {\bibfield  {journal} {\bibinfo  {journal} {\prl}\ }\textbf {\bibinfo
  {volume} {123}},\ \bibinfo {eid} {251101} (\bibinfo {year} {2019})},\ \Eprint
  {https://arxiv.org/abs/1906.09994}  { arXiv:1906.09994
  [astro-ph.HE]}\BibitemShut {NoStop}%
\bibitem [{\citenamefont {{DeRocco}}\ and\ \citenamefont
  {{Graham}}(2019)}]{2019PhRvL.123y1102D}%
  \BibitemOpen
  \bibfield  {author} {\bibinfo {author} {\bibfnamefont {W.}~\bibnamefont
  {{DeRocco}}}\ and\ \bibinfo {author} {\bibfnamefont {P.~W.}\ \bibnamefont
  {{Graham}}},\ }\href {https://doi.org/10.1103/PhysRevLett.123.251102}
  {\bibfield  {journal} {\bibinfo  {journal} {\prl}\ }\textbf {\bibinfo
  {volume} {123}},\ \bibinfo {eid} {251102} (\bibinfo {year} {2019})},\ \Eprint
  {https://arxiv.org/abs/1906.07740}  { arXiv:1906.07740
  [astro-ph.CO]}\BibitemShut {NoStop}%
\bibitem [{\citenamefont {{Keith}}\ and\ \citenamefont
  {{Hooper}}(2021)}]{2021PhRvD.104f3033K}%
  \BibitemOpen
  \bibfield  {author} {\bibinfo {author} {\bibfnamefont {C.}~\bibnamefont
  {{Keith}}}\ and\ \bibinfo {author} {\bibfnamefont {D.}~\bibnamefont
  {{Hooper}}},\ }\href {https://doi.org/10.1103/PhysRevD.104.063033} {\bibfield
   {journal} {\bibinfo  {journal} {\prd}\ }\textbf {\bibinfo {volume} {104}},\
  \bibinfo {eid} {063033} (\bibinfo {year} {2021})},\ \Eprint
  {https://arxiv.org/abs/2103.08611}  { arXiv:2103.08611
  [astro-ph.CO]}\BibitemShut {NoStop}%
\bibitem [{\citenamefont {{Barrau}}\ \emph {et~al.}(2002)\citenamefont
  {{Barrau}} \emph {et~al.}}]{2002A&A...388..676B}%
  \BibitemOpen
  \bibfield  {author} {\bibinfo {author} {\bibfnamefont {A.}~\bibnamefont
  {{Barrau}}} \emph {et~al.},\ }\href
  {https://doi.org/10.1051/0004-6361:20020313} {\bibfield  {journal} {\bibinfo
  {journal} {\aap}\ }\textbf {\bibinfo {volume} {388}},\ \bibinfo {pages} {676}
  (\bibinfo {year} {2002})},\ \Eprint {https://arxiv.org/abs/astro-ph/0112486}
  { arXiv:astro-ph/0112486 [astro-ph]}\BibitemShut {NoStop}%
\bibitem [{\citenamefont {{Abe}}\ \emph {et~al.}(2017)\citenamefont {{Abe}}
  \emph {et~al.}}]{2017AdSpR..60..806A}%
  \BibitemOpen
  \bibfield  {author} {\bibinfo {author} {\bibfnamefont {K.}~\bibnamefont
  {{Abe}}} \emph {et~al.} (\bibinfo {collaboration} {BESS}),\ }\href
  {https://doi.org/10.1016/j.asr.2016.11.004} {\bibfield  {journal} {\bibinfo
  {journal} {\AdSpR}\ }\textbf {\bibinfo {volume} {60}},\ \bibinfo {pages}
  {806} (\bibinfo {year} {2017})}\BibitemShut {NoStop}%
\bibitem [{\citenamefont {{Golden}}\ \emph {et~al.}(1979)\citenamefont
  {{Golden}} \emph {et~al.}}]{1979PhRvL..43.1196G}%
  \BibitemOpen
  \bibfield  {author} {\bibinfo {author} {\bibfnamefont {R.~L.}\ \bibnamefont
  {{Golden}}} \emph {et~al.},\ }\href
  {https://doi.org/10.1103/PhysRevLett.43.1196} {\bibfield  {journal} {\bibinfo
   {journal} {\prl}\ }\textbf {\bibinfo {volume} {43}},\ \bibinfo {pages}
  {1196} (\bibinfo {year} {1979})}\BibitemShut {NoStop}%
\bibitem [{\citenamefont {{Buffington}}\ \emph {et~al.}(1981)\citenamefont
  {{Buffington}}, \citenamefont {{Schindler}},\ and\ \citenamefont
  {{Pennypacker}}}]{1981ApJ...248.1179B}%
  \BibitemOpen
  \bibfield  {author} {\bibinfo {author} {\bibfnamefont {A.}~\bibnamefont
  {{Buffington}}}, \bibinfo {author} {\bibfnamefont {S.~M.}\ \bibnamefont
  {{Schindler}}},\ and\ \bibinfo {author} {\bibfnamefont {C.~R.}\ \bibnamefont
  {{Pennypacker}}},\ }\href {https://doi.org/10.1086/159247} {\bibfield
  {journal} {\bibinfo  {journal} {\apj}\ }\textbf {\bibinfo {volume} {248}},\
  \bibinfo {pages} {1179} (\bibinfo {year} {1981})}\BibitemShut {NoStop}%
\bibitem [{\citenamefont {{Turner}}(1982)}]{1982Natur.297..379T}%
  \BibitemOpen
  \bibfield  {author} {\bibinfo {author} {\bibfnamefont {M.~S.}\ \bibnamefont
  {{Turner}}},\ }\href {https://doi.org/10.1038/297379a0} {\bibfield  {journal}
  {\bibinfo  {journal} {\nat}\ }\textbf {\bibinfo {volume} {297}},\ \bibinfo
  {pages} {379} (\bibinfo {year} {1982})}\BibitemShut {NoStop}%
\bibitem [{\citenamefont {{Yoshimura}}\ \emph {et~al.}(1995)\citenamefont
  {{Yoshimura}} \emph {et~al.}}]{1995PhRvL..75.3792Y}%
  \BibitemOpen
  \bibfield  {author} {\bibinfo {author} {\bibfnamefont {K.}~\bibnamefont
  {{Yoshimura}}} \emph {et~al.},\ }\href
  {https://doi.org/10.1103/PhysRevLett.75.3792} {\bibfield  {journal} {\bibinfo
   {journal} {\prl}\ }\textbf {\bibinfo {volume} {75}},\ \bibinfo {pages}
  {3792} (\bibinfo {year} {1995})}\BibitemShut {NoStop}%
\bibitem [{\citenamefont {{Maki}}\ \emph {et~al.}(1996)\citenamefont {{Maki}},
  \citenamefont {{Mitsui}},\ and\ \citenamefont
  {{Orito}}}]{1996PhRvL..76.3474M}%
  \BibitemOpen
  \bibfield  {author} {\bibinfo {author} {\bibfnamefont {K.}~\bibnamefont
  {{Maki}}}, \bibinfo {author} {\bibfnamefont {T.}~\bibnamefont {{Mitsui}}},\
  and\ \bibinfo {author} {\bibfnamefont {S.}~\bibnamefont {{Orito}}},\ }\href
  {https://doi.org/10.1103/PhysRevLett.76.3474} {\bibfield  {journal} {\bibinfo
   {journal} {\prl}\ }\textbf {\bibinfo {volume} {76}},\ \bibinfo {pages}
  {3474} (\bibinfo {year} {1996})},\ \Eprint
  {https://arxiv.org/abs/astro-ph/9601025}  { arXiv:astro-ph/9601025
  [astro-ph]}\BibitemShut {NoStop}%
\bibitem [{\citenamefont {{Mitsui}}\ \emph {et~al.}(1996)\citenamefont
  {{Mitsui}}, \citenamefont {{Maki}},\ and\ \citenamefont
  {{Orito}}}]{1996PhLB..389..169M}%
  \BibitemOpen
  \bibfield  {author} {\bibinfo {author} {\bibfnamefont {T.}~\bibnamefont
  {{Mitsui}}}, \bibinfo {author} {\bibfnamefont {K.}~\bibnamefont {{Maki}}},\
  and\ \bibinfo {author} {\bibfnamefont {S.}~\bibnamefont {{Orito}}},\ }\href
  {https://doi.org/10.1016/S0370-2693(96)01363-9} {\bibfield  {journal}
  {\bibinfo  {journal} {\plb}\ }\textbf {\bibinfo {volume} {389}},\ \bibinfo
  {pages} {169} (\bibinfo {year} {1996})},\ \Eprint
  {https://arxiv.org/abs/astro-ph/9608123}  { arXiv:astro-ph/9608123
  [astro-ph]}\BibitemShut {NoStop}%
\bibitem [{\citenamefont {{Sj{\"o}strand}}(1994)}]{1994CoPhC..82...74S}%
  \BibitemOpen
  \bibfield  {author} {\bibinfo {author} {\bibfnamefont {T.}~\bibnamefont
  {{Sj{\"o}strand}}},\ }\href {https://doi.org/10.1016/0010-4655(94)90132-5}
  {\bibfield  {journal} {\bibinfo  {journal} {\CPC}\ }\textbf {\bibinfo
  {volume} {82}},\ \bibinfo {pages} {74} (\bibinfo {year} {1994})}\BibitemShut
  {NoStop}%
\bibitem [{\citenamefont {{Yoshimura}}(2001)}]{2001AdSpR..27..693Y}%
  \BibitemOpen
  \bibfield  {author} {\bibinfo {author} {\bibfnamefont {K.}~\bibnamefont
  {{Yoshimura}}},\ }\href {https://doi.org/10.1016/S0273-1177(01)00111-9}
  {\bibfield  {journal} {\bibinfo  {journal} {\AdSpR}\ }\textbf {\bibinfo
  {volume} {27}},\ \bibinfo {pages} {693} (\bibinfo {year} {2001})}\BibitemShut
  {NoStop}%
\bibitem [{\citenamefont {{Abe}}\ \emph {et~al.}(2012)\citenamefont {{Abe}}
  \emph {et~al.}}]{2012PhRvL.108e1102A}%
  \BibitemOpen
  \bibfield  {author} {\bibinfo {author} {\bibfnamefont {K.}~\bibnamefont
  {{Abe}}} \emph {et~al.} (\bibinfo {collaboration} {BESS}),\ }\href
  {https://doi.org/10.1103/PhysRevLett.108.051102} {\bibfield  {journal}
  {\bibinfo  {journal} {\prl}\ }\textbf {\bibinfo {volume} {108}},\ \bibinfo
  {eid} {051102} (\bibinfo {year} {2012})},\ \Eprint
  {https://arxiv.org/abs/1107.6000}  { arXiv:1107.6000
  [astro-ph.HE]}\BibitemShut {NoStop}%
\bibitem [{\citenamefont {{Yamamoto}}\ \emph {et~al.}(2013)\citenamefont
  {{Yamamoto}} \emph {et~al.}}]{2013AdSpR..51..227B}%
  \BibitemOpen
  \bibfield  {author} {\bibinfo {author} {\bibfnamefont {A.}~\bibnamefont
  {{Yamamoto}}} \emph {et~al.} (\bibinfo {collaboration} {BESS}),\ }\href
  {https://doi.org/10.1016/j.asr.2011.07.012} {\bibfield  {journal} {\bibinfo
  {journal} {\AdSpR}\ }\textbf {\bibinfo {volume} {51}},\ \bibinfo {pages}
  {227} (\bibinfo {year} {2013})}\BibitemShut {NoStop}%
\bibitem [{\citenamefont {{Aramaki}}\ \emph {et~al.}(2014)\citenamefont
  {{Aramaki}} \emph {et~al.}}]{2014APh....59...12A}%
  \BibitemOpen
  \bibfield  {author} {\bibinfo {author} {\bibfnamefont {T.}~\bibnamefont
  {{Aramaki}}} \emph {et~al.},\ }\href
  {https://doi.org/10.1016/j.astropartphys.2014.03.011} {\bibfield  {journal}
  {\bibinfo  {journal} {\APh}\ }\textbf {\bibinfo {volume} {59}},\ \bibinfo
  {pages} {12} (\bibinfo {year} {2014})},\ \Eprint
  {https://arxiv.org/abs/1401.8245}  { arXiv:1401.8245
  [astro-ph.HE]}\BibitemShut {NoStop}%
\bibitem [{\citenamefont {{von Doetinchem}}\ \emph {et~al.}(2020)\citenamefont
  {{von Doetinchem}} \emph {et~al.}}]{2020JCAP...08..035V}%
  \BibitemOpen
  \bibfield  {author} {\bibinfo {author} {\bibfnamefont {P.}~\bibnamefont {{von
  Doetinchem}}} \emph {et~al.},\ }\href
  {https://doi.org/10.1088/1475-7516/2020/08/035} {\bibfield  {journal}
  {\bibinfo  {journal} {\jcap}\ }\textbf {\bibinfo {volume} {2020}},\ \bibinfo
  {eid} {035} (\bibinfo {year} {2020})},\ \Eprint
  {https://arxiv.org/abs/2002.04163}  { arXiv:2002.04163
  [astro-ph.HE]}\BibitemShut {NoStop}%
\bibitem [{\citenamefont {{Barrau}}\ \emph
  {et~al.}(2003{\natexlab{b}})\citenamefont {{Barrau}}, \citenamefont
  {{Blais}}, \citenamefont {{Boudoul}},\ and\ \citenamefont
  {{Polarski}}}]{2003PhLB..551..218B}%
  \BibitemOpen
  \bibfield  {author} {\bibinfo {author} {\bibfnamefont {A.}~\bibnamefont
  {{Barrau}}}, \bibinfo {author} {\bibfnamefont {D.}~\bibnamefont {{Blais}}},
  \bibinfo {author} {\bibfnamefont {G.}~\bibnamefont {{Boudoul}}},\ and\
  \bibinfo {author} {\bibfnamefont {D.}~\bibnamefont {{Polarski}}},\ }\href
  {https://doi.org/10.1016/S0370-2693(02)03060-5} {\bibfield  {journal}
  {\bibinfo  {journal} {\plb}\ }\textbf {\bibinfo {volume} {551}},\ \bibinfo
  {pages} {218} (\bibinfo {year} {2003}{\natexlab{b}})},\ \Eprint
  {https://arxiv.org/abs/astro-ph/0210149}  { arXiv:astro-ph/0210149
  [astro-ph]}\BibitemShut {NoStop}%
\bibitem [{\citenamefont {{Barrau}}\ \emph
  {et~al.}(2003{\natexlab{c}})\citenamefont {{Barrau}} \emph
  {et~al.}}]{2003A&A...398..403B}%
  \BibitemOpen
  \bibfield  {author} {\bibinfo {author} {\bibfnamefont {A.}~\bibnamefont
  {{Barrau}}} \emph {et~al.},\ }\href
  {https://doi.org/10.1051/0004-6361:20021588} {\bibfield  {journal} {\bibinfo
  {journal} {\aap}\ }\textbf {\bibinfo {volume} {398}},\ \bibinfo {pages} {403}
  (\bibinfo {year} {2003}{\natexlab{c}})},\ \Eprint
  {https://arxiv.org/abs/astro-ph/0207395}  { arXiv:astro-ph/0207395
  [astro-ph]}\BibitemShut {NoStop}%
\bibitem [{\citenamefont {{Herms}}\ \emph {et~al.}(2017)\citenamefont
  {{Herms}}, \citenamefont {{Ibarra}}, \citenamefont {{Vittino}},\ and\
  \citenamefont {{Wild}}}]{2017JCAP...02..018H}%
  \BibitemOpen
  \bibfield  {author} {\bibinfo {author} {\bibfnamefont {J.}~\bibnamefont
  {{Herms}}}, \bibinfo {author} {\bibfnamefont {A.}~\bibnamefont {{Ibarra}}},
  \bibinfo {author} {\bibfnamefont {A.}~\bibnamefont {{Vittino}}},\ and\
  \bibinfo {author} {\bibfnamefont {S.}~\bibnamefont {{Wild}}},\ }\href
  {https://doi.org/10.1088/1475-7516/2017/02/018} {\bibfield  {journal}
  {\bibinfo  {journal} {\jcap}\ }\textbf {\bibinfo {volume} {2017}},\ \bibinfo
  {eid} {018} (\bibinfo {year} {2017})},\ \Eprint
  {https://arxiv.org/abs/1610.00699}  { arXiv:1610.00699
  [astro-ph.HE]}\BibitemShut {NoStop}%
\bibitem [{\citenamefont {{Bugaev}}\ \emph {et~al.}(2009)\citenamefont
  {{Bugaev}} \emph {et~al.}}]{2009arXiv0906.3182B}%
  \BibitemOpen
  \bibfield  {author} {\bibinfo {author} {\bibfnamefont {E.~V.}\ \bibnamefont
  {{Bugaev}}} \emph {et~al.},\ }\href@noop {} {\bibfield  {journal} {\bibinfo
  {journal} {\arxiv}\ } (\bibinfo {year} {2009})},\ \Eprint
  {https://arxiv.org/abs/0906.3182}  { arXiv:0906.3182
  [astro-ph.CO]}\BibitemShut {NoStop}%
\bibitem [{\citenamefont {{Porter}}(1978)}]{1978IrAJ...13..173P}%
  \BibitemOpen
  \bibfield  {author} {\bibinfo {author} {\bibfnamefont {N.~A.}\ \bibnamefont
  {{Porter}}},\ }\href@noop {} {\bibfield  {journal} {\bibinfo  {journal}
  {Irish Astron.~J.}\ }\textbf {\bibinfo {volume} {13}},\ \bibinfo {pages}
  {173} (\bibinfo {year} {1978})}\BibitemShut {NoStop}%
\bibitem [{\citenamefont {{Porter}}\ and\ \citenamefont
  {{Weekes}}(1978)}]{1978MNRAS.183..205P}%
  \BibitemOpen
  \bibfield  {author} {\bibinfo {author} {\bibfnamefont {N.~A.}\ \bibnamefont
  {{Porter}}}\ and\ \bibinfo {author} {\bibfnamefont {T.~C.}\ \bibnamefont
  {{Weekes}}},\ }\href {https://doi.org/10.1093/mnras/183.2.205} {\bibfield
  {journal} {\bibinfo  {journal} {\mnras}\ }\textbf {\bibinfo {volume} {183}},\
  \bibinfo {pages} {205} (\bibinfo {year} {1978})}\BibitemShut {NoStop}%
\bibitem [{\citenamefont {{Alexandreas}}\ \emph {et~al.}(1993)\citenamefont
  {{Alexandreas}} \emph {et~al.}}]{1993PhRvL..71.2524A}%
  \BibitemOpen
  \bibfield  {author} {\bibinfo {author} {\bibfnamefont {D.~E.}\ \bibnamefont
  {{Alexandreas}}} \emph {et~al.},\ }\href
  {https://doi.org/10.1103/PhysRevLett.71.2524} {\bibfield  {journal} {\bibinfo
   {journal} {\prl}\ }\textbf {\bibinfo {volume} {71}},\ \bibinfo {pages}
  {2524} (\bibinfo {year} {1993})}\BibitemShut {NoStop}%
\bibitem [{\citenamefont {{Semikoz}}(1994)}]{1994ApJ...436..254S}%
  \BibitemOpen
  \bibfield  {author} {\bibinfo {author} {\bibfnamefont {D.~V.}\ \bibnamefont
  {{Semikoz}}},\ }\href {https://doi.org/10.1086/174898} {\bibfield  {journal}
  {\bibinfo  {journal} {\apj}\ }\textbf {\bibinfo {volume} {436}},\ \bibinfo
  {pages} {254} (\bibinfo {year} {1994})}\BibitemShut {NoStop}%
\bibitem [{\citenamefont {{Ukwatta}}\ \emph {et~al.}(2016)\citenamefont
  {{Ukwatta}} \emph {et~al.}}]{2016APh....80...90U}%
  \BibitemOpen
  \bibfield  {author} {\bibinfo {author} {\bibfnamefont {T.~N.}\ \bibnamefont
  {{Ukwatta}}} \emph {et~al.},\ }\href
  {https://doi.org/10.1016/j.astropartphys.2016.03.007} {\bibfield  {journal}
  {\bibinfo  {journal} {\APh}\ }\textbf {\bibinfo {volume} {80}},\ \bibinfo
  {pages} {90} (\bibinfo {year} {2016})},\ \Eprint
  {https://arxiv.org/abs/1510.04372}  { arXiv:1510.04372
  [astro-ph.HE]}\BibitemShut {NoStop}%
\bibitem [{\citenamefont {{Porter}}(1996)}]{1996SSRv...75...67P}%
  \BibitemOpen
  \bibfield  {author} {\bibinfo {author} {\bibfnamefont {N.~A.}\ \bibnamefont
  {{Porter}}},\ }\href {https://doi.org/10.1007/BF00195025} {\bibfield
  {journal} {\bibinfo  {journal} {\ssr}\ }\textbf {\bibinfo {volume} {75}},\
  \bibinfo {pages} {67} (\bibinfo {year} {1996})}\BibitemShut {NoStop}%
\bibitem [{\citenamefont {{Doro}}\ \emph {et~al.}(2021)\citenamefont {{Doro}},
  \citenamefont {{S{\'a}nchez-Conde}},\ and\ \citenamefont
  {{H{\"u}tten}}}]{2021arXiv211101198D}%
  \BibitemOpen
  \bibfield  {author} {\bibinfo {author} {\bibfnamefont {M.}~\bibnamefont
  {{Doro}}}, \bibinfo {author} {\bibfnamefont {M.~A.}\ \bibnamefont
  {{S{\'a}nchez-Conde}}},\ and\ \bibinfo {author} {\bibfnamefont
  {M.}~\bibnamefont {{H{\"u}tten}}},\ }\href@noop {} {\bibfield  {journal}
  {\bibinfo  {journal} {\arxiv}\ } (\bibinfo {year} {2021})},\ \Eprint
  {https://arxiv.org/abs/2111.01198}  { 2111.01198 [astro-ph.HE]}\BibitemShut
  {NoStop}%
\bibitem [{\citenamefont {{Barrau}}(2000)}]{2000APh....12..269B}%
  \BibitemOpen
  \bibfield  {author} {\bibinfo {author} {\bibfnamefont {A.}~\bibnamefont
  {{Barrau}}},\ }\href {https://doi.org/10.1016/S0927-6505(99)00103-6}
  {\bibfield  {journal} {\bibinfo  {journal} {\APh}\ }\textbf {\bibinfo
  {volume} {12}},\ \bibinfo {pages} {269} (\bibinfo {year} {2000})},\ \Eprint
  {https://arxiv.org/abs/astro-ph/9907347}  { arXiv:astro-ph/9907347
  [astro-ph]}\BibitemShut {NoStop}%
\bibitem [{\citenamefont {{Greisen}}(1966)}]{1966PhRvL..16..748G}%
  \BibitemOpen
  \bibfield  {author} {\bibinfo {author} {\bibfnamefont {K.}~\bibnamefont
  {{Greisen}}},\ }\href {https://doi.org/10.1103/PhysRevLett.16.748} {\bibfield
   {journal} {\bibinfo  {journal} {\prl}\ }\textbf {\bibinfo {volume} {16}},\
  \bibinfo {pages} {748} (\bibinfo {year} {1966})}\BibitemShut {NoStop}%
\bibitem [{\citenamefont {{Zatsepin}}\ and\ \citenamefont
  {{Kuz'min}}(1966)}]{1966JETPL...4...78Z}%
  \BibitemOpen
  \bibfield  {author} {\bibinfo {author} {\bibfnamefont {G.~T.}\ \bibnamefont
  {{Zatsepin}}}\ and\ \bibinfo {author} {\bibfnamefont {V.~A.}\ \bibnamefont
  {{Kuz'min}}},\ }\href@noop {} {\bibfield  {journal} {\bibinfo  {journal}
  {\JETPL}\ }\textbf {\bibinfo {volume} {4}},\ \bibinfo {pages} {78} (\bibinfo
  {year} {1966})},\ \bibinfo {note} {[\ZhETF~\textbf{4}, 114
  (1966)]}\BibitemShut {NoStop}%
\bibitem [{\citenamefont {{Takeda}}\ \emph {et~al.}(1998)\citenamefont
  {{Takeda}} \emph {et~al.}}]{1998PhRvL..81.1163T}%
  \BibitemOpen
  \bibfield  {author} {\bibinfo {author} {\bibfnamefont {M.}~\bibnamefont
  {{Takeda}}} \emph {et~al.} (\bibinfo {collaboration} {AGASA}),\ }\href
  {https://doi.org/10.1103/PhysRevLett.81.1163} {\bibfield  {journal} {\bibinfo
   {journal} {\prl}\ }\textbf {\bibinfo {volume} {81}},\ \bibinfo {pages}
  {1163} (\bibinfo {year} {1998})},\ \Eprint
  {https://arxiv.org/abs/astro-ph/9807193}  { arXiv:astro-ph/9807193
  [astro-ph]}\BibitemShut {NoStop}%
\bibitem [{\citenamefont {{Baker}}\ and\ \citenamefont
  {{Thamm}}(2021)}]{2021arXiv210510506B}%
  \BibitemOpen
  \bibfield  {author} {\bibinfo {author} {\bibfnamefont {M.~J.}\ \bibnamefont
  {{Baker}}}\ and\ \bibinfo {author} {\bibfnamefont {A.}~\bibnamefont
  {{Thamm}}},\ }\href@noop {} {\bibfield  {journal} {\bibinfo  {journal}
  {\arxiv}\ } (\bibinfo {year} {2021})},\ \Eprint
  {https://arxiv.org/abs/2105.10506}  { arXiv:2105.10506 [hep-ph]}\BibitemShut
  {NoStop}%
\bibitem [{\citenamefont {{Jelley}}\ \emph {et~al.}(1977)\citenamefont
  {{Jelley}}, \citenamefont {{Baird}},\ and\ \citenamefont
  {{Omongain}}}]{1977Natur.267..499J}%
  \BibitemOpen
  \bibfield  {author} {\bibinfo {author} {\bibfnamefont {J.~V.}\ \bibnamefont
  {{Jelley}}}, \bibinfo {author} {\bibfnamefont {G.~A.}\ \bibnamefont
  {{Baird}}},\ and\ \bibinfo {author} {\bibfnamefont {E.}~\bibnamefont
  {{Omongain}}},\ }\href {https://doi.org/10.1038/267499a0} {\bibfield
  {journal} {\bibinfo  {journal} {\nat}\ }\textbf {\bibinfo {volume} {267}},\
  \bibinfo {pages} {499} (\bibinfo {year} {1977})}\BibitemShut {NoStop}%
\bibitem [{\citenamefont {{Von Korff}}\ \emph {et~al.}(2013)\citenamefont {{Von
  Korff}} \emph {et~al.}}]{2013ApJ...767...40V}%
  \BibitemOpen
  \bibfield  {author} {\bibinfo {author} {\bibfnamefont {J.}~\bibnamefont {{Von
  Korff}}} \emph {et~al.},\ }\href {https://doi.org/10.1088/0004-637X/767/1/40}
  {\bibfield  {journal} {\bibinfo  {journal} {\apj}\ }\textbf {\bibinfo
  {volume} {767}},\ \bibinfo {eid} {40} (\bibinfo {year} {2013})},\ \Eprint
  {https://arxiv.org/abs/1211.1338}  { arXiv:1211.1338
  [astro-ph.IM]}\BibitemShut {NoStop}%
\bibitem [{\citenamefont {{Cline}}\ \emph
  {et~al.}(1999{\natexlab{b}})\citenamefont {{Cline}}, \citenamefont
  {{Matthey}},\ and\ \citenamefont {{Otwinowski}}}]{1999ApJ...527..827C}%
  \BibitemOpen
  \bibfield  {author} {\bibinfo {author} {\bibfnamefont {D.~B.}\ \bibnamefont
  {{Cline}}}, \bibinfo {author} {\bibfnamefont {C.}~\bibnamefont {{Matthey}}},\
  and\ \bibinfo {author} {\bibfnamefont {S.}~\bibnamefont {{Otwinowski}}},\
  }\href {https://doi.org/10.1086/308094} {\bibfield  {journal} {\bibinfo
  {journal} {\apj}\ }\textbf {\bibinfo {volume} {527}},\ \bibinfo {pages} {827}
  (\bibinfo {year} {1999}{\natexlab{b}})},\ \Eprint
  {https://arxiv.org/abs/astro-ph/9905346}  { arXiv:astro-ph/9905346
  [astro-ph]}\BibitemShut {NoStop}%
\bibitem [{\citenamefont {{Belyanin}}\ \emph
  {et~al.}(1998{\natexlab{a}})\citenamefont {{Belyanin}}, \citenamefont
  {{Kocharovsky}},\ and\ \citenamefont {{Kocharovsky
  v.}}}]{1998R&QE...41...22B}%
  \BibitemOpen
  \bibfield  {author} {\bibinfo {author} {\bibfnamefont {A.~A.}\ \bibnamefont
  {{Belyanin}}}, \bibinfo {author} {\bibfnamefont {V.~V.}\ \bibnamefont
  {{Kocharovsky}}},\ and\ \bibinfo {author} {\bibfnamefont {V.}~\bibnamefont
  {{Kocharovsky v.}}},\ }\href {https://doi.org/10.1007/BF02676709} {\bibfield
  {journal} {\bibinfo  {journal} {Radiophys.~Quantum Electron.}\ }\textbf
  {\bibinfo {volume} {41}},\ \bibinfo {pages} {22} (\bibinfo {year}
  {1998}{\natexlab{a}})}\BibitemShut {NoStop}%
\bibitem [{\citenamefont {{Belyanin}}\ \emph
  {et~al.}(1998{\natexlab{b}})\citenamefont {{Belyanin}}, \citenamefont
  {{Kocharovsky}},\ and\ \citenamefont {{Kocharovsky}}}]{1998AdSpR..22.1111B}%
  \BibitemOpen
  \bibfield  {author} {\bibinfo {author} {\bibfnamefont {A.~A.}\ \bibnamefont
  {{Belyanin}}}, \bibinfo {author} {\bibfnamefont {V.~V.}\ \bibnamefont
  {{Kocharovsky}}},\ and\ \bibinfo {author} {\bibfnamefont {V.~V.}\
  \bibnamefont {{Kocharovsky}}},\ }\href
  {https://doi.org/10.1016/S0273-1177(98)00204-X} {\bibfield  {journal}
  {\bibinfo  {journal} {\AdSpR}\ }\textbf {\bibinfo {volume} {22}},\ \bibinfo
  {pages} {1111} (\bibinfo {year} {1998}{\natexlab{b}})}\BibitemShut {NoStop}%
\bibitem [{\citenamefont {{Czerny}}\ \emph {et~al.}(2011)\citenamefont
  {{Czerny}}, \citenamefont {{Janiuk}}, \citenamefont {{Cline}},\ and\
  \citenamefont {{Otwinowski}}}]{2011NewA...16...33C}%
  \BibitemOpen
  \bibfield  {author} {\bibinfo {author} {\bibfnamefont {B.}~\bibnamefont
  {{Czerny}}}, \bibinfo {author} {\bibfnamefont {A.}~\bibnamefont {{Janiuk}}},
  \bibinfo {author} {\bibfnamefont {D.~B.}\ \bibnamefont {{Cline}}},\ and\
  \bibinfo {author} {\bibfnamefont {S.}~\bibnamefont {{Otwinowski}}},\ }\href
  {https://doi.org/10.1016/j.newast.2010.06.001} {\bibfield  {journal}
  {\bibinfo  {journal} {\na}\ }\textbf {\bibinfo {volume} {16}},\ \bibinfo
  {pages} {33} (\bibinfo {year} {2011})},\ \Eprint
  {https://arxiv.org/abs/1006.1470}  { arXiv:1006.1470
  [astro-ph.HE]}\BibitemShut {NoStop}%
\bibitem [{\citenamefont {{Schroedter}}\ \emph {et~al.}(2009)\citenamefont
  {{Schroedter}} \emph {et~al.}}]{2009APh....31..102S}%
  \BibitemOpen
  \bibfield  {author} {\bibinfo {author} {\bibfnamefont {M.}~\bibnamefont
  {{Schroedter}}} \emph {et~al.},\ }\href
  {https://doi.org/10.1016/j.astropartphys.2008.12.002} {\bibfield  {journal}
  {\bibinfo  {journal} {\APh}\ }\textbf {\bibinfo {volume} {31}},\ \bibinfo
  {pages} {102} (\bibinfo {year} {2009})},\ \Eprint
  {https://arxiv.org/abs/0812.0546}  { arXiv:0812.0546 [astro-ph]}\BibitemShut
  {NoStop}%
\bibitem [{\citenamefont {{Daghigh}}\ and\ \citenamefont
  {{Kapusta}}(2002)}]{2002PhRvD..65f4028D}%
  \BibitemOpen
  \bibfield  {author} {\bibinfo {author} {\bibfnamefont {R.}~\bibnamefont
  {{Daghigh}}}\ and\ \bibinfo {author} {\bibfnamefont {J.}~\bibnamefont
  {{Kapusta}}},\ }\href {https://doi.org/10.1103/PhysRevD.65.064028} {\bibfield
   {journal} {\bibinfo  {journal} {\prd}\ }\textbf {\bibinfo {volume} {65}},\
  \bibinfo {eid} {064028} (\bibinfo {year} {2002})},\ \Eprint
  {https://arxiv.org/abs/gr-qc/0109090}  { arXiv:gr-qc/0109090
  [gr-qc]}\BibitemShut {NoStop}%
\bibitem [{\citenamefont {{Porter}}\ and\ \citenamefont
  {{Weekes}}(1977)}]{1977ApJ...212..224P}%
  \BibitemOpen
  \bibfield  {author} {\bibinfo {author} {\bibfnamefont {N.~A.}\ \bibnamefont
  {{Porter}}}\ and\ \bibinfo {author} {\bibfnamefont {T.~C.}\ \bibnamefont
  {{Weekes}}},\ }\href {https://doi.org/10.1086/155040} {\bibfield  {journal}
  {\bibinfo  {journal} {\apj}\ }\textbf {\bibinfo {volume} {212}},\ \bibinfo
  {pages} {224} (\bibinfo {year} {1977})}\BibitemShut {NoStop}%
\bibitem [{\citenamefont {{Fegan}}\ \emph {et~al.}(1978)\citenamefont
  {{Fegan}}, \citenamefont {{McBreen}}, \citenamefont {{Obrien}},\ and\
  \citenamefont {{Osullivan}}}]{1978Natur.271..731F}%
  \BibitemOpen
  \bibfield  {author} {\bibinfo {author} {\bibfnamefont {D.~J.}\ \bibnamefont
  {{Fegan}}}, \bibinfo {author} {\bibfnamefont {B.}~\bibnamefont {{McBreen}}},
  \bibinfo {author} {\bibfnamefont {D.}~\bibnamefont {{Obrien}}},\ and\
  \bibinfo {author} {\bibfnamefont {C.}~\bibnamefont {{Osullivan}}},\ }\href
  {https://doi.org/10.1038/271731a0} {\bibfield  {journal} {\bibinfo  {journal}
  {\nat}\ }\textbf {\bibinfo {volume} {271}},\ \bibinfo {pages} {731} (\bibinfo
  {year} {1978})}\BibitemShut {NoStop}%
\bibitem [{\citenamefont {{Porter}}\ and\ \citenamefont
  {{Weekes}}(1979)}]{1979Natur.277..199P}%
  \BibitemOpen
  \bibfield  {author} {\bibinfo {author} {\bibfnamefont {N.~A.}\ \bibnamefont
  {{Porter}}}\ and\ \bibinfo {author} {\bibfnamefont {T.~C.}\ \bibnamefont
  {{Weekes}}},\ }\href {https://doi.org/10.1038/277199a0} {\bibfield  {journal}
  {\bibinfo  {journal} {\nat}\ }\textbf {\bibinfo {volume} {277}},\ \bibinfo
  {pages} {199} (\bibinfo {year} {1979})}\BibitemShut {NoStop}%
\bibitem [{\citenamefont {{Bhat}}\ \emph {et~al.}(1980)\citenamefont {{Bhat}}
  \emph {et~al.}}]{1980Natur.284..433B}%
  \BibitemOpen
  \bibfield  {author} {\bibinfo {author} {\bibfnamefont {P.~N.}\ \bibnamefont
  {{Bhat}}} \emph {et~al.},\ }\href {https://doi.org/10.1038/284433a0}
  {\bibfield  {journal} {\bibinfo  {journal} {\nat}\ }\textbf {\bibinfo
  {volume} {284}},\ \bibinfo {pages} {433} (\bibinfo {year}
  {1980})}\BibitemShut {NoStop}%
\bibitem [{\citenamefont {{Bhat}}\ \emph {et~al.}(1982)\citenamefont {{Bhat}}
  \emph {et~al.}}]{1982MNRAS.199.1007B}%
  \BibitemOpen
  \bibfield  {author} {\bibinfo {author} {\bibfnamefont {P.~N.}\ \bibnamefont
  {{Bhat}}} \emph {et~al.},\ }\href {https://doi.org/10.1093/mnras/199.4.1007}
  {\bibfield  {journal} {\bibinfo  {journal} {\mnras}\ }\textbf {\bibinfo
  {volume} {199}},\ \bibinfo {pages} {1007} (\bibinfo {year}
  {1982})}\BibitemShut {NoStop}%
\bibitem [{\citenamefont {{Linton}}\ \emph {et~al.}(2006)\citenamefont
  {{Linton}} \emph {et~al.}}]{2006JCAP...01..013L}%
  \BibitemOpen
  \bibfield  {author} {\bibinfo {author} {\bibfnamefont {E.~T.}\ \bibnamefont
  {{Linton}}} \emph {et~al.} (\bibinfo {collaboration} {Whipple}),\ }\href
  {https://doi.org/10.1088/1475-7516/2006/01/013} {\bibfield  {journal}
  {\bibinfo  {journal} {\jcap}\ }\textbf {\bibinfo {volume} {2006}},\ \bibinfo
  {eid} {013} (\bibinfo {year} {2006})}\BibitemShut {NoStop}%
\bibitem [{\citenamefont {{Petkov}}\ \emph {et~al.}(2008)\citenamefont
  {{Petkov}} \emph {et~al.}}]{2008AstL...34..509P}%
  \BibitemOpen
  \bibfield  {author} {\bibinfo {author} {\bibfnamefont {V.~B.}\ \bibnamefont
  {{Petkov}}} \emph {et~al.},\ }\href
  {https://doi.org/10.1134/S106377370808001X} {\bibfield  {journal} {\bibinfo
  {journal} {\astl}\ }\textbf {\bibinfo {volume} {34}},\ \bibinfo {pages} {509}
  (\bibinfo {year} {2008})},\ \Eprint {https://arxiv.org/abs/0808.3093}  {
  arXiv:0808.3093 [astro-ph]}\BibitemShut {NoStop}%
\bibitem [{\citenamefont {{Cassanyes}}(2015)}]{Cassanyes:2015wpr}%
  \BibitemOpen
  \bibfield  {author} {\ }\bibinfo {author} {\bibfnamefont {M.}~\bibnamefont
  {{Cassanyes}}},\ \href@noop {} {Master's thesis} (\bibinfo {year}
  {2015})\BibitemShut {NoStop}%
\bibitem [{\citenamefont {{Archambault}}(2017)}]{2017ICRC...35..691A}%
  \BibitemOpen
  \bibfield  {author} {\bibinfo {author} {\bibfnamefont {S.}~\bibnamefont
  {{Archambault}}} (\bibinfo {collaboration} {VERITAS}),\ }in\ \href@noop {}
  {\emph {\bibinfo {booktitle} {35th International Cosmic Ray Conference
  (ICRC2017)}}},\ \bibinfo {series} {Proc.~Sci.}, Vol.\ \bibinfo {volume}
  {301}\ (\bibinfo {year} {2017})\ p.\ \bibinfo {pages} {691},\ \Eprint
  {https://arxiv.org/abs/1709.00307} {arXiv:1709.00307 [astro-ph.HE]}
  \BibitemShut {NoStop}%
\bibitem [{\citenamefont {{Kumar}}(2019)}]{2019ICRC...36..719K}%
  \BibitemOpen
  \bibfield  {author} {\bibinfo {author} {\bibfnamefont {S.}~\bibnamefont
  {{Kumar}}} (\bibinfo {collaboration} {VERITAS}),\ }in\ \href@noop {} {\emph
  {\bibinfo {booktitle} {36th International Cosmic Ray Conference
  (ICRC2019)}}},\ \bibinfo {series} {Proc.~Sci.}, Vol.~\bibinfo {volume} {36}\
  (\bibinfo {year} {2019})\ p.\ \bibinfo {pages} {719},\ \Eprint
  {https://arxiv.org/abs/1909.01171} {arXiv:1909.01171 [astro-ph.HE]}
  \BibitemShut {NoStop}%
\bibitem [{\citenamefont {{Tavernier}}\ \emph {et~al.}(2021)\citenamefont
  {{Tavernier}}, \citenamefont {{Brun}}, \citenamefont {{Glicenstein}},\ and\
  \citenamefont {{Marandon}}}]{HESS:2021rto}%
  \BibitemOpen
  \bibfield  {author} {\bibinfo {author} {\bibfnamefont {T.}~\bibnamefont
  {{Tavernier}}}, \bibinfo {author} {\bibfnamefont {F.}~\bibnamefont {{Brun}}},
  \bibinfo {author} {\bibfnamefont {J.~F.}\ \bibnamefont {{Glicenstein}}},\
  and\ \bibinfo {author} {\bibfnamefont {V.}~\bibnamefont {{Marandon}}}
  (\bibinfo {collaboration} {HESS}),\ }in\ \href
  {https://doi.org/10.22323/1.395.0518} {\emph {\bibinfo {booktitle} {37th
  International Cosmic Ray Conference (ICRC2021)}}},\ \bibinfo {series and
  number} {Proc.~Sci.}\ (\bibinfo {year} {2021})\ p.\ \bibinfo {pages}
  {518}\BibitemShut {NoStop}%
\bibitem [{\citenamefont {{Abdo}}\ \emph {et~al.}(2015)\citenamefont {{Abdo}}
  \emph {et~al.}}]{2015APh....64....4A}%
  \BibitemOpen
  \bibfield  {author} {\bibinfo {author} {\bibfnamefont {A.~A.}\ \bibnamefont
  {{Abdo}}} \emph {et~al.} (\bibinfo {collaboration} {Milagro}),\ }\href
  {https://doi.org/10.1016/j.astropartphys.2014.10.007} {\bibfield  {journal}
  {\bibinfo  {journal} {Astroparticle Physics}\ }\textbf {\bibinfo {volume}
  {64}},\ \bibinfo {pages} {4} (\bibinfo {year} {2015})},\ \Eprint
  {https://arxiv.org/abs/1407.1686}  { arXiv:1407.1686
  [astro-ph.HE]}\BibitemShut {NoStop}%
\bibitem [{\citenamefont {{Albert}}\ \emph {et~al.}(2020)\citenamefont
  {{Albert}} \emph {et~al.}}]{2020JCAP...04..026A}%
  \BibitemOpen
  \bibfield  {author} {\bibinfo {author} {\bibfnamefont {A.}~\bibnamefont
  {{Albert}}} \emph {et~al.} (\bibinfo {collaboration} {HAWC}),\ }\href
  {https://doi.org/10.1088/1475-7516/2020/04/026} {\bibfield  {journal}
  {\bibinfo  {journal} {\jcap}\ }\textbf {\bibinfo {volume} {2020}},\ \bibinfo
  {eid} {026} (\bibinfo {year} {2020})},\ \Eprint
  {https://arxiv.org/abs/1911.04356}  { arXiv:1911.04356
  [astro-ph.HE]}\BibitemShut {NoStop}%
\bibitem [{\citenamefont {{L{\'o}pez-Coto}}\ \emph {et~al.}(2021)\citenamefont
  {{L{\'o}pez-Coto}}, \citenamefont {{Doro}}, \citenamefont {{de Angelis}},
  \citenamefont {{Mariotti}},\ and\ \citenamefont
  {{Harding}}}]{2021JCAP...08..040L}%
  \BibitemOpen
  \bibfield  {author} {\bibinfo {author} {\bibfnamefont {R.}~\bibnamefont
  {{L{\'o}pez-Coto}}}, \bibinfo {author} {\bibfnamefont {M.}~\bibnamefont
  {{Doro}}}, \bibinfo {author} {\bibfnamefont {A.}~\bibnamefont {{de
  Angelis}}}, \bibinfo {author} {\bibfnamefont {M.}~\bibnamefont
  {{Mariotti}}},\ and\ \bibinfo {author} {\bibfnamefont {J.~P.}\ \bibnamefont
  {{Harding}}},\ }\href {https://doi.org/10.1088/1475-7516/2021/08/040}
  {\bibfield  {journal} {\bibinfo  {journal} {\jcap}\ }\textbf {\bibinfo
  {volume} {2021}},\ \bibinfo {eid} {040} (\bibinfo {year} {2021})},\ \Eprint
  {https://arxiv.org/abs/2103.16895}  { arXiv:2103.16895
  [astro-ph.HE]}\BibitemShut {NoStop}%
\bibitem [{\citenamefont {{Ackermann}}\ \emph {et~al.}(2018)\citenamefont
  {{Ackermann}} \emph {et~al.}}]{2018ApJ...857...49A}%
  \BibitemOpen
  \bibfield  {author} {\bibinfo {author} {\bibfnamefont {M.}~\bibnamefont
  {{Ackermann}}} \emph {et~al.},\ }\href
  {https://doi.org/10.3847/1538-4357/aaac7b} {\bibfield  {journal} {\bibinfo
  {journal} {\apj}\ }\textbf {\bibinfo {volume} {857}},\ \bibinfo {eid} {49}
  (\bibinfo {year} {2018})}\BibitemShut {NoStop}%
\bibitem [{\citenamefont {{Smith}}\ \emph {et~al.}(2013)\citenamefont {{Smith}}
  \emph {et~al.}}]{2013APh....45...56S}%
  \BibitemOpen
  \bibfield  {author} {\bibinfo {author} {\bibfnamefont {M.~W.~E.}\
  \bibnamefont {{Smith}}} \emph {et~al.},\ }\href
  {https://doi.org/10.1016/j.astropartphys.2013.03.003} {\bibfield  {journal}
  {\bibinfo  {journal} {\APh}\ }\textbf {\bibinfo {volume} {45}},\ \bibinfo
  {pages} {56} (\bibinfo {year} {2013})},\ \Eprint
  {https://arxiv.org/abs/1211.5602}  { arXiv:1211.5602
  [astro-ph.HE]}\BibitemShut {NoStop}%
\bibitem [{\citenamefont {{Te{\v{s}}i{\'c}}}(2015)}]{2015ICRC...34..328T}%
  \BibitemOpen
  \bibfield  {author} {\bibinfo {author} {\bibfnamefont {G.}~\bibnamefont
  {{Te{\v{s}}i{\'c}}}},\ }in\ \href {https://doi.org/10.22323/1.236.0328}
  {\emph {\bibinfo {booktitle} {34th International Cosmic Ray Conference
  (ICRC2015)}}},\ \bibinfo {series} {Proc.~Sci.}, Vol.~\bibinfo {volume} {34}\
  (\bibinfo {year} {2015})\ p.\ \bibinfo {pages} {328}\BibitemShut {NoStop}%
\bibitem [{\citenamefont {{Petkov}}\ \emph {et~al.}(2019)\citenamefont
  {{Petkov}}, \citenamefont {{Bugaev}},\ and\ \citenamefont
  {{Klimai}}}]{2019arXiv191201317P}%
  \BibitemOpen
  \bibfield  {author} {\bibinfo {author} {\bibfnamefont {V.~B.}\ \bibnamefont
  {{Petkov}}}, \bibinfo {author} {\bibfnamefont {E.~V.}\ \bibnamefont
  {{Bugaev}}},\ and\ \bibinfo {author} {\bibfnamefont {P.~A.}\ \bibnamefont
  {{Klimai}}},\ }\href@noop {} {\bibfield  {journal} {\bibinfo  {journal}
  {\arxiv}\ } (\bibinfo {year} {2019})},\ \Eprint
  {https://arxiv.org/abs/1912.01317}  { 1912.01317 [astro-ph.HE]}\BibitemShut
  {NoStop}%
\bibitem [{\citenamefont {{Dave}}\ and\ \citenamefont
  {{Taboada}}(2019)}]{2019ICRC...36..863D}%
  \BibitemOpen
  \bibfield  {author} {\bibinfo {author} {\bibfnamefont {P.}~\bibnamefont
  {{Dave}}}\ and\ \bibinfo {author} {\bibfnamefont {I.}~\bibnamefont
  {{Taboada}}} (\bibinfo {collaboration} {IceCube}),\ }in\ \href@noop {} {\emph
  {\bibinfo {booktitle} {36th International Cosmic Ray Conference
  (ICRC2019)}}},\ \bibinfo {series} {Proc.~Sci.}, Vol.~\bibinfo {volume} {36}\
  (\bibinfo {year} {2019})\ p.\ \bibinfo {pages} {863},\ \Eprint
  {https://arxiv.org/abs/1908.05403} {arXiv:1908.05403 [astro-ph.HE]}
  \BibitemShut {NoStop}%
\bibitem [{\citenamefont {{Trofimenko}}(1990)}]{1990Ap&SS.168..277T}%
  \BibitemOpen
  \bibfield  {author} {\bibinfo {author} {\bibfnamefont {A.~P.}\ \bibnamefont
  {{Trofimenko}}},\ }\href {https://doi.org/10.1007/BF00636873} {\bibfield
  {journal} {\bibinfo  {journal} {\apss}\ }\textbf {\bibinfo {volume} {168}},\
  \bibinfo {pages} {277} (\bibinfo {year} {1990})}\BibitemShut {NoStop}%
\bibitem [{\citenamefont {{Dokuchaev}}\ and\ \citenamefont
  {{Volkova}}(1995)}]{1995SSRv...74..467D}%
  \BibitemOpen
  \bibfield  {author} {\bibinfo {author} {\bibfnamefont {V.~I.}\ \bibnamefont
  {{Dokuchaev}}}\ and\ \bibinfo {author} {\bibfnamefont {L.~V.}\ \bibnamefont
  {{Volkova}}},\ }\href {https://doi.org/10.1007/BF00751434} {\bibfield
  {journal} {\bibinfo  {journal} {\ssr}\ }\textbf {\bibinfo {volume} {74}},\
  \bibinfo {pages} {467} (\bibinfo {year} {1995})}\BibitemShut {NoStop}%
\bibitem [{\citenamefont {{Acevedo}}\ \emph {et~al.}(2021)\citenamefont
  {{Acevedo}}, \citenamefont {{Bramante}}, \citenamefont {{Goodman}},
  \citenamefont {{Kopp}},\ and\ \citenamefont
  {{Opferkuch}}}]{2021JCAP...04..026A}%
  \BibitemOpen
  \bibfield  {author} {\bibinfo {author} {\bibfnamefont {J.~F.}\ \bibnamefont
  {{Acevedo}}}, \bibinfo {author} {\bibfnamefont {J.}~\bibnamefont
  {{Bramante}}}, \bibinfo {author} {\bibfnamefont {A.}~\bibnamefont
  {{Goodman}}}, \bibinfo {author} {\bibfnamefont {J.}~\bibnamefont {{Kopp}}},\
  and\ \bibinfo {author} {\bibfnamefont {T.}~\bibnamefont {{Opferkuch}}},\
  }\href {https://doi.org/10.1088/1475-7516/2021/04/026} {\bibfield  {journal}
  {\bibinfo  {journal} {\jcap}\ }\textbf {\bibinfo {volume} {2021}},\ \bibinfo
  {eid} {026} (\bibinfo {year} {2021})},\ \Eprint
  {https://arxiv.org/abs/2012.09176}  { arXiv:2012.09176 [hep-ph]}\BibitemShut
  {NoStop}%
\bibitem [{\citenamefont {{Scholtz}}\ and\ \citenamefont
  {{Unwin}}(2020)}]{2020PhRvL.125e1103S}%
  \BibitemOpen
  \bibfield  {author} {\bibinfo {author} {\bibfnamefont {J.}~\bibnamefont
  {{Scholtz}}}\ and\ \bibinfo {author} {\bibfnamefont {J.}~\bibnamefont
  {{Unwin}}},\ }\href {https://doi.org/10.1103/PhysRevLett.125.051103}
  {\bibfield  {journal} {\bibinfo  {journal} {\prl}\ }\textbf {\bibinfo
  {volume} {125}},\ \bibinfo {eid} {051103} (\bibinfo {year} {2020})},\ \Eprint
  {https://arxiv.org/abs/1909.11090}  { arXiv:1909.11090 [hep-ph]}\BibitemShut
  {NoStop}%
\bibitem [{\citenamefont {{Arbey}}\ and\ \citenamefont
  {{Auffinger}}(2020)}]{2020arXiv200602944A}%
  \BibitemOpen
  \bibfield  {author} {\bibinfo {author} {\bibfnamefont {A.}~\bibnamefont
  {{Arbey}}}\ and\ \bibinfo {author} {\bibfnamefont {J.}~\bibnamefont
  {{Auffinger}}},\ }\href@noop {} {\bibfield  {journal} {\bibinfo  {journal}
  {\arxiv}\ } (\bibinfo {year} {2020})},\ \Eprint
  {https://arxiv.org/abs/2006.02944}  { arXiv:2006.02944 [gr-qc]}\BibitemShut
  {NoStop}%
\bibitem [{\citenamefont {{Page}}\ \emph {et~al.}(2008)\citenamefont {{Page}},
  \citenamefont {{Carr}},\ and\ \citenamefont
  {{MacGibbon}}}]{2008PhRvD..78f4044P}%
  \BibitemOpen
  \bibfield  {author} {\bibinfo {author} {\bibfnamefont {D.~N.}\ \bibnamefont
  {{Page}}}, \bibinfo {author} {\bibfnamefont {B.~J.}\ \bibnamefont {{Carr}}},\
  and\ \bibinfo {author} {\bibfnamefont {J.~H.}\ \bibnamefont {{MacGibbon}}},\
  }\href {https://doi.org/10.1103/PhysRevD.78.064044} {\bibfield  {journal}
  {\bibinfo  {journal} {\prd}\ }\textbf {\bibinfo {volume} {78}},\ \bibinfo
  {eid} {064044} (\bibinfo {year} {2008})},\ \Eprint
  {https://arxiv.org/abs/0709.2381}  { arXiv:0709.2381 [astro-ph]}\BibitemShut
  {NoStop}%
\bibitem [{\citenamefont {{Sobrinho}}\ and\ \citenamefont
  {{Augusto}}(2014)}]{2014MNRAS.441.2878S}%
  \BibitemOpen
  \bibfield  {author} {\bibinfo {author} {\bibfnamefont {J.~L.~G.}\
  \bibnamefont {{Sobrinho}}}\ and\ \bibinfo {author} {\bibfnamefont
  {P.}~\bibnamefont {{Augusto}}},\ }\href
  {https://doi.org/10.1093/mnras/stu786} {\bibfield  {journal} {\bibinfo
  {journal} {\mnras}\ }\textbf {\bibinfo {volume} {441}},\ \bibinfo {pages}
  {2878} (\bibinfo {year} {2014})},\ \Eprint {https://arxiv.org/abs/1406.1785}
  { arXiv:1406.1785 [astro-ph.CO]}\BibitemShut {NoStop}%
\bibitem [{\citenamefont {{Nambu}}\ and\ \citenamefont
  {{Noda}}(2022)}]{2022PhRvD.105d5022N}%
  \BibitemOpen
  \bibfield  {author} {\bibinfo {author} {\bibfnamefont {Y.}~\bibnamefont
  {{Nambu}}}\ and\ \bibinfo {author} {\bibfnamefont {S.}~\bibnamefont
  {{Noda}}},\ }\href {https://doi.org/10.1103/PhysRevD.105.045022} {\bibfield
  {journal} {\bibinfo  {journal} {\prd}\ }\textbf {\bibinfo {volume} {105}},\
  \bibinfo {eid} {045022} (\bibinfo {year} {2022})},\ \Eprint
  {https://arxiv.org/abs/2109.07044}  { arXiv:2109.07044 [gr-qc]}\BibitemShut
  {NoStop}%
\bibitem [{\citenamefont {{Geringer-Sameth}}\ and\ \citenamefont
  {{Koushiappas}}(2012)}]{2012MNRAS.425..862G}%
  \BibitemOpen
  \bibfield  {author} {\bibinfo {author} {\bibfnamefont {A.}~\bibnamefont
  {{Geringer-Sameth}}}\ and\ \bibinfo {author} {\bibfnamefont {S.~M.}\
  \bibnamefont {{Koushiappas}}},\ }\href
  {https://doi.org/10.1111/j.1365-2966.2012.21139.x} {\bibfield  {journal}
  {\bibinfo  {journal} {\mnras}\ }\textbf {\bibinfo {volume} {425}},\ \bibinfo
  {pages} {862} (\bibinfo {year} {2012})},\ \Eprint
  {https://arxiv.org/abs/1012.1873}  { arXiv:1012.1873
  [astro-ph.HE]}\BibitemShut {NoStop}%
\bibitem [{\citenamefont {{Green}}(2001)}]{2001PhRvD..65b7301G}%
  \BibitemOpen
  \bibfield  {author} {\bibinfo {author} {\bibfnamefont {A.~M.}\ \bibnamefont
  {{Green}}},\ }\href {https://doi.org/10.1103/PhysRevD.65.027301} {\bibfield
  {journal} {\bibinfo  {journal} {\prd}\ }\textbf {\bibinfo {volume} {65}},\
  \bibinfo {eid} {027301} (\bibinfo {year} {2001})},\ \Eprint
  {https://arxiv.org/abs/astro-ph/0105253}  { arXiv:astro-ph/0105253
  [astro-ph]}\BibitemShut {NoStop}%
\bibitem [{\citenamefont {{Cai}}\ \emph {et~al.}(2021)\citenamefont {{Cai}},
  \citenamefont {{Chen}}, \citenamefont {{Ding}},\ and\ \citenamefont
  {{Wang}}}]{2021arXiv211210422C}%
  \BibitemOpen
  \bibfield  {author} {\bibinfo {author} {\bibfnamefont {Y.-F.}\ \bibnamefont
  {{Cai}}}, \bibinfo {author} {\bibfnamefont {C.}~\bibnamefont {{Chen}}},
  \bibinfo {author} {\bibfnamefont {Q.}~\bibnamefont {{Ding}}},\ and\ \bibinfo
  {author} {\bibfnamefont {Y.}~\bibnamefont {{Wang}}},\ }\href@noop {}
  {\bibfield  {journal} {\bibinfo  {journal} {\arxiv}\ } (\bibinfo {year}
  {2021})},\ \Eprint {https://arxiv.org/abs/2112.10422}  { arXiv:2112.10422
  [astro-ph.CO]}\BibitemShut {NoStop}%
\bibitem [{\citenamefont {{Mosbech}}\ and\ \citenamefont
  {{Picker}}(2022)}]{2022arXiv220305743M}%
  \BibitemOpen
  \bibfield  {author} {\bibinfo {author} {\bibfnamefont {M.~R.}\ \bibnamefont
  {{Mosbech}}}\ and\ \bibinfo {author} {\bibfnamefont {Z.~S.~C.}\ \bibnamefont
  {{Picker}}},\ }\href@noop {} {\bibfield  {journal} {\bibinfo  {journal}
  {\arxiv}\ } (\bibinfo {year} {2022})},\ \Eprint
  {https://arxiv.org/abs/2203.05743}  { arXiv:2203.05743
  [astro-ph.HE]}\BibitemShut {NoStop}%
\bibitem [{\citenamefont {{Mukhopadhyay}}\ \emph {et~al.}(2022)\citenamefont
  {{Mukhopadhyay}}, \citenamefont {{Majumdar}},\ and\ \citenamefont
  {{Halder}}}]{2022arXiv220313008M}%
  \BibitemOpen
  \bibfield  {author} {\bibinfo {author} {\bibfnamefont {U.}~\bibnamefont
  {{Mukhopadhyay}}}, \bibinfo {author} {\bibfnamefont {D.}~\bibnamefont
  {{Majumdar}}},\ and\ \bibinfo {author} {\bibfnamefont {A.}~\bibnamefont
  {{Halder}}},\ }\href@noop {} {\bibfield  {journal} {\bibinfo  {journal}
  {\arxiv}\ } (\bibinfo {year} {2022})},\ \Eprint
  {https://arxiv.org/abs/2203.13008}  { arXiv:2203.13008
  [astro-ph.CO]}\BibitemShut {NoStop}%
\bibitem [{\citenamefont {{Lehmann}}\ \emph {et~al.}(2018)\citenamefont
  {{Lehmann}}, \citenamefont {{Profumo}},\ and\ \citenamefont
  {{Yant}}}]{2018JCAP...04..007L}%
  \BibitemOpen
  \bibfield  {author} {\bibinfo {author} {\bibfnamefont {B.~V.}\ \bibnamefont
  {{Lehmann}}}, \bibinfo {author} {\bibfnamefont {S.}~\bibnamefont
  {{Profumo}}},\ and\ \bibinfo {author} {\bibfnamefont {J.}~\bibnamefont
  {{Yant}}},\ }\href {https://doi.org/10.1088/1475-7516/2018/04/007} {\bibfield
   {journal} {\bibinfo  {journal} {\jcap}\ }\textbf {\bibinfo {volume}
  {2018}},\ \bibinfo {eid} {007} (\bibinfo {year} {2018})},\ \Eprint
  {https://arxiv.org/abs/1801.00808}  { arXiv:1801.00808
  [astro-ph.CO]}\BibitemShut {NoStop}%
\bibitem [{\citenamefont {{Chambers}}\ \emph {et~al.}(1997)\citenamefont
  {{Chambers}}, \citenamefont {{Hiscock}},\ and\ \citenamefont
  {{Taylor}}}]{1997PhRvL..78.3249C}%
  \BibitemOpen
  \bibfield  {author} {\bibinfo {author} {\bibfnamefont {C.~M.}\ \bibnamefont
  {{Chambers}}}, \bibinfo {author} {\bibfnamefont {W.~A.}\ \bibnamefont
  {{Hiscock}}},\ and\ \bibinfo {author} {\bibfnamefont {B.}~\bibnamefont
  {{Taylor}}},\ }\href {https://doi.org/10.1103/PhysRevLett.78.3249} {\bibfield
   {journal} {\bibinfo  {journal} {\prl}\ }\textbf {\bibinfo {volume} {78}},\
  \bibinfo {pages} {3249} (\bibinfo {year} {1997})},\ \Eprint
  {https://arxiv.org/abs/gr-qc/9703018}  { arXiv:gr-qc/9703018
  [gr-qc]}\BibitemShut {NoStop}%
\bibitem [{\citenamefont {{Taylor}}\ \emph {et~al.}(1998)\citenamefont
  {{Taylor}}, \citenamefont {{Chambers}},\ and\ \citenamefont
  {{Hiscock}}}]{1998PhRvD..58d4012T}%
  \BibitemOpen
  \bibfield  {author} {\bibinfo {author} {\bibfnamefont {B.~E.}\ \bibnamefont
  {{Taylor}}}, \bibinfo {author} {\bibfnamefont {C.~M.}\ \bibnamefont
  {{Chambers}}},\ and\ \bibinfo {author} {\bibfnamefont {W.~A.}\ \bibnamefont
  {{Hiscock}}},\ }\href {https://doi.org/10.1103/PhysRevD.58.044012} {\bibfield
   {journal} {\bibinfo  {journal} {\prd}\ }\textbf {\bibinfo {volume} {58}},\
  \bibinfo {eid} {044012} (\bibinfo {year} {1998})},\ \Eprint
  {https://arxiv.org/abs/gr-qc/9801044}  { arXiv:gr-qc/9801044
  [gr-qc]}\BibitemShut {NoStop}%
\bibitem [{\citenamefont {{Calz{\`a}}}\ \emph {et~al.}(2021)\citenamefont
  {{Calz{\`a}}}, \citenamefont {{March-Russell}},\ and\ \citenamefont
  {{Rosa}}}]{2021arXiv211013602C}%
  \BibitemOpen
  \bibfield  {author} {\bibinfo {author} {\bibfnamefont {M.}~\bibnamefont
  {{Calz{\`a}}}}, \bibinfo {author} {\bibfnamefont {J.}~\bibnamefont
  {{March-Russell}}},\ and\ \bibinfo {author} {\bibfnamefont {J.~G.}\
  \bibnamefont {{Rosa}}},\ }\href@noop {} {\bibfield  {journal} {\bibinfo
  {journal} {\arxiv}\ } (\bibinfo {year} {2021})},\ \Eprint
  {https://arxiv.org/abs/2110.13602}  { arXiv:2110.13602
  [astro-ph.CO]}\BibitemShut {NoStop}%
\bibitem [{\citenamefont {{Arbey}}\ \emph
  {et~al.}(2020{\natexlab{c}})\citenamefont {{Arbey}}, \citenamefont
  {{Auffinger}},\ and\ \citenamefont {{Silk}}}]{2020MNRAS.494.1257A}%
  \BibitemOpen
  \bibfield  {author} {\bibinfo {author} {\bibfnamefont {A.}~\bibnamefont
  {{Arbey}}}, \bibinfo {author} {\bibfnamefont {J.}~\bibnamefont
  {{Auffinger}}},\ and\ \bibinfo {author} {\bibfnamefont {J.}~\bibnamefont
  {{Silk}}},\ }\href {https://doi.org/10.1093/mnras/staa765} {\bibfield
  {journal} {\bibinfo  {journal} {\mnras}\ }\textbf {\bibinfo {volume} {494}},\
  \bibinfo {pages} {1257} (\bibinfo {year} {2020}{\natexlab{c}})},\ \Eprint
  {https://arxiv.org/abs/1906.04196}  { arXiv:1906.04196
  [astro-ph.CO]}\BibitemShut {NoStop}%
\bibitem [{\citenamefont {{Hofmann}}\ \emph {et~al.}(2016)\citenamefont
  {{Hofmann}}, \citenamefont {{Barausse}},\ and\ \citenamefont
  {{Rezzolla}}}]{2016ApJ...825L..19H}%
  \BibitemOpen
  \bibfield  {author} {\bibinfo {author} {\bibfnamefont {F.}~\bibnamefont
  {{Hofmann}}}, \bibinfo {author} {\bibfnamefont {E.}~\bibnamefont
  {{Barausse}}},\ and\ \bibinfo {author} {\bibfnamefont {L.}~\bibnamefont
  {{Rezzolla}}},\ }\href {https://doi.org/10.3847/2041-8205/825/2/L19}
  {\bibfield  {journal} {\bibinfo  {journal} {\apjl}\ }\textbf {\bibinfo
  {volume} {825}},\ \bibinfo {eid} {L19} (\bibinfo {year} {2016})},\ \Eprint
  {https://arxiv.org/abs/1605.01938}  { arXiv:1605.01938 [gr-qc]}\BibitemShut
  {NoStop}%
\bibitem [{\citenamefont {{Doctor}}\ \emph {et~al.}(2021)\citenamefont
  {{Doctor}}, \citenamefont {{Farr}},\ and\ \citenamefont
  {{Holz}}}]{2021ApJ...914L..18D}%
  \BibitemOpen
  \bibfield  {author} {\bibinfo {author} {\bibfnamefont {Z.}~\bibnamefont
  {{Doctor}}}, \bibinfo {author} {\bibfnamefont {B.}~\bibnamefont {{Farr}}},\
  and\ \bibinfo {author} {\bibfnamefont {D.~E.}\ \bibnamefont {{Holz}}},\
  }\href {https://doi.org/10.3847/2041-8213/ac0334} {\bibfield  {journal}
  {\bibinfo  {journal} {\apjl}\ }\textbf {\bibinfo {volume} {914}},\ \bibinfo
  {eid} {L18} (\bibinfo {year} {2021})},\ \Eprint
  {https://arxiv.org/abs/2103.04001}  { arXiv:2103.04001
  [astro-ph.HE]}\BibitemShut {NoStop}%
\bibitem [{\citenamefont {{Randall}}\ and\ \citenamefont
  {{Sundrum}}(1999{\natexlab{a}})}]{1999PhRvL..83.3370R}%
  \BibitemOpen
  \bibfield  {author} {\bibinfo {author} {\bibfnamefont {L.}~\bibnamefont
  {{Randall}}}\ and\ \bibinfo {author} {\bibfnamefont {R.}~\bibnamefont
  {{Sundrum}}},\ }\href {https://doi.org/10.1103/PhysRevLett.83.3370}
  {\bibfield  {journal} {\bibinfo  {journal} {\prl}\ }\textbf {\bibinfo
  {volume} {83}},\ \bibinfo {pages} {3370} (\bibinfo {year}
  {1999}{\natexlab{a}})},\ \Eprint {https://arxiv.org/abs/hep-ph/9905221}  {
  arXiv:hep-ph/9905221 [hep-ph]}\BibitemShut {NoStop}%
\bibitem [{\citenamefont {{Randall}}\ and\ \citenamefont
  {{Sundrum}}(1999{\natexlab{b}})}]{1999PhRvL..83.4690R}%
  \BibitemOpen
  \bibfield  {author} {\bibinfo {author} {\bibfnamefont {L.}~\bibnamefont
  {{Randall}}}\ and\ \bibinfo {author} {\bibfnamefont {R.}~\bibnamefont
  {{Sundrum}}},\ }\href {https://doi.org/10.1103/PhysRevLett.83.4690}
  {\bibfield  {journal} {\bibinfo  {journal} {\prl}\ }\textbf {\bibinfo
  {volume} {83}},\ \bibinfo {pages} {4690} (\bibinfo {year}
  {1999}{\natexlab{b}})},\ \Eprint {https://arxiv.org/abs/hep-th/9906064}  {
  arXiv:hep-th/9906064 [hep-th]}\BibitemShut {NoStop}%
\bibitem [{\citenamefont {{Arkani-Hamed}}\ \emph {et~al.}(1998)\citenamefont
  {{Arkani-Hamed}}, \citenamefont {{Dimopoulos}},\ and\ \citenamefont
  {{Dvali}}}]{1998PhLB..429..263A}%
  \BibitemOpen
  \bibfield  {author} {\bibinfo {author} {\bibfnamefont {N.}~\bibnamefont
  {{Arkani-Hamed}}}, \bibinfo {author} {\bibfnamefont {S.}~\bibnamefont
  {{Dimopoulos}}},\ and\ \bibinfo {author} {\bibfnamefont {G.}~\bibnamefont
  {{Dvali}}},\ }\href {https://doi.org/10.1016/S0370-2693(98)00466-3}
  {\bibfield  {journal} {\bibinfo  {journal} {\plb}\ }\textbf {\bibinfo
  {volume} {429}},\ \bibinfo {pages} {263} (\bibinfo {year} {1998})},\ \Eprint
  {https://arxiv.org/abs/hep-ph/9803315}  { arXiv:hep-ph/9803315
  [hep-ph]}\BibitemShut {NoStop}%
\bibitem [{\citenamefont {{Antoniadis}}\ \emph {et~al.}(1998)\citenamefont
  {{Antoniadis}}, \citenamefont {{Arkani-Hamed}}, \citenamefont
  {{Dimopoulos}},\ and\ \citenamefont {{Dvali}}}]{1998PhLB..436..257A}%
  \BibitemOpen
  \bibfield  {author} {\bibinfo {author} {\bibfnamefont {I.}~\bibnamefont
  {{Antoniadis}}}, \bibinfo {author} {\bibfnamefont {N.}~\bibnamefont
  {{Arkani-Hamed}}}, \bibinfo {author} {\bibfnamefont {S.}~\bibnamefont
  {{Dimopoulos}}},\ and\ \bibinfo {author} {\bibfnamefont {G.}~\bibnamefont
  {{Dvali}}},\ }\href {https://doi.org/10.1016/S0370-2693(98)00860-0}
  {\bibfield  {journal} {\bibinfo  {journal} {\plb}\ }\textbf {\bibinfo
  {volume} {436}},\ \bibinfo {pages} {257} (\bibinfo {year} {1998})},\ \Eprint
  {https://arxiv.org/abs/hep-ph/9804398}  { arXiv:hep-ph/9804398
  [hep-ph]}\BibitemShut {NoStop}%
\bibitem [{\citenamefont {{Arkani-Hamed}}\ \emph {et~al.}(1999)\citenamefont
  {{Arkani-Hamed}}, \citenamefont {{Dimopoulos}},\ and\ \citenamefont
  {{Dvali}}}]{1999PhRvD..59h6004A}%
  \BibitemOpen
  \bibfield  {author} {\bibinfo {author} {\bibfnamefont {N.}~\bibnamefont
  {{Arkani-Hamed}}}, \bibinfo {author} {\bibfnamefont {S.}~\bibnamefont
  {{Dimopoulos}}},\ and\ \bibinfo {author} {\bibfnamefont {G.}~\bibnamefont
  {{Dvali}}},\ }\href {https://doi.org/10.1103/PhysRevD.59.086004} {\bibfield
  {journal} {\bibinfo  {journal} {\prd}\ }\textbf {\bibinfo {volume} {59}},\
  \bibinfo {eid} {086004} (\bibinfo {year} {1999})},\ \Eprint
  {https://arxiv.org/abs/hep-ph/9807344}  { arXiv:hep-ph/9807344
  [hep-ph]}\BibitemShut {NoStop}%
\bibitem [{\citenamefont {{Myers}}\ and\ \citenamefont
  {{Perry}}(1986)}]{1986AnPhy.172..304M}%
  \BibitemOpen
  \bibfield  {author} {\bibinfo {author} {\bibfnamefont {R.~C.}\ \bibnamefont
  {{Myers}}}\ and\ \bibinfo {author} {\bibfnamefont {M.~J.}\ \bibnamefont
  {{Perry}}},\ }\href {https://doi.org/10.1016/0003-4916(86)90186-7} {\bibfield
   {journal} {\bibinfo  {journal} {\AnPhy}\ }\textbf {\bibinfo {volume}
  {172}},\ \bibinfo {pages} {304} (\bibinfo {year} {1986})}\BibitemShut
  {NoStop}%
\bibitem [{\citenamefont {{Kanti}}(2004)}]{2004IJMPA..19.4899K}%
  \BibitemOpen
  \bibfield  {author} {\bibinfo {author} {\bibfnamefont {P.}~\bibnamefont
  {{Kanti}}},\ }\href {https://doi.org/10.1142/S0217751X04018324} {\bibfield
  {journal} {\bibinfo  {journal} {\IJMPA}\ }\textbf {\bibinfo {volume} {19}},\
  \bibinfo {pages} {4899} (\bibinfo {year} {2004})},\ \Eprint
  {https://arxiv.org/abs/hep-ph/0402168}  { arXiv:hep-ph/0402168
  [hep-ph]}\BibitemShut {NoStop}%
\bibitem [{\citenamefont {{Kanti}}\ and\ \citenamefont
  {{Winstanley}}(2015)}]{2015qabh.book..229K}%
  \BibitemOpen
  \bibfield  {author} {\bibinfo {author} {\bibfnamefont {P.}~\bibnamefont
  {{Kanti}}}\ and\ \bibinfo {author} {\bibfnamefont {E.}~\bibnamefont
  {{Winstanley}}},\ }\bibfield  {title} {\bibinfo {title} {{Hawking Radiation
  from Higher-Dimensional Black Holes}},\ }in\ \href
  {https://doi.org/10.1007/978-3-319-10852-0\_8} {\emph {\bibinfo {booktitle}
  {Quantum Aspects of Black Holes}}}\ (\bibinfo {year} {2015})\ pp.\ \bibinfo
  {pages} {229--265},\ \Eprint {https://arxiv.org/abs/1402.3952}
  {arXiv:1402.3952 [hep-th]} \BibitemShut {NoStop}%
\bibitem [{\citenamefont {{Kanti}}(2009)}]{2009LNP...769..387K}%
  \BibitemOpen
  \bibfield  {author} {\bibinfo {author} {\bibfnamefont {P.}~\bibnamefont
  {{Kanti}}},\ }\bibfield  {title} {\bibinfo {title} {{Black Holes at the Large
  Hadron Collider}},\ }in\ \href
  {https://doi.org/10.1007/978-3-540-88460-6\_10} {\emph {\bibinfo {booktitle}
  {Physics of Black Holes}}},\ Vol.\ \bibinfo {volume} {769}\ (\bibinfo {year}
  {2009})\ p.\ \bibinfo {pages} {387},\ \Eprint
  {https://arxiv.org/abs/0802.2218} {arXiv:0802.2218 [hep-th]} \BibitemShut
  {NoStop}%
\bibitem [{\citenamefont {{Aliferis}}\ \emph {et~al.}(2015)\citenamefont
  {{Aliferis}}, \citenamefont {{Kofinas}},\ and\ \citenamefont
  {{Zarikas}}}]{2015PhRvD..91d5002A}%
  \BibitemOpen
  \bibfield  {author} {\bibinfo {author} {\bibfnamefont {G.}~\bibnamefont
  {{Aliferis}}}, \bibinfo {author} {\bibfnamefont {G.}~\bibnamefont
  {{Kofinas}}},\ and\ \bibinfo {author} {\bibfnamefont {V.}~\bibnamefont
  {{Zarikas}}},\ }\href {https://doi.org/10.1103/PhysRevD.91.045002} {\bibfield
   {journal} {\bibinfo  {journal} {\prd}\ }\textbf {\bibinfo {volume} {91}},\
  \bibinfo {eid} {045002} (\bibinfo {year} {2015})},\ \Eprint
  {https://arxiv.org/abs/1406.6215}  { arXiv:1406.6215 [hep-ph]}\BibitemShut
  {NoStop}%
\bibitem [{\citenamefont {{Aliferis}}\ and\ \citenamefont
  {{Zarikas}}(2021)}]{2021PhRvD.103b3509A}%
  \BibitemOpen
  \bibfield  {author} {\bibinfo {author} {\bibfnamefont {G.}~\bibnamefont
  {{Aliferis}}}\ and\ \bibinfo {author} {\bibfnamefont {V.}~\bibnamefont
  {{Zarikas}}},\ }\href {https://doi.org/10.1103/PhysRevD.103.023509}
  {\bibfield  {journal} {\bibinfo  {journal} {\prd}\ }\textbf {\bibinfo
  {volume} {103}},\ \bibinfo {eid} {023509} (\bibinfo {year} {2021})},\ \Eprint
  {https://arxiv.org/abs/2006.13621}  { arXiv:2006.13621 [gr-qc]}\BibitemShut
  {NoStop}%
\bibitem [{\citenamefont {{Clancy}}\ \emph {et~al.}(2003)\citenamefont
  {{Clancy}}, \citenamefont {{Guedens}},\ and\ \citenamefont
  {{Liddle}}}]{2003PhRvD..68b3507C}%
  \BibitemOpen
  \bibfield  {author} {\bibinfo {author} {\bibfnamefont {D.}~\bibnamefont
  {{Clancy}}}, \bibinfo {author} {\bibfnamefont {R.}~\bibnamefont
  {{Guedens}}},\ and\ \bibinfo {author} {\bibfnamefont {A.~R.}\ \bibnamefont
  {{Liddle}}},\ }\href {https://doi.org/10.1103/PhysRevD.68.023507} {\bibfield
  {journal} {\bibinfo  {journal} {\prd}\ }\textbf {\bibinfo {volume} {68}},\
  \bibinfo {eid} {023507} (\bibinfo {year} {2003})},\ \Eprint
  {https://arxiv.org/abs/astro-ph/0301568}  { arXiv:astro-ph/0301568
  [astro-ph]}\BibitemShut {NoStop}%
\bibitem [{\citenamefont {{Sendouda}}\ \emph {et~al.}(2003)\citenamefont
  {{Sendouda}}, \citenamefont {{Nagataki}},\ and\ \citenamefont
  {{Sato}}}]{2003PhRvD..68j3510S}%
  \BibitemOpen
  \bibfield  {author} {\bibinfo {author} {\bibfnamefont {Y.}~\bibnamefont
  {{Sendouda}}}, \bibinfo {author} {\bibfnamefont {S.}~\bibnamefont
  {{Nagataki}}},\ and\ \bibinfo {author} {\bibfnamefont {K.}~\bibnamefont
  {{Sato}}},\ }\href {https://doi.org/10.1103/PhysRevD.68.103510} {\bibfield
  {journal} {\bibinfo  {journal} {\prd}\ }\textbf {\bibinfo {volume} {68}},\
  \bibinfo {eid} {103510} (\bibinfo {year} {2003})},\ \Eprint
  {https://arxiv.org/abs/astro-ph/0309170}  { arXiv:astro-ph/0309170
  [astro-ph]}\BibitemShut {NoStop}%
\bibitem [{\citenamefont {{Sendouda}}\ \emph {et~al.}(2005)\citenamefont
  {{Sendouda}}, \citenamefont {{Kohri}}, \citenamefont {{Nagataki}},\ and\
  \citenamefont {{Sato}}}]{2005PhRvD..71f3512S}%
  \BibitemOpen
  \bibfield  {author} {\bibinfo {author} {\bibfnamefont {Y.}~\bibnamefont
  {{Sendouda}}}, \bibinfo {author} {\bibfnamefont {K.}~\bibnamefont {{Kohri}}},
  \bibinfo {author} {\bibfnamefont {S.}~\bibnamefont {{Nagataki}}},\ and\
  \bibinfo {author} {\bibfnamefont {K.}~\bibnamefont {{Sato}}},\ }\href
  {https://doi.org/10.1103/PhysRevD.71.063512} {\bibfield  {journal} {\bibinfo
  {journal} {\prd}\ }\textbf {\bibinfo {volume} {71}},\ \bibinfo {eid} {063512}
  (\bibinfo {year} {2005})},\ \Eprint {https://arxiv.org/abs/astro-ph/0408369}
  { arXiv:astro-ph/0408369 [astro-ph]}\BibitemShut {NoStop}%
\bibitem [{\citenamefont {{Jain}}\ and\ \citenamefont
  {{Panda}}(2005)}]{2005ICRC....9...33J}%
  \BibitemOpen
  \bibfield  {author} {\bibinfo {author} {\bibfnamefont {P.}~\bibnamefont
  {{Jain}}}\ and\ \bibinfo {author} {\bibfnamefont {S.}~\bibnamefont
  {{Panda}}},\ }in\ \href@noop {} {\emph {\bibinfo {booktitle} {29th
  International Cosmic Ray Conference (ICRC29), Volume 9}}},\ \bibinfo {series}
  {Proc.~Sci.}, Vol.~\bibinfo {volume} {9}\ (\bibinfo {year} {2005})\
  p.~\bibinfo {pages} {33},\ \Eprint {https://arxiv.org/abs/astro-ph/0509324}
  {arXiv:astro-ph/0509324 [astro-ph]} \BibitemShut {NoStop}%
\bibitem [{\citenamefont {{Kavic}}\ \emph {et~al.}(2008)\citenamefont
  {{Kavic}}, \citenamefont {{Simonetti}}, \citenamefont {{Cutchin}},
  \citenamefont {{Ellingson}},\ and\ \citenamefont
  {{Patterson}}}]{2008JCAP...11..017K}%
  \BibitemOpen
  \bibfield  {author} {\bibinfo {author} {\bibfnamefont {M.}~\bibnamefont
  {{Kavic}}}, \bibinfo {author} {\bibfnamefont {J.~H.}\ \bibnamefont
  {{Simonetti}}}, \bibinfo {author} {\bibfnamefont {S.~E.}\ \bibnamefont
  {{Cutchin}}}, \bibinfo {author} {\bibfnamefont {S.~W.}\ \bibnamefont
  {{Ellingson}}},\ and\ \bibinfo {author} {\bibfnamefont {C.~D.}\ \bibnamefont
  {{Patterson}}},\ }\href {https://doi.org/10.1088/1475-7516/2008/11/017}
  {\bibfield  {journal} {\bibinfo  {journal} {\jcap}\ }\textbf {\bibinfo
  {volume} {2008}},\ \bibinfo {eid} {017} (\bibinfo {year} {2008})},\ \Eprint
  {https://arxiv.org/abs/0801.4023}  { arXiv:0801.4023 [astro-ph]}\BibitemShut
  {NoStop}%
\bibitem [{\citenamefont {{Alexeyev}}\ \emph {et~al.}(2015)\citenamefont
  {{Alexeyev}}, \citenamefont {{Rannu}}, \citenamefont {{Dyadina}},
  \citenamefont {{Latosh}},\ and\ \citenamefont
  {{Turyshev}}}]{2015JETP..120..966A}%
  \BibitemOpen
  \bibfield  {author} {\bibinfo {author} {\bibfnamefont {S.~O.}\ \bibnamefont
  {{Alexeyev}}}, \bibinfo {author} {\bibfnamefont {K.~A.}\ \bibnamefont
  {{Rannu}}}, \bibinfo {author} {\bibfnamefont {P.~I.}\ \bibnamefont
  {{Dyadina}}}, \bibinfo {author} {\bibfnamefont {B.~N.}\ \bibnamefont
  {{Latosh}}},\ and\ \bibinfo {author} {\bibfnamefont {S.~G.}\ \bibnamefont
  {{Turyshev}}},\ }\href {https://doi.org/10.1134/S1063776115060011} {\bibfield
   {journal} {\bibinfo  {journal} {Soviet Journal of Experimental and
  Theoretical Physics}\ }\textbf {\bibinfo {volume} {120}},\ \bibinfo {pages}
  {966} (\bibinfo {year} {2015})},\ \Eprint {https://arxiv.org/abs/1501.04217}
  { arXiv:1501.04217 [gr-qc]}\BibitemShut {NoStop}%
\bibitem [{\citenamefont {{Johnson}}(2020)}]{2020JCAP...09..046J}%
  \BibitemOpen
  \bibfield  {author} {\bibinfo {author} {\bibfnamefont {G.}~\bibnamefont
  {{Johnson}}},\ }\href {https://doi.org/10.1088/1475-7516/2020/09/046}
  {\bibfield  {journal} {\bibinfo  {journal} {\jcap}\ }\textbf {\bibinfo
  {volume} {2020}},\ \bibinfo {eid} {046} (\bibinfo {year} {2020})},\ \Eprint
  {https://arxiv.org/abs/2005.07467}  { arXiv:2005.07467
  [astro-ph.CO]}\BibitemShut {NoStop}%
\bibitem [{\citenamefont {{MacGibbon}}(1987)}]{1987Natur.329..308M}%
  \BibitemOpen
  \bibfield  {author} {\bibinfo {author} {\bibfnamefont {J.~H.}\ \bibnamefont
  {{MacGibbon}}},\ }\href {https://doi.org/10.1038/329308a0} {\bibfield
  {journal} {\bibinfo  {journal} {\nat}\ }\textbf {\bibinfo {volume} {329}},\
  \bibinfo {pages} {308} (\bibinfo {year} {1987})}\BibitemShut {NoStop}%
\bibitem [{\citenamefont {{Barrow}}\ \emph {et~al.}(1992)\citenamefont
  {{Barrow}}, \citenamefont {{Copeland}},\ and\ \citenamefont
  {{Liddle}}}]{1992PhRvD..46..645B}%
  \BibitemOpen
  \bibfield  {author} {\bibinfo {author} {\bibfnamefont {J.~D.}\ \bibnamefont
  {{Barrow}}}, \bibinfo {author} {\bibfnamefont {E.~J.}\ \bibnamefont
  {{Copeland}}},\ and\ \bibinfo {author} {\bibfnamefont {A.~R.}\ \bibnamefont
  {{Liddle}}},\ }\href {https://doi.org/10.1103/PhysRevD.46.645} {\bibfield
  {journal} {\bibinfo  {journal} {\prd}\ }\textbf {\bibinfo {volume} {46}},\
  \bibinfo {pages} {645} (\bibinfo {year} {1992})}\BibitemShut {NoStop}%
\bibitem [{\citenamefont {{Torres}}(2013)}]{2013PhRvD..87l3514T}%
  \BibitemOpen
  \bibfield  {author} {\bibinfo {author} {\bibfnamefont {R.}~\bibnamefont
  {{Torres}}},\ }\href {https://doi.org/10.1103/PhysRevD.87.123514} {\bibfield
  {journal} {\bibinfo  {journal} {\prd}\ }\textbf {\bibinfo {volume} {87}},\
  \bibinfo {eid} {123514} (\bibinfo {year} {2013})}\BibitemShut {NoStop}%
\bibitem [{\citenamefont {{Dvali}}\ \emph {et~al.}(2020)\citenamefont
  {{Dvali}}, \citenamefont {{Eisemann}}, \citenamefont {{Michel}},\ and\
  \citenamefont {{Zell}}}]{2020PhRvD.102j3523D}%
  \BibitemOpen
  \bibfield  {author} {\bibinfo {author} {\bibfnamefont {G.}~\bibnamefont
  {{Dvali}}}, \bibinfo {author} {\bibfnamefont {L.}~\bibnamefont {{Eisemann}}},
  \bibinfo {author} {\bibfnamefont {M.}~\bibnamefont {{Michel}}},\ and\
  \bibinfo {author} {\bibfnamefont {S.}~\bibnamefont {{Zell}}},\ }\href
  {https://doi.org/10.1103/PhysRevD.102.103523} {\bibfield  {journal} {\bibinfo
   {journal} {\prd}\ }\textbf {\bibinfo {volume} {102}},\ \bibinfo {eid}
  {103523} (\bibinfo {year} {2020})},\ \Eprint
  {https://arxiv.org/abs/2006.00011}  { arXiv:2006.00011 [hep-th]}\BibitemShut
  {NoStop}%
\bibitem [{\citenamefont {{Green}}(1999)}]{1999PhRvD..60f3516G}%
  \BibitemOpen
  \bibfield  {author} {\bibinfo {author} {\bibfnamefont {A.~M.}\ \bibnamefont
  {{Green}}},\ }\href {https://doi.org/10.1103/PhysRevD.60.063516} {\bibfield
  {journal} {\bibinfo  {journal} {\prd}\ }\textbf {\bibinfo {volume} {60}},\
  \bibinfo {eid} {063516} (\bibinfo {year} {1999})},\ \Eprint
  {https://arxiv.org/abs/astro-ph/9903484}  { arXiv:astro-ph/9903484
  [astro-ph]}\BibitemShut {NoStop}%
\bibitem [{\citenamefont {{Sahlen}}(2003)}]{2003astro.ph..4478S}%
  \BibitemOpen
  \bibfield  {author} {\bibinfo {author} {\bibfnamefont {M.}~\bibnamefont
  {{Sahlen}}},\ }\href@noop {} {\bibfield  {journal} {\bibinfo  {journal}
  {arXiv e-prints}\ } (\bibinfo {year} {2003})},\ \Eprint
  {https://arxiv.org/abs/astro-ph/0304478}  { astro-ph/0304478
  [astro-ph]}\BibitemShut {NoStop}%
\bibitem [{\citenamefont {{Bell}}\ and\ \citenamefont
  {{Volkas}}(1999)}]{1999PhRvD..59j7301B}%
  \BibitemOpen
  \bibfield  {author} {\bibinfo {author} {\bibfnamefont {N.~F.}\ \bibnamefont
  {{Bell}}}\ and\ \bibinfo {author} {\bibfnamefont {R.~R.}\ \bibnamefont
  {{Volkas}}},\ }\href {https://doi.org/10.1103/PhysRevD.59.107301} {\bibfield
  {journal} {\bibinfo  {journal} {\prd}\ }\textbf {\bibinfo {volume} {59}},\
  \bibinfo {eid} {107301} (\bibinfo {year} {1999})},\ \Eprint
  {https://arxiv.org/abs/astro-ph/9812301}  { arXiv:astro-ph/9812301
  [astro-ph]}\BibitemShut {NoStop}%
\bibitem [{\citenamefont {{Bernal}}\ \emph
  {et~al.}(2021{\natexlab{a}})\citenamefont {{Bernal}}, \citenamefont
  {{Hajkarim}},\ and\ \citenamefont {{Xu}}}]{2021PhRvD.104g5007B}%
  \BibitemOpen
  \bibfield  {author} {\bibinfo {author} {\bibfnamefont {N.}~\bibnamefont
  {{Bernal}}}, \bibinfo {author} {\bibfnamefont {F.}~\bibnamefont
  {{Hajkarim}}},\ and\ \bibinfo {author} {\bibfnamefont {Y.}~\bibnamefont
  {{Xu}}},\ }\href {https://doi.org/10.1103/PhysRevD.104.075007} {\bibfield
  {journal} {\bibinfo  {journal} {\prd}\ }\textbf {\bibinfo {volume} {104}},\
  \bibinfo {eid} {075007} (\bibinfo {year} {2021}{\natexlab{a}})},\ \Eprint
  {https://arxiv.org/abs/2107.13575}  { arXiv:2107.13575 [hep-ph]}\BibitemShut
  {NoStop}%
\bibitem [{\citenamefont {{Bernal}}\ \emph
  {et~al.}(2021{\natexlab{b}})\citenamefont {{Bernal}}, \citenamefont
  {{Perez-Gonzalez}}, \citenamefont {{Xu}},\ and\ \citenamefont
  {{Zapata}}}]{2021PhRvD.104l3536B}%
  \BibitemOpen
  \bibfield  {author} {\bibinfo {author} {\bibfnamefont {N.}~\bibnamefont
  {{Bernal}}}, \bibinfo {author} {\bibfnamefont {Y.~F.}\ \bibnamefont
  {{Perez-Gonzalez}}}, \bibinfo {author} {\bibfnamefont {Y.}~\bibnamefont
  {{Xu}}},\ and\ \bibinfo {author} {\bibfnamefont {{\'O}.}~\bibnamefont
  {{Zapata}}},\ }\href {https://doi.org/10.1103/PhysRevD.104.123536} {\bibfield
   {journal} {\bibinfo  {journal} {\prd}\ }\textbf {\bibinfo {volume} {104}},\
  \bibinfo {eid} {123536} (\bibinfo {year} {2021}{\natexlab{b}})},\ \Eprint
  {https://arxiv.org/abs/2110.04312}  { arXiv:2110.04312 [hep-ph]}\BibitemShut
  {NoStop}%
\bibitem [{\citenamefont {{Morrison}}\ \emph {et~al.}(2019)\citenamefont
  {{Morrison}}, \citenamefont {{Profumo}},\ and\ \citenamefont
  {{Yu}}}]{2019JCAP...05..005M}%
  \BibitemOpen
  \bibfield  {author} {\bibinfo {author} {\bibfnamefont {L.}~\bibnamefont
  {{Morrison}}}, \bibinfo {author} {\bibfnamefont {S.}~\bibnamefont
  {{Profumo}}},\ and\ \bibinfo {author} {\bibfnamefont {Y.}~\bibnamefont
  {{Yu}}},\ }\href {https://doi.org/10.1088/1475-7516/2019/05/005} {\bibfield
  {journal} {\bibinfo  {journal} {\jcap}\ }\textbf {\bibinfo {volume} {2019}},\
  \bibinfo {eid} {005} (\bibinfo {year} {2019})},\ \Eprint
  {https://arxiv.org/abs/1812.10606}  { arXiv:1812.10606
  [astro-ph.CO]}\BibitemShut {NoStop}%
\bibitem [{\citenamefont {{Khlopov}}\ \emph {et~al.}(2006)\citenamefont
  {{Khlopov}}, \citenamefont {{Barrau}},\ and\ \citenamefont
  {{Grain}}}]{2006CQGra..23.1875K}%
  \BibitemOpen
  \bibfield  {author} {\bibinfo {author} {\bibfnamefont {M.~Y.}\ \bibnamefont
  {{Khlopov}}}, \bibinfo {author} {\bibfnamefont {A.}~\bibnamefont
  {{Barrau}}},\ and\ \bibinfo {author} {\bibfnamefont {J.}~\bibnamefont
  {{Grain}}},\ }\href {https://doi.org/10.1088/0264-9381/23/6/004} {\bibfield
  {journal} {\bibinfo  {journal} {\CQG}\ }\textbf {\bibinfo {volume} {23}},\
  \bibinfo {pages} {1875} (\bibinfo {year} {2006})},\ \Eprint
  {https://arxiv.org/abs/astro-ph/0406621}  { arXiv:astro-ph/0406621
  [astro-ph]}\BibitemShut {NoStop}%
\bibitem [{\citenamefont {{Kitabayashi}}(2022)}]{2022PTEP.2022c3B02K}%
  \BibitemOpen
  \bibfield  {author} {\bibinfo {author} {\bibfnamefont {T.}~\bibnamefont
  {{Kitabayashi}}},\ }\href {https://doi.org/10.1093/ptep/ptac025} {\bibfield
  {journal} {\bibinfo  {journal} {\PTEP}\ }\textbf {\bibinfo {volume} {2022}},\
  \bibinfo {eid} {033B02} (\bibinfo {year} {2022})},\ \Eprint
  {https://arxiv.org/abs/2107.11692}  { arXiv:2107.11692 [hep-ph]}\BibitemShut
  {NoStop}%
\bibitem [{\citenamefont {{Lennon}}\ \emph {et~al.}(2018)\citenamefont
  {{Lennon}}, \citenamefont {{March-Russell}}, \citenamefont
  {{Petrossian-Byrne}},\ and\ \citenamefont {{Tillim}}}]{2018JCAP...04..009L}%
  \BibitemOpen
  \bibfield  {author} {\bibinfo {author} {\bibfnamefont {O.}~\bibnamefont
  {{Lennon}}}, \bibinfo {author} {\bibfnamefont {J.}~\bibnamefont
  {{March-Russell}}}, \bibinfo {author} {\bibfnamefont {R.}~\bibnamefont
  {{Petrossian-Byrne}}},\ and\ \bibinfo {author} {\bibfnamefont
  {H.}~\bibnamefont {{Tillim}}},\ }\href
  {https://doi.org/10.1088/1475-7516/2018/04/009} {\bibfield  {journal}
  {\bibinfo  {journal} {\jcap}\ }\textbf {\bibinfo {volume} {2018}},\ \bibinfo
  {eid} {009} (\bibinfo {year} {2018})},\ \Eprint
  {https://arxiv.org/abs/1712.07664}  { arXiv:1712.07664 [hep-ph]}\BibitemShut
  {NoStop}%
\bibitem [{\citenamefont {{Allahverdi}}\ \emph {et~al.}(2018)\citenamefont
  {{Allahverdi}}, \citenamefont {{Dent}},\ and\ \citenamefont
  {{Osinski}}}]{2018PhRvD..97e5013A}%
  \BibitemOpen
  \bibfield  {author} {\bibinfo {author} {\bibfnamefont {R.}~\bibnamefont
  {{Allahverdi}}}, \bibinfo {author} {\bibfnamefont {J.}~\bibnamefont
  {{Dent}}},\ and\ \bibinfo {author} {\bibfnamefont {J.}~\bibnamefont
  {{Osinski}}},\ }\href {https://doi.org/10.1103/PhysRevD.97.055013} {\bibfield
   {journal} {\bibinfo  {journal} {\prd}\ }\textbf {\bibinfo {volume} {97}},\
  \bibinfo {eid} {055013} (\bibinfo {year} {2018})},\ \Eprint
  {https://arxiv.org/abs/1711.10511}  { arXiv:1711.10511
  [astro-ph.CO]}\BibitemShut {NoStop}%
\bibitem [{\citenamefont {{Samanta}}\ and\ \citenamefont
  {{Urban}}(2021)}]{2021arXiv211204836S}%
  \BibitemOpen
  \bibfield  {author} {\bibinfo {author} {\bibfnamefont {R.}~\bibnamefont
  {{Samanta}}}\ and\ \bibinfo {author} {\bibfnamefont {F.~R.}\ \bibnamefont
  {{Urban}}},\ }\href@noop {} {\bibfield  {journal} {\bibinfo  {journal}
  {\arxiv}\ } (\bibinfo {year} {2021})},\ \Eprint
  {https://arxiv.org/abs/2112.04836}  { arXiv:2112.04836 [hep-ph]}\BibitemShut
  {NoStop}%
\bibitem [{\citenamefont {{Hooper}}\ \emph {et~al.}(2019)\citenamefont
  {{Hooper}}, \citenamefont {{Krnjaic}},\ and\ \citenamefont
  {{McDermott}}}]{2019JHEP...08..001H}%
  \BibitemOpen
  \bibfield  {author} {\bibinfo {author} {\bibfnamefont {D.}~\bibnamefont
  {{Hooper}}}, \bibinfo {author} {\bibfnamefont {G.}~\bibnamefont
  {{Krnjaic}}},\ and\ \bibinfo {author} {\bibfnamefont {S.~D.}\ \bibnamefont
  {{McDermott}}},\ }\href {https://doi.org/10.1007/JHEP08(2019)001} {\bibfield
  {journal} {\bibinfo  {journal} {\jhep}\ }\textbf {\bibinfo {volume} {2019}},\
  \bibinfo {eid} {1} (\bibinfo {year} {2019})},\ \Eprint
  {https://arxiv.org/abs/1905.01301}  { arXiv:1905.01301 [hep-ph]}\BibitemShut
  {NoStop}%
\bibitem [{\citenamefont {{Gondolo}}\ \emph {et~al.}(2020)\citenamefont
  {{Gondolo}}, \citenamefont {{Sandick}},\ and\ \citenamefont {{Shams Es
  Haghi}}}]{2020PhRvD.102i5018G}%
  \BibitemOpen
  \bibfield  {author} {\bibinfo {author} {\bibfnamefont {P.}~\bibnamefont
  {{Gondolo}}}, \bibinfo {author} {\bibfnamefont {P.}~\bibnamefont
  {{Sandick}}},\ and\ \bibinfo {author} {\bibfnamefont {B.}~\bibnamefont
  {{Shams Es Haghi}}},\ }\href {https://doi.org/10.1103/PhysRevD.102.095018}
  {\bibfield  {journal} {\bibinfo  {journal} {\prd}\ }\textbf {\bibinfo
  {volume} {102}},\ \bibinfo {eid} {095018} (\bibinfo {year} {2020})},\ \Eprint
  {https://arxiv.org/abs/2009.02424}  { arXiv:2009.02424 [hep-ph]}\BibitemShut
  {NoStop}%
\bibitem [{\citenamefont {{Bernal}}\ and\ \citenamefont
  {{Zapata}}(2021{\natexlab{a}})}]{2021JCAP...03..007B}%
  \BibitemOpen
  \bibfield  {author} {\bibinfo {author} {\bibfnamefont {N.}~\bibnamefont
  {{Bernal}}}\ and\ \bibinfo {author} {\bibfnamefont {{\'O}.}~\bibnamefont
  {{Zapata}}},\ }\href {https://doi.org/10.1088/1475-7516/2021/03/007}
  {\bibfield  {journal} {\bibinfo  {journal} {\jcap}\ }\textbf {\bibinfo
  {volume} {2021}},\ \bibinfo {eid} {007} (\bibinfo {year}
  {2021}{\natexlab{a}})},\ \Eprint {https://arxiv.org/abs/2010.09725}  {
  arXiv:2010.09725 [hep-ph]}\BibitemShut {NoStop}%
\bibitem [{\citenamefont {{Bernal}}\ and\ \citenamefont
  {{Zapata}}(2021{\natexlab{b}})}]{2021JCAP...03..015B}%
  \BibitemOpen
  \bibfield  {author} {\bibinfo {author} {\bibfnamefont {N.}~\bibnamefont
  {{Bernal}}}\ and\ \bibinfo {author} {\bibfnamefont {{\'O}.}~\bibnamefont
  {{Zapata}}},\ }\href {https://doi.org/10.1088/1475-7516/2021/03/015}
  {\bibfield  {journal} {\bibinfo  {journal} {\jcap}\ }\textbf {\bibinfo
  {volume} {2021}},\ \bibinfo {eid} {015} (\bibinfo {year}
  {2021}{\natexlab{b}})},\ \Eprint {https://arxiv.org/abs/2011.12306}  {
  arXiv:2011.12306 [astro-ph.CO]}\BibitemShut {NoStop}%
\bibitem [{\citenamefont {{Bernal}}\ and\ \citenamefont
  {{Zapata}}(2021{\natexlab{c}})}]{2021PhLB..81536129B}%
  \BibitemOpen
  \bibfield  {author} {\bibinfo {author} {\bibfnamefont {N.}~\bibnamefont
  {{Bernal}}}\ and\ \bibinfo {author} {\bibfnamefont {{\'O}.}~\bibnamefont
  {{Zapata}}},\ }\href {https://doi.org/10.1016/j.physletb.2021.136129}
  {\bibfield  {journal} {\bibinfo  {journal} {\plb}\ }\textbf {\bibinfo
  {volume} {815}},\ \bibinfo {eid} {136129} (\bibinfo {year}
  {2021}{\natexlab{c}})},\ \Eprint {https://arxiv.org/abs/2011.02510}  {
  arXiv:2011.02510 [hep-ph]}\BibitemShut {NoStop}%
\bibitem [{\citenamefont {{Li}}\ and\ \citenamefont
  {{Liao}}(2022)}]{2022arXiv220314443L}%
  \BibitemOpen
  \bibfield  {author} {\bibinfo {author} {\bibfnamefont {T.}~\bibnamefont
  {{Li}}}\ and\ \bibinfo {author} {\bibfnamefont {J.}~\bibnamefont {{Liao}}},\
  }\href@noop {} {\bibfield  {journal} {\bibinfo  {journal} {\arxiv}\ }
  (\bibinfo {year} {2022})},\ \Eprint {https://arxiv.org/abs/2203.14443}  {
  arXiv:2203.14443 [hep-ph]}\BibitemShut {NoStop}%
\bibitem [{\citenamefont {{Calabrese}}\ \emph
  {et~al.}(2022{\natexlab{b}})\citenamefont {{Calabrese}}, \citenamefont
  {{Chianese}}, \citenamefont {{Fiorillo}},\ and\ \citenamefont
  {{Saviano}}}]{2022arXiv220317093C}%
  \BibitemOpen
  \bibfield  {author} {\bibinfo {author} {\bibfnamefont {R.}~\bibnamefont
  {{Calabrese}}}, \bibinfo {author} {\bibfnamefont {M.}~\bibnamefont
  {{Chianese}}}, \bibinfo {author} {\bibfnamefont {D.~F.~G.}\ \bibnamefont
  {{Fiorillo}}},\ and\ \bibinfo {author} {\bibfnamefont {N.}~\bibnamefont
  {{Saviano}}},\ }\href@noop {} {\bibfield  {journal} {\bibinfo  {journal}
  {\arxiv}\ } (\bibinfo {year} {2022}{\natexlab{b}})},\ \Eprint
  {https://arxiv.org/abs/2203.17093}  { arXiv:2203.17093 [hep-ph]}\BibitemShut
  {NoStop}%
\bibitem [{\citenamefont {{Calabrese}}\ \emph
  {et~al.}(2022{\natexlab{c}})\citenamefont {{Calabrese}}, \citenamefont
  {{Chianese}}, \citenamefont {{Fiorillo}},\ and\ \citenamefont
  {{Saviano}}}]{2022PhRvD.105b1302C}%
  \BibitemOpen
  \bibfield  {author} {\bibinfo {author} {\bibfnamefont {R.}~\bibnamefont
  {{Calabrese}}}, \bibinfo {author} {\bibfnamefont {M.}~\bibnamefont
  {{Chianese}}}, \bibinfo {author} {\bibfnamefont {D.~F.~G.}\ \bibnamefont
  {{Fiorillo}}},\ and\ \bibinfo {author} {\bibfnamefont {N.}~\bibnamefont
  {{Saviano}}},\ }\href {https://doi.org/10.1103/PhysRevD.105.L021302}
  {\bibfield  {journal} {\bibinfo  {journal} {\prd}\ }\textbf {\bibinfo
  {volume} {105}},\ \bibinfo {eid} {L021302} (\bibinfo {year}
  {2022}{\natexlab{c}})},\ \Eprint {https://arxiv.org/abs/2107.13001}  {
  arXiv:2107.13001 [hep-ph]}\BibitemShut {NoStop}%
\bibitem [{\citenamefont {{Carr}}\ \emph
  {et~al.}(2021{\natexlab{c}})\citenamefont {{Carr}}, \citenamefont
  {{K{\"u}hnel}},\ and\ \citenamefont {{Visinelli}}}]{2021MNRAS.506.3648C}%
  \BibitemOpen
  \bibfield  {author} {\bibinfo {author} {\bibfnamefont {B.}~\bibnamefont
  {{Carr}}}, \bibinfo {author} {\bibfnamefont {F.}~\bibnamefont
  {{K{\"u}hnel}}},\ and\ \bibinfo {author} {\bibfnamefont {L.}~\bibnamefont
  {{Visinelli}}},\ }\href {https://doi.org/10.1093/mnras/stab1930} {\bibfield
  {journal} {\bibinfo  {journal} {\mnras}\ }\textbf {\bibinfo {volume} {506}},\
  \bibinfo {pages} {3648} (\bibinfo {year} {2021}{\natexlab{c}})},\ \Eprint
  {https://arxiv.org/abs/2011.01930}  { arXiv:2011.01930
  [astro-ph.CO]}\BibitemShut {NoStop}%
\end{thebibliography}%
	
\end{document}